\documentclass[a4paper,fleqn]{cas-sc}
\usepackage[numbers]{natbib}
\usepackage{multirow}
\usepackage{bm}

\def\tsc#1{\csdef{#1}{\textsc{\lowercase{#1}}\xspace}}
\tsc{WGM}
\tsc{QE}
\tsc{EP}
\tsc{PMS}
\tsc{BEC}
\tsc{DE}
\newcommand{\be}{\begin{equation}}
\newcommand{\ee}{\end{equation}}
\newcommand{\bea}{\begin{eqnarray}}
\newcommand{\eea}{\end{eqnarray}}
\newcommand{\eg}{\textit{e.g.}}

\newcommand{\ie}{\textit{i.e.}}
\newcommand{\raa}{R_{\mathrm{AA}}}

\newcommand{\lsim}{\lesssim}

\newcommand{\Ds}{{\cal D}_s}
\newcommand{\nn}{\nonumber\\}
\newcommand{\pa}{\partial} 
\newcommand{\pT}{p_{\rm T}}
\newcommand{\Tpc}{T_{\rm pc}}
\newcommand{\sqrts}{\sqrt{s_{\rm NN}}}
\newcommand{\Ncoll}{N_{\rm coll}}

\begin{document}
\let\WriteBookmarks\relax
\def\floatpagepagefraction{1}
\def\textpagefraction{.001}

\shorttitle{Charm and Bottom Hadrons in Hot Hadronic Matter}

\shortauthors{Santosh K. Das {\it et al.} }

\title [mode = title]{Charm and Bottom Hadrons in Hot Hadronic Matter}                      
%\tnotemark[1,2]

%\tnotetext[1]{This document is the results of the research
%   project funded by the National Science Foundation.}

%\tnotetext[2]{The second title footnote which is a longer text matter
%  to fill through the whole text width and overflow into
%  another line in the footnotes area of the first page.}

%[type=editor,
%\author[1,3]{CV Radhakrishnan}[type=editor,
%                        auid=000,bioid=1,
%                        prefix=Sir,
%                        role=Researcher,
%                        orcid=0000-0001-7511-2910]

%\cormark[1]


%\ead{cvr_1@tug.org.in}

%\ead[url]{www.cvr.cc, cvr@sayahna.org}

%\credit{Conceptualization of this study, Methodology, Software}

%\affiliation[1]{organization={Elsevier B.V.},
%    addressline={Radarweg 29}, 
%    city={Amsterdam},
    % citysep={}, % Uncomment if no comma needed between city and postcode
%    postcode={1043 NX}, 
%    country={The Netherlands}}

\author[1]{Santosh K. Das}

\affiliation[1]{organization={School of Physical Sciences, Indian Institute of Technology Goa},
    city={Farmagudi, Ponda},
    % citysep={}, % Uncomment if no comma needed between city and postcode
   postcode={403401}, 
%    state={Trivandrum},
  country={India}}

\author[2]{Juan M. Torres-Rincon}

\affiliation[2]{organization={Departament de F\'isica Qu\`antica i Astrof\'isica and Institut de Ci\`encies del Cosmos (ICCUB), Facultat de F\'isica,  Universitat de Barcelona},
 addressline={Mart\'i i Franqu\`es 1}, 
 city={Barcelona},
  % citysep={}, % Uncomment if no comma needed between city and postcode
    postcode={08028}, 
%    state={Trivandrum},
   country={Spain}}

%[style=chinese]

\author[3]{Ralf Rapp}
%[%
%   role=Co-ordinator,
%   suffix=Jr,
%\ead{rapp@comp.tamu.edu}
%\ead[URL]{}

%\credit{Data curation, Writing - Original draft preparation}

\affiliation[3]{organization={Cyclotron Institute and Department of Physics and Astronomy, Texas A\&M University},
    addressline={MS 3666}, 
   city={College Station},
    % citysep={}, % Uncomment if no comma needed between city and postcode
    postcode={77843-3366}, 
    state={Texas},
   country={U.S.A.}}

%\author%
%[1,3]
%{Rishi T.}
%\cormark[2]
%\fnmark[1,3]
%\ead{rishi@stmdocs.in}
%\ead[URL]{www.stmdocs.in}

%\affiliation[3]{organization={STM Document Engineering Pvt Ltd.},
%    addressline={Mepukada}, 
%    city={Malayinkil},
    % citysep={}, % Uncomment if no comma needed between city and postcode
%    postcode={695571}, 
%    state={Trivandrum},
%    country={India}}

%\cortext[cor2]{Principal corresponding author}

%\fntext[fn1]{This is the first author footnote. but is common to third
%  author as well.}
%\fntext[fn2]{Another author footnote, this is a very long footnote and
%  it should be a really long footnote. But this footnote is not yet
%  sufficiently long enough to make two lines of footnote text.}

%\nonumnote{This note has no numbers. In this work we demonstrate $a_b$
%  the formation Y\_1 of a new type of polariton on the interface
%  between a cuprous oxide slab and a polystyrene micro-sphere placed
%  on the slab.
% }

\begin{abstract}
Heavy quarks, and the hadrons containing them, are excellent probes of the QCD medium formed in high-energy heavy-ion collisions, as they provide essential information on the transport properties of the medium and how quarks color-neutralize into hadrons. Large theoretical and phenomenological efforts have been dedicated thus far to assess the diffusion of charm and bottom quarks in the quark-gluon plasma and their subsequent hadronization into heavy-flavor (HF) hadrons. However, the fireball formed in heavy-ion collisions also features an extended hadronic phase, and therefore any quantitative analysis of experimental observables needs to account for the rescattering of charm and bottom hadrons. This is further reinforced by the presence of a QCD cross-over transition and the notion that the interaction strength is maximal in the vicinity of the pseudo-critical temperature.  We review existing approaches for evaluating the interactions of open HF hadrons in a hadronic heat bath and the pertinent results for scattering amplitudes, spectral functions and transport coefficients. While most of the work to date has focused on $D$-mesons, we also discuss excited states as well as HF baryons and the bottom sector. Both the HF hadro-chemistry and bottom observables will play a key role in future experimental measurements. We also conduct a survey of transport calculations in heavy-ion collisions that have included effects of hadronic HF diffusion and assess its impact on various observables.

%This template helps you to create a properly formatted \LaTeX\ manuscript.
%\noindent\texttt{\textbackslash begin{abstract}} \dots 
%\texttt{\textbackslash end{abstract}} and
%\verb+\begin{keyword}+ \verb+...+ \verb+\end{keyword}+ 
%which contain the abstract and keywords respectively. 
%\noindent Each keyword shall be separated by a \verb+\sep+ command.
\end{abstract}

% \includegraphics{figs/grabs.pdf}

%\item Research highlights item 1
%\item Research highlights item 2
%\item Research highlights item 3

% Keywords
\begin{keywords}
\sep Heavy-flavor transport \sep Hadronic matter \sep Ultra-relativistic heavy-ion collisions \sep Effective field theories 
\end{keywords}

\maketitle
\tableofcontents
\newpage
%%%%%%%%%%%%%%%%%%%%%%%%%%%%%%%%%%%%%%%%5
\section{Introduction and motivation}
%%%%%%%%%%%%%%%%%%%%%%%%%%%%%%%%%%%%%%%%
Ultra-relativistic heavy-ion collisions (URHICs) provide the unique opportunity to study hot QCD matter akin to the one which filled the early Universe, just a few microseconds after the Big Bang. One of the main goals of these campaigns is to determine the transport properties of the medium and acquire a microscopic understanding of the underlying interactions and the emerging many-body phenomena. The heavy charm ($c$) and bottom ($b$) quarks play a central role in these objectives~\cite{Rapp:2018qla,Dong:2019byy,Dong:2019unq,He:2022ywp}. Since their masses are much larger than the typical temperatures reached in heavy-ion collisions, their diffusion through the fireball can serve as a ``Brownian marker'' of their coupling strength to the medium. In addition, the conversion of heavy quarks into various heavy-flavor (HF) hadrons in the transition from the quark-gluon plasma (QGP) to hadronic matter offers valuable insights into mechanisms of hadronization. Subsequently, HF hadrons continue to diffuse through the hadronic phase until the fireball medium has become dilute enough for them to decouple from the environment.
In the present review we will focus on the in-medium properties of open HF hadrons, \ie, hadrons that carry a non-zero charm-quark or bottom-quark number. The in-medium physics of quarkonia, which involves their own specifics, will not be addressed here (see, \eg, Refs.~\cite{Rapp:2008tf,Mocsy:2013syh,Rothkopf:2019ipj,Andronic:2024oxz} for comprehensive overviews). Furthermore, our discussion will be mostly geared toward hot hadronic matter with a small net-baryon content, as produced in URHICs. Complementary information about in-medium $D$-meson properties can be obtained by studying HF hadrons in cold nuclei via suitable production experiments, both with hadron and electron beams, which is beyond the scope of the present review, see, \eg, Refs.~\cite{Li:2020sru,Hosaka:2016ypm}.

Experimental measurements over the last twenty years at the Relativistic Heavy-Ion Collider (RHIC) and the Large Hadron Collider (LHC) have covered a large variety of open HF observables in heavy-ion collisions, mostly focusing on the so-called nuclear modification factor, $\raa$, and elliptic flow coefficient, $v_2$. The former quantifies the deviation of the particle spectra (or yields) in nucleus-nucleus (AA) collisions from those in $pp$ collisions, while the $v_2$ quantifies the azimuthal asymmetry of the transverse-momentum ($\pT$) spectra usually associated with a collective fireball expansion in non-central collisions (or, at high $\pT$, with path-length differences when propagating through an elliptic fireball).

Pioneering HF measurements were performed at RHIC using electrons from semi-leptonic HF hadron decays; a large reduction of the $\raa$ in 200\,GeV Au+Au collisions was observed, accompanied by a substantial positive $v_2$ coefficient~\cite{PHENIX:2006iih}. These observations already indicated a large interaction strength of the (predominant) charm quarks in the QCD medium, as well as the importance of recombination processes in their hadronization, \ie, rather late in the medium evolution~\cite{vanHees:2005wb}. Given the fact that the QCD transition at vanishing quark chemical potential is a crossover, this already implies that effects from the hadronic phase, which were not taken into account in the calculations at the time, might be significant. State-of-the-art data for $D$-meson production in 5\,TeV Pb+Pb collisions at the LHC have now reached a precision that enabled, through model comparisons, quantitative constraints on one of the key transport parameters, \ie, the spatial HF diffusion coefficient~\cite{Rapp:2018qla,ALICE:2021rxa}, $\Ds$. 
The constraints amount to a range of values of 1.5~$\le (2\pi T)\Ds \le \,4.5$  for charm quarks in the QGP at temperatures near the pseudo-critical one, 
$T_{\rm pc}\simeq 160$\,MeV~\cite{Aoki:2009sc,HotQCD:2018pds}, including a commonly used scaling by the thermal wavelength of the medium to render a dimensionless quantity. This is tantalizingly close to a lower bound of $\sim$1 conjectured to exist in the strong-coupling limit of conformal quantum field theory~\cite{Policastro:2002se}.  
Some of the transport model calculations in this extraction account for the hadronic diffusion effects and predict a significant contribution to the final observed $v_2$, at the level of 
$\sim$10-30\%~\cite{Song:2015sfa,He:2019vgs}. Again, the fact that one expects a minimum of the (scaled) transport coefficients in the vicinity of the transition temperature reiterates the importance of accounting for hadronic diffusion in any future precision assessments of the HF transport properties from heavy-ion data.   

As mentioned above, the conversion of a universal heavy-quark (HQ) distribution function, after diffusion through the QGP, into color-neutral particles offers unique insights into hadronization mechanisms. However, this implies that one has sufficient control over the transport properties of the individual hadron species, such as $D$, $D_s$ or charm-baryons which diffuse through the subsequent hot hadronic medium. In addition, even the measurement of ground-state hadrons (which are not subject to strong decays) requires an understanding of feed-down effects, \ie, contributions from the strong (and electromagnetic) decays of excited states, such as $D^*\to D + \pi$.  This affects not only the kinematics of the ground-state hadrons, but also their abundances. For example, recent discoveries of a large ratio for the yield of charm baryons to charm mesons, $\Lambda_c/D^0$, of up to $\sim$0.5 at low $\pT$ in hadronic and heavy-ion collisions at RHIC and the LHC~\cite{STAR:2019ank,ALICE:2021bib,CMS:2019uws} (which is a factor of near 5 larger than observed in $e^+e^-$ annihilation) suggest the presence of a large number of excited charm baryons~\cite{He:2019tik} (likewise in the bottom sector~\cite{He:2022tod}). If so, these resonances can provide a substantial source of interaction strength in hadronic matter and thus, in turn,  have a significant impact on the kinematic distributions of the HF daughter particles. All these features call for a systematic assessment of in-medium properties of HF hadrons. Large interaction strengths will then require an underlying quantum many-body theory that accounts for, and ultimately predicts, the spectral properties of the charm degrees of freedom in hadronic matter.     

Last but not least, one should keep in mind the plans for measuring HF observables in heavy-ion collisions at lower center-of-mass energies, where the role of the hadronic phase, relative to the QGP, will become more important. Pertinent measurements already exist from a RHIC beam energy scan at 39 and 62\,GeV~\cite{STAR:2014yia}, and may also be conducted with much improved precision in fixed-target configurations at the LHC using the LHCb detector~\cite{BoenteGarcia:2024sjf}. There are very promising efforts toward even lower energies at the SPS with the NA60+ experiment~\cite{NA60:2022sze}, which opens up exciting opportunities to study transport properties in QCD matter at relatively large baryon chemical potentials. 

The remainder of this review is organized as follows. In Sec.~\ref{sec:charm} we discuss the in-medium properties of $D$ mesons, starting with a brief introduction to HQ symmetries (Sec.~\ref{ssec:hqet}), followed by the construction of vacuum interactions with light hadrons (Sec.~\ref{ssec:Dint}), and their applications to assess in-medium $D$-meson spectral properties (Sec.~\ref{ssec:Dmed}) and transport coefficients (Sec.~\ref{ssec:Dtrans}). In Sec.~\ref{sec:bottom-baryons} we collect the same, albeit much more sparse information on the interactions (Sec.~\ref{ssec:Bint}) and transport coefficients (Sec.~\ref{ssec:Btrans}) of $B$-mesons, as well as baryons containing a heavy quark (Secs.~\ref{ssec:baryon-int} and \ref{ssec:baryon-trans}). 
In Sec.~\ref{sec:pheno} we turn to applications of heavy-hadron diffusion through the hadronic phase of the fireballs created in URHICs; 
we scrutinize the sensitivity of the $D$-meson, $B$-meson and heavy-baryon observables, specifically their nuclear modification factor and elliptic flow, to contributions from hadronic transport as computed in available models and in the context of experimental data from RHIC and the LHC in Secs.~\ref{ssec:D-obs} and \ref{ssec:B+baryon-obs}, elaborate on HF chemistry observables including baryon-to-meson ratios in Sec.~\ref{ssec:chem}, and provide a critical assessment of the current state-of affairs of the hadronic HF diffusion coefficient, put into context with results for the QGP, in Sec.~\ref{ssec:assess}. 
We summarize and conclude in Sec.~\ref{sec:sum}.

%%%%%%%%%%%%%%%%%%%%%%%%%%%%%%%%%%%%%%%%%%%%%
\section{$D$-meson interactions and transport coefficients in hot matter}
\label{sec:charm}
%%%%%%%%%%%%%%%%%%%%%%%%%%%%%%%%%%%%%%%%%%%%%%%%
Open HF hadrons offer distinct opportunities to study various aspects of hadronic physics. For example, heavy-light mesons, being made up of one heavy ($c$ or $b$) and one light valence quark, can be used as rather clean probes of the interactions of a light quark with its environment, such as their coupling to QCD vacuum condensates or mean-fields in nuclear matter~\cite{Hayashigaki:2000es,Saito:2005rv,Hosaka:2016ypm}. Alternatively, the large mass of heavy quarks creates the possibility to use heavy hadrons as probes of the transport properties of hadronic matter, by invoking the concept of a Brownian motion in a medium of much lighter constituents~\cite{Rapp:2009my}. This aspect will be the focus of the present review, where the canonical experimental context is the nearly net-baryon free medium produced in URHICs at RHIC and the LHC. The hadronic phase emerges once the fireball formed in these reactions has passed through a possible QGP phase and has converted back into hadronic degrees of freedom.
Within the medium of confined particles, heavy hadrons interact with other mesons and baryons along their trajectory and until they decouple and ultimately reach the experimental detectors. Since pions (followed by other light mesons like kaons and $\eta$'s) are the most abundant species in the system due to their small mass, and since $D$-mesons are the most abundant HF hadrons, this section will focus on their interactions. 
 
The scenario we have in mind is that of a locally thermalized thermal bath, composed of light particles (pion, kaons...), in which $D$-mesons might not yet be fully equilibrated, but, owing to the preceding partonic evolution, are typically not very far away from equilibrium. This sets the stage for a diffusion description using Brownian motion with a sequence of thermal random kicks with small moment transfer in an evolving thermal medium. This can be realized in Langevin simulations of a Fokker-Planck equation, where the microphysics is encoded in effective transport parameters (equivalent of Wilson coefficients in the effective-field-theory terminology). These are inputs to the equations of motion and depend on temperature, baryon density and the momentum of the Brownian particle. In the Fokker-Planck framework, these are a set of drag and momentum-space diffusion coefficients (which, in principle, are related via the fluctuation-dissipation theorem).

In the case of charm (the lightest heavy flavor), the difference between the heavy-particle mass and the temperature of the system might not be well separated in some cases. If so, the expansion of in the $T/M$ ratio underlying Fokker-Planck may not be accurate anymore, and one might want to take recourse to a Boltzmann-like equation. In particular, this could happen for charm quarks in the deconfined phase of heavy-ion collisions where the early temperatures can reach a substantial fraction of the charm-quark mass (say, $T \simeq 0.5$\,GeV), and thus corrections to the Fokker-Planck approximation become noticeable~\cite{Das:2013kea,Rapp:2018qla}. However, there is also a price to pay in that the Boltzmann equation is a semiclassical approach that is based on the assumption of well-defined quasiparticles in the collision term; this entails limitations for the description of strongly coupled systems where the collisional widths of the medium constituents become comparable to their masses (or kinetic energies). This scenario is quite possibly realized for the ``strongly coupled QGP'' (sQGP) at temperatures up to at least 2 times above the pseudo-critical one~\cite{Liu:2016ysz}. 
If, on the other hand, the mass of the Brownian particle remains significantly larger than its collisional width, \ie, if it remains a good quasiparticle, one can encode the strong-coupling properties of the medium in the calculation of the transport coefficients by accounting for the pertinent spectral functions of the light partons, including collisional widths which are comparable to their energies~\cite{Liu:2018syc}.
For $D$ mesons and $\Lambda_c$ baryons (and certainly for bottom hadrons) in hadronic matter, the hierarchy required for a Fokker-Planck (or Langevin) description is generally well satisfied~\cite{Tolos:2016slr},  since the charm-hadron masses are larger than the $c$-quark mass and the temperatures in the hadronic medium are smaller than in the QGP. On the other hand, the collisional width of light hadrons in hadronic matter can still be quite substantial leading to nontrivial spectral functions; \eg, the collisional broadening of the $\rho$ meson is expected to reach several hundred MeV at temperatures of around 150\,MeV~\cite{Rapp:1999ej} which is supported by dilepton measurements~\cite{NA60:2006ymb}. This suggests that also the hadronic medium is strongly coupled, and a classical Boltzmann approach for $D$-meson interactions with light hadrons may not be reliable. In this case, one can still include the off-shell effects of the medium particles in the transport coefficients or in the heavy-light scattering amplitude, see, \eg, Ref.~\cite{Torres-Rincon:2023qll} for an off-shell application with $D$ mesons.
 
\begin{figure}[!t] 
\centering
\includegraphics[scale=0.70]{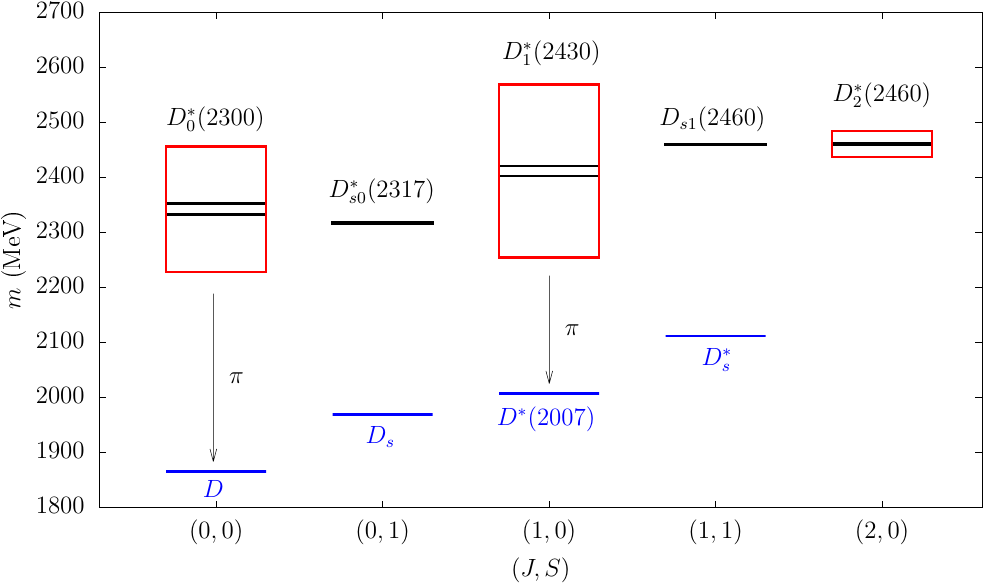}
\caption{Low-lying experimentally identified states in the open charm meson sector~\cite{pdg2024,ParticleDataGroup:2022pth}, as functions of their angular momentum $J$, and their strangeness content $S$. States represented with blue color are ground states, black boxes show their masses with uncertainties, and the red boxes give a measure of their decay widths. Total decay widths less than $\Gamma = 5$ MeV are not shown, since the states effectively live during all the URHIC lifetime.}
\label{fig:spectrum}
\end{figure}

A reliable description of transport processes requires good control over the microphysics that figures in the evaluation of the transport coefficients, as a function of temperature and baryon density. Heavy-quark symmetry enables the use of effective-field theory (EFT) to construct the interaction between the heavy and light hadrons. 
Since at moderate temperature, the main population of the bath consists of light (pseudo-) Goldstone bosons, $\pi$, $K$, $\bar{K}$ and $\eta$, an effective chiral Lagrangian can be constructed with a well-defined power counting. This Lagrangian can be extended to account for the dynamics of $D$ mesons. A summary of the low-lying open charm-meson states that will figure in the description later in the text is shown in Fig.~\ref{fig:spectrum}; both the ground states as well as several excited states are displayed, with their properties are taken from Refs.~\cite{pdg2024,ParticleDataGroup:2022pth}.

The remainder of this section is organized as follows:
In Sec.~\ref{ssec:hqet} we start with a brief historical perspective on the construction of EFTs incorporating both chiral and HQ symmetry, including developments prior to 2011 which were dedicated to hadron spectroscopy and are unrelated to in-medium physics.
In Sec.~\ref{ssec:Dint} we focus on the applications to $D$-meson interactions with light hadrons, by combining chiral perturbation theory ($\chi$PT) with HQ spin symmetry (HQSS) into heavy-meson $\chi$PT for the low-energy regime (Sec.~\ref{sssec:sym}), and unitarized approaches necessary to incorporate effects of resonance formation as well as interactions with excited states in the hadron resonance gas (Sec.~\ref{sssec:resonances}).
In Sec.~\ref{ssec:Dmed} we discuss how these interactions are implemented to evaluate in-medium $D$-meson properties, \ie,  their spectral properties (Sec.~\ref{sssec:spectral}) and scattering amplitudes at finite temperature (Sec.~\ref{sssec:D-ampl}), including the issue of self-consistency.
In Sec.~\ref{ssec:Dtrans} we turn to the calculations of $D$-meson transport coefficients in hot matter, specifically the drag and momentum diffusion coefficients needed in Langevin simulations (Sec.~\ref{sssec:Ddrag}), the spatial diffusion coefficient   (Sec.~\ref{sssec:spatial}), and effects of finite baryon chemical potential (Sec.~\ref{sssec:finite-mu}).   

%%%%%%%%%%%%%%%%%%%%%%%%%%%%%%%%%%%%%%%%%%%%%%%%%%%%%%%%%%%55
\subsection{Heavy-quark symmetry and effective theories} 
\label{ssec:hqet}
%%%%%%%%%%%%%%%%%%%%%%%%%%%%%%%%%%%%%%%%%%%%%%%%%%%%%%%%%%%%%%%%%%

A starting point for an effective theory description of $D$-meson scattering off light hadrons can be based on symmetry considerations in the limits of heavy and light masses of the $D$ mesons and pseudo-Goldstone bosons, respectively. The effective approach is termed Heavy-Quark Effective Theory (HQET)~\cite{Isgur:1989vq,Eichten:1989zv,Georgi:1990um,Grinstein:1990mj}. This effective theory exploits that the heavy-hadron mass is the largest scale in the system (dominating over the typical exchanged momentum and the nonperturbative QCD scale, $\Lambda_{QCD}$), in which case one can expand the Lagrangian in inverse powers of the mass.
Assuming that individual collisions take place instantaneously, so that any thermal effect can be neglected in this process, we concentrate on the vacuum EFT description of heavy-light interactions.
Initial works defining the HQET Lagrangian for the description of heavy hadrons in the HQ mass limit were carried out in the early 1990's~\cite{Isgur:1989vq,Eichten:1989zv,Georgi:1990um,Grinstein:1990mj}, following the philosophy of effective field approaches pioneered by Weinberg~\cite{Weinberg:1978kz}. 
Soon thereafter, a systematic Lagrangian incorporating HQ and chiral symmetries at lowest order was developed in Refs.~\cite{Yan:1992gz, Burdman:1992gh,Wise:1992hn} and applied to strong and semileptonic decays of heavy hadrons.

Next-to-leading-order (NLO) effects were soon incorporated following the HQET power counting and came along with unknown low-energy coefficients (not fixed by symmetry principles)~\cite{Cheng:1993gc,Cheng:1993kp}. The NLO effects refer to $1/m_Q$ terms in the Lagrangian, while the chiral Lagrangian is already expanded up to NLO to incorporate finite masses of the pseudo-Goldstone bosons. The effective Lagrangian was later used to compute genuine scattering amplitudes at NLO in the chiral expansion for collisions of heavy mesons off light mesons~\cite{Hofmann:2003je}, and predicted new resonant states. In such a task an extension to higher energies is required, by applying a unitarization method. While this method breaks the pre-established power counting it restores unitarity in the scattering amplitudes, allowing for the emergence of bound and/or resonant states (by dynamically generating poles of the $T$ matrix in the complex energy plane). This path was exploited with the LO effective Lagrangian in the charm sector~\cite{Kolomeitsev:2003ac}, predicting some of the states that will be relevant later. A description of both charm and bottom sectors was possible using the HQ flavor symmetry (HQFS)~\cite{Guo:2006fu}. The HQFS is a result of the expansion of the HQET Lagrangian and its power counting. Since at lowest-order the Lagrangian does not depend on the heavy-hadron mass (it is ${\cal O} (1/m_H^0)$), the interaction is blind to the heavy flavor, and the physics of charmed and bottomed states are the same.
The inclusion of the vector $1^-$ channel can be done thanks to the HQSS~\cite{Guo:2006rp}. This symmetry is also a result of the lowest-order HQET Lagrangian, which does not depend on the HQ spin orientation. Therefore, the interaction is the same for the heavy pseudoscalar state ($D$ meson) and its vector partner ($D^*$ meson).
Subsequently, the methodology was extended up to chiral NLO~\cite{Hofmann:2003je,Guo:2008gp,Guo:2009ct,Geng:2010vw}. These papers established a solid effective framework to understand the heavy-meson interactions with pseudo-Goldstone bosons.
Alternative EFT descriptions use the Weinberg-Tomozawa interaction (extending vector meson exchange at low energies into the pseudoscalar sector), and a broken SU(4) flavor symmetric Lagrangian~\cite{Gamermann:2006nm,Gamermann:2007fi}.

%%%%%%%%%%%%%%%%%%%%%%%%%%%%%%%%%%%%%%%%%%%%%
\subsection{$D$-meson interactions with light hadrons}
\label{ssec:Dint}
%%%%%%%%%%%%%%%%%%%%%%%%%%%%%%%%%%%%%%%%%%%%

The development of the HQET and based on a separation of scales given the HQ mass limit to define a sensible power counting in the Lagrangian, eventually gave rise to approaches in which the same hierarchies are deployed for applications to the phenomenology of URHICs. The systematic study of the processes that govern the dynamics and equilibration of $D$-mesons flourished in 2011, when four independent works appeared over a short period of time~\cite{Laine:2011is,He:2011yi,Ghosh:2011bw,Abreu:2011ic}. In all these works the dynamics of $D$-mesons were used---under different approximations---to evaluate their transport properties for temperatures typical of URHICs after hadronization.
  
In Refs.~\cite{Laine:2011is,Ghosh:2011bw} HQ and chiral symmetry have been combined to obtain an effective Lagrangian at lowest order from which the drag and diffusion coefficients  were extracted. In Refs.~\cite{He:2011yi, Abreu:2011ic} the emphasis was on resonance interactions---introduced either explicitly~\cite{He:2011yi} or generated dynamically~\cite{Abreu:2011ic}---which extended the validity of the scattering amplitudes to higher energies. These were used to compute the same coefficients as well as the spatial diffusion coefficient, $\Ds$. In the next subsections we will elaborate on these (and later) calculations, commenting on their main differences and the implications for the pertinent results. 

%%%%%%%%%%%%%%%%%%%%%%%%%%%%%%%%%%%% 
\subsubsection{Low-energy region: chiral and heavy-quark symmetries}
\label{sssec:sym}
%%%%%%%%%%%%%%%%%%%%%%%%%%%%%%%%%%%%%
To account for the dynamics of heavy mesons immersed into a thermal bath of pseudo-Goldstone bosons one can expect that both HQSS and chiral symmetry are key to set a hierarchy of scales able to address the problem of $D$-meson propagation. In Ref.~\cite{Laine:2011is}, this double expansion (used at the Lagrangian level as the Heavy Meson Chiral Perturbation Theory) was utilized to compute the heavy-meson diffusion coefficients. The dominant interaction of the heavy meson was considered to be a contact interaction with pions, $H \pi \rightarrow H \pi$. Subleading terms were not considered for the resulting transport coefficient (we will come back to them in Sec.~\ref{ssec:Dtrans}), and the validity limits for the temperature were quoted in the range $m_\pi/\pi \sim 30 \textrm{ MeV} \ll T \ll 80 \textrm{ MeV}$. As will be clarified in the subsequent section, the use of lowest-order interactions eventually leads to an overestimate of the interaction strength with an associated underestimate of the values of the diffusion coefficient $\Ds$.

The work in Ref.~\cite{Ghosh:2011bw} also used effective interactions combining chiral and HQ symmetries. A more extensive study of the transport coefficients as a function of temperature was performed, evaluating all coefficients appearing in the Fokker-Planck equation. The EFT used was based on previous developments of Ref.~\cite{Geng:2010vw}, where the Lagrangian was expanded at leading order (LO) in both chiral and HQ expansions. Moreover, a full set of pseudo-Goldstone bosons was considered ($\pi$, $K$, $\bar{K}$, $\eta$), interacting through contact terms plus vector-meson ($D^*,D_s^*$) exchanges. The latter were assumed to be stable states, \ie, their widths were neglected, which may become unreliable if large collisional widths of the vector mesons develop~\cite{Rapp:1999ej}. 
In addition, an effective model for $D$-meson interactions with nucleons was taken from a gauged SU(4) Lagrangian~\cite{Liu:2002vw}. The width of the heavy-vector meson is again crucial since it has a large effect on the determination of the $D\pi D^*$ coupling due to its near-threshold decay kinematics, introducing additional uncertainty.

In these works~\cite{Laine:2011is,Ghosh:2011bw}, the interactions are restricted to low energies, given the power counting of the EFT. The chiral expansion is the most restrictive one since the Goldstone boson momenta should, in principle, satisfy $k/\Lambda_\chi \ll 1$ (which can be thought of as the typical momentum exchanged in the collisions), with $\Lambda_\chi = 4 \pi f_\pi \simeq 1$ GeV the chiral scale. This constraint limits the range of temperatures where reliable results can be obtained since the typical momenta of the particle in the bath scale with $T$. This is a well-known feature for finite-temperature applications $\chi$PT~\cite{Schenk:1993ru}. For example, the $\pi\pi$ interaction described by the $\chi$PT expansion cannot capture the $\rho$-meson peak, whose presence starts to become important already at rather moderates temperatures, estimated to be around 100\,MeV~\cite{Schenk:1993ru}. The argument can be translated to our heavy-light systems. For the $D$ meson-light meson interaction at $T=100$ MeV, the average (relativistic) thermal pion energy is $\langle E_\pi \rangle = m_\pi + 3/2 T \simeq 300$ MeV, while the one for $D$-mesons is $\langle E_D \rangle \simeq 2$ GeV. If such two mesons collide head on, the center-of-mass energy of the pair is $\sqrt{s} \simeq 2.3$ GeV, which is close to the nominal PDG mass of the $D_0^*(2300)$ resonance, which is already the next higher resonance in the $D\pi$ channel (after the narrow $D^*(2010)$), with a vacuum  width of $\Gamma=229$ MeV. Since the EFT cannot describe these resonances the validity of Heavy-Meson Chiral Perturbation Theory is effectively restricted to $T \ll 100$ MeV, and transport coefficients start picking up uncertainties due to an uncontrolled increase of the scattering amplitudes (unless some effective cutoff or form factor is used to tame this increase). A more quantitative discussion of this issue will be given in Sec.~\ref{ssec:Dtrans}.

%%%%%%%%%%%%%%%%%%%%%%%%%%%%%%%%%%%%%
\subsubsection{Resonance region}
\label{sssec:resonances}
%%%%%%%%%%%%%%%%%%%%%%%%%%%%%%%%%%%%%%
Hadronic matter at temperatures above $T\simeq100$ MeV features a rapidly increasing number of resonance states. For example, at the typical chemical freeze-out temperature in URHICs, $T_{\rm chem} \simeq 160$\,MeV, the density of $\rho$ mesons alone is close to 1/3 of that of pions (implying that the pions from $\rho$ decays contribute more than half of the ``direct'' pions to the finally observed yield). At the same time, the hadron resonance gas (HRG) model is known to give a rather good description of the basic thermodynamics quantities of QCD matter as computed in lattice QCD (LQCD)~\cite{Ratti:2018ksb}.  For a reliable description of $D$-meson transport properties in this regime it is therefore mandatory to extent the description of heavy-hadron interactions to include resonant scattering channels, especially since the pertinent poles (in the complex plane) usually lead to a large interaction strength. 
In addition, perturbative interactions obtained from an EFT, when truncated at a given order in the momentum power expansion, introduce a violation of unitary bounds for the scattering amplitudes~\cite{dobado2012effective}.

In the literature, different ways of introducing the effect of resonant $D$-meson interactions have been pursued to control the energy dependence of the scattering amplitudes, and consequently, the temperature dependence of the transport coefficients. All methods will necessarily break the original power counting of the EFT, but they restore the exact unitarity condition of the scattering matrix elements and achieve an improved  description of the hadron content of the thermal bath. In the following, we cover the unitarization methods based on $T$-matrix equations (Sec.~\ref{ssssec:unitarity})  and the explicit introduction of resonant interactions guided by the HRG (Sec.~\ref{ssssec:hrg}).

%%%%%%%%%%%%%%%%%%%%%%%%%%%%%%%%%%%%%%%%%%%%%%%%%%%%%%%%%%%%%%%%%%%%%
\paragraph{Dynamically-generated resonances: unitarization models.}
\label{ssssec:unitarity}
%%%%%%%%%%%%%%%%%%%%%%%%%%%%%%%%%%%%%%%%%%%%%%%%%%%%%%%%%%%%%%%%%%%%%%%%
A popular line of investigation to improve the description of the EFT at moderate energies without introducing extra degrees of freedom is the application of unitarization techniques. The idea is to construct a scattering amplitude, $T$, from the perturbative term, $V$, given by the EFT at a certain order. The unitarity condition for the ${\cal S}$-matrix is translated to the ${\cal T}$-operator (${\cal S}=\mathbb{1}- i{\cal T}$) as
\begin{equation}
 {\cal S}^\dag {\cal S} = \mathbb{1} \qquad \rightarrow \qquad 
2 \textrm{Im } {\cal T} = - {\cal T}^\dag {\cal T} \ . \label{eq:unitarity}
\end{equation}
Any truncation of the EFT produces scattering amplitudes which fulfills the unitary condition Eq.~\eqref{eq:unitarity} only perturbatively. However, this is enough for the cross section to break physical bounds at moderate energies. 

\begin{figure}[!ht]
\centering
\includegraphics[scale=0.45]{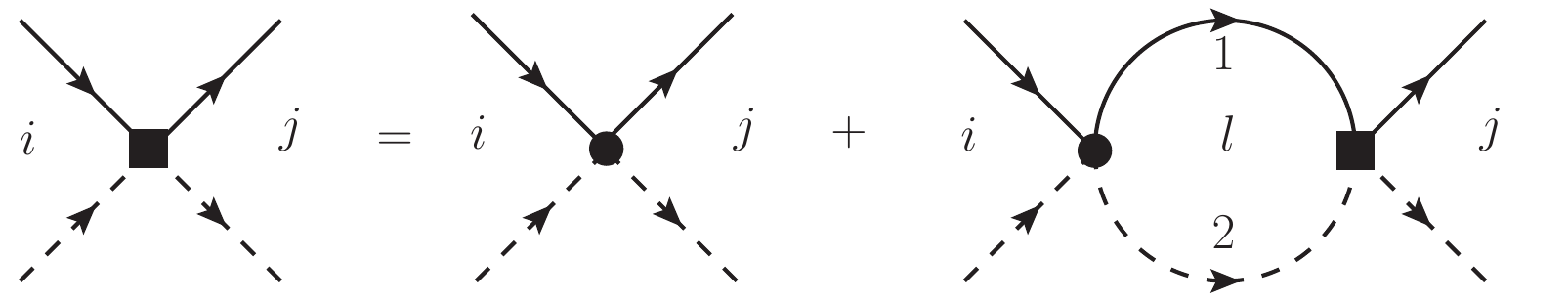}
\caption{Diagrammatic representation of the $T$-matrix Eq.~\eqref{eq:Tmatrix}, where the $T$-matrix element (square) is obtained from the integral equation using the perturbative potential $V$ (circle) as kernel.}
\label{fig:Tmatrix}
\end{figure}

For two-body scattering, the perturbative amplitude, or potential $V$, can be used as the kernel of the Bethe-Salpeter equation for the $T$-matrix elements,
\be 
T_{ij} (P, k_i,k_j)=V_{ij} (P,k_i,k_j) + i \sum_l \int \frac{d^4q}{(2\pi)^4} V_{il} (P,k_i,q) \Delta_1(q) \Delta_2(P-q) T_{lj} (P,q,k_j) \ , 
\label{eq:Tmatrix}
\ee
where $i,j$ denote the initial and final scattering channels, $k_i$ is the relative momentum of the particles in the channel $i$, $P$ is the total momentum of the pair, and $\Delta_{1,2}$ are individual meson propagators for the particles $1$ and $2$ that are involved in the intermediate channel, $l$. A pictorial representation of Eq.~\eqref{eq:Tmatrix} is given in Fig.~\ref{fig:Tmatrix}, where the square vertex represents the $T_{ij}$ and the filled circle the $V_{ij}$.

The kernel of the $T$-matrix Eq.~\eqref{eq:Tmatrix} is the perturbative potential, $V_{ij}$, calculated at a given order in the original EFT, which is then iterated to obtain the $T$ matrix. In the so-called {\it on-shell factorization scheme}~\cite{Oller:1997ti,Oset:1997it}, the off-shell contribution of the potentials $V_{ij}$ is encoded in the renormalization of masses and vertices, and only the on-shell part of the potential is used in the Bethe-Salpeter equation, leading to,
\be 
T_{ij} (P, k_i,k_j)=V_{ij} (P,k_i,k_j) +  \sum_l V_{il} (P,k_i,k_j) G_l(P) T_{lj} (P,k_i,k_j) \ ,
\label{eq:Tmatrix2}
\ee
with
\be G_l(P)=i\int \frac{d^4q}{(2\pi)^4} \Delta_1 (q) \Delta_2 (P-q) \ 
\ee
being the so-called {\it loop function}, or two-particle propagator function. Eq.~\eqref{eq:Tmatrix2} is now an algebraic equation which can be solved much easier than Eq.~\eqref{eq:Tmatrix}, especially for coupled-channel cases where inelastic collisions are taken into account. 
In the center-of-mass frame of the collision where the total pair momentum vanishes, $P=( \sqrt{s},0,0,0)$, ones has
\be 
T_{ij} (s)=V_{ij} (s) +  \sum_l V_{il} (s) G_l(s) T_{lj} (s) \ . 
\ee
The unitarity condition, Eq.~\eqref{eq:unitarity}, for the $T$-matrix element amounts to the optical theorem for coupled channels,
\be 
\textrm{Im } T_{ij} (s)= \sum_l T^*_{i \rightarrow l} \ \textrm{Im } G_l(s) \ T_{l \rightarrow j}(s) \ 
\ee
with
\be 
\textrm{Im }G_l(s)= \frac{q(s)}{4\pi \sqrt{s}}=\frac{\lambda^{1/2}(s,m_{1}^2,m_{2}^2)}{8\pi s} = \frac{\sqrt{[s-(m_1+m_2)^2][s-(m_1-m_2)^2]}}{8\pi s}\ , 
\ee
where $q(s)$ is the CM momentum in the loop, $m_1$ and $m_2$ are the masses of the two particles propagating in the loop of channel $a$, and $\lambda(a,b,c)$ is the triangular (or K\"all\'en) function. Notice that for real $V_{ij}$ the optical theorem can now be satisfied thanks to the imaginary part of the loop function, which is nonzero in the physical range of $s$, \ie, above the unitary threshold.

The restoration of the exact unitarity condition of the scattering amplitude has important consequences. First of all, for complex values of the energy, $z=\sqrt{s}$,  non-analyticities in the form of poles might appear in the different Riemann sheets (RSs) of the resulting $T(z)$ function. Those appearing on the real axis of the first RS are associated with bound states, since they appear below the elastic threshold. Poles appearing in higher Riemann sheets (typically on the second one, \ie, the one directly connected to the real energy axis) are associated with resonances, since the poles appear with an imaginary energy argument, $\textrm{Im } z_{\textrm{pole}} \neq 0$. While these generated states appear in the non-physical complex energy plane, their effect is felt on the real energy axis above threshold, and can potentially dominate the cross section over a certain energy range. Therefore, for the calculation of scattering processes and the computation of transport coefficients such states are very relevant.

\begin{figure}[!ht] 
\centering
\includegraphics[scale=0.27]{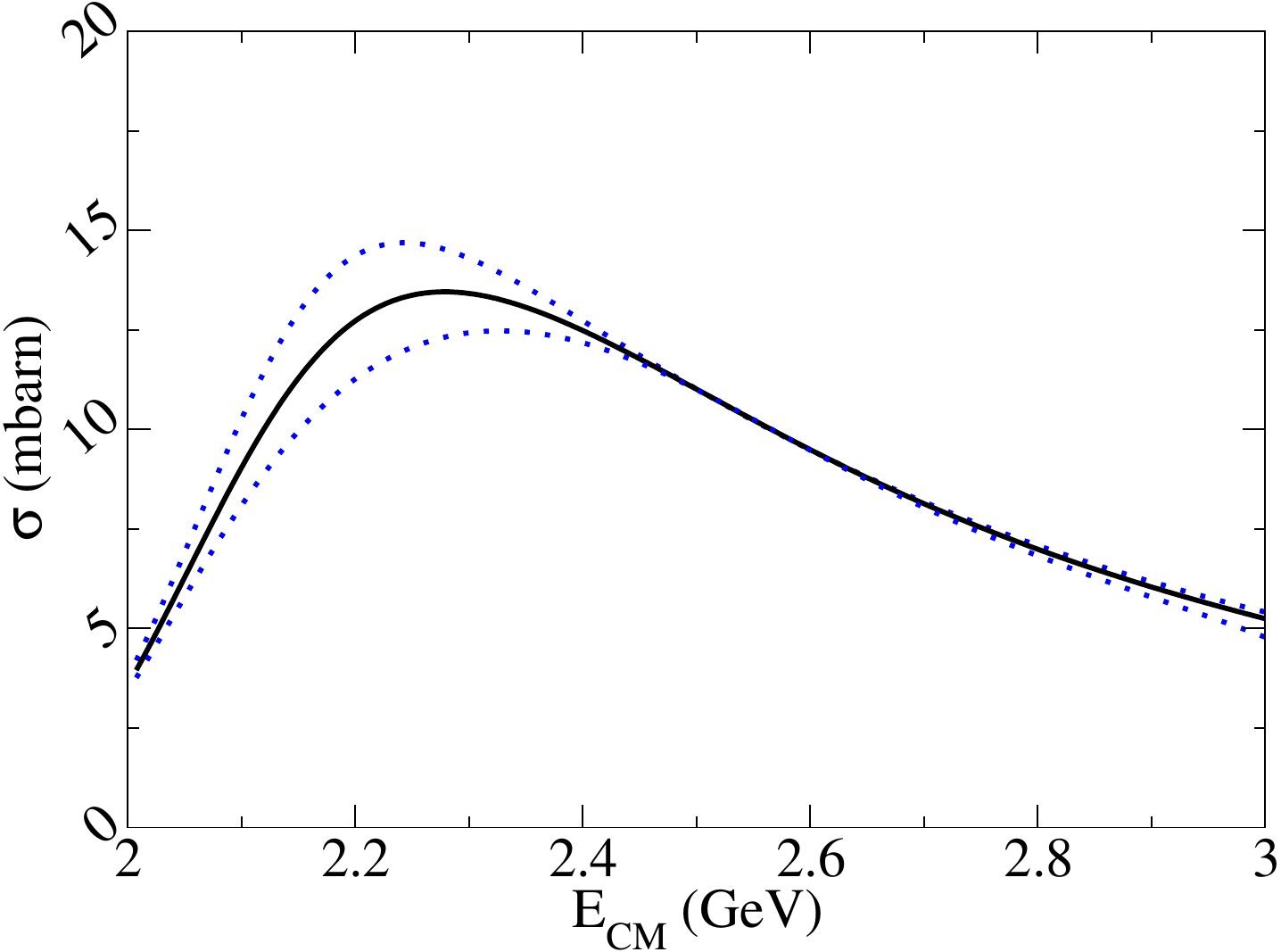}
\includegraphics[scale=0.27]{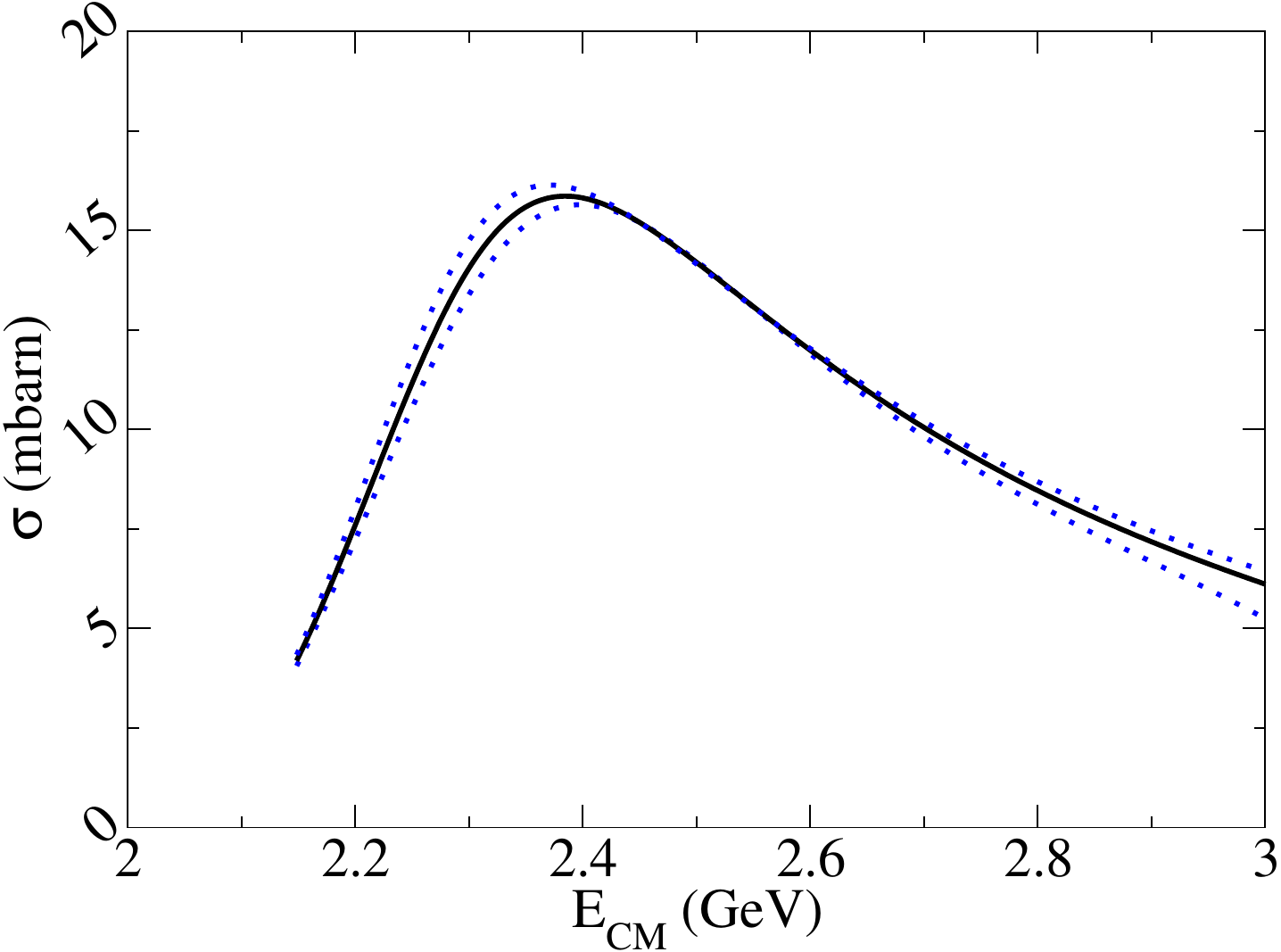}
\caption{$I=1/2$ total cross sections of $D\pi$ (left panel) and $D^* \pi$ (right panel) scattering as functions of the center-of-mass energy of the collision. The resonant peaks are respectively identified  with the $D_0(2300)$ and $D_1(2420)$ states~\cite{ParticleDataGroup:2022pth}. The dotted lines serve as error bands of the calculation due to the uncertainties in the low-energy coefficients of the EFT. Adapted from Ref.~\cite{Abreu:2011ic}.}
\label{fig:XsecAbreu2011}
\end{figure}

In the context of $D$-meson interactions with light hadrons, this idea was put forward in Ref.~\cite{Abreu:2011ic} where the same EFT incorporating chiral and HQ symmetries was used to construct the perturbative amplitudes, $V_{ij}$. The chiral expansion was taken up to NLO, but the interaction covered only $D$-meson interactions with $\pi$'s (in subsequent extensions of this work the whole set of pseudoscalar light mesons, as well as $N$ and $\Delta$ baryons, were incorporated in a coupled channel basis~\cite{Tolos:2013kva}).
The $T$-matrix Eq.~\eqref{eq:Tmatrix2} was then solved using the NLO potentials, to generate the final unitary scattering amplitudes. Using HQFS the interactions of the heavy vector meson $D^*$ with pions were also constructed. The final $D\pi$ and $D^* \pi$ cross sections in the isospin $I=1/2$ channel of Ref.~\cite{Abreu:2011ic} are summarized in Fig.~\ref{fig:XsecAbreu2011} as a function of the center-of-mass energy, $\sqrt{s}=E_{CM}$ .
The qualitative behavior of the two cross sections is similar (due to the HQFS also the magnitudes are similar), and one observes the typical resonant behavior. For the $S$-wave $J=0$ case the peak has been ascribed to the $D_0(2300)$ resonant and for the $J=1$ (also $S$-wave) case with the heavy-quark spin partner, the $D_1(2420)$. While the unitarization method generates such peaks with the appropriate quantum numbers and approximately correct masses and widths, the freedom of selecting the NLO low-energy constants (LECs) helps to fix the parameters to reproduce the experimental values. The dotted lines in the plots of Fig.~\ref{fig:XsecAbreu2011} define an error band due to the natural variation of the low-energy constants. The use of these scattering amplitudes to compute the transport coefficient~\cite{Abreu:2011ic} will be detailed in Sec.~\ref{ssec:Dtrans}.

In Ref.~\cite{Tolos:2013kva} the remaining set of light mesons ($K,\bar{K},\eta$), as well as nucleons and $\Delta$ baryons, were added. The more complete use of coupled channels in the $T$-matrix led to an improvement in the LECs, in agreement with LQCD data~\cite{Liu:2012zya}, which appeared soon before~\cite{Tolos:2013kva}. The resulting cross sections covered many elastic and inelastic channels (see Ref.~\cite{Song:2015sfa} for concrete examples) and they were used in transport model simulations of hadrons in the Parton-Hadron String Dynamics (PHSD) code, see Sec.~\ref{sssec_phsd}.

\begin{figure}[h]
\centering
\includegraphics[scale=0.5]{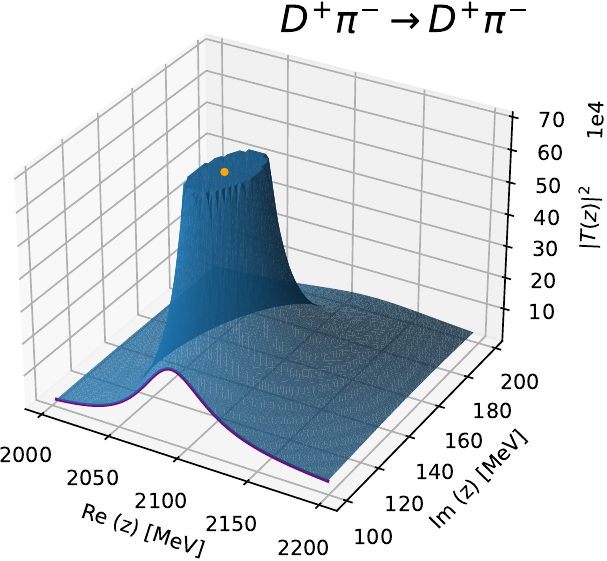}
\includegraphics[scale=0.5]{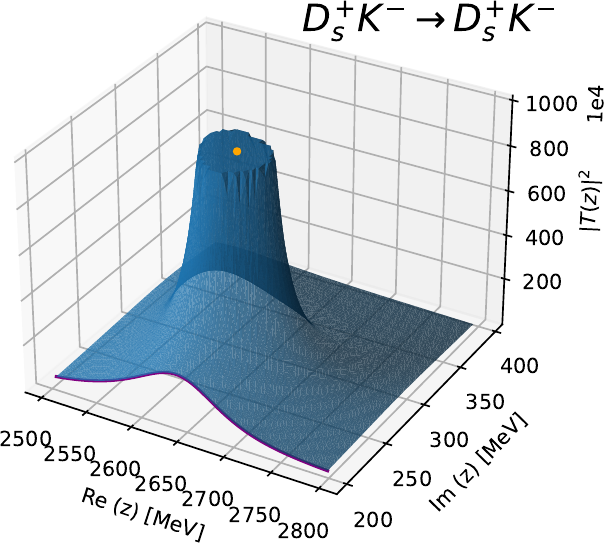}
\includegraphics[scale=0.5]{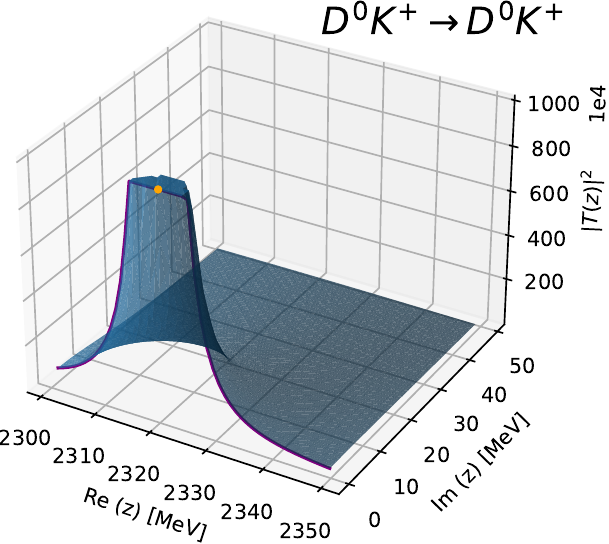}
\caption{Lower pole (left panel) and higher pole (center panel) of the $D_0(2300)$ state, obtained upon unitarization of the $D\pi$ scattering. Pole of the $D_s^{*0}(2317)$ state (right panel) obtained upon unitarization of the $DK$ scattering~\cite{Montana:2020lfi,Montana:2020vjg}.}
\label{fig:D0poles}
\end{figure}
The search for poles in complex energy plane was systematized in Refs.~\cite{Montana:2020lfi,Montana:2020vjg} for $D$ and $D^*$ where constraints from recent LQCD data~\cite{Guo:2018tjx} on the values of the LECs were used.
The results of the pole search are summarized in Fig.~\ref{fig:D0poles}. In the left panel we show the $|T|^2$ of the $D^+ \pi^-$ elastic scattering in the complex plane on the second Riemann sheet. A pole is found with a sizable decay width (imaginary part). In the center panel we show the $D_s^+K^-$ coupled channel at higher energies where another pole is found, also with a large imaginary part. These two poles share the same quantum numbers and the scattering channels are coupled in the $T$-matrix equation. Recent studies~\cite{Kolomeitsev:2003ac,Guo:2006fu,Guo:2009ct,Albaladejo:2016lbb} have determined that these 2 poles correspond to a single state denoted as $D^*_0(2300)$~\cite{ParticleDataGroup:2022pth} which are present in all coupled channels but with different strengths (couplings). See Ref.~\cite{Meissner:2020khl} for further discussion of these structures, also in different hadron systems. Therefore, depending on the particular scattering process, the cross section can reflect in a more pronounced way one pole or another. Finally, in the right panel of Fig.~\ref{fig:D0poles}, the first Riemann sheet of the $D^0K^+$ process reveals a pole without imaginary part. It is identified with the $D_{s0}^* (2317)$ state, which is below the elastic threshold, but that would appear as an enhancement in the cross section above to it.

In summary of the case of $D$-meson--light-meson interactions, the unitarization method allows to uncover a rather rich set of channels with resonances and bound states that are dynamically generated and provide substantial contributions to the $D$-meson cross section at the typical energies that URHICs temperatures $T \simeq 150$ MeV are probing.

%%%%%%%%%%%%%%%%%%%%%%%%%%%%%%%%%%%%%%%%%%
\paragraph{Hadron resonance gas}
\label{ssssec:hrg}
%%%%%%%%%%%%%%%%%%%%%%%%%%%%%%%%%%%%%%%%%%%
The first attempt to systematically incorporate resonant $D$-meson interactions within a resonance gas of light hadrons into the evaluation of its transport properties was carried out in Ref.~\cite{He:2011yi} (for a pion gas an initial calculation was done in Ref.~\cite{Fuchs:2004fh}). 
The interactions have been constructed using Breit-Wigner parameterizations for both bound and scattering states including microscopic widths based on partial-wave interactions and thresholds as appropriate. 
The $D$-meson interactions were implemented in a finite-temperature gas at vanishing chemical potential including scattering off $\pi,K,\eta, \rho,\omega,K^*,N,\bar{N}, \Delta, \bar{\Delta}$, covering a considerable set of hadrons. Each channel has been studied separately according to the most prominent resonances that appear at the energies of interest.

For interactions with pions the $D_0^*(2300)$ ($D_0^*$(2308) at that time), the $D^*(2010)$ (the vector HQ partner of the $D$ meson), and the narrow $D_2^*(2460)$ state were accounted for. The Breit-Wigner parameters (masses and widths) were estimated using empirical widths and decay branchings in the pertinent channels.  The interaction with kaons was taken from the unitarized EFT result of Ref.~\cite{Kolomeitsev:2003ac} which accounts for the predominant $D_{s0}^* (2317)$ bound state below threshold, not directly visible but causing the cross section to be enhanced close to the $DK$ threshold (similar to what has been discussed in the previous section). The $D\eta$ interaction was taken from the same reference. Interactions with vector mesons were based on the poles found by a unitarized EFT using hidden-gauge formalism with SU(4) chiral symmetry~\cite{Gamermann:2007fi}, and interactions with baryons were taken from the result of unitarized EFT constructed 
in Refs.~\cite{Lutz:2005vx} (interactions with nucleons and anti-nucleons) and \cite{Hofmann:2003je} (with $\Delta$ anti-/baryons ).

\begin{figure} 
\centering
\includegraphics[scale=0.29]{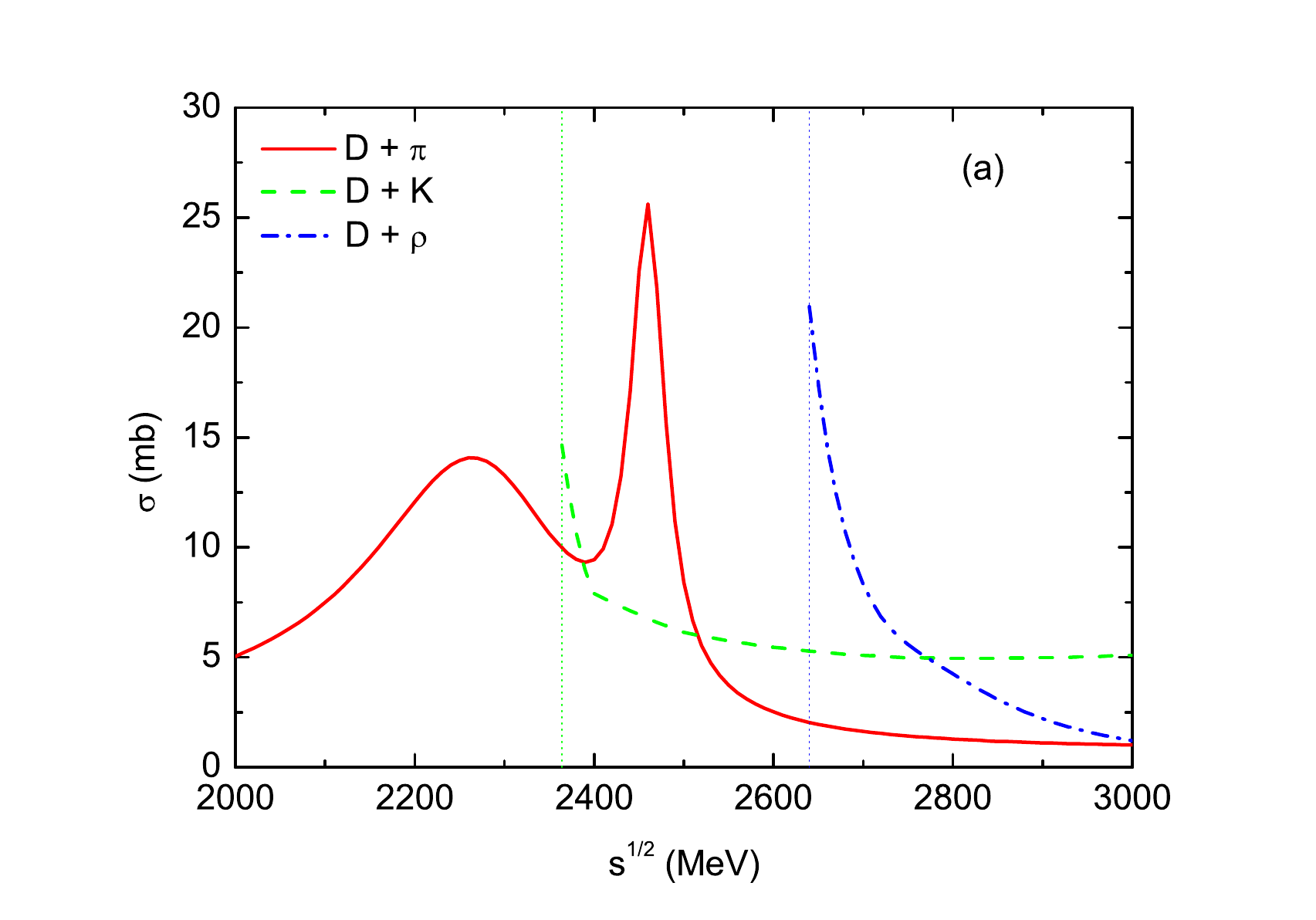}
\hspace{-0.5cm}
\includegraphics[scale=0.29]{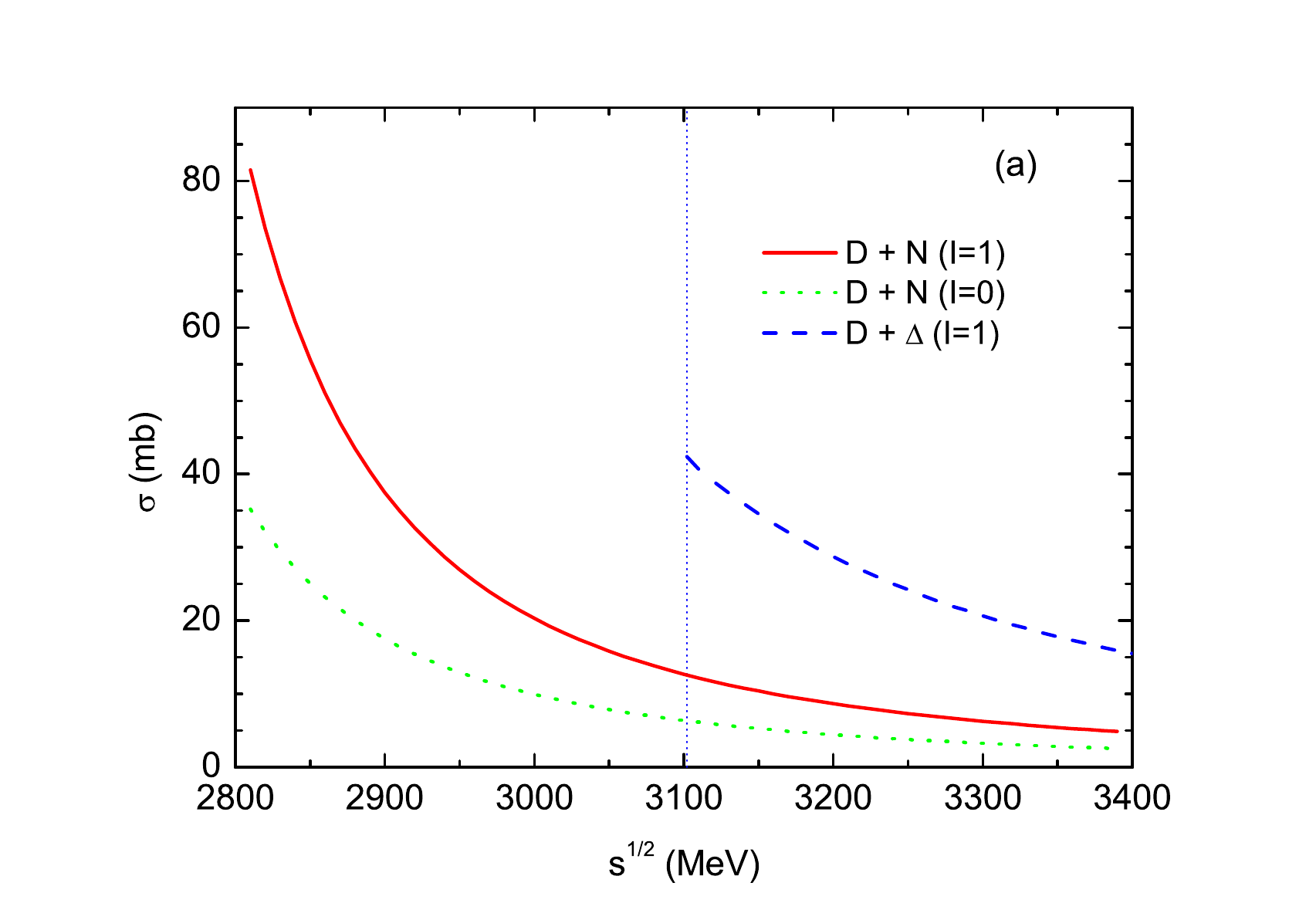}
\caption{Left panel: Total cross section of $D\pi$, $DK$ and $D\rho$ processes as functions of the center-of-mass energy of the meson pair. Right panel: $DN$ and $D\Delta$ total cross sections as function of the center-of-mass energy. Both results are obtained using the resonance gas model. Figures adapted from Ref.~\cite{He:2011yi}.}
\label{fig:Xsec_He}
\end{figure}

\begin{figure} 
\centering
\includegraphics[scale=0.5]{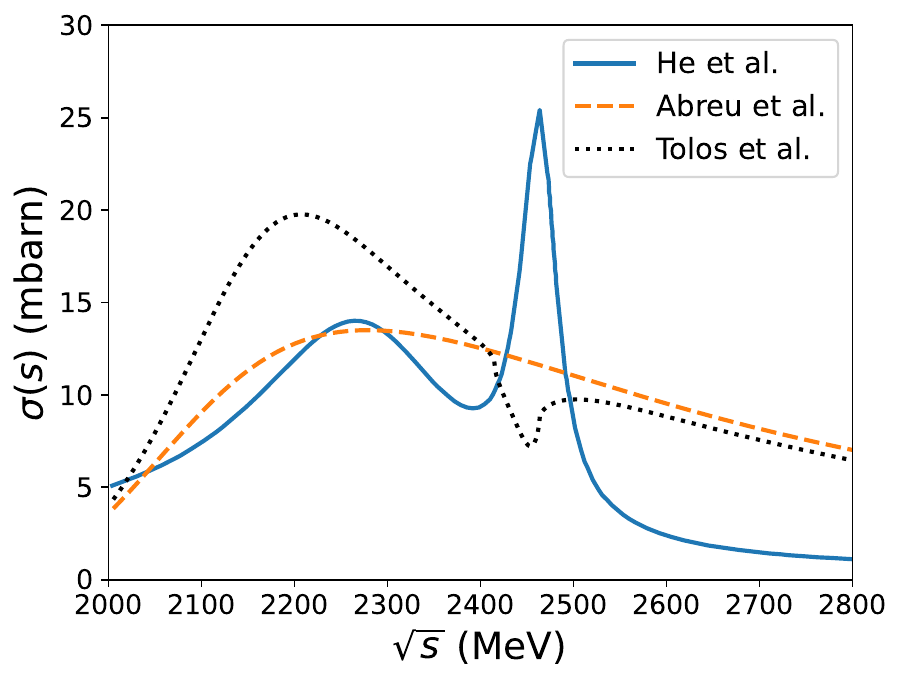}
\caption{Total (isospin averaged) cross section of the $D\pi$ scattering comparing unitarization and resonance gas models. Abreu {\it et al.}~\cite{Abreu:2011ic} uses exclusive $S-$wave $D\pi \rightarrow D\pi$ interaction in which the $D_0^*(2300)$ bump is seen. Tolos {\it et al.}~\cite{Tolos:2013kva} incorporates the $D\pi \leftrightarrow D\eta$ coupled channel (the dip at $\sqrt{s}=2450$ MeV signals the opening of the $D \eta$ channel). He {\it et al.}~\cite{He:2011yi} uses a resonance model in which the  $D_0^*(2300)$ and the $D_2^*$ are included (the $D_2^*(2460)$ resonance is not generated in the previous two approaches, since it can only decay into $D\pi$ in a relative $D$-wave).}
\label{figXsec_comparison}
\end{figure}

In Fig.~\ref{fig:Xsec_He} we display a summary of the total cross sections of (most of) the channels evaluated in Ref.~\cite{He:2011yi}; one notices the prominence of the aforementioned resonances, or threshold enhancements originating from the presence of bound states. Again, we stress that these features are of quite an opposite tendency  compared to perturbative EFT approaches, where an increase of the cross section with energy is usually obtained. The influence of these cross sections on the transport coefficients~\cite{He:2011yi} will be elaborated in Sec.~\ref{ssec:Dtrans}.

Subsequent works have also combined the two methods (unitarized and explicit resonant interactions) to address $D$-meson cross sections. Reference~\cite{Ozvenchuk:2014rpa} contains combined results from unitarized EFT~\cite{Tolos:2013kva} and the resonance gas model~\cite{He:2011yi} for interactions of $D$ mesons with mesons and nucleons, as well as results from the hidden local gauge SU(4) EFT for the interaction with $\rho$ mesons~\cite{Lin:2000jp} (without unitarization but with the use of form factors to control the ultraviolet behavior of the scattering amplitudes).  

In Fig.~\ref{figXsec_comparison},
we present a compilation of the mostly used $D\pi$ cross sections after averaging in the two isospin channels ($I=1/2$ and $I=3/2$). While, on average, the absolute value of the different cross sections are alike, their details are different. The effective theory used in Abreu {\it et al.}~\cite{Abreu:2011ic} works only with the $L=0$ partial wave in the diagonal $D\pi \rightarrow D\pi$ channel. The $D_0^*(2300)$ state is generated dynamically and appears as a very broad resonance. In Tolos {\it et al.}~\cite{Tolos:2013kva} the previous calculation is extended to coupled channels and improves the determination of the model parameters. The $D_0^*(2300)$ peak remains, but it is slightly shifted to lower energies and threshold cusp effects do appear. Finally, the calculation of He {\it et al.} in Ref.~\cite{He:2011yi} employs a resonance model for the $D\pi$ scattering in which the resonances are introduced explicitly. The peak of the  $D_0^*(2300)$ is accompanied by the narrow $D_2^*(2460)$ state, which is absent in the previous calculations since it appears in the $D\pi$ system in the $L=2$ partial wave.

It is probably fair to say that the fine details of the cross sections are not essential for the description of decay widths or transport coefficients, since the latter are calculated by performing averages of the differential cross sections over the thermal momentum distributions of the heat bath particles. For example, a very narrow resonance is not likely to contribute as much as a broad resonance that dominates the scattering cross section over a broader energy range (unless the value of the cross section of the former dominates over the second one).
On the other hand, a faithful description of the angular dependence of the scattering amplitudes (or differential cross sections) can be quite relevant to properly access the transport properties that depend on anisotropic momentum exchange~\cite{Plumari:2012ep,Hammelmann:2023fqw}. 
While the {\em collision} rate can in principle be computed from the total cross section, the {\em thermalization} rate (and other transport coefficients like the momentum diffusion coefficient), involve the differential cross section with an angular weighting.
Therefore, the results for an isotropic differential cross section versus a forward-peaked are generally very different even when the total cross section is the same in both cases.
This is, in fact, well known for the for the case of HQ diffusion in the QGP, where, \eg, perturbative one-gluon exchange amplitudes can  have fairly large cross sections but are rather inefficient for the transport coefficient since the angular distribution is strongly forward peaked and thus ineffective in isotropization. This was one of the original motivations of introducing heavy-light resonance interactions in the QGP~\cite{vanHees:2004gq,Rapp:2009my}.

%%%%%%%%%%%%%%%%%%%%%%%%%%%%%%%%%%%%%%%%%%%%%%%
\subsection{In-medium $D$-meson properties}
\label{ssec:Dmed}
%%%%%%%%%%%%%%%%%%%%%%%%%%%%%%%%%%%%%%%%%%%%%
In this subsection we will address the temperature and density dependence of the $D$-meson properties when immersed in a thermal gas of light hadronic constituents. Quantities like the mass, decay width and two-body scattering amplitudes acquire a dependence on temperature and density, and these might affect not only the propagation of the heavy particle but also their transport coefficients. Within the hadron gas description we will review the different approximations to calculate the in-medium spectral function of the $D$-meson in hot matter in Sec.~\ref{sssec:spectral}, briefly alluding to results at finite chemical potential in Sec.~\ref{sssec:chem}, and discuss
the temperature dependence of the $D$-meson scattering amplitudes in Sec.~\ref{sssec:D-ampl}.

%%%%%%%%%%%%%%%%%%%%%%%%%%%%%%%%%%%%%%%%%%%
\subsubsection{Spectral properties}
\label{sssec:spectral}
%%%%%%%%%%%%%%%%%%%%%%%%%%%%%%%%%%%%%%%%%%%
Interactions of the $D$ meson with particles of the bath modify its spectral properties, especially when temperatures and/or densities are high. These properties can be addressed using standard QFT techniques at finite temperature, \eg, the imaginary-time formalism~\cite{le2000thermal,kapusta2007finite,laine2016basics}, based on the scattering processes that have been described before in vacuum.

At finite temperature, several works have addressed the thermal spectral function of $D$ mesons. It can be defined as,
\be 
\rho (p_0,{\bm p})= -\frac{1}{\pi} \frac{ \textrm{Im } \Sigma^R (p_0,{\bm p})}{\left[p_0^2-{\bm p}^2-m_{D,0}^2- \textrm{Re }\Sigma^R(p_0,{\bm p}) \right]^2+ \left[\textrm{Im } \Sigma^R(p_0,{\bm p}) \right]^2} \ ,
\ee
where $m_{D,0}$ is the vacuum mass and $\Sigma^R(p_0,{\bm p})$ the retarded self-energy of the $D$ meson. The latter can be computed from the $D$-meson amplitudes discussed in the previous section by folding them over the thermal distribution of the pertinent medium particles.

\begin{figure}[th]
\centering
\includegraphics[scale=0.33]{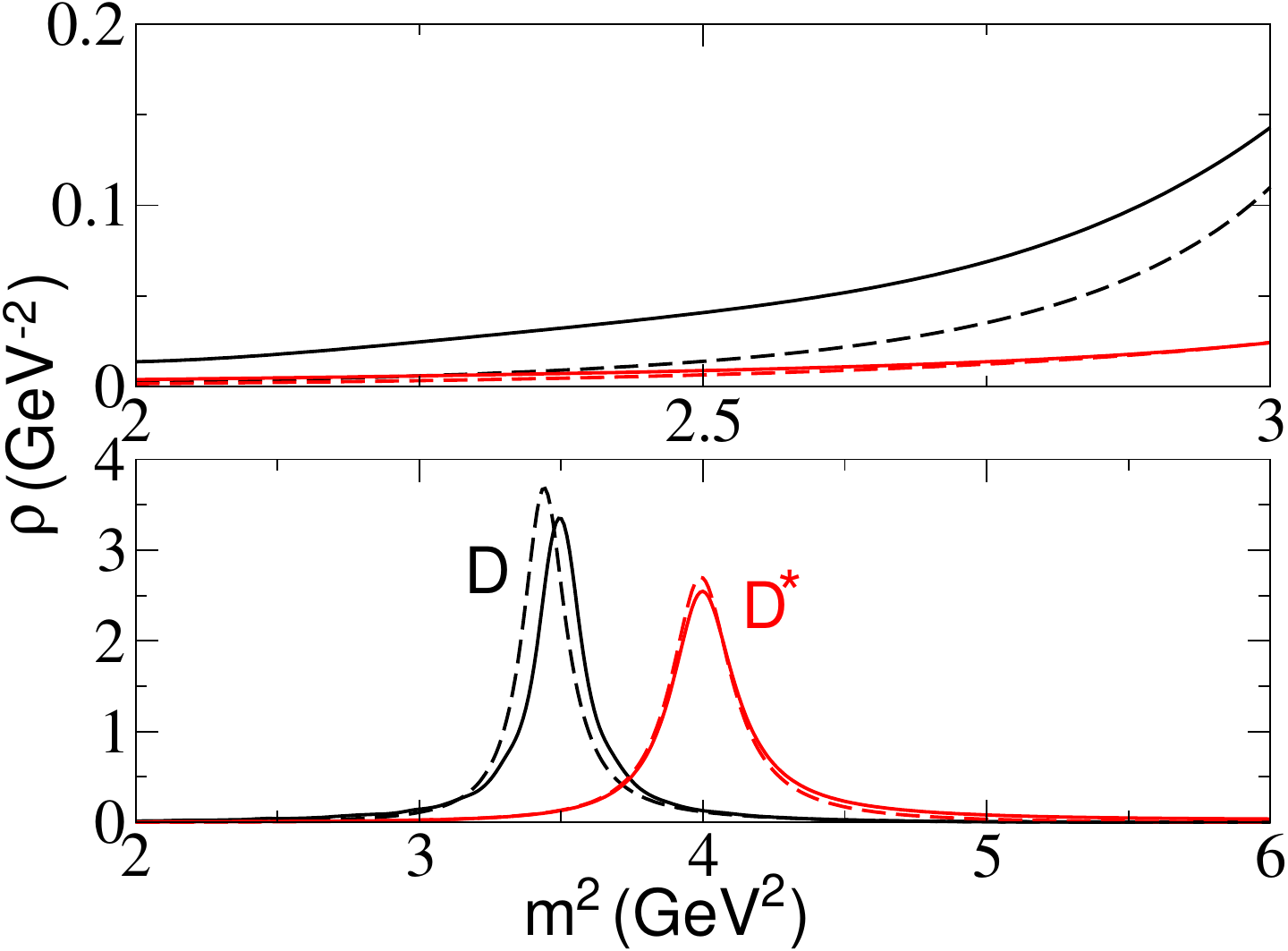}
\hspace{5mm}
\includegraphics[scale=0.34]{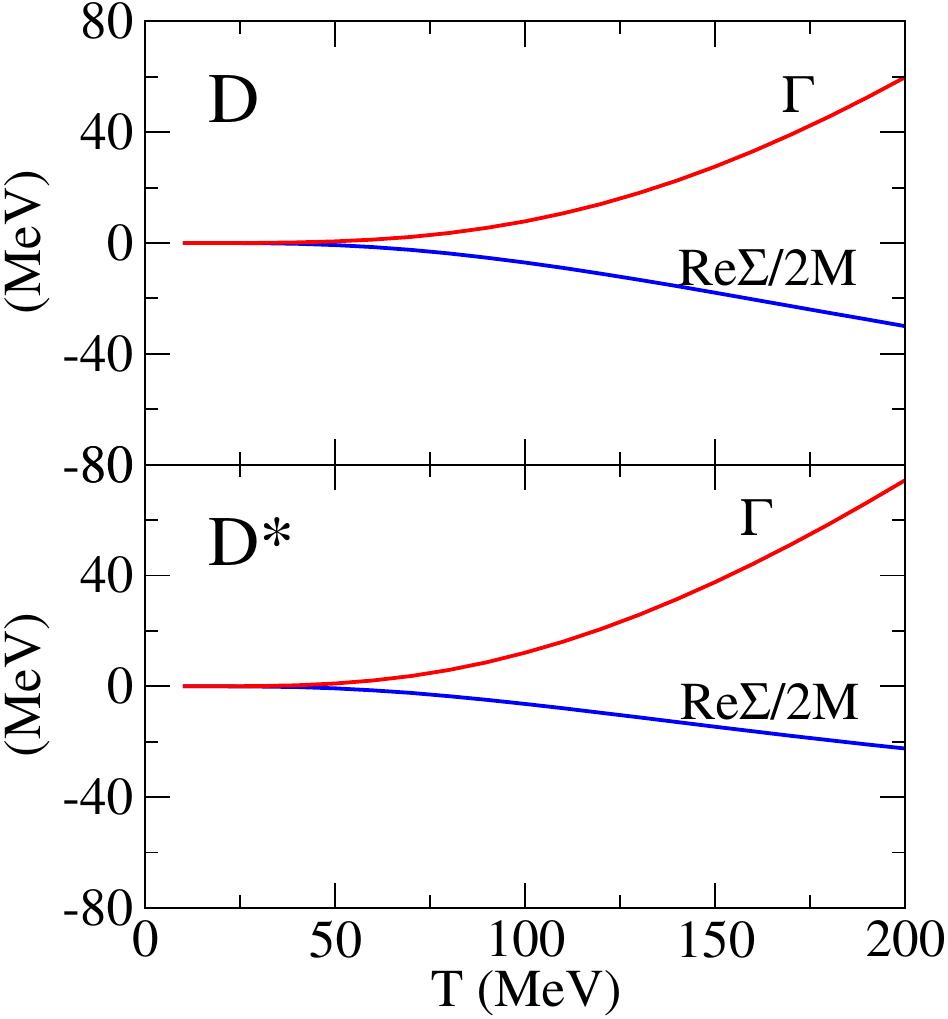}
\caption{In-medium $D$-meson properties in a pion gas at vanishing 3-momentum and $T=200$ MeV from Ref.~\cite{Fuchs:2004fh}; left panels: spectral functions for $D^*$ (red lines) and $D$ (black lines) at vanishing three-momentum where dashed and solid lines correspond to the lowest order and first step in a self consistent iteration; right panels: collisional width (red lines) and mass shift (blue lines), corresponding to the imaginary and real parts of the self energy, as a function of temperature for $D$ (upper panel) and $D^*$ (lower panel).    
Figures taken from Ref.~\cite{Fuchs:2004fh}}.
\label{fig:Dpi_fuchs}
\end{figure}
In Ref.~\cite{Fuchs:2004fh} resonant scattering amplitudes in a thermal pion gas have been used in the isospin $I=1/2$ channel of the $D\pi$ channels in the form of relativistic Breit-Wigner shapes.
In the left panel of Fig.~\ref{fig:Dpi_fuchs} we reproduce the resulting  spectral functions of the $D$ and $D^*$ mesons~\cite{Fuchs:2004fh}, plotted as a function the invariant mass squared, $m^2 = p_0^2-{\bm p}^2$  at a temperature of $T=200$\,MeV. Black and red lines correspond to $D$ and $D^*$ mesons, respectively, and solid (dashed) lines correspond to the zeroth order (first iteration) in the the self-consistent equations that relate the self-energy with the $D$-meson propagator and the scattering amplitude (see Sec.~\ref{sssec:D-ampl} for additional details). 
The spectral functions show a typical collisional broadening which is quantified as a function of temperature in the right panel of Fig.~\ref{fig:Dpi_fuchs}, reaching up to $\Gamma=60$\,MeV at the highest $T$=200\,MeV (generated by the imaginary part of the self-energy, $\Gamma=-{\rm Im}\Sigma/M$). The shift of the $D$-meson mass with respect to the vacuum one (corresponding to the  real part of $\Sigma/(2M)$)  is more clearly seen in right panel of Fig.~\ref{fig:Dpi_fuchs}, where the results for the self-energies of $D$ (upper panel) and $D^*$ (lower panel) are plotted as a function of temperature. At $T=150$ MeV a downward shift in the $D$ and $D^*$ masses of 20 MeV and 15 MeV, respectively, is found. The in-medium decay width of the states amounts to 25 MeV and 40 MeV for $D$ and $D^*$, respectively. In the latter case the width is also modified by its coupling to scalar and tensor $D$-meson excitations.

\begin{figure}[th]  
\centering
\includegraphics[scale=0.28]{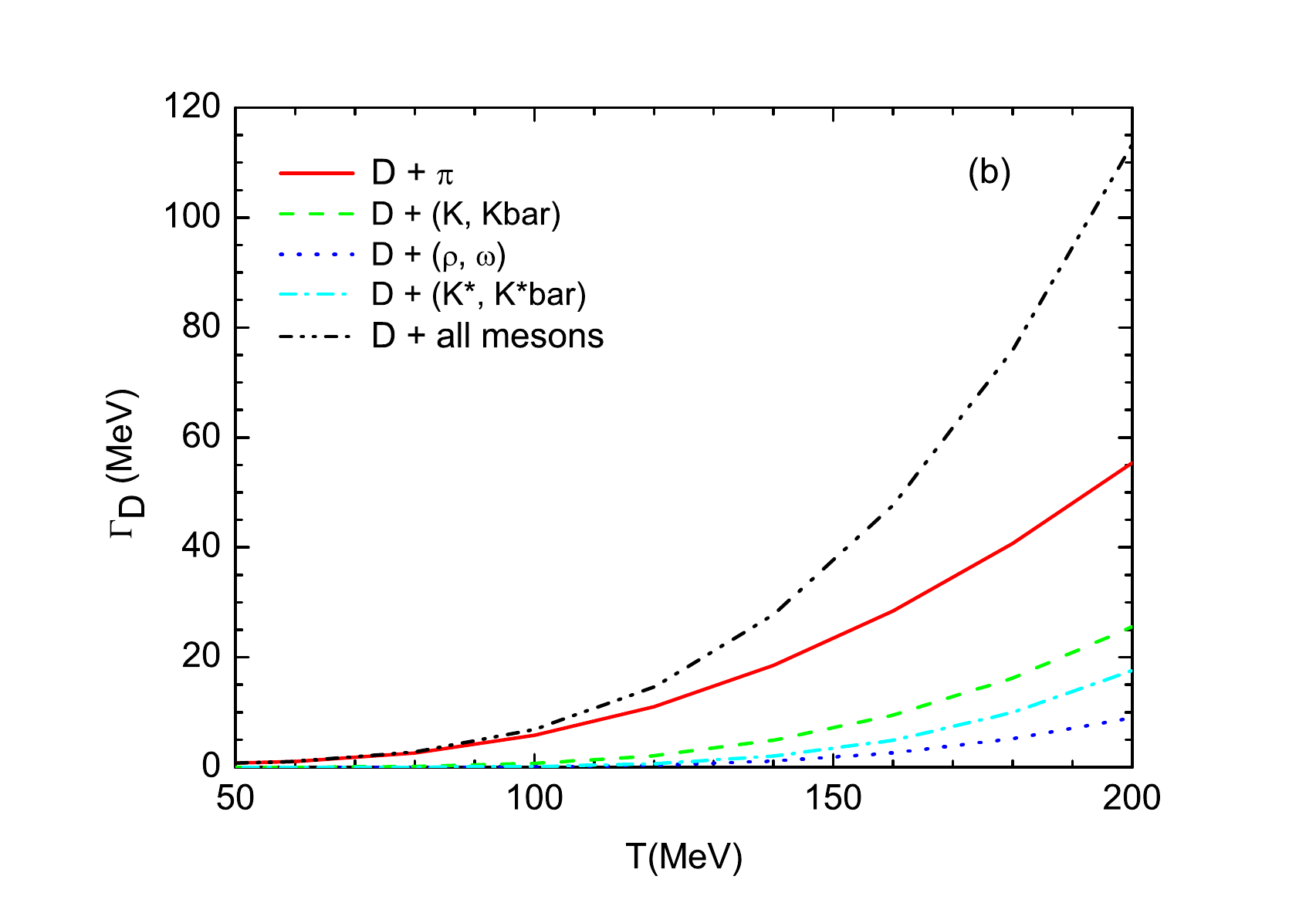}
\includegraphics[scale=0.28]{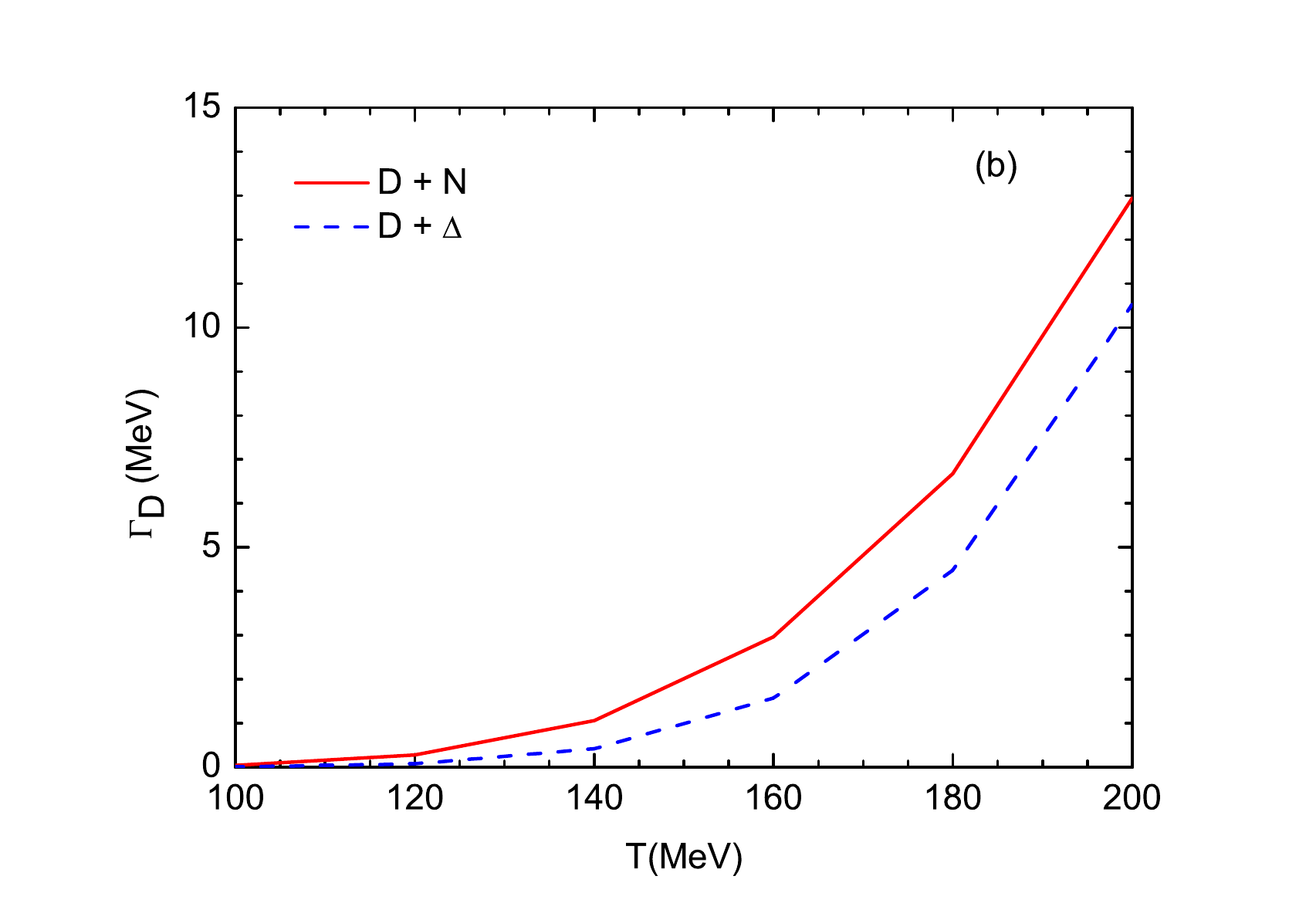}
\caption{Thermal decay width of $D$ mesons as a function of  temperature as computed in Ref.~\cite{He:2011yi}. The individual mesonic (left panel) and baryonic (right panel) contributions to the total decay width $\Gamma_D(T)$ are shown. Figures taken from~\cite{He:2011yi}.}
\label{fig:Dwidth_he}
\end{figure}
In Ref.~\cite{He:2011yi}, the results for the in-medium width of $D$ mesons in a hadron gas have also been calculated and are reproduced in Fig.~\ref{fig:Dwidth_he}, where the individual contributions from the light species are resolved (left panel: mesons; right panel: anti-/baryons). 
The width obtained from the interactions with pions is consistent with the results of Ref.~\cite{Fuchs:2004fh}, which is expected as approximations in both calculations are similar, \ie, based on a one-loop calculation of the self-energies using vacuum propagators. From Fig.~\ref{fig:Dwidth_he} it becomes clear that the contribution from other mesons cannot be neglected at moderate temperatures; \eg, at $T=150$ MeV the total $D$-meson width amounts to $\Gamma_D \simeq 40$ MeV, approximately 50\% larger than the pion gas result (and rapidly increasing beyond that point). In the right panel of Fig.~\ref{fig:Dwidth_he} the contribution of the $N$ and $\Delta$ baryons to the $D$-meson width is shown. The contribution is rather small due to the low density of baryon and anti-baryon states for the considered temperatures (in a gas at vanishing net baryon density), adding about 10\% to the meson gas result. However, due to the large set of excited baryons in the spectrum, additional effects can be expected from those, although empirical constraints are more difficult to obtain. No thermal mass shifts were reported in Ref.~\cite{He:2011yi}.

An effective field theory perspective was used in Ref.~\cite{Ghosh:2013xea} to evaluate the spectral properties of $D$ and $D^*$ mesons from interactions with the light pseudoscalar mesons, using the low-energy interactions from Ref.~\cite{Geng:2010vw}. The heavy-meson thermal self-energy is calculated to one loop level, including interactions mixing $D$ and $D^*$ mesons, and incorporating non-dispersive tadpoles (which only modify the real part of the meson self-energy) and dispersive diagrams (which introduce damping into the heavy-meson propagation). In particular, the analysis of the $D$ and $D^*$ meson self-energies exhibits the contribution of the Landau and unitary cuts. While no momentum-integrated thermal decay width $\Gamma(T)$ is reported in that work, values of $-{\rm Im} \Sigma (p_0,{\bm p})/M$ at $T=150$ MeV for a $D$ meson with momentum $|{\bm p}|=300$ MeV reach up to $15-20$ MeV, depending on the value of the energy $p_0$. This is somehow smaller but still comparable to the works discussed above.

\begin{figure}[ht]
\centering
\includegraphics[scale=0.82]{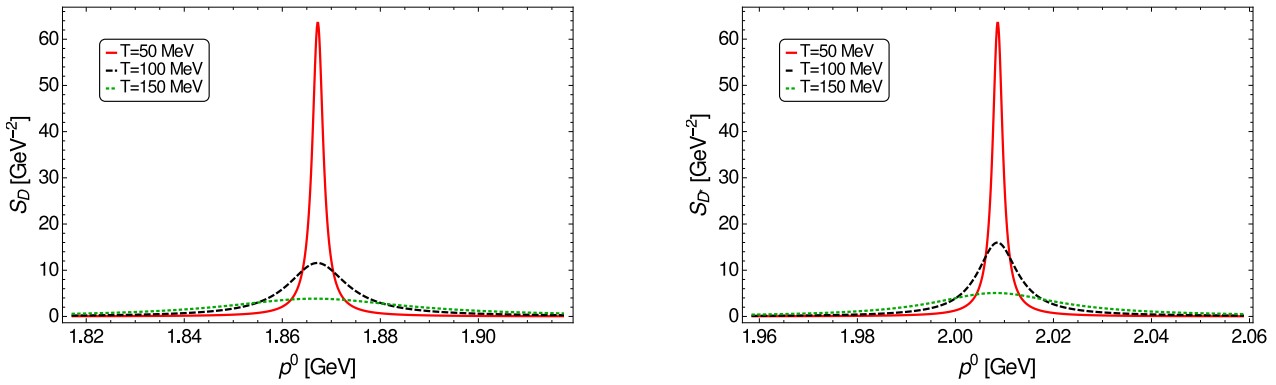}
\includegraphics[scale=1.18]{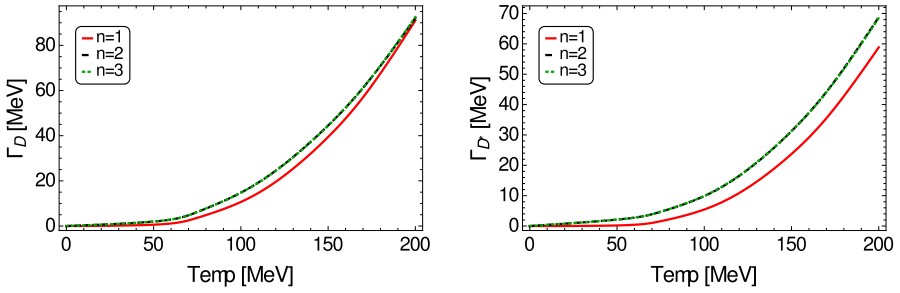}
\caption{In-medium properties of $D$ mesons (left panels) and $D^*$ mesons as functions of the temperature as calculated in Ref.~\cite{Cleven:2017fun}. Top panels: Spectral function of $D$ and $D^*$ mesons at rest ${\bm p}=0$ as functions of the energy $p^0$, for three different temperatures. Lower panels: Thermal decay width of $D$ and $D^*$ mesons as functions of the temperature, for three subsequent iterations in the self-consistent solution of the $T$-matrix equation (see text for details). Figures taken from Ref.~\cite{Cleven:2017fun}.}\label{fig:Dwidth_cleven}
\end{figure}

In Ref.~\cite{Cleven:2017fun} an EFT with unitarized amplitudes was applied to evaluate medium corrections. The interaction with pions is a contact interaction (Weinberg-Tomozawa term) from a chiral Lagrangian with SU(4) flavor extension. The $T$-matrix equation is solved with temperature dependent propagators, and the self-energy is computed via a one-loop diagram including medium dependence as well. The resulting $D$-meson spectral function is iterated in the $T$-matrix equation until convergence is achieved, thereby solving the self-consistency problem. As a simplifying assumption, the thermal mass shift of the $D$ meson is neglected, and only the thermal width from a the gas of pions was reported.
The pertinent spectral function and thermal widths are shown in upper and lower panels, respectively, of Fig.~\ref{fig:Dwidth_cleven}, for both $D$ (left panels) and $D^*$ mesons  (right panels) at vanishing 3-momentum. Both spectral functions exhibit a considerable broadening with temperature. In the lower panel, more quantitative results for the widths are displayed. The different lines correspond to the number, $n$, of iteration steps of the self-consistent set of equations. Self-consistency is reached rather quickly but leads to a noticeable increase of the width. At $T=150$ MeV the latter  reaches $\Gamma \simeq 40$ MeV for $D$-mesons  while  the one for vector mesons is $\Gamma_{D^*} \simeq 30$ MeV. These values are larger than the pion gas results of Refs.~\cite{Fuchs:2004fh,He:2011yi}, but only a small part of the discrepancy originates from the self-consistency. It may be rather attributed to the non-resonant contributions to the interactions provided by the 4-point vertices, as well as the Bose enhancement factor of the pion in the $D\pi$ loop.

\begin{figure}[!ht]
\centering
\includegraphics[scale=0.55]{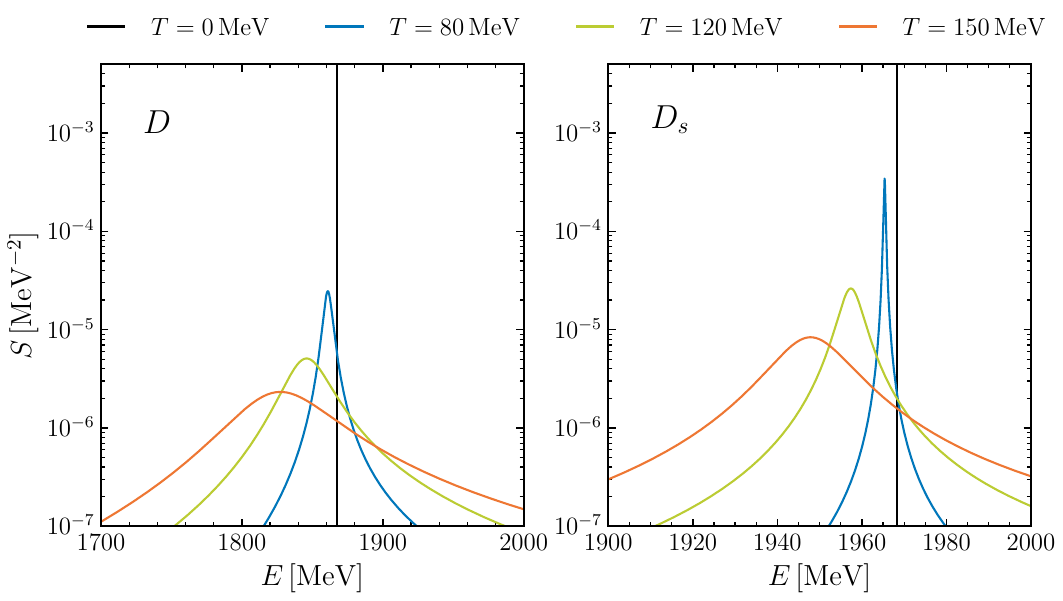}
\includegraphics[scale=0.55]{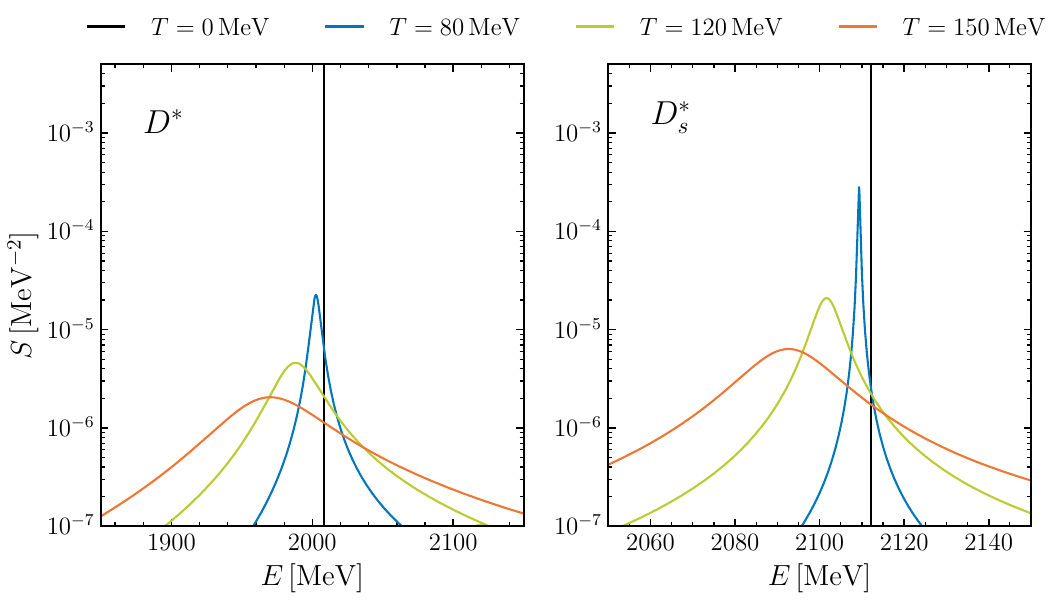}
\caption{Ground state spectral functions in the heavy-light meson sector taken from Refs.~\cite{Montana:2020lfi,Montana:2020vjg}. We present the spectral function of $D$ (top, left), $D_s$ (top, right), $D^*$ (bottom, left) and $D_s^*$ (bottom, right) mesons as functions of the temperature.}\label{fig:Dspec_montana}
\end{figure}
The solution of the self-consistent problem was reviewed and extended in Ref.~\cite{Montana:2020lfi} for the $D$ and $D_s$ mesons, and Ref.~\cite{Montana:2020vjg} for the $D^*$ and $D_s^*$ mesons. The EFT is the one described in Sec.~\ref{ssec:hqet} incorporating a double expansion in chiral and HQ symmetries, both broken by the finite meson masses. The scattering amplitudes are unitarized in medium, via the self-consistent solution of the single- and two-body propagators equations. The thermal bath is augmented with respect to~\cite{Cleven:2017fun} to account for $K$, 
$\bar{K}$ and $\eta$ mesons. 
The pertinent spectral functions are shown in Fig.~\ref{fig:Dspec_montana} for vanishing meson 3-momentum.
\begin{figure}[ht]
\centering
\includegraphics[scale=0.60]{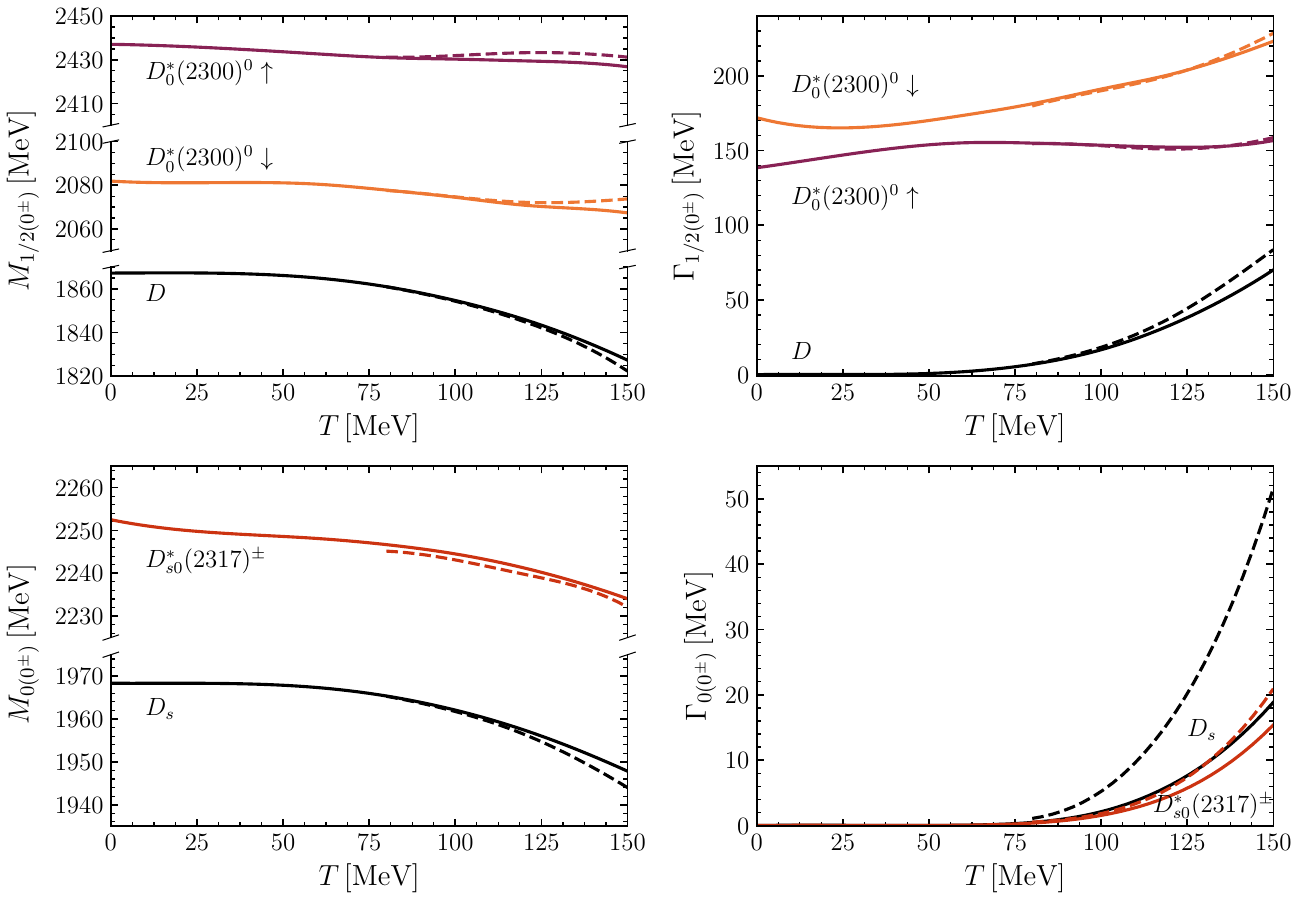}
\caption{Thermal masses (left column) and decay widths (right column) of several charmed meson states from Ref.~\cite{Montana:2020vjg}. Top panel: $D$-meson and the two poles of the $D_0^*(2300)$ state. Bottom panel: $D_s$ and $D_{s0}^* (2317)$ pole.}\label{fig:Dmeson_montana}
\end{figure}
In addition to a thermal broadening (the vacuum widths of the vector mesons are neglected in comparison with the acquired thermal widths) a significant attractive mass shift has been reported. At $T=150$ MeV the mass shifts for the $D,D^*,D_s,D_s^*$ amount to  $\Delta m=$-50 MeV, -48 MeV, -33 MeV, -30 MeV, respectively, as illustrated in the left panels of Fig.~\ref{fig:Dmeson_montana} where also the masses of the different generated states ($D_0^*(2300), D_{s0}^*(2317),D_1^*(2430),D_{s1}^*(2460)$) are displayed as a function of temperature, while in the right panels the thermal width are reproduced from~\cite{Montana:2020vjg}. For the $D$ meson the total width is $\Gamma_D=80$ MeV at $T=150$ MeV ($\Gamma_{D^*}=90$ MeV for the vector state). This value is more than twice as large as in the meson gas calculation of Ref.~\cite{He:2011yi}. Strange $D$ mesons, $D_s$ and $D_s^*$ acquire $\Gamma \simeq 60$ MeV at $T=150$ MeV. 
The reason of this increase has to do with the use of thermal scattering amplitudes in the self-consistent problem and the opening of a new scattering channel usually referred to as Landau cut (as opposed to the standard unitarity cut), which is not present in vacuum which allows for extra channels for the heavy particle to decay into. The effect of these processes is sizable, according to Ref.~\cite{Montana:2020vjg}. These will be further elaborated more in the following section.

In the recent work of Ref.~\cite{Braaten:2023vgs} the $D$ mesons are dressed by thermal pions using an effective theory at LO in the chiral power counting, and NLO in the heavy-meson mass expansion. This is a perturbative approach which does not incorporate resonant interactions. The thermal corrections in the $D$ and $D^*$ self-energies are Very small. At $T=130$ MeV the thermal-mass shift of $D$ ($D^*$) mesons is $\simeq +2.5$ MeV ($\simeq -0.5$ MeV), while the thermal-decay width acquired by these states is $\Gamma \simeq 100$ keV. These value are  much smaller than most of the results discussed above, presumably due to the perturbative character of the approach and the suppressed NLO HQ corrections.

\begin{figure} 
\centering
\includegraphics[scale=0.3]{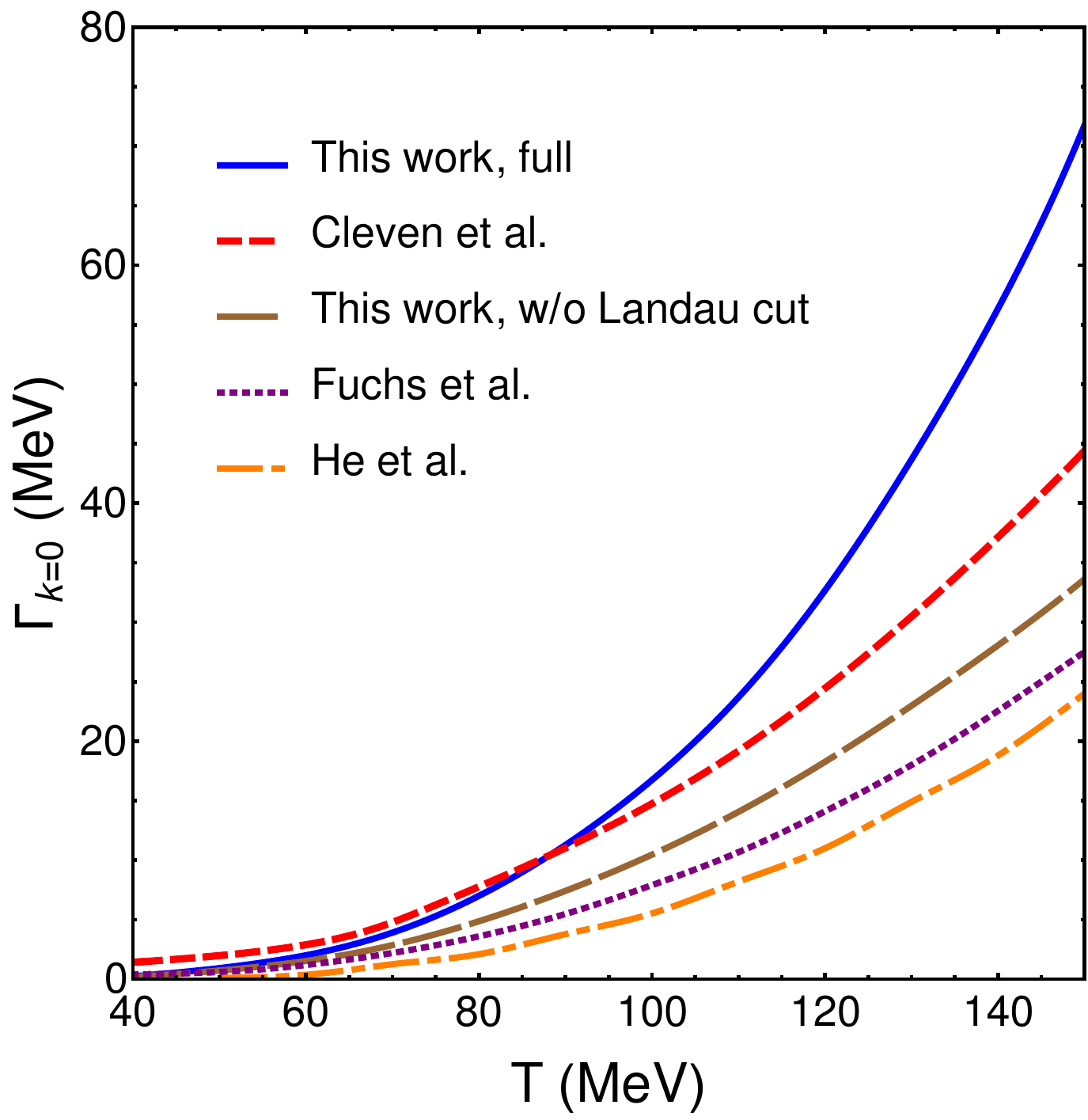}
\caption{Compilation of different results for the thermal decay width of a static ($k \rightarrow 0$) $D$ meson moving in a gas of pions as a function of the temperature. From top to bottom, the different results are taken from Refs.~\cite{Torres-Rincon:2021yga,Cleven:2017fun,Torres-Rincon:2021yga,Fuchs:2004fh,He:2011yi}. Figure taken from Ref.~\cite{Torres-Rincon:2021yga}.}
\label{fig:Dwidth_comparison}
\end{figure}

A summary of the different results for the collisional broadening of $D$ mesons in hot matter is compiled in Fig.~\ref{fig:Dwidth_comparison} as extracted from Ref.~\cite{Torres-Rincon:2021yga}, where for the sake of a consistent comparison  only the interaction with pions have been included.
We observe approximate agreement for works in a first iteration of the many-body problem (that is using vacuum amplitudes and masses) and a moderate increase of the widths if self-consistency is carried out. The results of Ref.~\cite{Torres-Rincon:2021yga} are about a factor two larger for most of the relevant temperature range due to additional thermal processes that originate from in-medium interaction associated with a ``Landau cut'' (whose effect will be detailed in Sec.~\ref{sssec:D-ampl}).

\begin{figure} 
\centering
\includegraphics[scale=0.75]{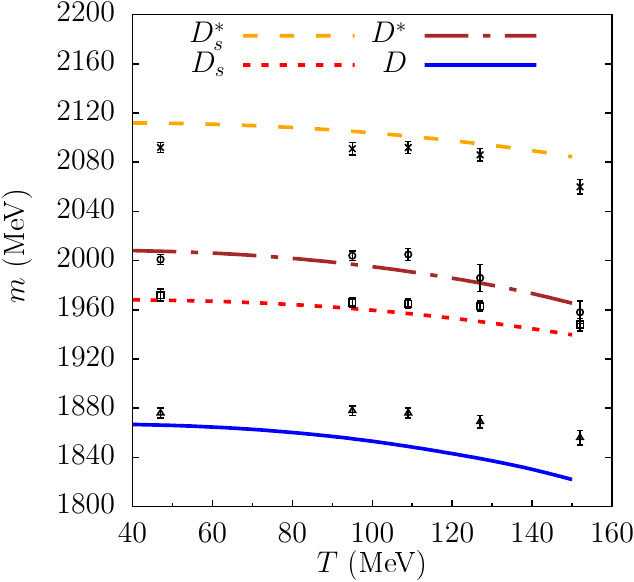}
\caption{Ground state thermal masses in the heavy-light meson sector ($D,D_sD^*D_s^*$) taken from Refs.~\cite{Montana:2020lfi,Montana:2020vjg}. Data points are the result from the LQCD calculation of Ref.~\cite{Aarts:2022krz} using $N_f=2+1$ flavors with vacuum pion mass of $m_\pi=239(1)$ MeV.}\label{fig:DmesonlatticeQCD}
\end{figure}

Finally, in Fig.~\ref{fig:DmesonlatticeQCD} we present LQCD results for the $D, D_s, D^*$ and $D_s^*$ meson masses as functions of the temperature, which where extracted from fits to Euclidean correlators~\cite{Aarts:2022krz}. The simulations implement $N_f=2+1$ dynamical quarks, where the strange quark mass is set to the physical point, but the light sector uses a pion mass of $m_\pi=239(1)$ MeV, at the smallest temperature. The temperature range spans $T=47-169$ MeV. This study found no discernible temperature dependence of any of the four states until $T=127$ MeV, and a reduction of the thermal masses, of around 20 to 40 MeV at $T=152$ MeV. In Fig.~\ref{fig:DmesonlatticeQCD} the LQCD results are plotted together with the calculations of Refs.~\cite{Montana:2020lfi,Montana:2020vjg}. Despite the quantitative difference of the results, presumably due to a pion mass that is slightly above the physical value in the LQCD simulations, the qualitative trend is rather similar.

%%%%%%%%%%%%%%%%%%%%%%%%%%%%%%%%%%%%%%%%%%%%%%%%%%
\subsubsection{Finite Chemical Potential}
\label{sssec:chem}
%%%%%%%%%%%%%%%%%%%%%%%%%%%%%%%%%%%%%%%%%%%%%%%%%%%
Even though this review is mostly dedicated to finite temperature effects we also give a brief account of in-medium effects at finite chemical potential, but refer to dedicated reviews of charm properties in nuclear matter, in, \eg, Refs.~\cite{Tolos:2013gta,Hosaka:2016ypm}.

Initial works found a significant mass shift effect of the $D$ meson in dense matter using a chiral effective hadronic model~\cite{Mishra:2003se} or QCD sum rules~\cite{Hayashigaki:2000es}.
Solving the self-consistent problem in which the dense medium affects the intermediate propagators can be addressed in the of the $G$-matrix formalism, which also accounts for Pauli-blocking which is particularly important at small temperatures and small nucleon energies. In Ref.~\cite{Tolos:2004yg} a separable, momentum-independent potential between mesons and baryons including coupled channels was employed for nuclear densities up to $\rho=1.5 \rho_0$ (where $\rho_0=0.16$\,fm${^{-3}}$ is  the nuclear saturation density). In Ref.~\cite{Tolos:2005ft} finite temperature effects up to $T=120$ MeV were considered. The $DN$ interaction gives rise to the generation of the $\Lambda_c (2593)$ resonance, and the spectral function of the $D$ meson at finite $T$ and $\rho$ features only a small mass shift ($\Delta m_D \simeq  10$ MeV) but a considerable broadening of $\Gamma = 50-150$ MeV.

Different interactions based on SU(4) symmetry were used in Ref.~\cite{Lutz:2005vx} to analyze both $D$ and $D_s$ states in medium;
a two-mode structure of the $D^+$ spectral function was found while the $D_s$ developed a resonance-hole loop which gives rise to a rather large broadening with strong 3-momentum dependence. Later, in Ref.~\cite{Mizutani:2006vq} based in $t$-channel (Weinberg-Tomozawa) vector meson exchange and subsequent unitarization, the two-mode structure for the $D$-meson was confirmed with an upwards shift of $20$ MeV at $\rho=\rho_0$ and $T=0$ in the $D$ meson quasiparticle mass and  a lower branch at about 200 MeV smaller energy. Additional temperature effects were introduced in Ref.~\cite{Tolos:2007vh}

The incorporation of HQSS in the interaction of $D$ and $D^*$ mesons at finite density was carried out in Ref.~\cite{Tolos:2009nn} leading to a SU(8) symmetry~\cite{Garcia-Recio:2008rjt}. Several coupled channels give rise to a number of dynamically generated $\Lambda_c$ and $\Sigma_c$ resonances. The in-medium results show a net attractive (repulsive) interaction for $D$ ($D^*$) meson producing a mass shift at $\rho=\rho_0$ of $\sim -15$ MeV (+20 MeV), plus a considerable broadening in both spectral functions with a rich structure of peaks due to the coupling of the $D,D^*$ mesons with various $\Lambda_c N^{-1}$ and $\Sigma_c N^{-1}$ states in the medium.

Finally, alternative approaches have been applied, for example a quark-based description in the Nambu-Jona--Lasinio model in Ref.~\cite{Blaschke:2011yv}, where the masses of $D^\pm$ mesons has been computed at finite temperature and density. The $D^+$ mass  slightly decreases at finite temperature until $\rho \simeq \rho_0$ and then increases. The $D^-$ mass increases due to a stronger Pauli blocking effect of the nucleons. The in-medium widths also become rather large, but in this case the decay process is into the quark constituent, since the NJL model does not feature confinement.

%%%%%%%%%%%%%%%%%%%%%%%%%%%%%%%%%%%%%%%%%%%%%%%%%%%%
\subsubsection{Thermal scattering amplitudes}
\label{sssec:D-ampl}
%%%%%%%%%%%%%%%%%%%%%%%%%%%%%%%%%%%%%%%%%%%%%%%%%%%%%%
The first step in evaluating in-medium properties is to utilize their vacuum interactions with a surrounding medium to evaluate their 1-body properties (self-energies and spectral function). In a self-consistent set up the in-medium propagators of the $D$ mesons propagators (and in principle also those of the light hadrons) give rise to modifications of their scattering amplitudes. The latter are the focus of this section.

At finite temperature the in-medium $T$-matrix equation was solved self-consistently in Refs.~\cite{Cleven:2017fun} and \cite{Montana:2020lfi}, using the above-mentioned EFT approaches. In Ref.~\cite{Torres-Rincon:2021yga} the different genuine finite-temperature effects were analyzed, and it was shown that in-medium interactions produce a substantial increase of the drag and momentum diffusion coefficients, as will be discussed in more detail in Sec.~\ref{ssec:Dtrans}. The genuine effects to be detailed in following consist of (i) off-shell propagation of particles, and (ii) opening of new scattering processes due to the Landau cut, being the latter the one responsible for the substantial increase of the scattering rate. 

\begin{figure}[ht]
\centering
\includegraphics[scale=0.65]{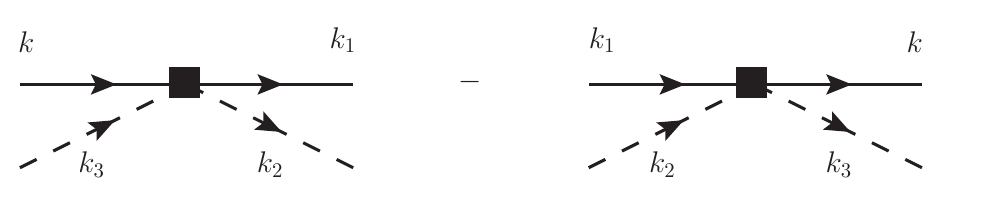}
\caption{Scattering diagrams corresponding to Eq.~\eqref{eq:GammaVacuum} showing the contribution to the decay width of a $D$ meson with momentum $k$ by collisions with light particles in the thermal medium. The $T$-matrix element (shown by the filled-square symbol) is computed in vacuum.}
\label{fig:gammavacuum}
\end{figure}

Let us start by revisiting the calculation of the $D$-meson decay width $\Gamma_k$ (transport parameters like friction or momentum diffusion coefficient carry a similar structure, with different momentum weights inside the integrals).
A more detailed study will be published elsewhere~\cite{gloria}. For a $D$ meson with energy and momentum $E_k,{\bm k}$ colliding with a light particle, say  $\pi$, from the medium, we label the scattering as $D(k),\pi(k_3) \leftrightarrow D(k_1),\pi(k_2)$. For a vacuum $T-$matrix the expression for the decay width of the $D$-meson with momentum $k$ has the standard form,
\begin{align}
\Gamma_{k}^{(\textrm{vac-$T$})}  & = \frac{1}{2E_k^D} \int \prod_{i=1}^3 
\frac{d^3 k_i }{(2\pi)^3} \frac{1}{2E^\pi_2} \frac{1}{2E^\pi_3} |T (E^D_k+E^\pi_3,\bm{k}+\bm{k}_3)|^2 (2\pi)^4 \delta^{(3)} ( \bm{k}+\bm{k}_3-\bm{k}_1-\bm{k}_2)
\nonumber \\
&\times \delta(E^D_k+E^\pi_3-E^D_1-E^\pi_2)  \left[ \tilde{f}^{(0)} (E^D_1) f^{(0)} (E^\pi_3)  \tilde{f}^{(0)} (E^\pi_2) -
f^{(0)} (E^D_1) f^{(0)} (E^\pi_2)  \tilde{f}^{(0)} (E^\pi_3)   \right]  \ , \label{eq:GammaVacuum} 
\end{align}
where $f^{(0)}$ are the equilibrium Bose-Einstein distribution function and $\tilde{f}^{(0)}=1+f^{(0)}$.
The interpretation of Eq.~(\ref{eq:GammaVacuum}) is standard: the first term represent an incoming pion with momentum (energy) of ${\bm k}_3 (E_3^\pi)$, carrying with it its pertinent Bose factor, $f^{(0)}$, 
while the outgoing $D$ and $\pi$ are subject to Bose enhancement factors (this scattering has a positive contribution to the decay width, since the $D$ meson with momentum $k$ disappears). In addition, one has the reverse binary reaction producing a final $D$ meson with energy $E_k^D$ out of a $D$ meson existing in the bath. This reverse process contributes with a negative sign to the decay width, since the $D$-meson with momentum $k$ is produced, rather than absorbed, in the reaction; however, it is typically  much suppressed due to the scarcity of heavy particles in the thermal bath. A pictorial representation is given in Fig.~\ref{fig:gammavacuum}.

A similar formula can be calculated when the interaction is self-consistently treated in medium. In Ref.~\cite{Torres-Rincon:2021yga} it was derived from the calculation of the imaginary part of the $D-$meson self-energy using the imaginary time (Matsubara) formalism (ITF), and alternatively, using the ``greater'' self-energy from the Kadanoff-Baym calculation for the case of an equilibrated thermal bath. Both methods provide the same result, which reads~\cite{Torres-Rincon:2021yga},
\begin{equation}
   \Gamma_{k}^{(\textrm{therm-$T$})} = \Gamma_k^{(U)} + \Gamma_k^{(L)} 
\end{equation}
with
\begin{align}
\Gamma_k^{(U)} & = \sum_{\lambda=\pm }\frac{\lambda}{2E^D_k} \frac{1}{\tilde{f}^{(0)}_k} \int dk_1^{0} \int \prod_{i=1}^3 \frac{d^3 k_i }{(2\pi)^3} \frac{1}{2E^\pi_2} \frac{1}{2E^\pi_3} |T (E^D_k+E^\pi_3,\bm{k}+\bm{k}_3)|^2 S_D(k^{0}_1,\bm{k}_1) \nonumber \\
&\times (2\pi)^4 \delta^{(3)} ( \bm{k}+\bm{k_}3-\bm{k}_1-\bm{k}_2) \delta(E^D_k+E^\pi_3-k^{0}_1- \lambda E^\pi_2) \tilde{f}^{(0)} (k^{0}_1) f^{(0)} (E^\pi_3)  \tilde{f}^{(0)} (\lambda E^\pi_2) \ ,      
\label{eq:gammaU} 
\end{align}
and
\begin{align}
\Gamma_k^{(L)} & = \sum_{\lambda=\pm} \frac{\lambda}{2E^D_k} \frac{1}{\tilde{f}^{(0)}_k}  \int dk_1^{0} \int \prod_{i=1}^3 \frac{d^3 k_i }{(2\pi)^3} \frac{1}{2E^\pi_2} \frac{1}{2E^\pi_3} |T (E^D_k-E^\pi_3,\bm{k}+\bm{k}_3)|^2 S_D(k_1^{0},\bm{k}_1) \nonumber \\
&\times (2\pi)^4 \delta^{(3)} ( \bm{k}+\bm{k}_3-\bm{k}_1-\bm{k}_2) \delta(E^D_k-E^\pi_3-k_1^{0}-\lambda E^\pi_2) \tilde{f}^{(0)} (k^{0}_1) \tilde{f}^{(0)} (E^\pi_3)  \tilde{f}^{(0)} (\lambda E^\pi_2)   \  \label{eq:gammaL} \ ,
\end{align}
where $S_D(k_1^{0},{\bm k}_1)$ is the spectral function of the $D$ meson in the bath, which is broadened due to its own interactions (light particles can also be off-shell, but this effect was neglected in Ref.~\cite{Torres-Rincon:2021yga}); consequently, the energy of the $D$ meson with momentum $k$ does not necessarily be on-shell, but we still denote it as $E_k$.

\begin{figure}[ht]
\centering
\includegraphics[scale=0.7]{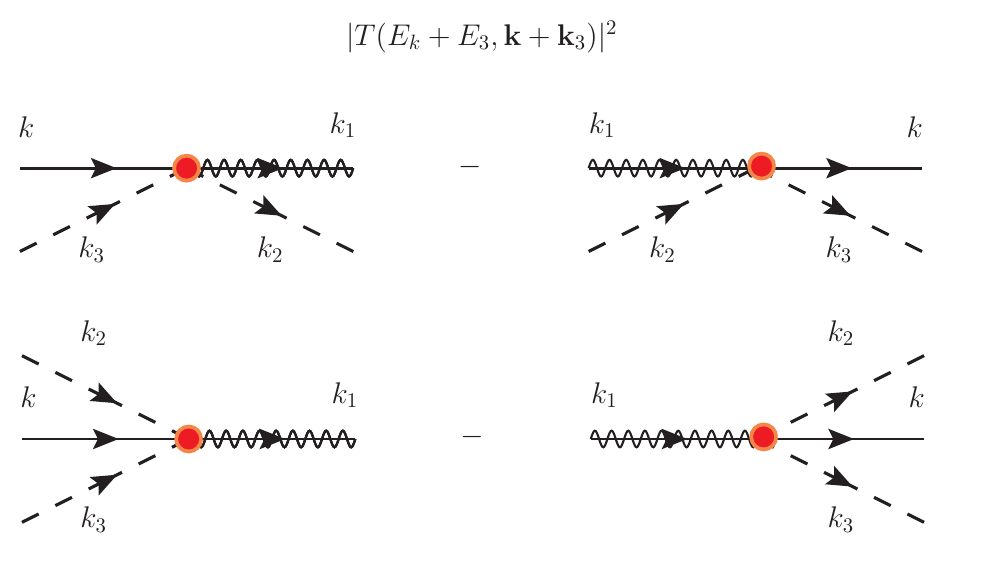}
\caption{Pictorial representation of Eq.~(\ref{eq:gammaUv2}) showing the contribution to the decay width of a $D$ meson with momentum $k$ by collisions with light particles in the thermal medium via $2 \leftrightarrow 2$ (upper diagrams) and $1\leftrightarrow 3$ processes (lower diagrams).}\label{fig:gammaU}
\end{figure}

The expressions Eq.~\eqref{eq:gammaU} and Eq.~\eqref{eq:gammaL}, shown in Ref.~\cite{Torres-Rincon:2021yga} do not have a direct interpretation. But one can make algebraic manipulations, using standard relations between the equilibrium distribution function to write it in an alternative form~\cite{gloria}. Eq.~\eqref{eq:gammaU} can be identically rewritten as
\begin{align}
\Gamma_k^{(U)} & = \frac{1}{2E_k} \int dk_1^0 \int \prod_{i=1}^3 \frac{d^3 k_i }{(2\pi)^3} \frac{1}{2E_2} \frac{1}{2E_3} |T (E_k+E_3,\bm{k}+\bm{k}_3)|^2 (2\pi)^4 \delta^{(3)} ( \bm{k}+\bm{k}_3-\bm{k}_1-\bm{k}_2) \nonumber \\
&\times \delta(E_k+E_3-k_1^0-E_2) S_D(k_1^0,\bm{k}_1) \left\{ \tilde{f}^{(0)} (k^0_1) f^{(0)} (E_3)  \tilde{f}^{(0)} (E_2) -
f^{(0)} (k^0_1) f^{(0)} (E_2)  \tilde{f}^{(0)} (E_3)   \right\}  \nonumber \\
& + \frac{1}{2E_k}  \int dk_1^0 \int \prod_{i=1}^3 \frac{d^3 k_i }{(2\pi)^3} \frac{1}{2E_2} \frac{1}{2E_3} |T (E_k+E_3,\bm{k}+\bm{k}_3)|^2  (2\pi)^4 \delta^{(3)} ( \bm{k}+\bm{k}_3-\bm{k}_1+\bm{k}_2) \nonumber \\
& \times \delta(E_k+E_3-k_1^0+E_2)S_D(k_1^0,\bm{k}_1)    \left\{ \tilde{f}^{(0)} (k^0_1) f^{(0)} (E_3)  f^{(0)} (E_2) -
f^{(0)} (k^0_1) \tilde{f}^{(0)} (E_2)  \tilde{f}^{(0)} (E_3)   \right\}  \ ,    \label{eq:gammaUv2} 
\end{align}
with a similar interpretation as before: the first term of Eq.~\eqref{eq:gammaUv2} is very similar to the expression in Eq.~\eqref{eq:GammaVacuum} and has the same interpretation. The difference is that the $D$ meson with ${\bm k}_1$ is off-shell, and its energy is integrated and weighted by the spectral function $S_D(k_1^0,{\bm k}_1)$. 
The difference of the second term in Eq.~\eqref{eq:gammaUv2}, as compared to the first one, is that the pion with ${\bm k}_2$ is changed from outgoing to an incoming particle in the direct term, and vice versa in the reverse one. Diagrammatically, this is understood as a $1 \leftrightarrow 3$ process which contributes to the decay width. This process is evidently only allowed because of the off shell nature of the $D$ meson, while in the practice it results in a rather small contribution~\cite{Torres-Rincon:2021yga}. The diagrams representing the terms in Eq.~\eqref{eq:gammaUv2} are depicted in Fig.~\ref{fig:gammaU}m where the solid-wiggled line represents an off-shell $D$ meson, whose quasi-particle energy should eventually be integrated over.

\begin{figure}[ht]
\centering
\includegraphics[scale=0.7]{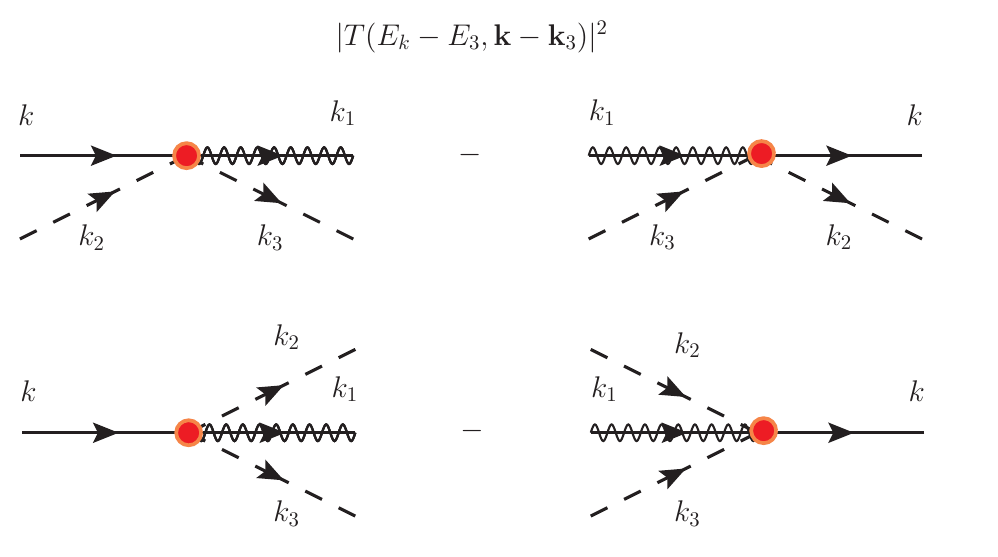}
\caption{Pictorial representation of Eq.~(\ref{eq:gammaLv2}) showing the contribution to the decay width of a $D$ meson with momentum $k$ by collisions with light particles in the thermal medium via $2 \leftrightarrow 2$ (upper diagrams) and $1\leftrightarrow 3$ processes (lower diagrams).}
\label{fig:gammaL}
\end{figure}

Similarly, Eq.~\eqref{eq:gammaL} can be identically rewritten as,
\begin{align} 
\Gamma_k^{(L)} & = \frac{1}{2E^D_k} \int dk_1^0 \int \prod_{i=1}^3 \frac{d^3 k_i }{(2\pi)^3} \frac{1}{2E^\pi_2} \frac{1}{2E^\pi_3} |T (E^D_k-E^\pi_3,\bm{k}-\bm{k}_3)|^2 (2\pi)^4 \delta^{(3)} ( \bm{k}-\bm{k}_3-\bm{k}_1+\bm{k}_2)  
\nonumber \\
& \times \delta(E^D_k-E^\pi_3-k_1^0+E^\pi_2) S_D(k_1^0,\bm{k}_1) \left\{ \tilde{f}^{(0)} (k^0_1) \tilde{f}^{(0)} (E^\pi_3)  f^{(0)} (E^\pi_2) - f^{(0)} (k^0_1) f^{(0)} (E^\pi_3)  \tilde{f}^{(0)} (E^\pi_2) \right\} 
\nonumber \\
+ &  \frac{1}{2E^D_k}  \int dk_1^0 \int \prod_{i=1}^3 \frac{d^3 k_i }{(2\pi)^3} \frac{1}{2E_2} \frac{1}{2E^\pi_3} |T (E^D_k-E^\pi_3,\bm{k}-\bm{k}_3)|^2 (2\pi)^4 \delta^{(3)} ( \bm{k}-\bm{k}_3-\bm{k}_1-\bm{k}_2) 
\nonumber \\
& \times \delta(E^D_k-E^\pi_3-k_1^0-E^\pi_2) S_D(k_1^0,\bm{k}_1) \left\{ \tilde{f}^{(0)} (k^0_1) \tilde{f}^{(0)} (E^\pi_3)  \tilde{f}^{(0)} (E^\pi_2) - f^{(0)} (k^0_1) f^{(0)} (E^\pi_3)  f^{(0)} (E^\pi_2) \right\}   \ . \label{eq:gammaLv2}
\end{align}
Following the energy-conserving delta-functions, as well as the statistical factors, one can interpret the two terms in a similar way than before. The first term represents a $2 \leftrightarrow 2$ scattering, while the second term is a $1 \leftrightarrow 3$ process, very much suppressed due to phase space and the narrow spectral function of the $D$ meson. The two processes (and their reversed) are plotted in Fig.~\ref{fig:gammaL}. The main difference with respect to the contribution of Eq.~\eqref{eq:gammaUv2} is that the $T$-matrix is evaluated at $(E_k^D-E_3^\pi, \bm{k}-\bm{k}_3)$, which is why we refer to it as the ``Landau cut'' contribution. In this kinematic range the scattering amplitude is nonzero only at finite temperature. Therefore the entire contribution of Eq.~\eqref{eq:gammaLv2} vanishes when the amplitude is taken in vacuum, but survives at finite temperature. 
Since the $2 \leftrightarrow 2$ scattering is not suppressed by phase space, it can have an important contribution to the decay width (and to the transport coefficients) which was quantified in Ref.~\cite{Torres-Rincon:2021yga}. It should be noted that the magnitude of these Landau cut contributions is sensitive to the way the vacuum terms are subtracted out of the purely thermal corrections. While in Ref.~\cite{Torres-Rincon:2021yga} it was checked that the Landau cut contribution vanishes at $T=0$ (as it should) there could be an ambiguity in the renormalization procedure, where a different choice of vacuum terms leads in a different result. This effect---which is likely the origin of the substantial increase of transport coefficients within this self-consistent approximation~\cite{Torres-Rincon:2021yga}---calls for a more detailed study which will be carried out in the future~\cite{gloria}.

For completeness, we briefly comment on various results at finite baryon density in relation to the scattering amplitudes. The effects of density in the scattering amplitude were considered in Ref.~\cite{Tolos:2004yg}, where the $DN$ amplitude is shown at nuclear saturation density and, more recently, in Ref.~\cite{Tolos:2009nn}, where the $DN \rightarrow DN$ and $D^*N \rightarrow D^*N$ amplitudes are shown as functions of baryon density. The dynamically generated states $\Lambda_c$ and $\Sigma_c$ shift only slightly since the increase due to Pauli blocking effect is partially compensated by the self-consistent inclusion of in-medium $D$ and $D^*$ meson self-energies.

%%%%%%%%%%%%%%%%%%%%%%%%%%%%%%%%%%%%%%%
\subsection{$D$-meson transport coefficients}
\label{ssec:Dtrans}
%%%%%%%%%%%%%%%%%%%%%%%%%%%%%%%%%%%%
In this section we will conduct quantitative comparisons of existing transport coefficients for $D$ mesons in hot hadron matter, aided by the detailed discussions of the underlying scattering amplitudes of the the preceding section. Specifically, we will address the momentum-space drag and diffusion coefficients in Sec.~\ref{sssec:Ddrag}, and the spatial diffusion coefficient in Sec.~\ref{sssec:spatial}. We will briefly allude to the effects of a finite baryon chemical potential in Sec.~\ref{sssec:finite-mu}.

%%%%%%%%%%%%%%%%%%%
\subsubsection{Drag force and momentum diffusion coefficients}
\label{sssec:Ddrag}
%%%%%%%%%%%%%%%%%%%%%%
The drag and momentum diffusion coefficients are key inputs to Langevin simulations as widely used in applications to RHIC phenomenology of HF transport, albeit mostly in the partonic phase. Here we focus on $D$-meson diffusion based on the interactions reviewed in Sec.~\ref{sec:charm}. The basic formalism to compute these coefficients is recapitulated in Appendix~\ref{app:DTR}. All approaches usually compute both the drag and diffusion coefficient. However, the two transport coefficients are not independent, but they are related through the fluctuation-dissipation theorem (FDT), expressed as $B(p) = T E_pA(p)$, where $A(p)$ and $B(p)$ represent the 3-momentum dependent drag and diffusion coefficients, respectively, and $E_p$ is the relativistic heavy-hadron energy.  
In practice, this relation is not always accurately satisfied, especially at high momentum, and one often rather uses the FDT to express $B(p)$ in terms of $A(p)$ (which has been found to be more reliable than vice versa)~\cite{Rapp:2018qla}. However, for small $p$ the FDT is usually well satisfied. For the sake of concise comparisons, we will therefore concentrate our discussion on the drag coefficient, at small momenta.

\begin{figure}[t] 
\centering
\includegraphics[scale=0.35]{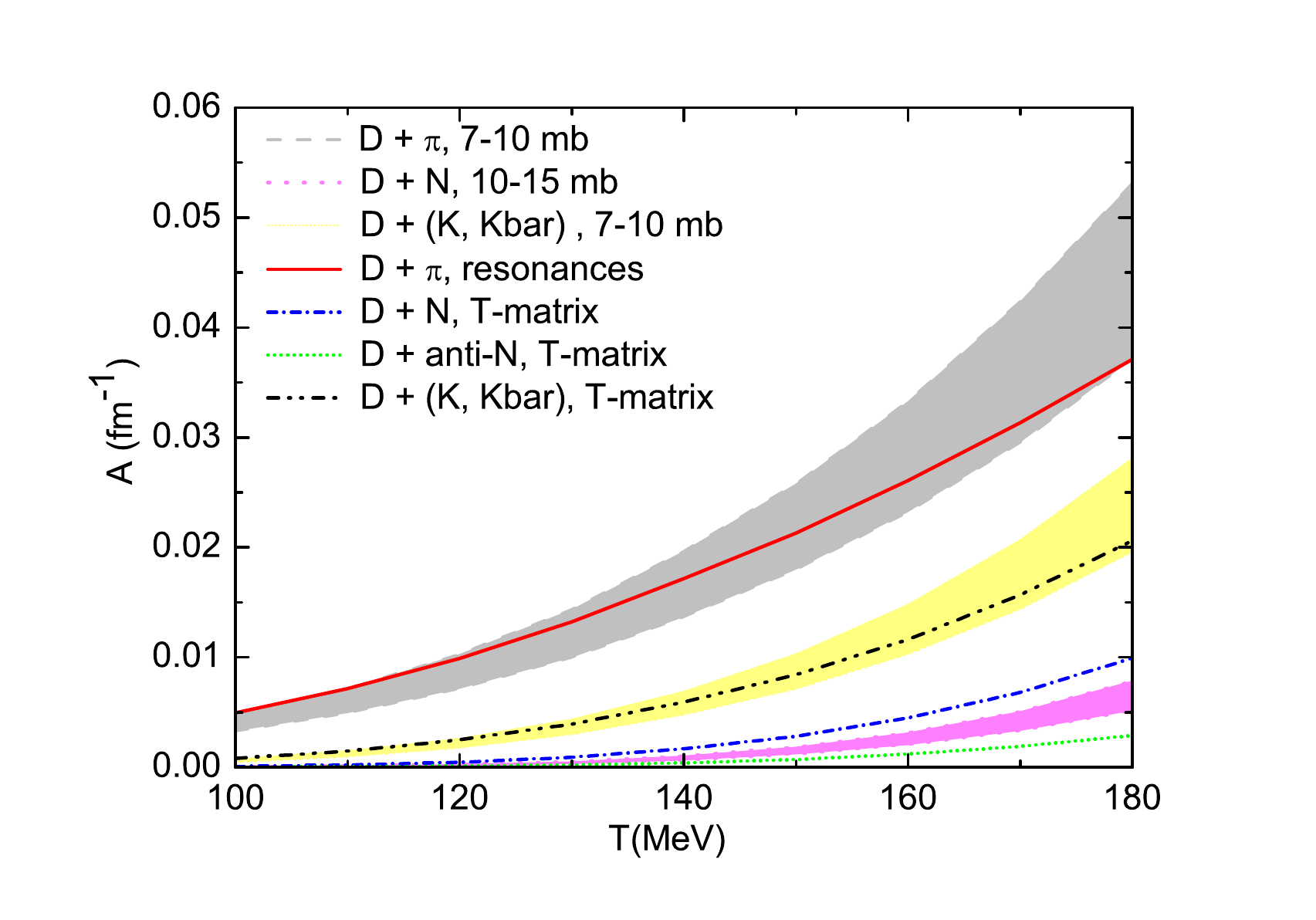}
\caption{The drag coefficient for $D$ mesons at momentum
$p$=0.1 GeV in a gas of pions (solid line), anti-/kaons (dash-double-dotted line), nucleons (dash-dotted line) and antinucleons (dotted line), as a function of temperature at vanishing baryon chemical potential using the scattering amplitudes of constructed in Ref.~\cite{He:2011yi}, where also the bands representing results obtained
with constant S$-$wave $D\pi$, $DK$ and $DN$ cross sections are shown. Figure adapted from Ref.~\cite{He:2011yi}.}
\label{fig:A_D-tamu}
\end{figure}
Let us start with the initial results for a pion gas from Ref.~\cite{Laine:2011is}, employing LO heavy-meson effective theory. They feature a rapid increase of the drag coefficient starting from rather low temperatures, growing as $T^6$.  As a consequence, it reaches rather large values already at temperatures well below $T=100$ MeV. As discussed above, the validity of these results is limited in temperature, probably below $T=50$ MeV. 

In the subsequent work~\cite{He:2011yi}, the drag and diffusion coefficients of $D$ mesons were determined through Breit-Wigner parameterized interactions with light mesons and baryons, as discussed in Sec.~\ref{ssssec:hrg}. The temperature dependence of the drag coefficient for a $D$-meson of momentum $p$ =0.1 GeV in a gas of pions (solid line), anti-/kaons (dash-double-dotted line), nucleons (dash-dotted line) and antinucleons (dotted line) using elastic scattering amplitudes is shown in the Fig.~\ref{fig:A_D-tamu} at vanishing chemical potentials. The largest contribution stems from pions with a temperature dependence that approximately scales with $T^3$. The authors also set up a schematic scenario where the $D\pi$ amplitude was replaced with a constant $S$-wave cross section, and found that a range of  $\sigma_{D\pi}^S=7-10$ mb can essentially recover the outcome of the microscopic calculations, although with a slightly stronger temperature dependence. To improve the estimation towards a more realistic hadron-resonance gas model, they additionally implemented $D$-meson scattering off $K$, $\eta$, $\rho$, $\omega$ and $K^*(892)$ in the meson sector and anti-/nucleons plus anti-/$\Delta$'s in the baryon sector. After pions, the other pseudo-Goldstone bosons, anti-/kaons and $\eta$'s, contribute the next-largest contributions, as much as 60\% of the $\pi$ contribution at the highest $T$. Following them are vector mesons (not shown), anti-/nucleons and anti-/$\Delta$'s, each at up to 35\% of the pion part. Also for these contributions, schematic cross sections following from the pion case using constituent-quark scaling, give a reasonable representation of the microscopic results.

In Ref.~\cite{Ghosh:2011bw}, the drag and diffusion coefficients of the $D$ meson in a hot hadronic medium comprised of pions, nucleons, kaons, and $\eta$'s, have been assessed within the framework of EFT. They evaluated the scattering amplitudes of the $D$ meson with the pseudo-Goldstone bosons ($\pi$, $K$ and $\eta$) through a covariant formulation of chiral perturbation theory, where the leading term incorporates $D^\ast$-meson exchanges along with a contact interaction~\cite{Geng:2010vw}. The $DN$ scattering amplitudes are determined by considering exchanges involving $\Lambda_c$ and $\Sigma_c$, utilizing the Lagrangian described in Ref.~\cite{Liu:2002vw}. Form factors are introduced at each interaction vertex to account for the finite size of the hadrons involved. In this calculation, the pion contribution clearly dominates, with the drag coefficient at, \eg, $T$=140\,MeV being about $\sim$40\%  larger than the one reported in Ref.~\cite{He:2011yi}, see, Fig.~\ref{fig:A_D-tamu}. The calculations done in Ref.~\cite{Ghosh:2011bw} do not incorporate unitarity, which could be a key reason for this discrepancy. On the other hand, the contributions of the other considered hadron species tend to be (significantly) smaller. 

In Ref.~\cite{Abreu:2011ic}, the drag and diffusion coefficients of the $D$ meson in a hot pion gas have been evaluated within an EFT framework which combines HQET to account for the $D$ meson, with $\chi$PT to describe the dynamics of pions. 
The authors utilize unitarization techniques to extend their calculations to higher temperatures and obtain a resonant cross section for the $D\pi$ scattering. 
%The amplitudes are unitarized to restore the unitarity of the S-matrix, which was compromised due to the truncation of the perturbative expansion. 
The magnitude of the drag coefficient obtained within this calculation is quite comparable to drag coefficient of $D$ mesons in a pion gas reported in Ref.~\cite{He:2011yi}.

\begin{figure} 
\centering
\includegraphics[scale=0.48]{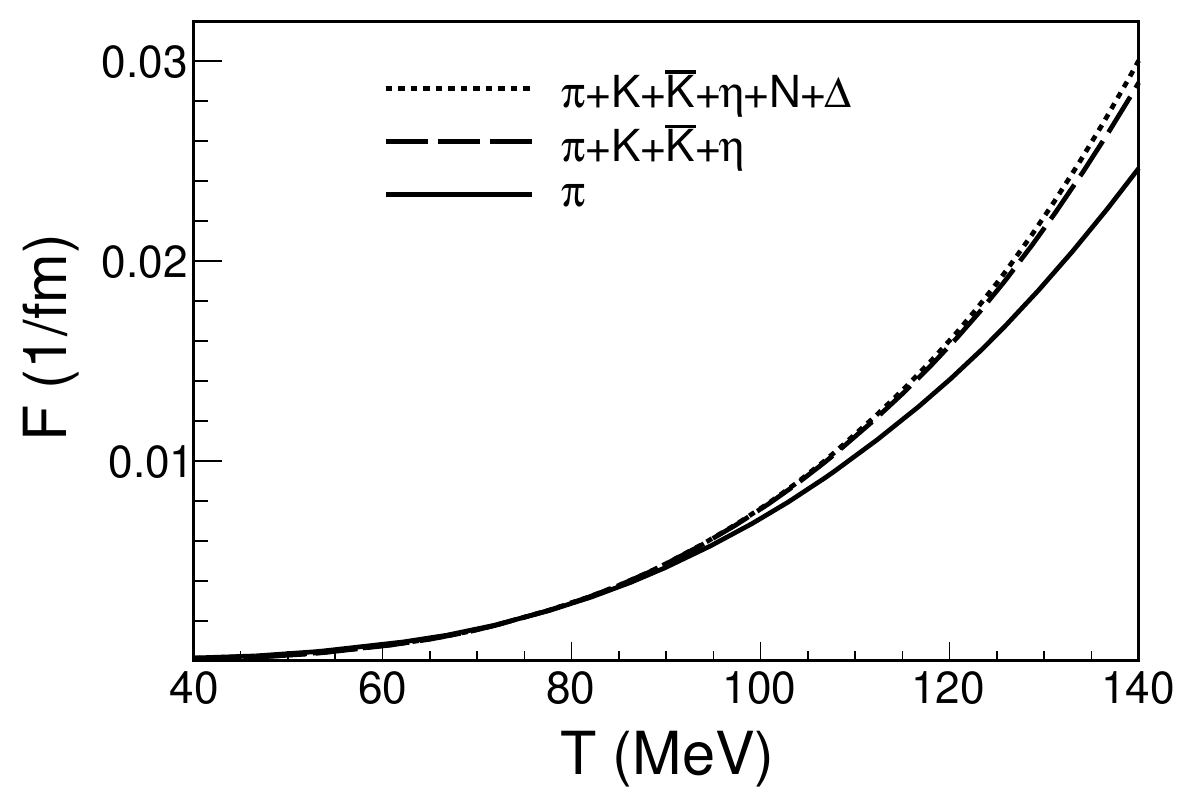}
\caption{Drag coefficient (denoted as $F$ in the plot instead of $A$) of a static $D$ meson as
a function of the temperature in a hadronic bath of light mesons, nucleons and $\Delta$ baryons~\cite{Tolos:2013kva}. The calculation uses scattering amplitudes obtained from a unitarization procedure incorporating potentials from EFTs that incorporate both chiral and HQ symmetries. Figure adapted from Ref.~\cite{Tolos:2013kva}.}
\label{fig:DDjl2}
\end{figure}
In Ref.~\cite{Tolos:2013kva}, the authors have extended the calculations of Ref.~\cite{Abreu:2011ic} for a pion gas to include the next relevant degrees of freedom, namely the complete set of pseudo-Goldstone bosons, kaons and $\eta$'s, along with baryonic degrees of freedom, \ie, nucleons and $\Delta$ baryons. All scattering amplitudes are derived from effective Lagrangians that adhere to chiral and HQ spin symmetries. Moreover, the amplitudes are subjected to unitarization to reestablish the unitarity of the {\cal S}-matrix. This technique helps to prevent the emergence of unphysical amplitudes and ensures a well-controlled energy dependence, which is crucial for describing the transport coefficients. In Fig.~\ref{fig:DDjl2} the variation of the drag coefficient with temperature is shown for different compositions of the hadronic heat bath. As evidenced by the plot, the interaction of $D$ mesons with pions dominates up to temperatures of at least $T=140$ MeV. Introducing kaons, antikaons and $\eta$ mesons slightly enhances the drag coefficient. However, the contribution from baryons to the drag coefficient is negligible at vanishing chemical potential.

\begin{figure}[t] 
\centering
\includegraphics[scale=0.35]{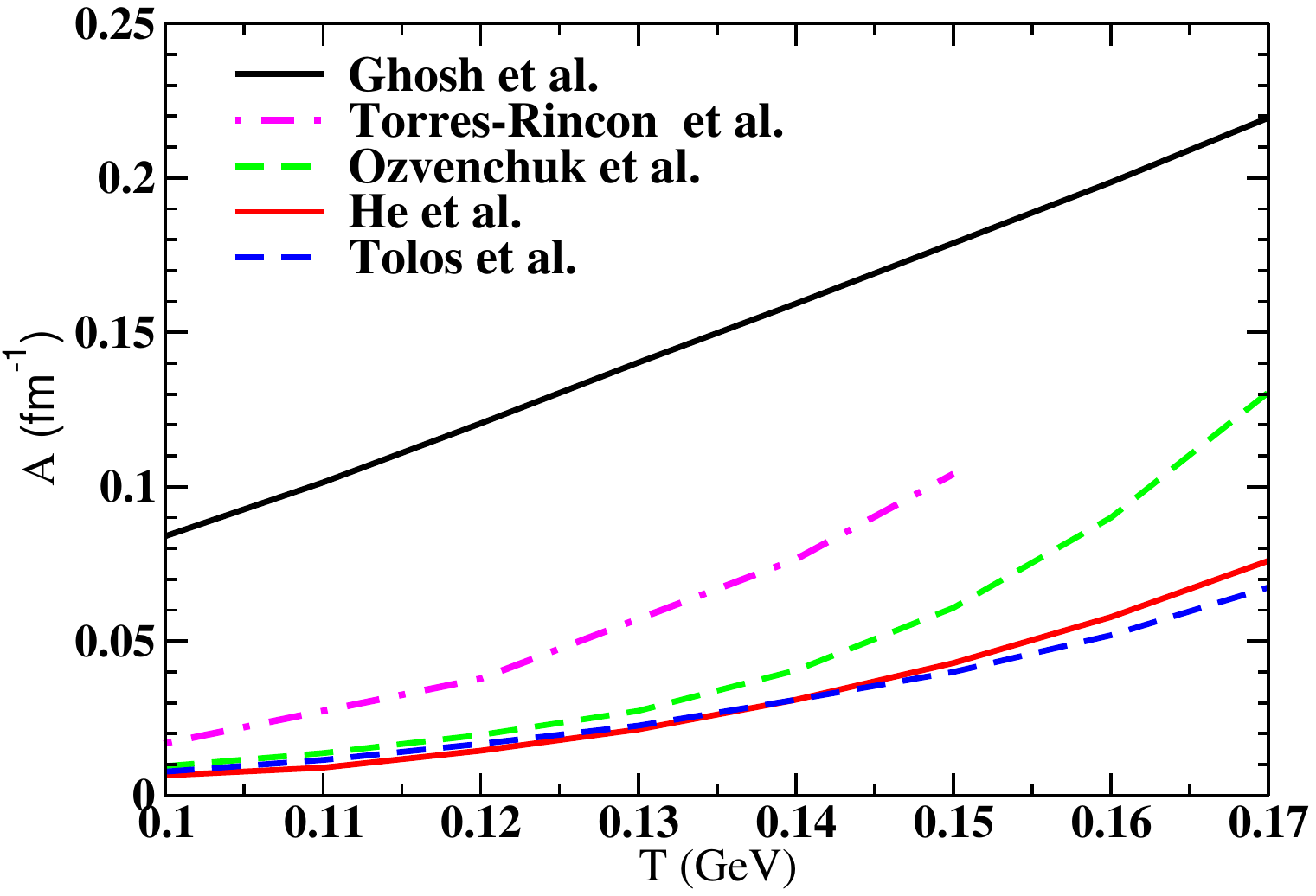}
\caption{Drag coefficient of $D$ mesons as
a function of temperature obtained within different models. The black solid line in the plot represents the results from Ref~\cite{Ghosh:2011bw}, while the model estimates from Ref. ~\cite{Torres-Rincon:2021yga} are depicted by the magenta dash-dotted line.  Additionally, the green dashed line  represents the result from Ref.~\cite{Ozvenchuk:2014rpa}, and the calculation from Ref.~\cite{He:2011yi} is presented by the red solid line. Results from Ref.~\cite{Tolos:2013kva} is presented by the blue dashed line. }
\label{fig:DDaic}
\end{figure}
In Ref.~\cite{Ozvenchuk:2014rpa}, the authors have further explored the topic by calculating the drag and diffusion coefficients of $D$ mesons as they propagate through hadronic matter. This analysis builds upon the model introduced in Ref.~\cite{Tolos:2013kva}. In this study, they incorporate higher states in the hadronic phase, which play a notable role at elevated temperatures. They have incorporated the $D\rho$ interaction~\cite{Lin:2000jp}, as well as mesons and baryons not listed by the particle data group, with the cross sections assumed to be constant, inspired by the methodology outlined in Ref.~\cite{He:2011yi}.
In Ref.~\cite{Torres-Rincon:2021yga,Montana:2023sft}, the $D$-meson transport coefficients were calculated by incorporating medium corrections for the interactions as discussed in Sec.~\ref{sssec:D-ampl}. They calculated the unitarized in-medium $D$-meson amplitudes~\cite{Montana:2020lfi,Montana:2020vjg} within a mesonic environment at finite temperature, with particular emphasis on incorporating off-shell effects. In-medium interactions introduce new kinematic domains that affect the $D$-meson properties, consequently impacting the transport coefficients.

In Fig.~\ref{fig:DDaic} we compile the dependence of the drag coefficient as a function of temperature as obtained using different models. Let us start by comparing the results obtained from Ref~\cite{Ozvenchuk:2014rpa} (green dashed line) to those of Ref.~\cite{Tolos:2013kva} (blue dashed line). Both calculations 
in principle employ identical baseline cross sections for $D$-meson rescattering with hadrons of the medium. The reason the drag coefficient obtained within the Ref.~\cite{Ozvenchuk:2014rpa} is larger than that of Ref.~\cite{Tolos:2013kva} is due to the inclusion of higher states in the hadronic medium, as they exert a significant influence at higher temperatures. The model estimates of the drag coefficient from Ref.~\cite{He:2011yi} based on Breit-Wigner amplitudes is depicted by the solid red line, and they turn out to be quite close to the results presented in Ref.~\cite{Tolos:2013kva}. The magenta dash-dotted line represents the result obtained within the calculation performed in Ref.~\cite{Torres-Rincon:2021yga,Montana:2023sft}. As discussed above, the in-medium interactions implemented in~\cite{Montana:2023sft} resulted in an increase of the drag coefficient by almost a factor of 2 compared to vacuum interaction~\cite{Tolos:2013kva}. The solid black line in Fig.~\ref{fig:DDaic} represents the drag coefficient obtained within the model presented in Ref.~\cite{Ghosh:2011bw}. It is quite a bit larger than all other model calculations, since the interactions leading to the drag coefficient reported in Ref.~\cite{Ghosh:2011bw} do not satisfy exact unitarity, which could potentially lead to an overestimation of the $D$-meson drag coefficient.

A few other attempts~\cite{Das:2011vba,Goswami:2023hdl} were also made to study the $D$-meson transport coefficients in the hot hadronic medium. In Ref.~\cite{Das:2011vba} they were evaluated in a hot hadronic medium consisting of pions, kaons and $\eta$'s using $\chi$PT through the scattering length~\cite{Liu:2009uz,Guo:2009ct}. The drag coefficient obtained within this approach up to next-to-next-to leading order (NNLO) is quite close to the results given in Ref.~\cite{Tolos:2013kva}. Recently, the $D$-meson transport coefficients have been evaluated 
in Ref.~\cite{Goswami:2023hdl} within a van der Waals hadron resonance gas model (VDWHRG)~\cite{Vovchenko:2016rkn}, including empirical short-range correlations in the HRG. A constant cross section was assumed for the drag coefficient as 
$A = \tau^{-1} = \sum_{j} n_{j} \langle \sigma_{Dj} v_{j} \rangle$, summing over many hadron species $j$, with number densities $n_{j}$, a pertinent cross section $\sigma_{Dj}$ and relative velocity $v_{Dj}$. The magnitude of the cross sections was adopted from Ref.~\cite{He:2011yi} as 10 (15)\,mb for $D$-meson scattering off mesons and baryons respectively. The large number of excited states in the HRG then leads to  a large friction coefficient, reaching $A\simeq 0.25$ fm$^{-1}$ at $T=160$ MeV and above $0.4$ fm$^{-1}$ at $T=180$ MeV. The latter converts into a spatial diffusion coefficient $\Ds (2\pi T)$ which is close to the quantum lower bound of one.  

%%%%%%%%%%%%%%%%%%%%%%%%%%%%%%%%%%%%%%%%%%%%%%%%%%%%%%
\subsubsection{Spatial diffusion coefficient}
\label{sssec:spatial}
%%%%%%%%%%%%%%%%%%%%%%%%%%%%%%%%%%%%%%%%%%%%%%%%%%%%%%%%%%
As elaborated in Appendix~\ref{app:DTR}, the spatial diffusion coefficient can be determined through the drag coefficient in the static limit, 
\bea \label{eq:ddx}
\Ds=\frac{T}{A (p \rightarrow 0) \ m_D}, 
\eea
where $m_D$ denotes the $D$-meson mass. Since the drag force $A(p)$, using nonrelativistic kinetic theory, varies inversely with the heavy mass $m_D$ cf. Eq.~\eqref{eq:A} (assuming that the transport cross section does not depend strongly on it), it is expected that the spatial diffusion coefficient, $\Ds$, will remain almost unaffected by changes in mass, at least, at lowest order. It is common to normalize the spatial diffusion coefficient by the thermal wavelength, represented as $1/(2\pi T)$, to render it a dimensionless quantity. 
In perturbation theory in the QGP, $2\pi T\Ds$ is directly related to the square of the inverse coupling constant, $1/\alpha_s^2$, which highlights its role as a measure of the (inverse) interaction strength in the medium (similar observations have been made in nonperturbative evaluations for HQ diffusion in the QGP~\cite{ZhanduoTang:2023pla}). Eq.~\eqref{eq:ddx} implies that if the drag coefficients are alike in two distinct models, then the spatial diffusion coefficient $\Ds$ will show similarity for the same $m_D$. One can also derive $\Ds$ from the momentum space diffusion coefficient using the Einstein relation $B=m_D A T$, where one finds $\Ds=T^2/B(p \rightarrow 0)$. In such scenario, there might be some discrepancies in the results obtained from these different approaches.

\begin{figure} 
\centering
\includegraphics[scale=0.35]{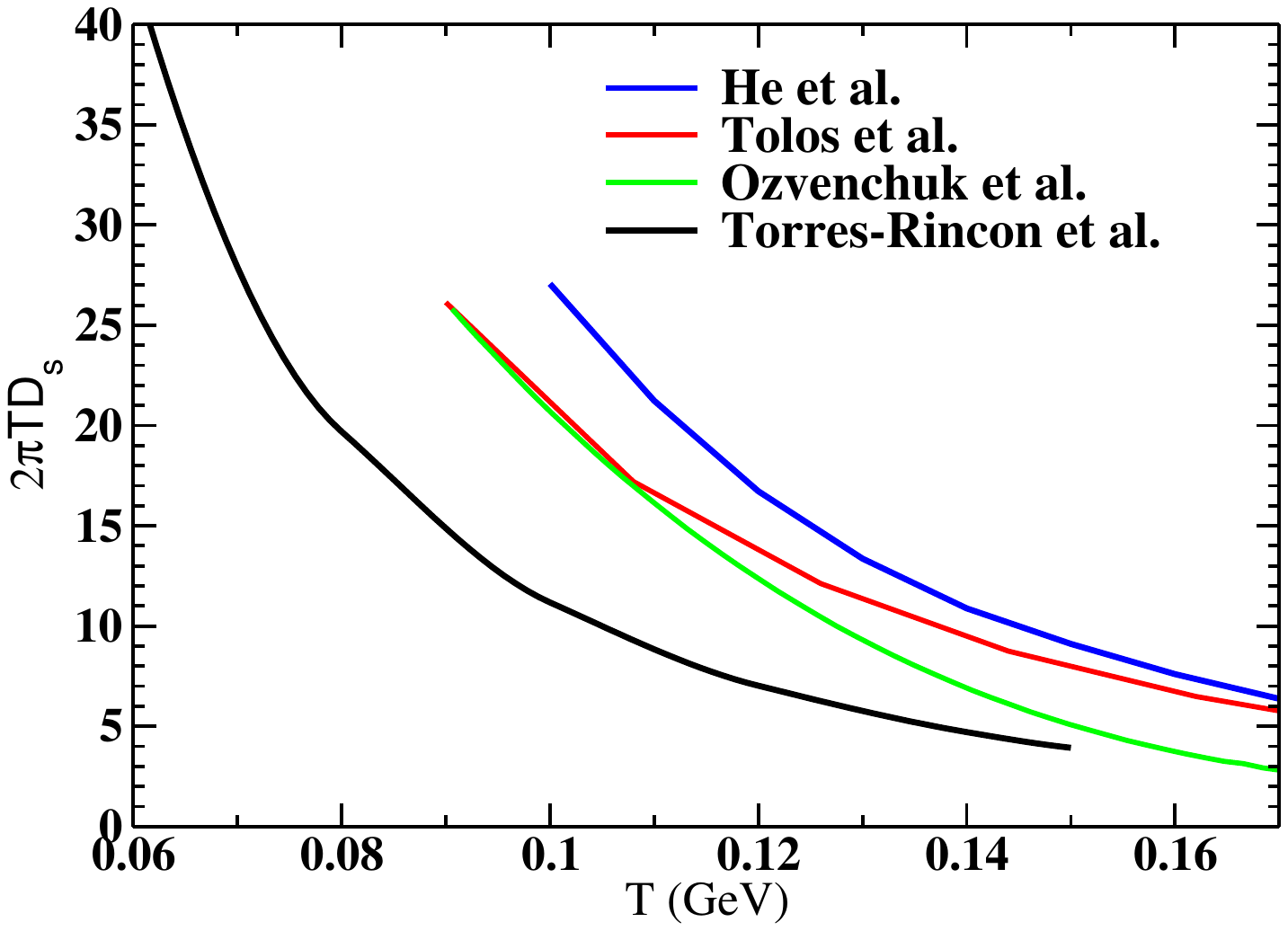}
\caption{The $D$-meson spatial diffusion coefficient, $2\pi T \Ds$, as
a function of temperature obtained within different models. The black solid line in the plot represents the results from Ref.~\cite{Torres-Rincon:2021yga}, while the model estimates from Ref.~\cite{Ozvenchuk:2014rpa} are depicted by the solid green line.  Additionally, the solid red line represents the result from Ref.~\cite{He:2011yi}, and the calculation from Ref.~\cite{Tolos:2013kva} is presented by the blue solid line.}
\label{fig:DDDx}
\end{figure}

Figure~\ref{fig:DDDx} illustrates the variation of  $2\pi T\Ds$ as a function of temperature from different model calculations.  The solid blue line represent the result from Ref.~\cite{He:2011yi} using empirical (Breit-Wigner) amplitudes. The $2\pi T\Ds$ from Ref.~\cite{Tolos:2013kva} is presented by the red solid line. As expected, the result obtained within the Ref.~\cite{Tolos:2013kva} is similar that of Ref.~\cite{He:2011yi}. The model estimates of the $\Ds$ from 
Ref.~\cite{Ozvenchuk:2014rpa}---depicted by the solid green line---is smaller than the result obtained within  Ref.~\cite{Tolos:2013kva} at larger temperatures but agrees well at temperatures below $T=120$ MeV where the HRG supports only few excited states. This is consistent with the findings of the Fig.~\ref{fig:DDaic}. 
The black solid line represent the result obtained within the calculation performed in Refs.~\cite{Torres-Rincon:2021yga,Montana:2023sft}. The in-medium interaction~\cite{Montana:2023sft} results in a decrease of $\Ds$ (increase of the drag coefficient) by almost a factor of 2  compared to vacuum interaction~\cite{Tolos:2013kva}.

%%%%%%%%%%%%%%%%%%%%%%%%%%%%%%%%
\subsubsection{Transport coefficients at finite baryon chemical potential}
\label{sssec:finite-mu}
%%%%%%%%%%%%%%%%%%%%%%%%%
\begin{figure}[t] 
\centering
\includegraphics[scale=0.48]{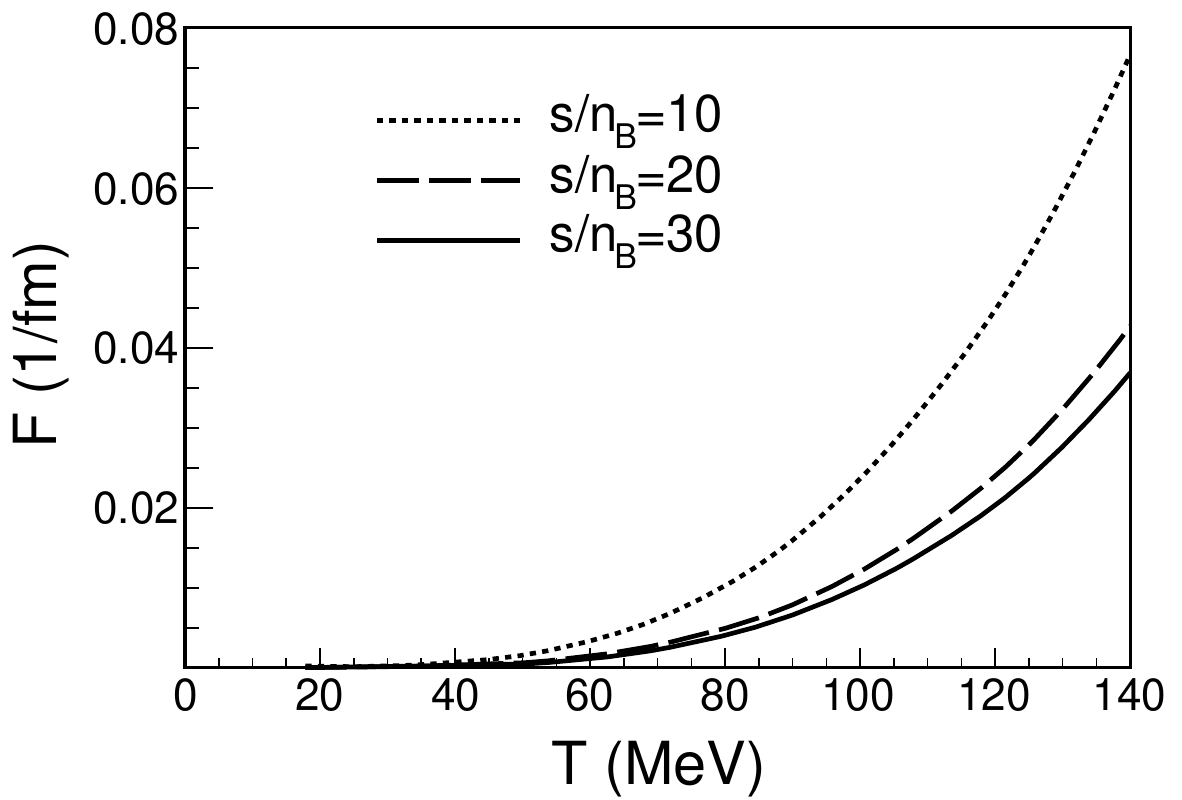}
\caption{Drag coefficient (denoted as $F$ in the plot instead of $A$) of $D$ mesons as a function of temperature at finite baryon chemical potential. The different lines correspond to different net baryon content, parameterized by different values of the entropy per (net) baryon. The lower the $s/n_B$ is, the higher the net baryon density of the gas. Figure taken from Ref.~\cite{Tolos:2013kva}.}
\label{fig:DDjl2b}
\end{figure}
For HICs at CM energies below the ones of the collider regime, a non-negligible net baryon density at mid-rapidity develops, increasing with decreasing $\sqrt{s_{NN}}$. Future efforts to study this matter are also directed towards exploring the transport coefficient of $D$ mesons under conditions of finite baryonic chemical potential~\cite{NA60:2022sze}. In Ref.~\cite{Tolos:2013kva}, the calculations were extended to a finite baryon chemical potential, considering heavy-ion collisions at future facilities such as FAIR or NICA, which are anticipated to operate center-of-mass energies roughly around 5-10\,AGeV. In that work, they examined the simplest scenario for isentropic trajectories in the QCD phase diagram as the fireball cools down, characterized by a constant ratio of entropy per (net)baryon, denoted by a fixed value of $s/n_B$. In particular, for FAIR conditions, the beam energy ranges from $E_{\rm lab} = 5-40$\,AGeV, corresponding to an approximate $s/n_B \sim 5-10$, with some variation depending on the thermal model employed~\cite{Bravina:2008ra}, while at SPS energies (up to $\sqrts\simeq20$\,GeV), the values can reach up to 30-40.  
They chose three values of $s/n_B=$10, 20, and 30, corresponding to chemical potentials, $\mu_B=$ 536, 361 and 286\,MeV, respectively, at a temperature of $T\simeq 160$ MeV. In Fig.~\ref{fig:DDjl2b}, the variation of the drag coefficient is shown as a function of temperature, exhibiting a strong increase which in part is driven by the simultaneous increase in baryon density at fixed $s/n_B$.
When increasing the latter, the system becomes more baryon-rich which is more pronounced as the specific entropy is lowered. Also note that baryons, with their relatively large masses, impart larger momentum changes in elastic collisions with the diffusing $D$ mesons than to the lighter pseudo-Goldstone bosons.
When comparing the results with those presented in Fig.~\ref{fig:DDjl2} one observes that the contributions from nucleons and $\Delta$ baryons significantly enhance the transport coefficients, especially toward the lower $s/n_B$ values relevant for the FAIR energy regime, see Refs.~\cite{Tolos:2013kva, Ozvenchuk:2014rpa} for further details.

It is worth noting that the contribution of baryon interactions to both the drag and diffusion coefficient exhibits an approximate simple scaling behavior with the fugacity~\cite{Tolos:2013kva},
\bea\label{eq:che}
A(T,\mu_B) \simeq A(T, \mu_B=0) \ e^{\mu_B/T}  \ ,  \nn
B(T,\mu_B) \simeq B(T, \mu_B=0) \ e^{\mu_B/T}  \ .
\eea
These relations can readily be obtained by assuming that classical statistics can be used instead of the Fermi-Dirac distribution for baryons.

%%%%%%%%%%%%%%%%%%%%%%%%%%%%%%%%%%
\section{$B$-mesons and heavy baryons in hot hadronic matter}
\label{sec:bottom-baryons}
%%%%%%%%%%%%%%%%%%%%%%%%%%%%
While most of the progress in open HF hadron systems has focused on $D$-meson interactions, heavier systems have also been considered in the literature. Heavy-quark flavor symmetry facilitates the extension of the dynamical description to the bottom sector, as the effective interaction Lagrangian remains independent of the heavy particle’s mass at leading order (LO) in 1/$m_Q$~\cite{Georgi:1990um}. The physical masses of the heavy hadrons explicitly break this symmetry but the formal interaction terms are identical at lowest order.

In this section, we will review the studies of $B$-meson properties in a hot confined medium (Sec.~\ref{ssec:Bint}) and the pertinent transport coefficients (Sec.~\ref{ssec:Btrans}), with emphasis on pointing out new aspects with respect to the charm case. We will also summarize results addressing heavy baryons, specifically single-charm and bottom-baryons $\Lambda_c$ and $\Lambda_b$, in a  thermal medium (Sec.~\ref{ssec:baryon-int}) and their transport coefficients (Sec.~\ref{ssec:baryon-trans}). 

%%%%%%%%%%%%%%%%%%%%%%%%%%%%%%%%%%%%%%%%%%%%%%%%%%%%%%%%%%%%%% 
\subsection{$B$-mesons interaction in hot hadronic matter}
\label{ssec:Bint}
%%%%%%%%%%%%%%%%%%%%%%%%%%%%%%%%%%%%%%%%%%%%%%%%%%%%%%%%%%
For systems in which the charm quark is replaced by a bottom quark\footnote{In the standard convention the equivalent of a $D$-meson would be a $\bar{B}$ meson, since it is the state carrying a $b$ quark. However, to ease the notation we will consider a $B$-meson instead. The mesonic interactions of the two states in a net baryon-free medium become the same.}, the HQ limit is more accurately satisfied. Therefore, an expansion in inverse powers of the HQ mass is expected to be an excellent guiding principle. Accounting for the chiral dynamics of the light degrees of freedom, the EFT used for $D$ mesons remains identical, at lowest order, to the one for $B$ mesons. Only the use of physical meson masses (as an explicit breaking of the HQ flavor symmetry) makes a difference in various observables.
  
In Sec.~\ref{ssec:hqet}, we have referenced several works incorporating the heavy-quark expansion to the $B$-meson dynamics. Some of the relevant literature that treats the interaction in vacuum can be found in Refs.~\cite{Kolomeitsev:2003ac,Guo:2006fu,Guo:2006rp,Flynn:2007ki,Liu:2009uz}.
In the context of URHICs, where a medium with temperature $T$ is considered, one of the first models for $B$-meson propagation in which the transport coefficients were estimated was elaborated in Ref.~\cite{Laine:2011is}. This work was already discussed in Sec.~\ref{sssec:sym}, and since the calculations were performed in the infinite HQ mass limit, the results for charm at LO can be directly transferred to the $B$-meson case (or any other particle with high enough mass). However as discussed before, the application of the results is rather limited in temperature, and it does not touch upon HF symmetry breaking effects. 

In Ref.~\cite{Das:2011vba} explicit calculations of transport coefficients of $B$-mesons were performed. Concerning the interaction with light degrees of freedom, this work utilizes the scattering lengths obtained in the ``covariant chiral perturbation theory'', inspired from the developments of Ref.~\cite{Geng:2010vw} and from the ``heavy-meson chiral perturbation theory''~\cite{Manohar:2000dt} taking the interactions up to NNLO~\cite{Liu:2009uz}. Since the perturbative interactions taken from these EFTs increase without limit with energy, the authors avoided unnaturally large scattering rates by using the values of the amplitudes at threshold (\ie, scattering lengths) for all energies. Since the validity of these amplitudes is restricted to energies close to threshold, the fact that the transport coefficients involve integrations over all energies of the therm-bath particles, a constant amplitude can mimic a saturated energy-dependent amplitude, in such a way that the average interaction has the correct order of magnitude.
  
In Ref.~\cite{Abreu:2012et} the same EFT of Ref.~\cite{Geng:2010vw} was used in analogy to the previous calculation for charm in Ref.~\cite{Abreu:2011ic} by the same authors, up to NLO in chiral expansion to account for the lowest-order diagrams in the HQ mass expansion. In that work the scattering amplitudes with $\pi,K,\bar{K}$ and $\eta$ mesons were unitarized which generated resonances that dominate the interaction. This could be done for both $B$ and $B^*$ mesons thanks to the HQ spin symmetry. It was found that the dominant role is played by the $J^P=0^+$ state at $\sqrt{s}=5534 + i105$\,MeV (compatible with results of Refs.~\cite{Kolomeitsev:2003ac,Guo:2006fu} in the same channel and identified with the HF partner of the $D_0^*(2300)$), together with a $J^P=1^+$ state at $\sqrt{s}=5587 +i122$ MeV (also compatible with Refs.~\cite{Kolomeitsev:2003ac,Guo:2006rp} and corresponding to the HF partner of the $D_1(2430)$). 

\begin{figure}[th]
\centering
\includegraphics[scale=0.4]{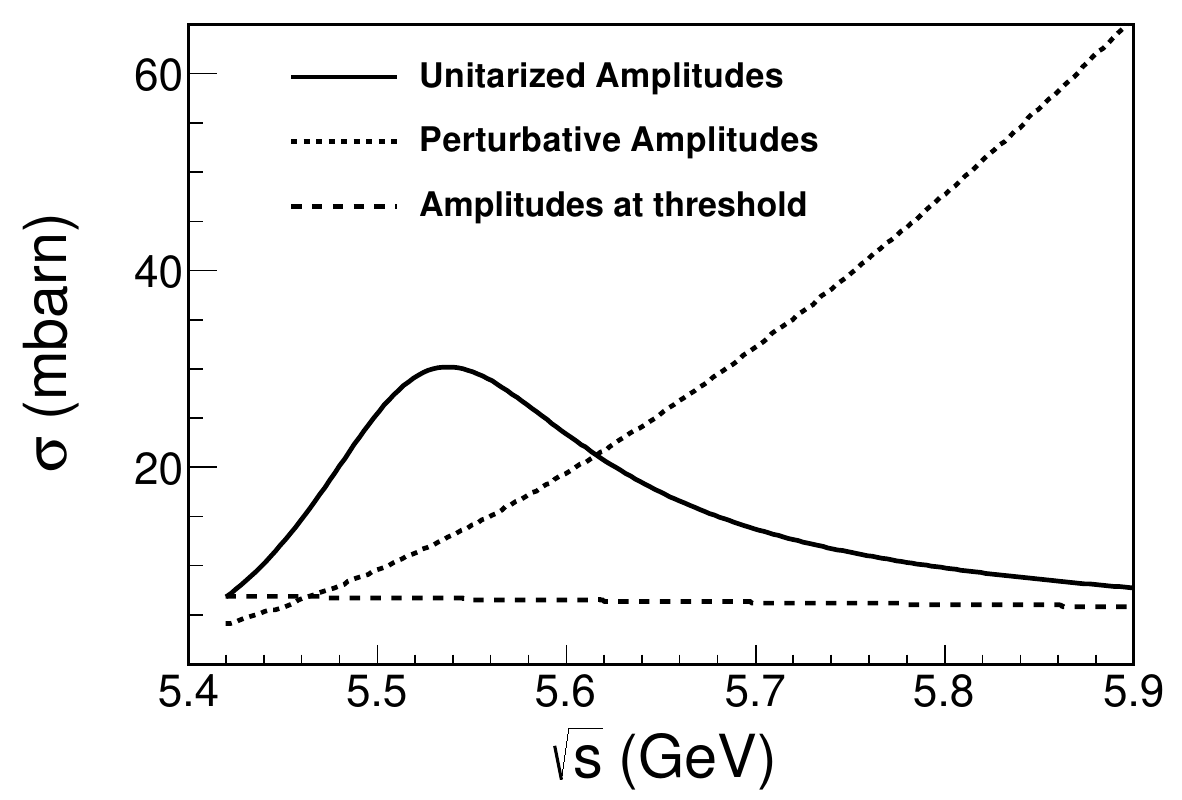}
\includegraphics[scale=0.4]{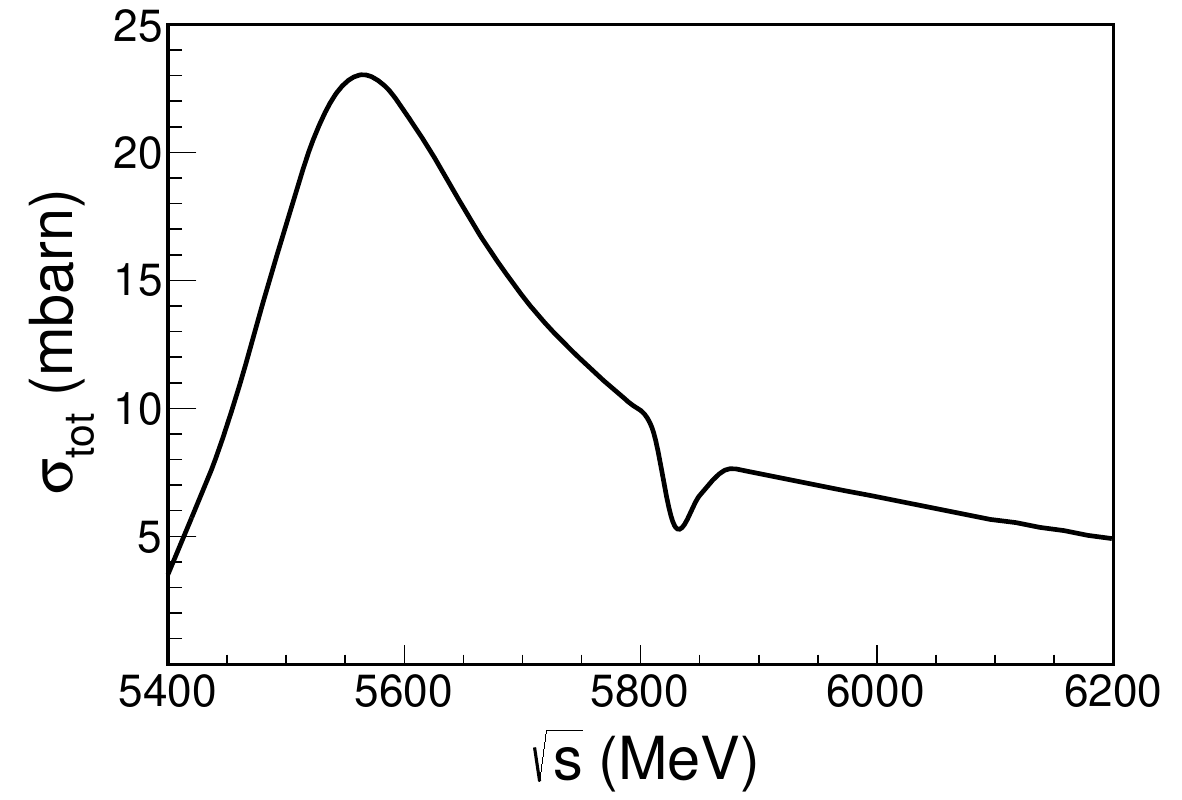}
\caption{Left panel: Comparison of the $B\pi$ total cross section as a function of the energy in the CM frame in different approximations: scattering amplitudes from the EFT (dotted line), from scattering amplitudes fixed at their threshold value (dashed line), and from scattering amplitudes after the on-shell unitarization procedure (solid line). Figure taken from Ref.~\cite{Abreu:2012et}. 
Right panel: Isospin averaged $B\pi$ cross section from Ref.~\cite{Torres-Rincon:2014ffa} where the full coupled-channel analysis was incorporated into the $T$-matrix equation. The dip around $\sqrt{s}=5830$ MeV is due to the opening of the $B_s K$ channel. Figure taken from~\cite{Torres-Rincon:2014ffa}.}
\label{fig:Bmesonxsec}
\end{figure}

In the left panel of Fig.~\ref{fig:Bmesonxsec} we display a comparison of the isospin averaged $B \pi$ cross section under different approximations, taken from Ref.~\cite{Abreu:2012et}.
The dotted line is the result of the EFT up to NLO for the $B\pi$ amplitude. Due to the absence of exact unitarity its validity is restricted to low energies. Upon unitarization of the perturbative amplitudes the cross section saturates and produces a resonant peak, which is shown by the solid line. Finally, taking the unitarized amplitude at threshold and using it for all energies one obtains the dashed line which is essentially constant with energy. The unitarized cross section differs significantly from that obtained using the threshold amplitude, resulting in different transport coefficients. However, at higher temperatures, this difference is partially reduced as momentum integrations over the thermal phase space of heat bath particles extend to center-of-mass energies well beyond the low-lying resonance structure.

In Ref.~\cite{Torres-Rincon:2014ffa}, the effective theory from~\cite{Abreu:2012et} was extended to incorporate coupled channels by including the $B_s$ states as well. Apart from several bound states, additional resonances were found together with the ones mentioned in Ref.~\cite{Abreu:2012et}. The results point to a double-pole structure of a $B^*_0$ state, similar to the one of the $D_0^*(2300)$. In Ref.~\cite{Torres-Rincon:2014ffa} the interactions with $N$ and $\Delta$ resonances were also considered by extending the calculation to the meson-baryon sector. Strictly speaking, the state considered was the $\bar{B}$ meson (the one with a valence $b$ quark) which interacts with baryons in a dense medium in an attractive manner. In the right panel of Fig.~\ref{fig:Bmesonxsec} we present the $B\pi$ isospin averaged cross section taken from~\cite{Torres-Rincon:2014ffa}. The effects of the coupled channels, as compared with the unitarized result in the left panel of the same figure, is a small reduction of the cross section, a mild shift of the peak position, and the presence of a distortion around $\sqrt{s}=5830$ MeV due to the opening of the $B_sK$ channel.   
In Ref.~\cite{Pathak:2014nfa}, the authors used a chiral effective model including mesons and baryons in which a Weinberg-Tomozawa interaction connects the heavy mesons with baryons. This accounts for density effects of the medium, which are taken into account together with temperature. No visible mass shift can be observed in the pure thermal case, but a systematic decrease of the thermal mass is seen when in conjunction with the increase of the baryon density.
  
Finally, the in-medium interactions of the $B, B_s$ and $B^*, B_s^*$ mesons were computed in a self-consistent scheme in Ref.~\cite{Montana:2023sft}, similar to the charm case carried out in Refs.~\cite{Montana:2020lfi,Montana:2020vjg}. The two states become quasiparticles at finite temperature with a mass and decay width due to the in-medium $T$-matrix interactions. The discussion in Sec.~\ref{ssec:Dmed} about the reduction and broadening of the meson thermal masses also applies to these states. At $T=150$ MeV, a mass reduction of around $30 (20)$ MeV and a thermal decay width of around $90 (30)$ MeV is found for $B$ ($B_s$) mesons.

%%%%%%%%%%%%%%%%%%%%%%%%%%%%%%%%%%%%%%%%%%%%%%%%%%%%
\subsection{$B$-meson transport coefficients}
\label{ssec:Btrans}
%%%%%%%%%%%%%%%%%%%%%%%%%%%%%%%%%%%%%%%%%%%%%%%%%%%%%
In this section, we compile existing results on the transport coefficients of $B$ mesons in hot hadron matter through quantitative comparisons. Our analysis will be based on the detailed discussions of the scattering amplitudes from the previous section, with a focus on  momentum-space drag and spatial diffusion coefficients. Although most research has focused on $D$ mesons, there have been some efforts to extend these studies to $B$ mesons. As with $D$ mesons in Sec.~\ref{sssec:Ddrag}, most approaches typically compute both the drag and diffusion coefficients. However, these two transport coefficients are not independent; they are related through the fluctuation-dissipation theorem (FDT). In this discussion, we will primarily emphasize the drag coefficient $A(p)$.

In Ref.~\cite{Das:2011vba}, as a first attempt, the drag and diffusion coefficients were  computed in a thermal bath of pions, kaons and $\eta$'s  using of scattering lengths~\cite{Liu:2009uz} as dynamical input up to NNLO in heavy-meson chiral perturbation theory with coupling constants taken from unquenched LQCD results~\cite{Ohki:2008py}. The utilization of the scattering length implies that the findings will be applicable primarily at low temperatures. 

In subsequent work~\cite{Abreu:2012et}, the drag and diffusion coefficients of $B$ mesons were determined in a hot hadronic medium consisting of pseudo-Goldstone bosons, \ie, pions, kaons, antikaons and $\eta$'s. The interactions of $B$ mesons and $B^*$ mesons were evaluated using NLO chiral perturbation theory, while adhering to the constraints imposed by HQ symmetry. To incorporate dynamically generated resonances and extend the analysis to higher temperatures,  standard unitarization techniques were employed.
This study extends the framework introduced in Ref.~\cite{Abreu:2011ic} to evaluate the drag and diffusion coefficients of $B$ mesons in a hot hadronic gas with a detailed analysis of the temperature and momentum dependencies of these coefficients. In Ref.~\cite{Torres-Rincon:2014ffa}, the authors delved deeper into the subject by investigating the scattering processes involving $\bar{B}$ mesons with light mesons and baryons like $N$ and $\Delta$. They adopted a unitarized approach with coupled channels, employing effective models consistent with chiral and HQ symmetries. We recall again that the $\bar{B}$ mesons are the bottom sector counterpart of $D$ mesons, containing one light antiquark. 
Their findings indicate that at zero baryon chemical potential, the primary contribution to the drag and diffusion coefficients arises from the interaction between $\bar{B}$ mesons and pions, simply because the latter are the prevalent degrees of freedom in the heat bath. 
Nevertheless, at finite baryon chemical potential, the contribution of baryons to the transport coefficients of $\bar{B}$ mesons becomes significant.  The interaction of both $B$ mesons and $\bar{B}$ mesons with pions, kaons, anti-kaons, and $\eta$ mesons results in identical transport coefficients.

In Ref.~\cite{Montana:2023sft}, the transport coefficients of $B$ mesons were computed by integrating medium corrections to the interactions, as detailed in Sec.~\ref{ssec:Bint}.
The in-medium unitarized amplitudes for $B$ mesons were computed within a mesonic environment at finite temperature, with particular attention to the inclusion of off-shell effects. This work builds upon their previous studies of Refs.~\cite{Montana:2020lfi,Montana:2020vjg}. The introduction of in-medium interactions introduces new kinematic domains (Landau cuts) that affect the properties of $B$ mesons and thus their transport coefficients.

\begin{figure}[th]
\centering
\includegraphics[scale=0.35]{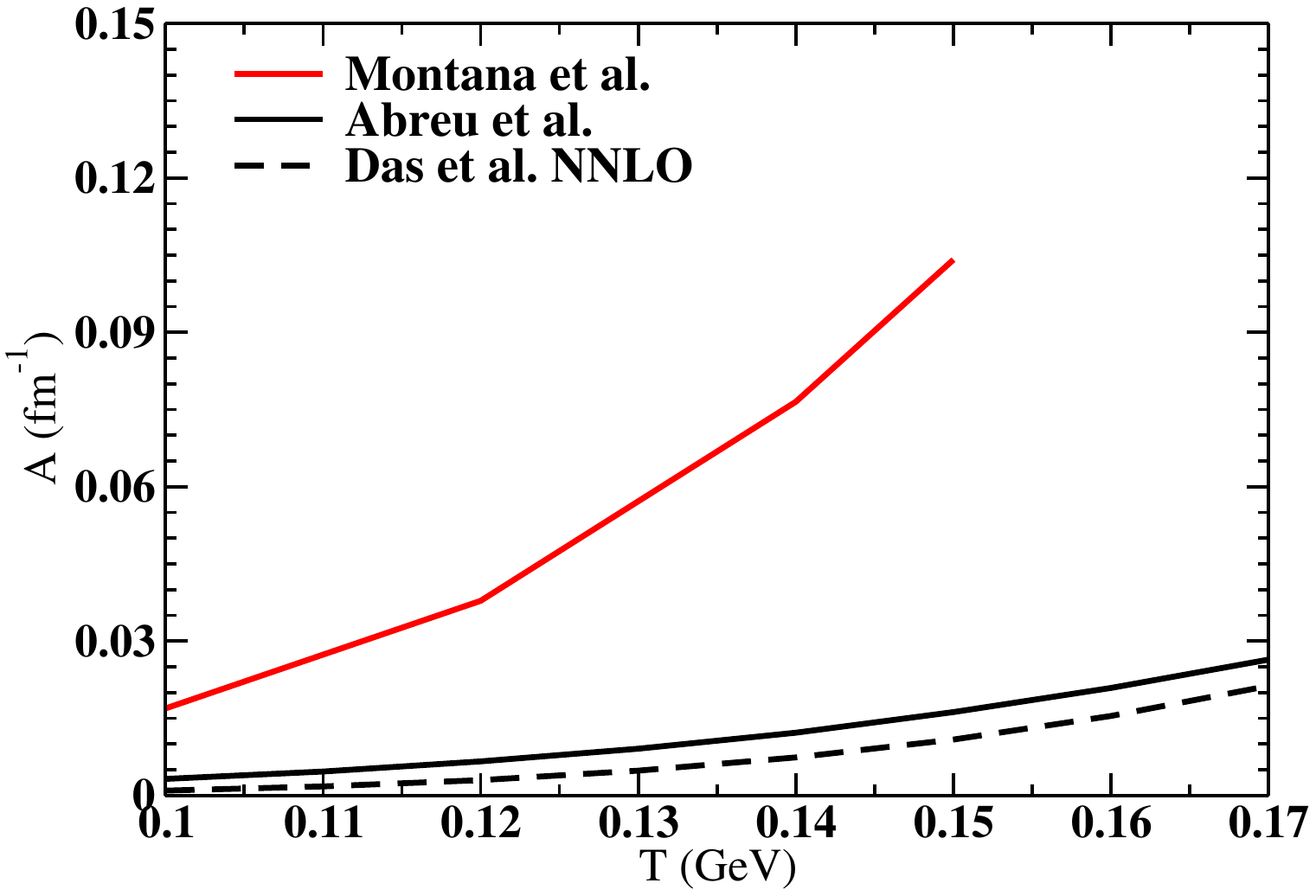}
\caption{Drag coefficient of $B$-mesons as
a function of temperature obtained within different models. The black solid line represents the results from Ref.~\cite{Abreu:2012et}, while the model estimates from Ref.~\cite{Montana:2023sft} are depicted by the red solid line.  Additionally, the black dashed line is the result from Ref.~\cite{Das:2011vba}.}
\label{fig:DDaic_B}
\end{figure}
In Fig.~\ref{fig:DDaic_B}, we compile the variation of the drag coefficients with temperature as obtained from the three models discussed above. The smaller magnitude of the drag coefficient obtained in Ref.~\cite{Das:2011vba} compared to that of Ref.~\cite{Abreu:2012et} can be attributed to the utilization of scattering lengths as inputs across the entire center-of-mass energy range.

The drag coefficient from Ref.~\cite{Montana:2023sft}, depicted by the red solid line, is substantially larger  compared to other results; this is attributed to the incorporation of medium corrections in the interactions as well as in the $B$-meson self-energy. As mentioned earlier, these corrections appear to be quite sensitive to the renormalization of the vacuum contribution of the scattering amplitudes, and need to be revisited in detail~\cite{gloria}.

\begin{figure}[th]
\centering
\includegraphics[scale=0.35]{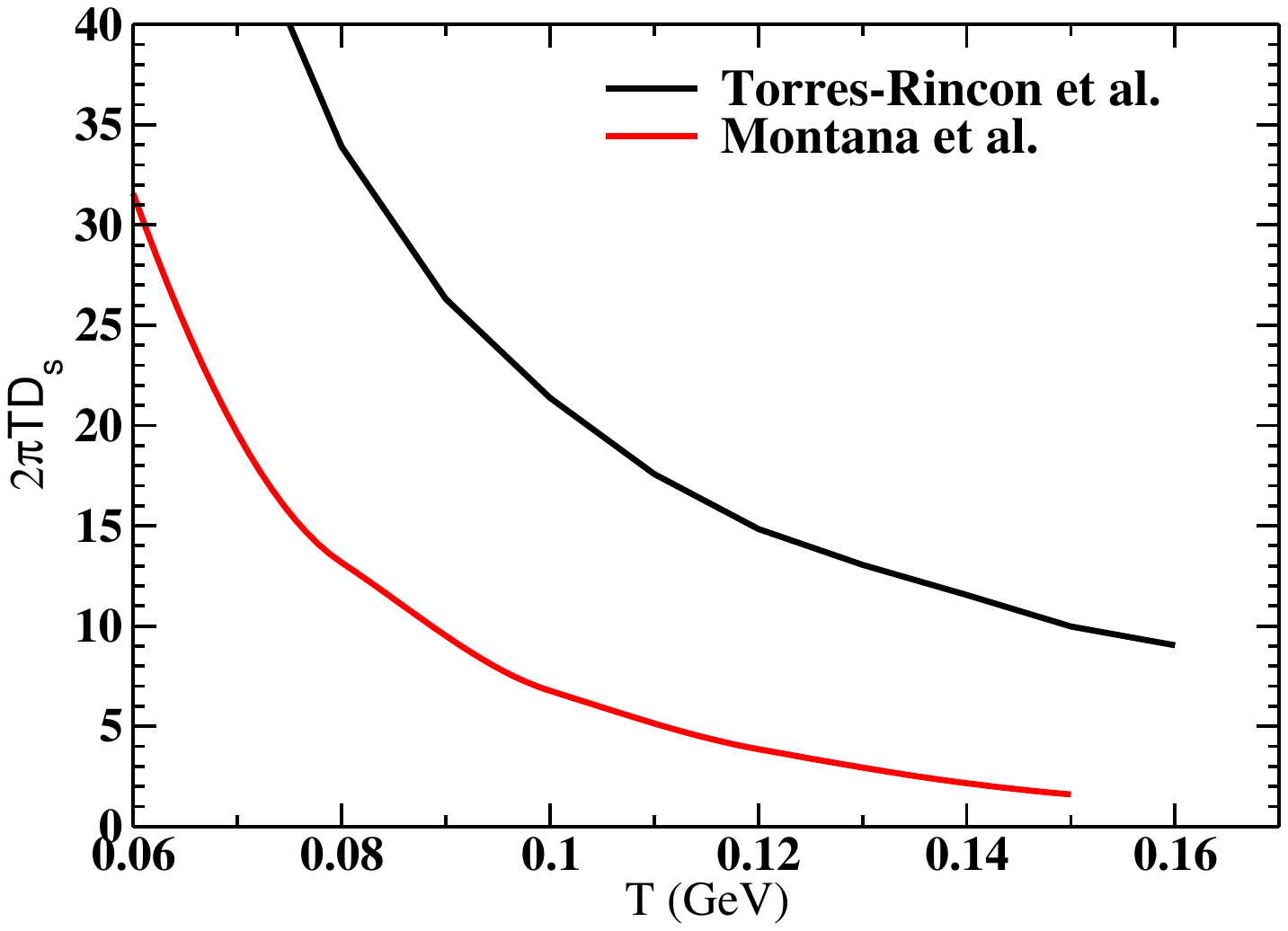}
\caption{The $B$-meson spatial diffusion coefficient, $2\pi T \Ds$, as
a function of temperature obtained within different models. The black solid line in the plot represents the results from a unitarized EFT for heavy+light mesons with vacuum scattering amplitudes~\cite{Torres-Rincon:2014ffa}, while the solid red line uses the same EFT but solves the self-consistent set of equations with in-medium interactions and $B$-meson self-energy~\cite{Montana:2023sft}.}
\label{fig:DDDx_B}
\end{figure}

Concerning the spatial diffusion coefficient, Fig.~\ref{fig:DDDx_B} illustrates the variation of $2\pi T\Ds$ of the $B$ mesons as a function of temperature, obtained using two different models. The solid black line represents the result obtained from Ref~\cite{Tolos:2016slr,Torres-Rincon:2014ffa} considering $B$ meson interaction with the pions, kaons, antikaons and $\eta$ mesons using an EFT that contains both chiral and HQ spin symmetries.
The solid red line represents the results obtained within Ref.~\cite{Montana:2023sft} considering the self-consistent medium corrections for the interactions. The incorporation of medium corrections into the interactions and the B-meson self-energy enhanced the drag coefficient, as
reported in Ref~\cite{Montana:2023sft}, resulting in a lower magnitude of $2\pi T\Ds$ compared to
the results presented in Ref.~\cite{Tolos:2016slr,Torres-Rincon:2014ffa}.

%%%%%%%%%%%%%%%%%%%%%%%%%%%%%%%%%%%%%%%%%%%%%%%%%%%%%%%%%%%%%%%%%%%%
\subsection{Heavy-baryon interactions in hot hadronic matter}
\label{ssec:baryon-int}
%%%%%%%%%%%%%%%%%%%%%%%%%%%%%%%%%%%%%%%%%%%%%%%%%%%%%%%%%%%%%%%%%%%%%
We now focus on the low-lying open-charm and  open-bottom baryons, $\Lambda_c$ and $\Lambda_b$, propagating through a thermal medium composed of light mesons ($\pi,K,\bar{K},\eta$). 

The interactions of these heavy baryons were studied in this context in Ref.~\cite{Tolos:2016slr},
using an effective model based on SU(6) $\times$ HQSS symmetry. The full symmetry group includes chiral symmetry (incorporating the pseudoscalar meson octet), spin symmetry (thus incorporating vector mesons), and HQ spin symmetry in the HF sector. Consequently, the Lagrangian contains 3+1 flavors, incorporating the mentioned mesons together with the $1/2^+$ and $3/2^+$ baryons. In the context of $D$-meson propagation, the same interaction was studied in Ref.~\cite{Tolos:2013kva} for the meson-baryon interaction with $C=1$ to describe the $DN$ resonant processes in the $S=0$ (strangeness) channel. Some of its coupled channels, $\Lambda_c \pi$  ($ S=0$,$I=1$) and  $\Lambda_c \eta$  ($ S=0$,$I=0$), together with other channels in the sector $S=\pm 1, I=1/2$, $\Lambda_c K$ and $\Lambda_c \bar{K}$, were taken together to study the $\Lambda_c$ case and extended to $\Lambda_b$ dynamics in Ref.~\cite{Tolos:2016slr} on equal footing.

The meson-baryon interactions are modelled by the exchange of vector resonances in the $t$-channel, in the limit where the exchanged momentum is much smaller than the vector masses. In this way the Weinberg-Tomozawa interaction gives a contact vertex (Born term) between the baryon and meson which reads
\be 
V^{IJS}_{ij} = D^{IJS}_{ij} \frac{2 \sqrt{s}- M_i-M_j}{4f_i f_j}  \sqrt{\frac{E_i+M_i}{2M_i}} \sqrt{\frac{E_j+M_j}{2M_j}} \ , 
\ee
where $i,j$ denote the incoming/outgoing channels, $D^{IJS}_{ij}$ are numerical coefficients describing the strength of the interaction, $M_i,E_i$ are the baryon mass and center-of-mass energy in the channel $i$, and $f_i$ is the meson decay constant in the channel $i$.
After on-shell unitarization of the perturbative amplitude a series of resonant states are dynamically generated. The most prominent one in the charm case is the $\Lambda_c(2595)$, while in the bottom sector, it is $\Lambda_b(5910)$. The generation of the $I=0$ $\Lambda_c(2595)$ state as a meson-baryon resonance was previously considered in similar effective models incorporating HQSS in Ref.~\cite{Garcia-Recio:2008rjt,Romanets:2012hm}, and the $\Lambda_b(5910)$ in Ref.~\cite{Garcia-Recio:2012lts}, as well as further states in other sectors. We refer the reader to a review~\cite{Tolos:2013gta} for a more comprehensive listing of dynamically generated states. 

\begin{figure} 
\centering
\includegraphics[scale=0.4]{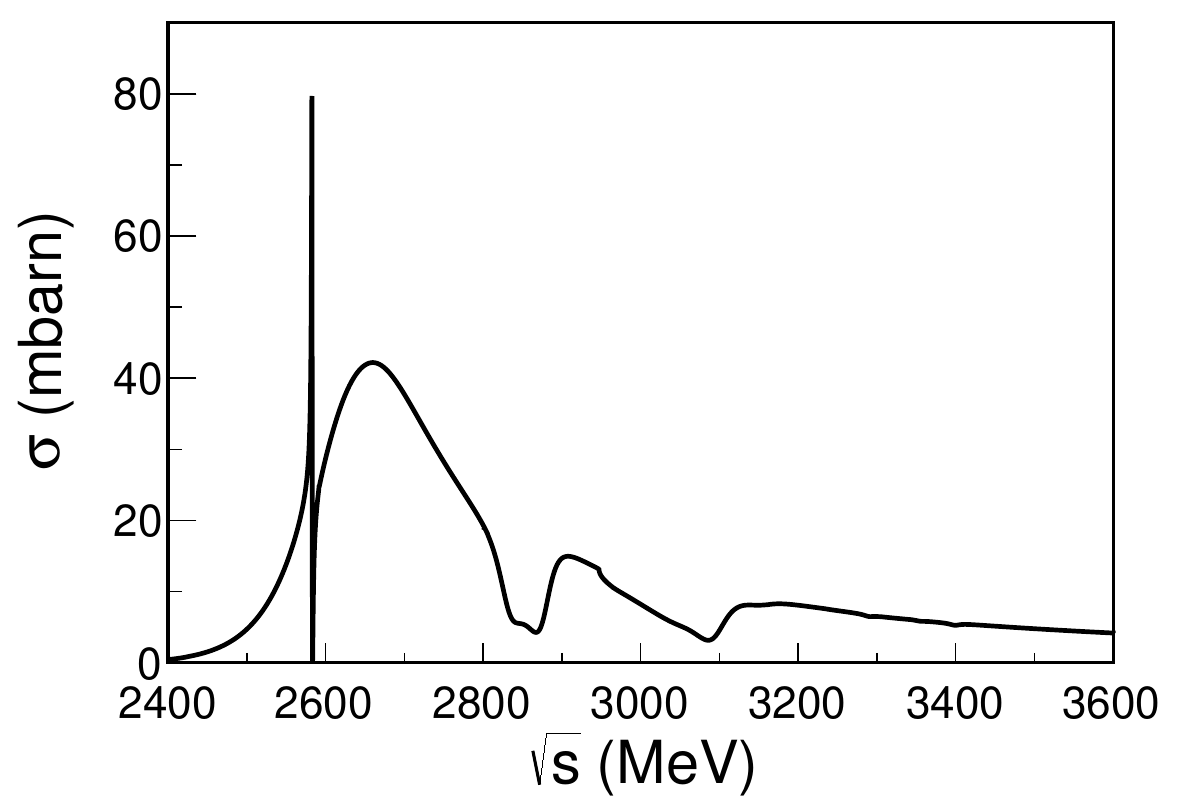}
\includegraphics[scale=0.4]{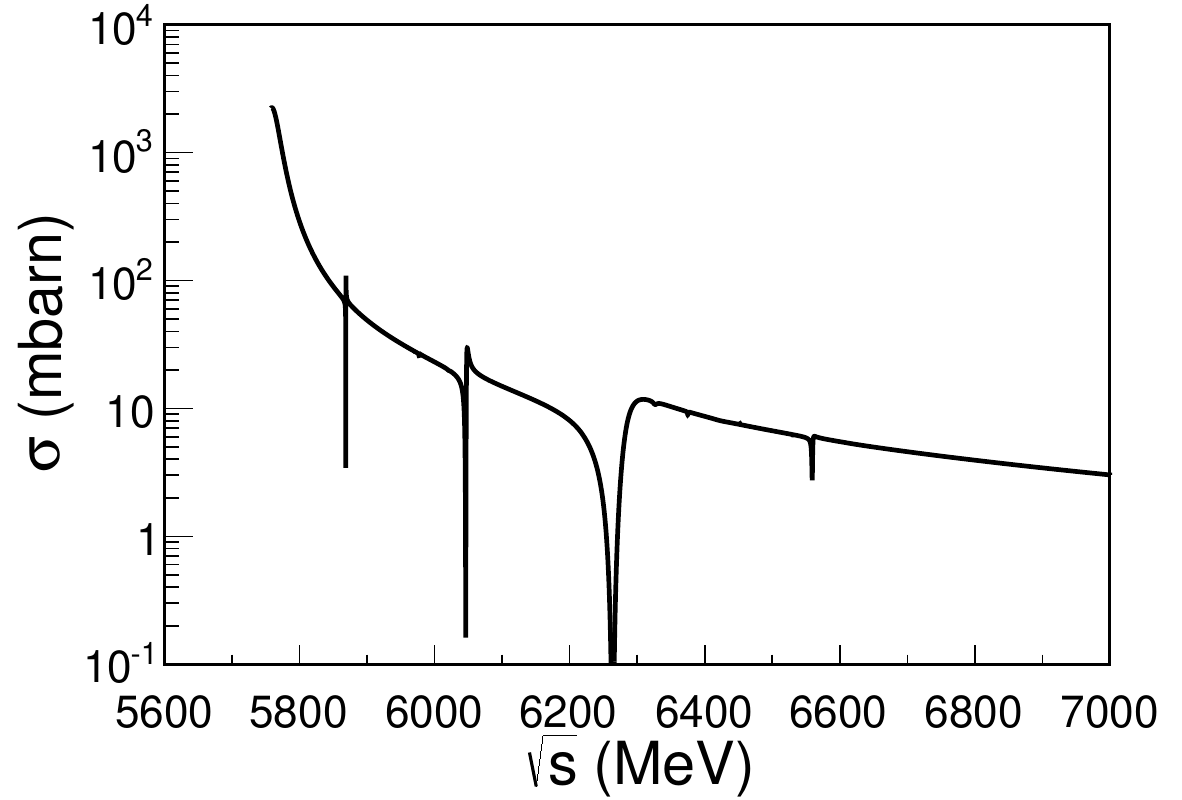}
\caption{Total cross section of $\Lambda_c \pi$ (left panel) and $\Lambda_b \pi$ (right panel) as functions of the CM total energy. Figure adapted from Ref.~\cite{Tolos:2016slr}.}
\label{fig:xsecLambdas}
\end{figure}

As an example of the cross sections obtained for  these states, we show in Fig.~\ref{fig:xsecLambdas} the $\Lambda_c \pi$ and $\Lambda_b \pi$ total cross section as functions of the total collision energy in the CM frame, taken from Ref.~\cite{Tolos:2016slr}. In the figure of the $\Lambda_c \pi$ one can observe several resonant structures which emerge with some threshold effect. The most prominent structure is related to the $\Lambda_c (2595)$ resonance. In the right panel, we show the $\Lambda_b \pi$ where a similar structure is expected (notice the semi-logarithmic axis in this case) with a strong effect due to the existence of energy thresholds of the different coupled channels. The large increase at low energies is likely due to a $\Sigma_b$ like ($I=1,J=1/2$) bound state below the $\Lambda_b \pi$ threshold~\cite{Torres-Rincon:2014ffa}. 
The transport coefficients of $\Lambda_c $ and $\Lambda_b$ obtained from these interactions are discussed in the following section.

%%%%%%%%%%%%%%%%%%%%%%%%%%%%%%%%
\subsection{Heavy-baryon transport coefficients}
\label{ssec:baryon-trans}
%%%%%%%%%%%%%%%%%%%%%%%%%%%%%%%%%%%
In this section, we will examine the existing transport coefficients for $\Lambda_c$ and $\Lambda_b$ in hot hadron matter by conducting quantitative comparisons. We will leverage the comprehensive discussions on scattering amplitudes from the previous section to facilitate our analysis. Similar to the earlier cases with $D$ and $B$ mesons, our discussion will be centered around the drag coefficient.

\begin{figure} 
\centering
\includegraphics[scale=0.27]{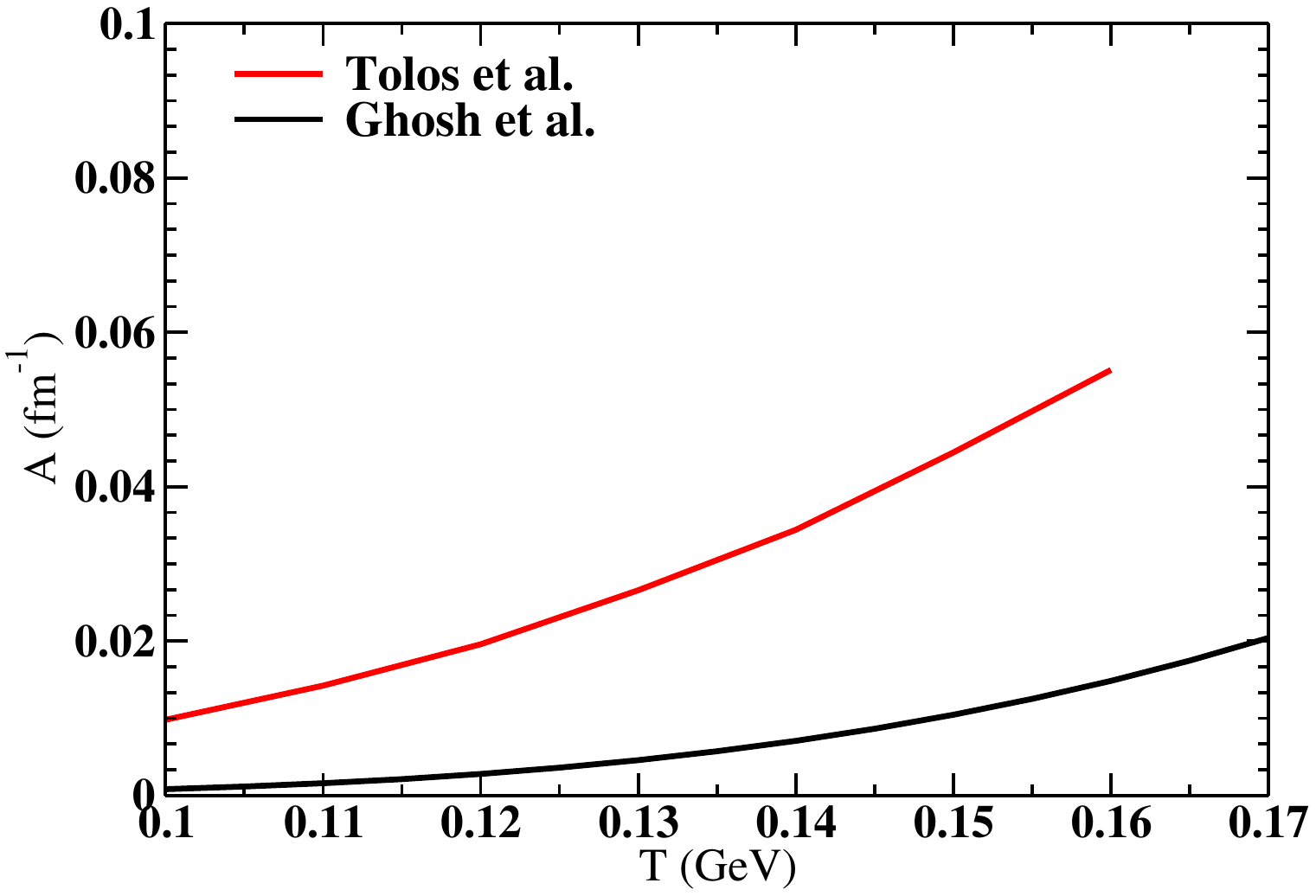}
\includegraphics[scale=0.27]{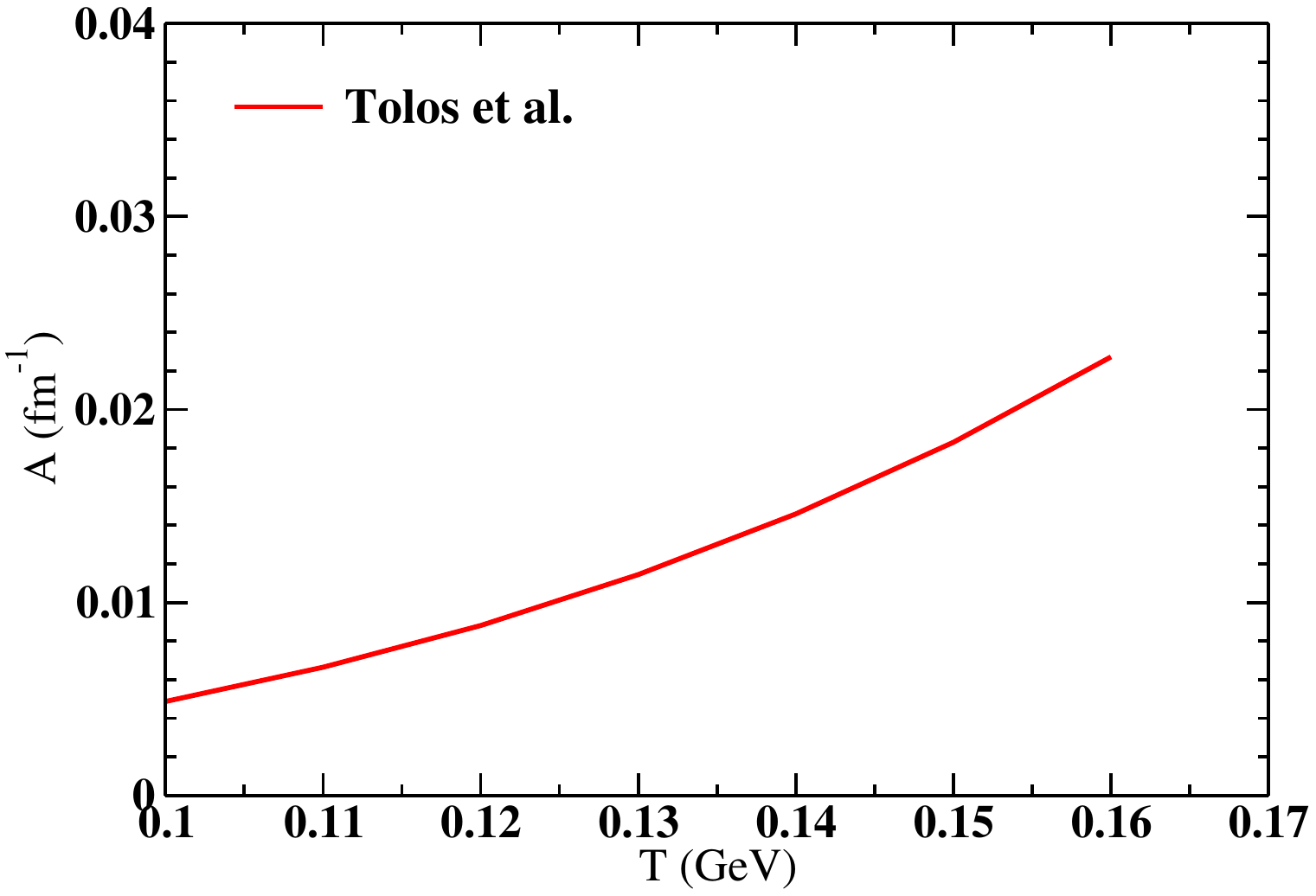}
\caption{Left panel: Drag coefficient of $\Lambda_c$ as a function of temperature obtained within different models. The black solid line represents the results from Ref.~\cite{Ghosh:2014oia}, while the model calculation from Ref.~\cite{Tolos:2016slr} are depicted by the red solid line. Right panel: Drag coefficient of $\Lambda_b$ as a function of temperature obtained from Ref.~\cite{Tolos:2016slr} is  depicted by the red solid line.}
\label{fig:DDaic_Lc}
\end{figure}

Ref.~\cite{Ghosh:2014oia} made a first attempt to compute the $\Lambda_c$ drag and diffusion coefficients in a thermal bath of pions, kaons and $\eta$'s  using  scattering lengths~\cite{Liu:2012uw} as dynamical input in heavy-meson chiral perturbation theory. 
The scattering length is associated with a range corresponding to maximum and minimum values, leading to a variation in the drag coefficient for $\Lambda_c$. Here, we present the average value of the drag coefficient. In subsequent work~\cite{Tolos:2016slr}, the drag and diffusion coefficients of $\Lambda_c$ were assessed using HQET in conjunction with $\chi$ PT. The authors employed unitarization techniques to extend their calculations to higher temperatures. 
The variation of the drag coefficients of $\Lambda_c$ as a function of temperature is shown in  the right panel of Fig.~\ref{fig:DDaic_Lc} obtained within the two different models discussed above.  The drag coefficient obtained in Ref.~\cite{Tolos:2016slr} is larger than that of the results obtained within Ref.~\cite{Ghosh:2014oia} employing scattering lengths.  Reference~\cite{Tolos:2016slr} made a first attempt to compute the drag and diffusion coefficients of $\Lambda_b$ in a thermal bath of pions, kaons and $\eta$ mesons using HQET in conjunction with $\chi$PT, also incorporating a unitarization method to extend the validity of the calculations to higher temperatures. The right panel of Fig.~\ref{fig:DDaic_Lc} shows the variation of the drag coefficient of $\Lambda_b$ obtained in Ref.~\cite{Tolos:2016slr}. However, the magnitude of the $\Lambda_b$ drag coefficient is smaller than that of the $\Lambda_c$ drag coefficient computed within the same model~\cite{Tolos:2016slr}.

\begin{figure} 
\centering
\includegraphics[scale=0.27]{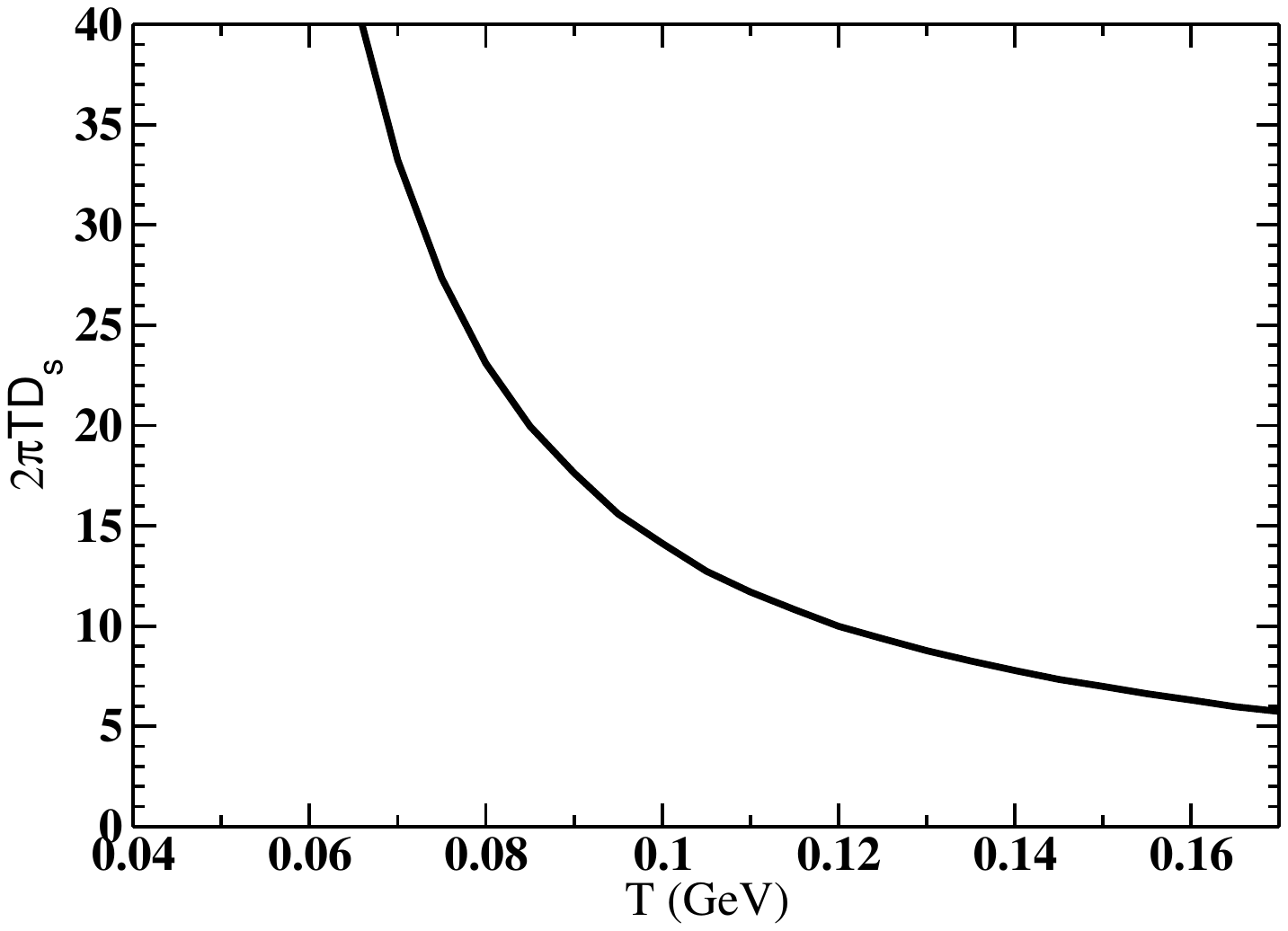}
\includegraphics[scale=0.27]{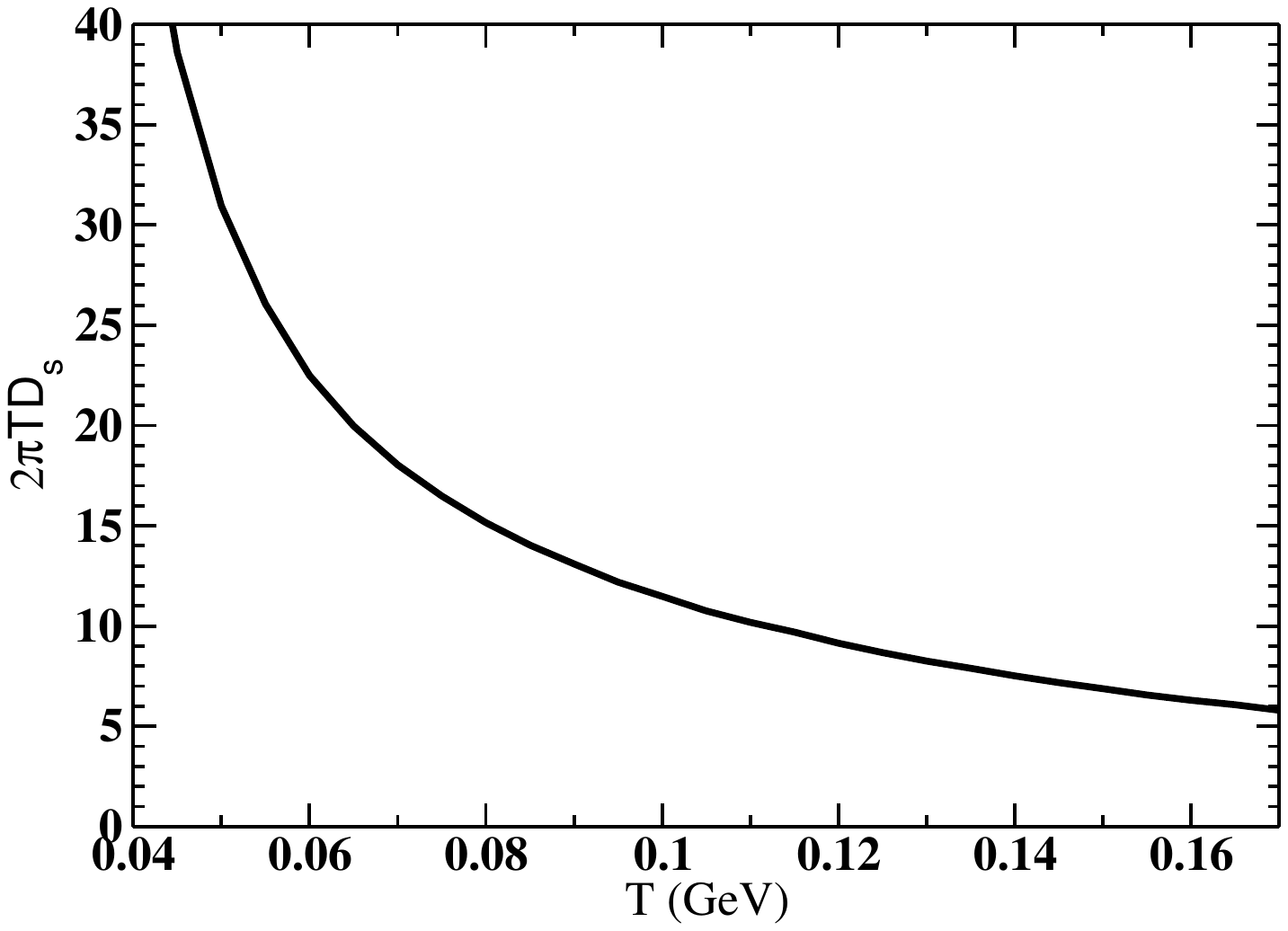}
\caption{ Left panel:  The $\Lambda_c$ spatial diffusion coefficient, $2\pi T \Ds$, as a function of temperature obtained within model calculation from Ref.~\cite{Tolos:2016slr}. Right panel:  The $\Lambda_b$ spatial diffusion coefficient, $2\pi T \Ds$, as a function of temperature obtained within model calculation from Ref.~\cite{Tolos:2016slr}.}
\label{fig:DDx_LcLb}
\end{figure}

The left panel of Fig.~\ref{fig:DDx_LcLb} illustrates the variation of $2\pi T\Ds$ of the $\Lambda_c$ as a function of temperature obtained within Ref.~\cite{Tolos:2016slr}. The right panel of Fig.~\ref{fig:DDx_LcLb} illustrates the variation of $2\pi T\Ds$ of the $\Lambda_b$ as a function of temperature obtained within Ref.~\cite{Tolos:2016slr}. As shown in Fig.~\ref{fig:DDx_LcLb}, the magnitude of $2\pi T\Ds$ for both the $\Lambda_c$ and $\Lambda_b$ are quite similar near the quark-hadron transition temperature.

\begin{figure} 
\centering
\includegraphics[scale=0.5]{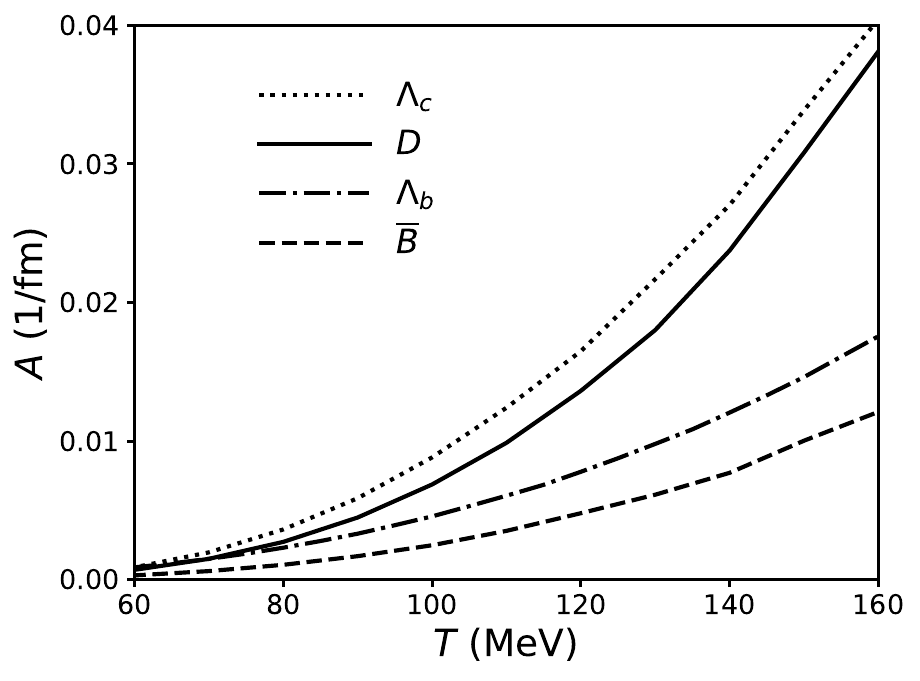}
\includegraphics[scale=0.5]{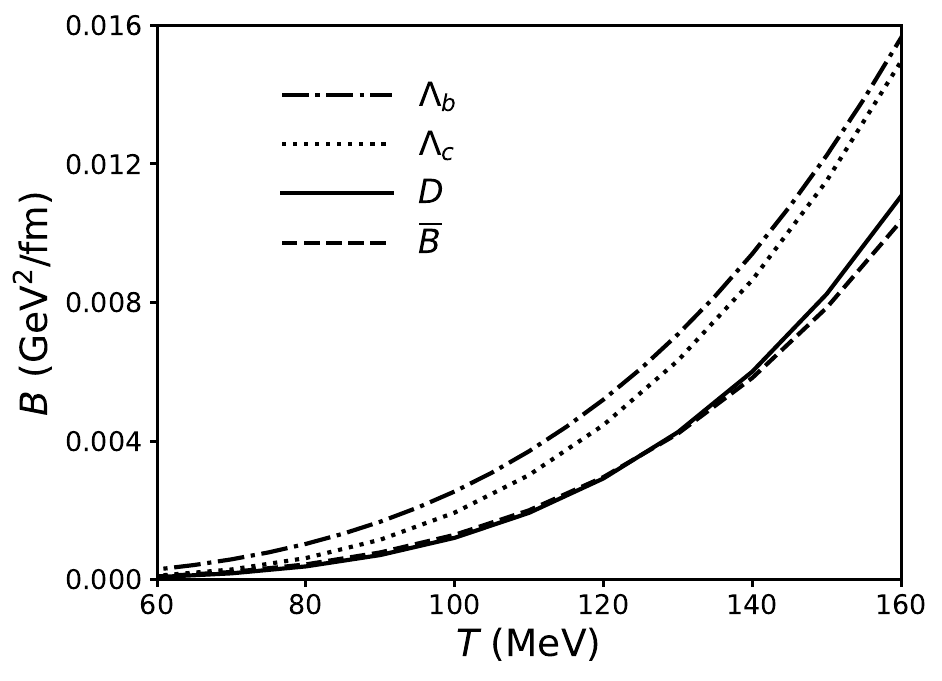}
\caption{The drag (left panel, here denoted as $F$ instead of $A$) and diffusion (right panel,  here denotes as $\Gamma_0$ instead of $B$) coefficients of $D,\bar{B},\Lambda_c,\Lambda_b$ hadrons  as a function of temperature at $p=100$ MeV in a gas of thermalized $\pi$ mesons.
Figures adapted from Ref.~\cite{Tolos:2016slr}.}\label{fig:LJSpi}
\end{figure}

In Secs.~\ref{sec:charm} and \ref{sec:bottom-baryons}, we have discussed the transport coefficients of $D$, $B$, $\Lambda_c$, and $\Lambda_b$ hadrons, which were obtained using various approaches within the hot hadronic medium. The magnitude of the transport coefficients varies across different models. However, it is notable that most of these models primarily concentrate on the transport coefficients of $D$ mesons.
Nevertheless, the literature does offer transport coefficients for all four hadrons---$D$, $B$, $\Lambda_c$ and $\Lambda_b$---within models based on unitarized baryon-meson and meson-meson interactions derived from EFTs that uphold chiral and heavy-quark symmetries. We will present a comparison of the magnitude of the transport coefficients for all four hadrons, as reported in the Ref.~\cite{Tolos:2016slr}.
 
In Fig.~\ref{fig:LJSpi}, the variation of drag coefficient (left panel) and diffusion coefficient (right panel, denoted as $\Gamma_0$ instead of $B$) coefficients of $D$, $B$, $\Lambda_c$, and $\Lambda_b$ hadrons depicted as a function of temperature in the thermal bath consisting solely of $\pi$. If one compares the drag and diffusion coefficients for the $D$ meson in Fig.~\ref{fig:LJSpi} with those for the $\Lambda_c$ reported in Ref.~\cite{Tolos:2016slr}, a striking similarity becomes apparent. This similarity can be comprehended in terms of the kinetic theory expressions of the drag and diffusion coefficients in the non-relativistic limit,
\begin{align}
 A(T) & \sim P \sigma \sqrt{\frac{m_l}{T}}\frac{1}{m_H} \ ,  \label{eq:A} \\
 B(T) & \sim P \sigma \sqrt{m_lT} \ , \label{eq:B}
\end{align}
where $m_H$ is the mass of the heavy hadron, $m_l$ is the mass of the light hadron part of the thermal bath, $\sigma$ is the interaction cross-section between the heavy and light hadrons, and $P$ is the pressure from the light-particle bath. The pressure exerted by the bath of light particles can be assumed to be the same for both $D$ mesons and $\Lambda_c$ baryons. If we use the drag and diffusion coefficients of the $D$ meson as a baseline and assume the validity of Eq.~(\ref{eq:B}), we can anticipate that the discrepancy between the diffusion coefficients of the $D$ and $\Lambda_c$ will be determined solely by the differences in their cross sections. The $\Lambda_c\pi$ curve has a similar qualitative trend than that of $D\pi$ ~\cite{Tolos:2016slr}. However, the thermally averaged cross sections for scattering of $\Lambda_c\pi$ is a bit larger than that of $D\pi$ in size. Therefore, the diffusion coefficients of $\Lambda_c$ are expected to be greater than those of the $D$ meson, as shown in the right panel of Fig.~\ref{fig:LJSpi}. Similarly, according to \cite{Tolos:2016slr}, the cross sections for $D\pi$ and $B\pi$ interactions are generally comparable in magnitude. Therefore, one can anticipate comparable $\Gamma_0$ coefficients, as illustrated in the right panel of Fig.~\ref{fig:LJSpi}. The right panel of Fig.~\ref{fig:LJSpi} shows that the diffusion coefficient of $\Lambda_b$ baryons is quite close to that of $\Lambda_c$. Despite the complexity of the $\Lambda_b\pi$ cross section, assuming Eq.~\eqref{eq:B} holds true, one can infer that the average cross section for $\Lambda_b\pi$ will be comparable to that of $\Lambda_c\pi$. 

On the contrary, as expressed in Eq.~(\ref{eq:A}), the drag coefficients are influenced by both the average cross section and the mass of the heavy hadrons. Beginning with $\Lambda_c$ compared to the $D$ meson, the $\Lambda_c\pi$ cross section is slightly larger than the $D\pi$ interaction, thereby leading to an increase in its drag coefficients. However, due to the larger mass of $\Lambda_c$ compared to that of the $D$ meson, its drag coefficient is reduced. As a result of these opposing effects, the drag coefficients of $\Lambda_c$ and $D$ mesons are quite similar.  The interaction cross-section of $B$ mesons with pions is similar to that of $D\pi$. Therefore, their drag coefficients should scale with the inverse of their masses. Therefore, one would expect the drag coefficient of the $B$ meson to be reduced by a factor of $m_B/m_D \sim 2.8$ compared to the drag coefficient of the $D$ meson. This expectation is fully consistent with the results shown in the left panel of Fig.~\ref{fig:LJSpi}. From the analysis of the $\Gamma_0$, we infer that the average cross section for $\Lambda_b\pi$ will be comparable to that of $\Lambda_c\pi$. Hence, one would expect the drag coefficient of the $\Lambda_b$  to be reduced by a factor of $m_{\Lambda_b}/m_{\Lambda_c}$ compared to the drag coefficient of the $\Lambda_c$ meson. The result depicted in the left panel of Fig.~\ref{fig:LJSpi} aligns with this expectation.

Examining the validity of the commonly used geometry scale in literature for computing spatial diffusion coefficient ($\Ds$)~\cite{Chen:2020lcn} would be an intriguing pursuit. As previously mentioned, the spatial diffusion coefficient can be calculated using the formula: $\Ds = \frac{T}{m_H A}$. Thus, $\Ds$ exhibits a relationship with $\sigma$ where it is inversely proportional to the interaction cross-section.
If we consider heavy hadrons as hard spheres with a radius $R$, then $\Ds$ is approximately inversely proportional to $R^2$. In this context, $R$ can be interpreted as the charge radius of heavy hadrons.
Therefore, the spatial diffusion coefficient for $\Lambda_c$ ($\Ds^{\Lambda_c}$) can be expressed in terms of the spatial diffusion coefficient for $D$ mesons ($\Ds^{D}$) as  $\Ds^{\Lambda_c}=\Ds^{D} (R_D^2/R_{\Lambda_c}^2)$. The sensitivity of the charge radius to the type of hadrons under consideration has been noted in literature ~\cite{Hwang:2001th}. Consequently, this geometric scaling approach may not be applicable universally across all hadrons, including the heavy hadrons discussed in this context.
As indicated in Eq.(\ref{eq:B}), the diffusion coefficient is directly linked to the cross-section $\sigma$, with $\Gamma$ being proportional to $\sigma$. The equation aligns with the findings depicted in the left panel of Fig.~\ref{fig:LJSpi}. If we substitute the cross-section with $R^2$, the radius, the relation becomes $\Gamma \sim R^2$. In this case, the scaling may not function in a similar manner. 
In the absence of the cross-section, the charge radius of the heavy hadron can provide a rough estimate of the diffusion coefficient, typically within an error bar of about a factor of 2.

%%%%%%%%%%%%%%%%%%%%%%%%%%%%%%%%%%%%%%%%%%%%%%%
\section{Heavy-flavor phenomenology in heavy-ion collisions}
\label{sec:pheno}
%%%%%%%%%%%%%%%%%%%%%%%%%%%%%%%%%%%%%%%%%%%%%%%
Observables related to single-inclusive HF hadron production in heavy-ion collisions (HICs) may be broadly categorized into two types: \\
\begin{itemize}
\item[(1)] Kinematic distributions of individual HF flavor hadrons, \ie, 
their $\pT$ spectra (perpendicular to the colliding beam axis) and pertinent azimuthal-angle distributions, as well as longitudinal-momentum distributions (along the beam axis, usually quantified via the rapidity variable, $y$); and 
\item[(2)] Heavy-flavor hadro-chemistry, usually investigated through the production ratios of different HF hadrons species (\eg, $D$, $D^*(2010)$, $D_s$, $\Lambda_c$, $\Sigma_c$),  either inclusive (\ie, momentum-integrated) or differential  in $\pT$ or rapidity $y$. 
\end{itemize}
Both observable types carry their own unique potential  (that we have briefly alluded to already in the introduction), and they are both affected by the hadronic phase, which is the main topic of this section. In addition to these ``direct'' HF observables, there is a third category in terms of ``indirect'' measurements via the HF decay products, most notably single-electron and -muon spectra arising from the semileptonic decays of HF hadrons (also refereed to as non-photonic leptons). While experimentally the semileptonic HF decay spectra can be rather well identified through displaced-vertex techniques, their theoretical interpretation is more involved since they are affected by both the kinetics and chemistry of the various HF hadrons decaying to them. For example, $D^+$ mesons, $D^0$ mesons and $\Lambda_c$ baryons have rather different semileptonic decay branchings  (16\% vs.~6.5\% vs.~4\%, respectively), which implies that the single-lepton yields strongly depend on (changes in) the HF hadro-chemistry~\cite{Martinez-Garcia:2007mzr}. In addition to single-inclusive observables, correlation measurements can provide further insights. For example, the angular decorrelation of two HF hadrons produced from the same $Q\bar Q$ of the original hard process (which tend to carry a back-to-back correlation) has been put forward as a means to distinguish radiative and collisional processes in the medium. A similar idea pertains to the diffusion of HF particles within a jet~\cite{CMS:2019jis,STAR:2019qbf}. More recently, it has become possible to study two-particle femtoscopic correlation functions of $D$ mesons with protons and light hadrons in the final state~\cite{ALICE:2022enj,ALICE:2024bhk}, which are sensitive to their final-state interactions.

Let us briefly summarize some of the main insights that have been obtained from HF phenomenology in  HICs to date, starting with the kinematic observables. 
The general expectation is that HQ interactions with the constituents of the thermal bath produced in HICs modify the momentum and angular distribution of the HF hadrons as compared to those produced in proton-proton (pp) collisions. 
In experiments, this effect is commonly quantified via  the nuclear modification factor, defined as
\begin{equation}
    \raa(\pT)=\frac{\frac{dN_{\rm AA}}{d^2\pT dy}}{\left\langle\Ncoll\right\rangle\frac{dN_{\rm pp}}{d^2\pT dy}} \ ,
\end{equation}
where $dN_{\rm AA}/d^2\pT dy$ is the $\pT$ spectrum of a given HF hadron measured in nucleus-nucleus (AA) collisions (usually within a given selection of collision centralities), $dN_{\rm pp}/d^2\pT dy$ is the transverse momentum distribution of the HF hadron measured  in pp collisions, and $\Ncoll$ is the number of nucleon-nucleon interactions in AA collisions. A value of $\raa$ = 1 indicates that  there is no medium or nuclear effect, and the AA collision just amounts to a superposition of independent binary nucleon-nucleon collisions. 
It is important to note that the concept of the $\raa$, which was originally introduced to quantify suppression of high-$\pT$ particle production in HICs as a consequence of their energy loss in the medium, extends all the way down to $\pT=0$ for HF particles. This is so because HQ pair production, due to the large HQ mass, is expected to be essentially constrained to initial binary NN collisions and thus scale with the binary-collision number at all $\pT$. This renders HF an extraordinary probe of the soft properties of the QCD medium, including a essential assessment of its transport parameters. As such, it also offers the opportunity to explicitly study the transition diffusive processes at low $\pT$ driven by elastic interactions to the energy loss regime of radiative processes at high $\pT$.

Another key observable measured in experiments is the HF hadron elliptic flow, defined as
\begin{equation}
    v_2(\pT)=\langle \cos \ (2\phi) \rangle = \Biggr \langle \frac{p_x^2-p_y^2}{p_x^2+p_y^2} \Biggr \rangle \ ,
\end{equation}
which is a measure of the anisotropy in the angular distribution of the HF hadron $\pT$ spectra. Here, $\phi$ is the azimuthal angle of the hadron's transverse momentum in the transverse plane defined relative to the impact parameter of the AA collision and the beam axis of the incoming nuclei.  In a non-central collision, the initial nuclear overlap zone in the transverse plane is spatially anisotropic in the coordinate space (almond- or football-shaped). If the medium thermalizes rapidly enough (which in practice is the case), it develops larger pressure gradients along the short ($x$) axis of the almond than its long ($y$) axis. As a consequence expansion, the initial spatial anisotropy will be converted into an anisotropy in momentum space that can be measured in the final hadron spectra. The key question in the present context then is in how far the heavy quarks (or HF hadrons in the hadronic phase) get ``dragged'' along with collective medium flow, rendering the HF elliptic flow an excellent measure of their coupling strength to (and thus the transport coefficients of) the QCD medium. Much less attention has been devoted to the directed flow characterized by an anisotropy coefficient, $v_1$. Specifically for  $D$-mesons~\cite{STAR:2019clv,ALICE:2019sgg}, $v_1$ observables have been proposed to study of  pre-equilibrium dynamics in the earliest aftermath of a non-central AA collision, including strong magnetic fields. 

Original expectations based on perturbative-QCD (pQCD) interactions of heavy quarks in the QGP led to predictions for an $\raa\sim 0.6$ for $D$-mesons, and $\sim$~0.8-0.9 for $B$ mesons at intermediate $\pT\simeq 5$\,GeV~\cite{Djordjevic:2005db} in central Au-Au ($\sqrt{s_{NN}} = 200$ MeV) collisions at RHIC, and the $v_2$ was predicted at a few percent, well below that measured for light hadrons~\cite{Armesto:2005mz}. Therefore, the first set of experimental results~\cite{STAR:2006btx,PHENIX:2005nhb,PHENIX:2006iih} came as a surprise, showing a strong modification of the $\raa$ and an elliptic flow much larger than the pQCD-based expectations. Since then, intense theoretical efforts have been conducted to understand this phenomenon, which turned to be even stronger in 5\,TeV Pb-Pb collisions at the LHC, and we refer the reader to several recent reviews on this topic~\cite{Dong:2019byy,Dong:2019unq,He:2022ywp}. 
While a complete description of HQ dynamics in heavy-ion collisions is quite complex and requires considering its propagation in the pre-equilibrium phase, QGP phase, through hadronization and the hadronic phase, it seems fair to say that strong nonperturbative HQ interactions in the QGP and through hadronization are required to get near to the experimental findings~\cite{Rapp:2018qla}. 

However, it remains a central goal of the ongoing heavy-ion programs and the associated theory efforts to quantify the HQ diffusion coefficients in QGP, in particular also their temperature and momentum dependencies, which go hand-in-hand when conducting phenomenological analyses. Clearly, the hadronic phase will have to play an significant part in these.

As for the hadro-chemical aspects, the measurement of various HF hadrons provides a unique opportunity to study how a universal distribution of heavy quarks after their diffusion through partonic matter converts into different hadrons, at all $\pT$. Also in this aspect, the hadronic phase can be expected to be relevant; for example, the $D^*$/$D$ ratio may well subject to change when quasielastic $D+\pi \to D^*$ interactions remain active in the hadronic evolution of the fireball (as has been observed, \eg, for the $\rho^0$/$\pi$ ratio~\cite{STAR:2003vqj,ALICE:2018qdv,Rapp:2003ar}). And while the $\Lambda_c/D$ ratio is commonly believed to be an excellent probe of recombination processes during hadronization~\cite{STAR:2019ank,ALICE:2018hbc}, it may also be sensitive to differences in the diffusion properties of these two hadrons in the hadronic phase. 

Future experimental programs at RHIC~\cite{Shi:2024eyk} and the LHC~\cite{ALICE:2023bsp,CMS:2018bxx} place a large emphasis on HF observables, in particular precision $v_2$ measurements across different HF hadron species, notably also in the bottom sector.
The NA60+ experiment has been proposed at the CERN SPS with energies ranging from  $\sqrts$ = 5-7\,GeV per nucleon pair with a major focus on precision measurements of HQ production, where the hadronic phase is expected to be even more important.

The remainder of this section is organized as follows: In Sec.~\ref{ssec:D-obs} we revisit calculations of hadronic rescattering in AA collisions and their effects on the kinematic distributions of $D$-mesons as reflected in the $\raa$ and $v_2$ observables, and likewise for $B$ mesons and HF baryons in Sec.~\ref{ssec:B+baryon-obs}. In Sec.~\ref{ssec:chem} we review aspects of hadro-chemical observables in the HF sector and conjecture on possible roles of hadronic rescattering which has received very little attention thus far. In Sec.~\ref{ssec:assess} we finish with a critical assessment of the significance of hadronic HF interactions in heavy-ion collisions and in the broader context of QCD matter research.  

%%%%%%%%%%%%%%%%%%%%%%%%%%%%%%%%%%
\subsection{Hadronic effects on $D$-meson observables}
\label{ssec:D-obs}
%%%%%%%%%%%%%%%%%%%%%%%%%%%%%%%%%%%%%%%
In HICs heavy quarks are primarily generated in hard processes that can be studied through pQCD calculations. Consequently, their initial momentum distribution can be theoretically evaluated and verified by experiment. The initial HQ spectra are generally of power-law type and thus harder than what one would expect in a thermalized medium (with some caveat due to the collective flow of the medium). 
The HQ interactions with the constituents of the bulk matter drive their spectra toward local thermal equilibrium, but since the thermalization time is delayed by a factor of order $m_Q/T$ they generally do not reach full equilibrium (which is precisely why they are an excellent gauge of the interaction strength with the medium). 
The standard approaches to study the time evolution of the HQ spectra in the QGP are the relativistic Langevin equation~\cite{Svetitsky:1987gq,GolamMustafa:1997id,Moore:2004tg,vanHees:2004gq,vanHees:2005wb} and the Boltzmann transport equation~\cite{Gossiaux:2008jv,Uphoff:2012gb,Uphoff:2011ad,Das:2013kea}. The input to the Langevin approach are the drag and diffusion coefficients discussed in the preceding sections, while for the Boltzmann equation HQ cross sections with light quarks and gluons serve as the inputs. 
Accurately simulating the dynamics of heavy quarks at the collision energies of RHIC and LHC also relies on providing a realistic space-time evolution of the expanding medium. Hydrodynamic or transport simulations are usually employed to simulate the evolving QGP phase, which can describe the measured collective properties of the system. 
Once the temperature of the QGP phase reaches the one of the quark-hadron cross-over transition, the heavy quarks will hadronize and continue to interact with the surrounding hadronic phase consisting of pions, kaons, $\eta$ mesons and nucleons, as well as their excited states. 
The QGP evolution and hadronization of HF particles can thus be viewed as an initial condition to their hadronic diffusion.

In Sec.~\ref{sssec_qgp} we will first review basic features of the time evolution of the HQ spectra and their elliptic flow that are pertinent to the subsequent hadronic evolution of $D$ mesons. We then analyze the results of various model implementations, specifically the Langevin calculations of the Texas A\&M University group (TAMU) in Sec.~\ref{sssec_tamu}, the parton-hadron string dynamics calculations (PHSD) in Sec.~\ref{sssec_phsd}, the results of the Nantes group who combine Boltzmann transport in the QGP with Langevin simulations in the hadronic phase in Sec.~\ref{sssec_nantes}, and the Duke University model which utilizes Langevin transport in the QGP and Boltzmann transport within the UrQMD model for the hadronic phase in Sec.~\ref{sssec_duke}. In Sec.~\ref{ssec:femtoscopy} we briefly discuss rather recent developments utilizing hadronic 2-body correlation functions to assess final-state interactions, specifically for the case of $D\pi$ and $DK$ channels.

%%%%%%%%%%%%%%%%%%%%%%%%%%%%%%%%%%%%%%%
\subsubsection{Evolution of HQ spectra and elliptic flow in the QGP}
\label{sssec_qgp}
%%%%%%%%%%%%%%%%%%%%%%%%%%%%%%%%%%%%%%%
\begin{figure}[thb]
\centering
\includegraphics[scale=0.65]{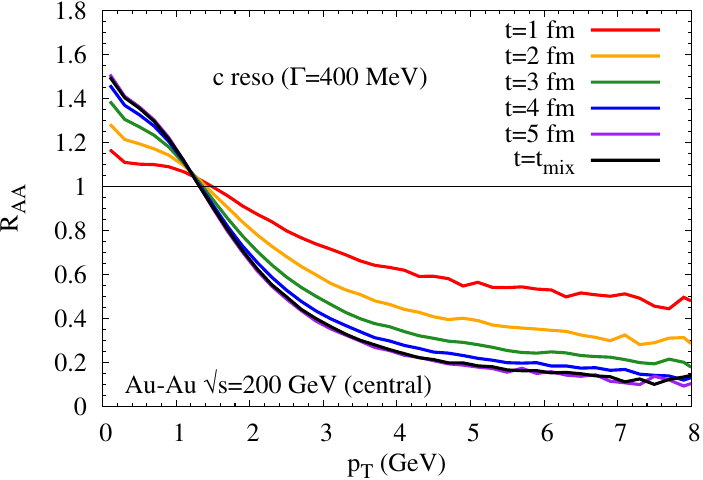}
\includegraphics[scale=0.65]{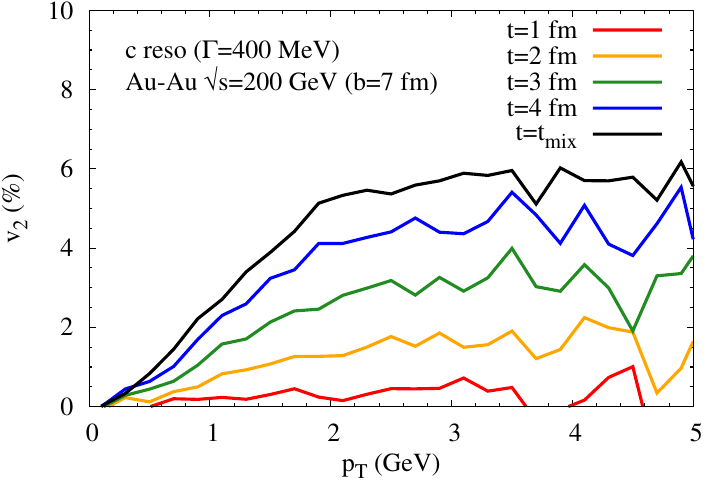}
\caption{The time evolution of the nuclear modification factor (left panel) and the elliptic flow (right panel) of charm quarks at RHIC, as computed within the TAMU transport approach. Figures adapted from Ref.~\cite{Rapp:2008qc}.}\label{fig:TTRaav2}
\end{figure}

To characterize HQ transport in the QGP phase, a quantitative assessment of heavy-hadron transport in the hadronic phase is required. To understand the possible impact of the hadronic phase on various experimental observables, the nuclear modification factor, $\raa$, and elliptic flow, $v_2$, it is instructive to understand the generic time evolution of both observables. This is shown in Fig.~\ref{fig:TTRaav2} for the resonance+pQCD model in the fireball evolution~\cite{vanHees:2005wb}, taken from the review in Ref.~\cite{Rapp:2008qc}. One finds that the $\raa$ of charm quarks is predominantly built up during the early stages of medium expansion, characterized by relatively high temperatures and energy densities, and tends to saturated within $t$=4-5~fm/$c$. This indicates that the $\raa$ may not be very sensitive to further changes in the hadronic phase following the QGP phase. However, the $v_2$ develops in the later stages of the evolution. Early on, the bulk medium does not carry significant $v_2$, as it requires time to develop it, with a time typical scale comparable to the medium size characterizing the initial spatial anisotropy. 
Subsequently, the bulk-matter $v_2$ in momentum space will be transferred to the heavy quarks due to its interaction with the thermal bath. This suggests that the time evolution of the HF $v_2$ will still be significant in the hadronic phase, even if the transport coefficients are not very large since the surrounding bulk-matter $v_2$ is substantial. 
Hence, the time evolution of the HF $\raa$ and $v_2$ is responsive to rather different stages of the fireball expansion, contrary to the strong correlation implied by the experimental results.  This will have important consequences for the interpretation of the experimental data to disentangle different energy loss models~\cite{PHENIX:2006iih,Das:2015ana} and to understand the impact of the hadronic phase on $\raa$ and $v_2$. 

\begin{figure}[thb] 
\centering
\includegraphics[scale=0.29]{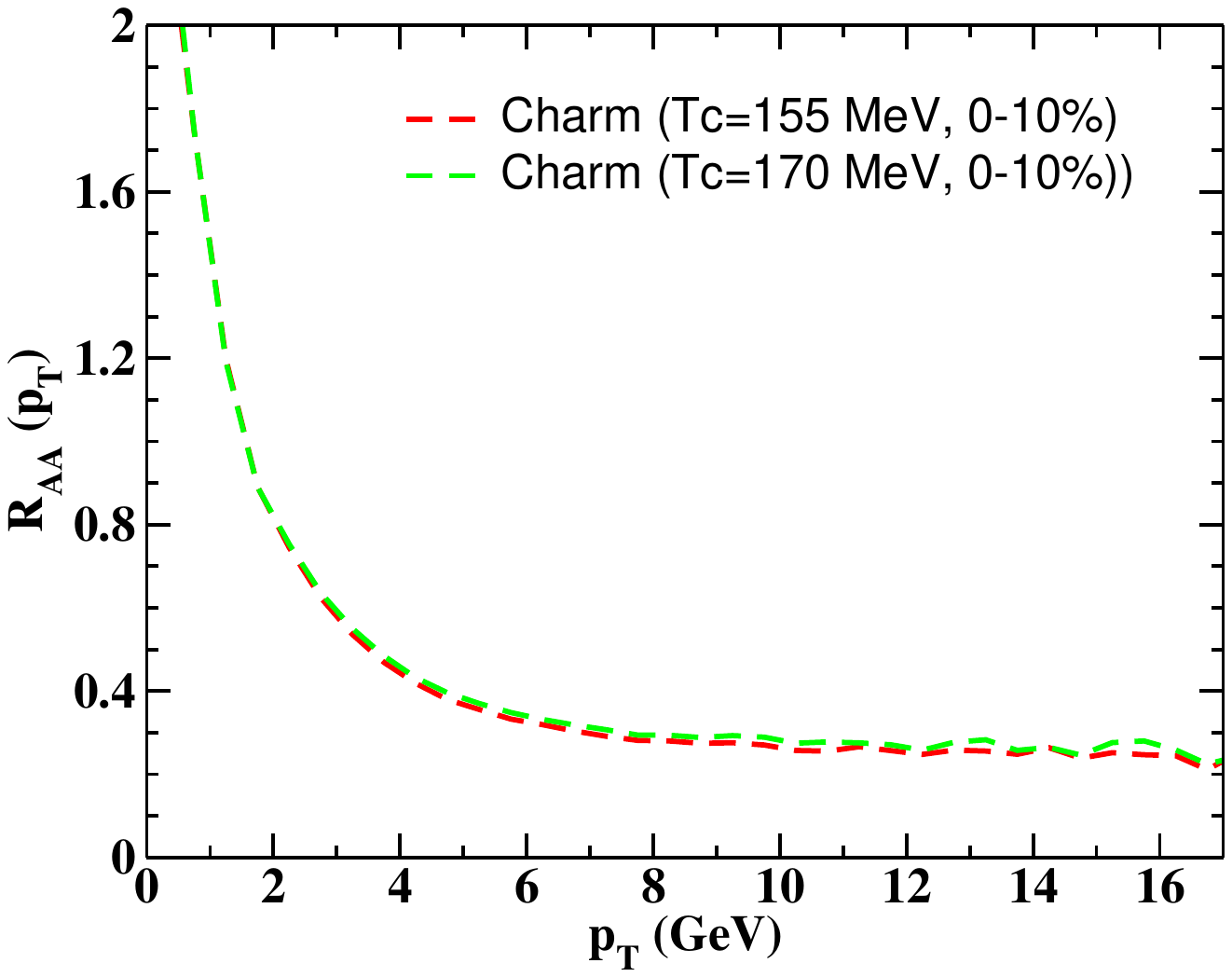}
\includegraphics[scale=0.29]{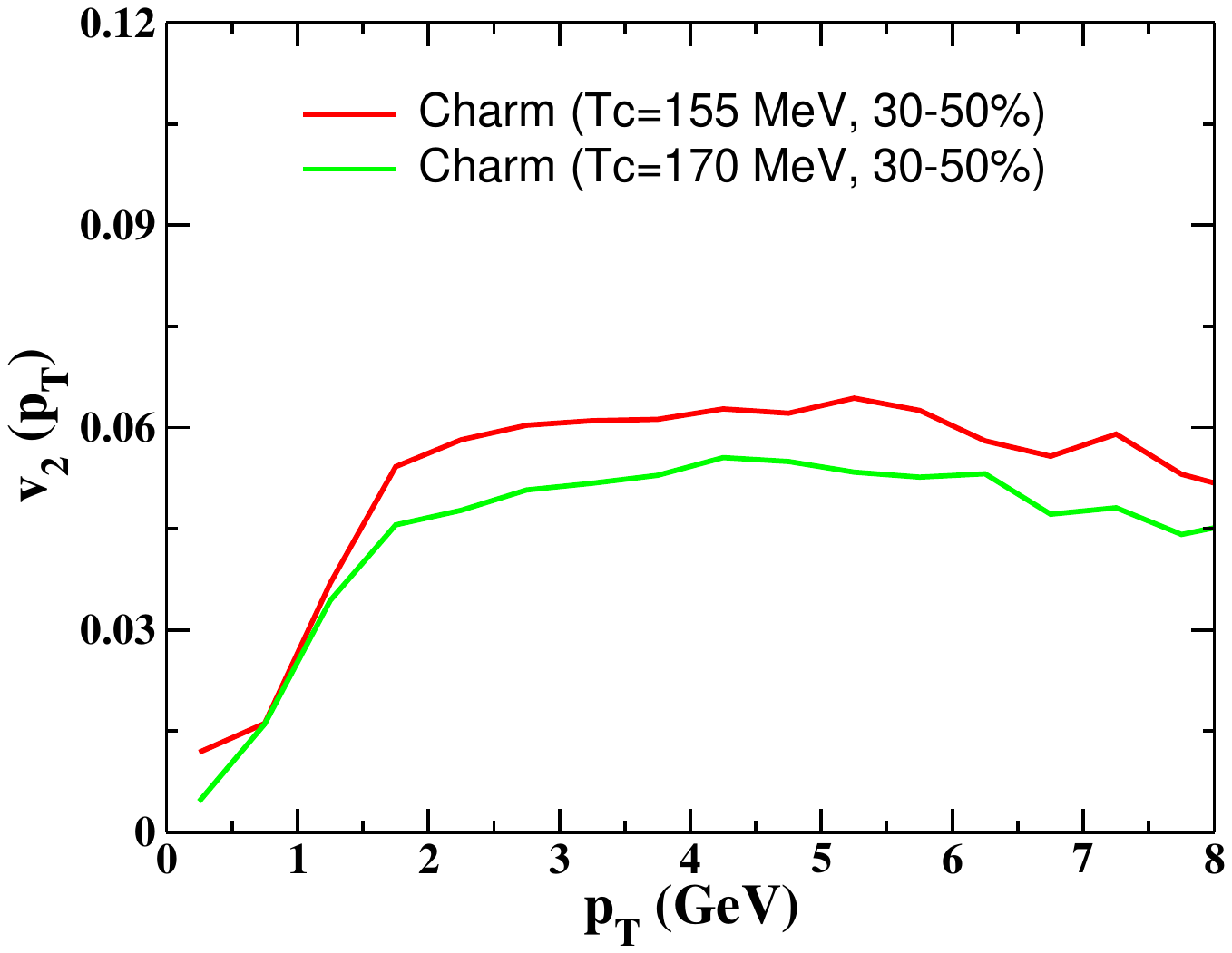}
\caption{Nuclear modification factor (left panel) and elliptic flow (right panel) of charm quarks at LHC, as computed within the Catania transport approach using pQCD*5 interaction; red and green lines correspond to the results at $T_{pc}=155$ MeV and $T_{pc}=170$ MeV, respectively. Figures adapted from Ref.~\cite{Rapp:2018qla}.}
\label{fig:raac}
\end{figure}

To further motivate the possible impact of hadronic phase, Fig.~\ref{fig:raac} depicts the impact of varying the pseudo-critical temperature, $\Tpc$ on the final spectra and $v_2$ of charm quarks as a function of $\pT$. This study utilizes the pQCD*5 interaction~\cite{Rapp:2018qla} within the Catania Langevin transport framework~\cite{Scardina:2017ipo}, examining two settings with varying critical temperatures: $\Tpc$=155 MeV (red lines) and $\Tpc$=170 MeV (green lines). As shown in Fig.~\ref{fig:raac}, the $\raa$ remains relatively unchanged, while the $v_2$ experiences a noticeable increase of up to 20$\%$ when the evolution is extended to lower temperatures. This finding aligns with prior studies~\cite{Rapp:2008zq, Das:2015ana}, which identify the suppression seen in the $\raa$ as primarily driven by density, with its greatest impact observed during the initial stages of the fireball. Conversely, the transfer of $v_2$ from the medium to heavy quarks demonstrates its highest effectiveness when the fireball's $v_2$ is large, typically occurring in the later phases of evolution closer to $\Tpc$. This effect becomes more pronounced when the coupling strength of the medium reaches a maximum near $\Tpc$. The pQCD*5 interaction model employed in this study does not feature an enhanced strength near $\Tpc$. Consequently, the anticipated increase in HQ $v_2$ near $\Tpc$ is expected to be even more noticeable in situations involving nonperturbative interactions~\cite{vanHees:2007me}. Again, this is suggestive for a further development of the HF $v_2$ in the hadronic phase, which in the following four section  will be investigated quantitatively in various model implementations.

A similar study has been carried out in Ref.~\cite{Inghirami:2018vqd} at FAIR/SPS energies, in Au-Au collisions with a beam energy of 25\,GeV (corresponding to $\sqrts\simeq7$\,GeV). A coarse-graining procedure of the UrQMD transport model has been carried out to obtain local temperatures and baryochemical potentials (alternatively, ideal hydrodynamic simulations initialized by UrQMD have also been considered). These results have then been employed as a background medium for Langevin simulations of $c$-quarks in a QGP (using the nonperturbative resonance model of Ref.~\cite{vanHees:2004gq}) and $D$ mesons in hadronic matter (using the coefficients from the EFT calculations of Ref.~\cite{Torres-Rincon:2014ffa}), both including the effects of finite baryo-chemical potentials, and connected with a coalescence/fragmentation model for hadronization. It was found that the $D$-meson $v_2$ right after hadronization varies by up to a factor of $\sim$2-3 at $\pT\simeq1.5$\,GeV when varying the hadronization temperature from $T$=160\,MeV down to 145 or 130\,MeV. On the other hand, the $\raa$ was much less affected, although it showed an overall interesting effect of becoming very large toward the kinematic limit in pp collisions at this collision energy, $\pT^{\rm pp,max} \simeq 1.6-2$\,GeV.

%%%%%%%%%%%%%%%%%%%%%%%%%%%%%%%%%%
\subsubsection{TAMU model}
\label{sssec_tamu}
%%%%%%%%%%%%%%%%%%%%%%%%%%%%%%%%%%%%%%%
\begin{figure}[thb]
\centering
\includegraphics[scale=0.28]{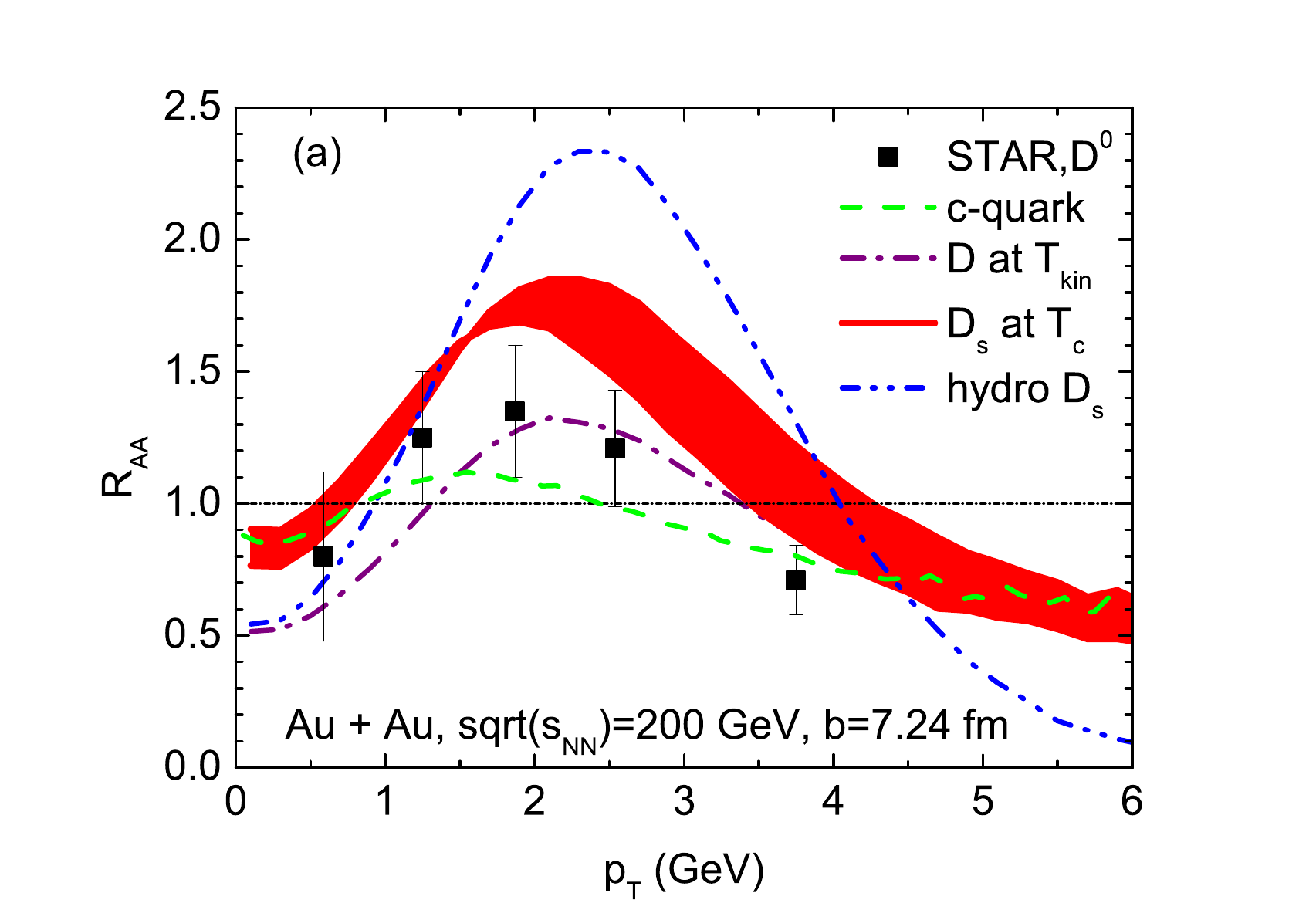}
\includegraphics[scale=0.28]{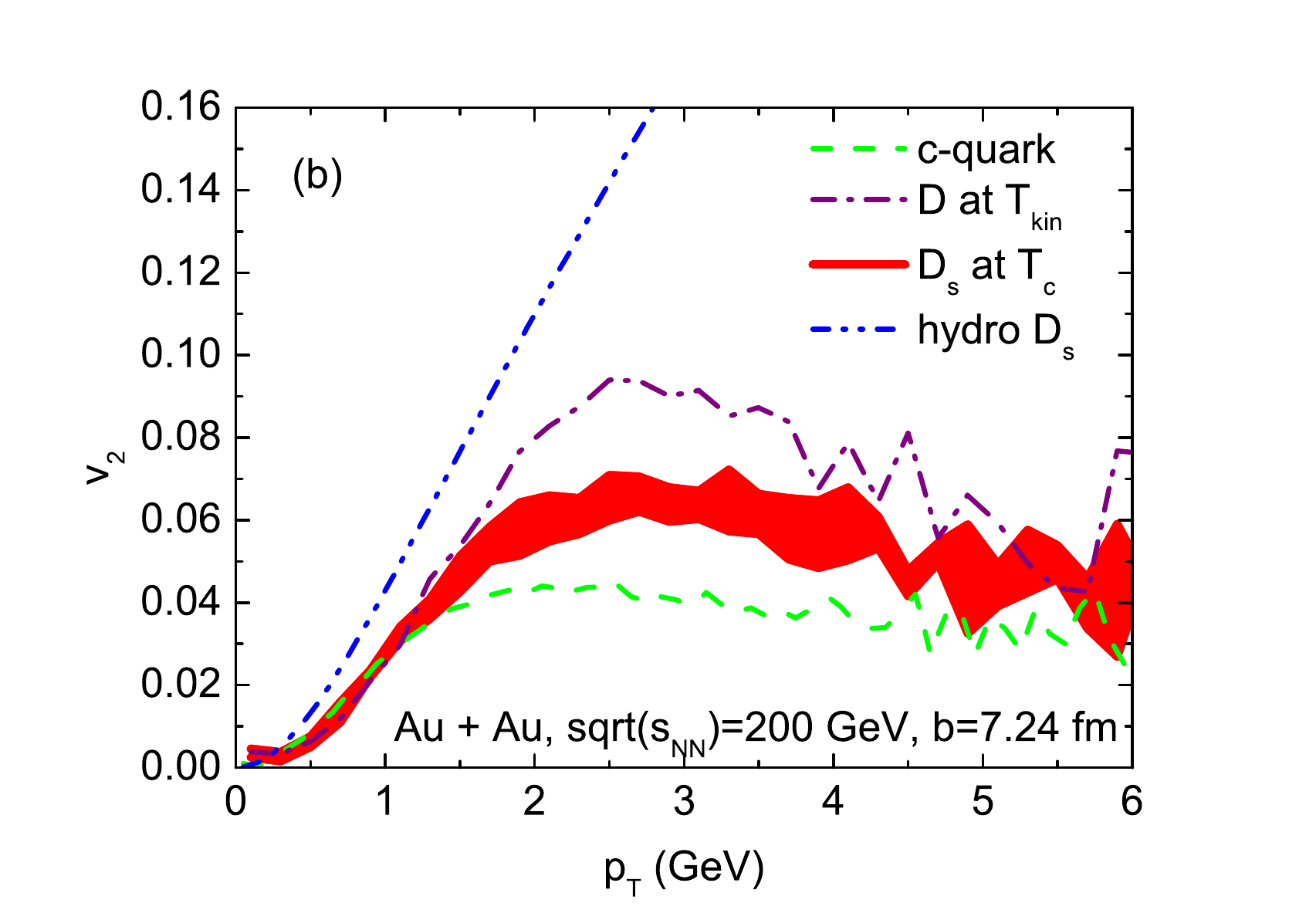}
\caption{Nuclear modification factor (left panel) and elliptic flow (right panel) of charm quarks (green dashed line), $D$-mesons (purple dashed-dotted line) and $D_s$-mesons (red line) at RHIC, as computed within the TAMU transport approach; the blue dashed double-dotted line represents the equilibrium limit for $D_s$ mesons in the hydrodynamic medium. The red uncertainty band in the nuclear modification factor is due to the inclusion or omission of a Cronin effect in the initial charm spectra. The red uncertainty band in the elliptic flow is due to the inclusion or neglect of diffusion effects in the hadronization window. Figures adapted from Ref.~\cite{He:2012df}.}
\label{fig:THRaav2_RHIC}
\end{figure}
One of the first attempts to characterize the impact of the hadronic phase was reported in Ref.~\cite{He:2012df} (see also Ref.~\cite{Das:2013lra} for an early more schematic study). In accordance with the widely accepted and empirically supported idea, the authors argued that multi-strange hadrons decouple not only chemically but also kinetically~\cite{STAR:2008bgi,He:2010vw} in the proximity of the hadronization transition; the main argument for this is that these hadrons largely lack the resonant cross sections (like $\pi N\to\Delta\to\pi N$ or $\pi K\to K^*\to\pi K$) that are abundantly operative in the light sector; a similar decoupling is then expected for $D_s$ meson. Therefore, by calculating the impact of the hadronic diffusion on the $D$ meson, a quantitative assessment of the hadronic transport coefficient can be achieved through a comparison of observables with the $D_s$. In this study, the dynamics of charm quarks was studied at RHIC energy using relativistic Langevin simulations. 
During the QGP phase, they utilized nonperturbative $T$-matrix calculations~\cite{vanHees:2007me,Riek:2010fk} for the HQ transport coefficients, with input potentials approximated by HQ internal energies computed in thermal LQCD. The resulting $T$-matrices reveal resonant states near the threshold in mesonic ($D$ and $D^*$) and color-antitriplet diquark channels close to the pseudo-critical temperature. These resonances  substantially augment the interaction strength, by a factor of around $\sim$5 compared to the pQCD Born amplitudes. The space-time evolution of the medium was approximated through boost-invariant ideal hydrodynamics. After undergoing diffusion through the QGP, the charm-quark distributions are converted into charmed hadrons by utilizing the resonance recombination model (RRM)~\cite{Ravagli:2007xx}, wherein thermal light and strange quarks combine into $D$ and $D_s$ mesons on the hydrodynamic hypersurface at a critical temperature of $\Tpc$=170\,MeV)~\cite{He:2011qa}. 
The strength of the interaction of the $D$ meson with light mesons and baryons has been assessed using existing microscopic models for $D$-hadron scattering~\cite{He:2011yi}. These models are constrained by considerations of chiral symmetry and vacuum spectroscopy, as described in Sec.~\ref{sec:charm} for $D$ mesons, and the pertinent transport coefficients are displayed in Fig.~\ref{fig:A_D-tamu}.

Figure~\ref{fig:THRaav2_RHIC} (left panel) displays the variation of $\raa$ of the $D$ and $D_s$ mesons in semicentral Au+Au collisions at RHIC; 
the results for the $D$-meson $\raa$  are consistent with STAR data in the  0-80\% centrality class~\cite{Zhang:2011uva}. A significant enhancement in the $D_s$-meson $\raa$ was found, which, however, is associated with a chemistry effect in the hadronization process due to charm-quark recombination with the enhanced strangeness abundance in Au+Au collisions, relative to pp. To provide further illustration of this phenomenon, the results are compared to the $\raa$ for charm quarks at $\Tpc$, representing the spectra of $D$ and $D_s$ mesons if coalescence were not present and only $\delta$-function fragmentation was applied. 
The impact of coalescence is evident, notably diminishing beyond $\pT$ values of approximately 5 GeV where fragmentation takes precedence, leading to the convergence of the  $\raa$ for $D$, $D_s$, and charm quarks. It was furthermore found that hadronic diffusion has a minimal impact on the $\raa$ of $D$ mesons (and was absent by construction for the $D_s$), attributed to a compensatory effect arising from the simultaneous decrease in temperature and increase in the flow of the medium.
Fig.~\ref{fig:THRaav2_RHIC} (right panel) displays the $\pT$ dependence of the $v_2$ of the $D$ and $D_s$ meson in semicentral Au+Au collisions at RHIC. While the charm-quark diffusion in the QGP imparts a sizable $v_2$ reaching up to approximately 4.5\%, coalescence with thermal light and strange quarks enhances this value by about 50 $\%$ for both $D$ and $D_s$ mesons. However, even as the $D_s$ spectra undergo freeze-out after hadronization, the coupling of $D$ meson to the hadronic medium persists~\cite{He:2011yi}. This coupling enables the $D$-meson to pick up more of the elliptic flow of the bulk medium generated during the QGP phase, leading to an additional increase in $v_2$ by approximately 30 $\%$ due to its interaction in the hadronic phase (dash-dotted line compared to the upper end of the red band for $D_s$ mesons in Fig.~\ref{fig:THRaav2_RHIC}). 
These findings support the expectations based on Fig.~\ref{fig:TTRaav2}, demonstrating that $D$-meson $\raa$ remains largely unaffected by interactions in the hadronic phase, while, in contrast, the impact on $v_2$ is notable, being sensitive to the later stages of the evolution with a significant enhancement generated by interactions in the hadronic phase.
More quantitatively, the $v_2$ splitting between $D$ and $D_s$, inspired by the early freeze-out of multistrange hadrons in the underlying hydro-evolution, has been suggested as a measure of the $v_2$ generated in the hadronic evolution, remaining essentially independent of the modeling of QGP diffusion and hadronization. Even in cases where the $D_s$ undergoes re-interactions in the hadronic phase, the $v_2$ splitting between $D$ and $D_s$ continues to serve as a lower limit for the transport coefficient in the hadronic medium. 

\begin{figure} 
\centering
\includegraphics[scale=0.28]{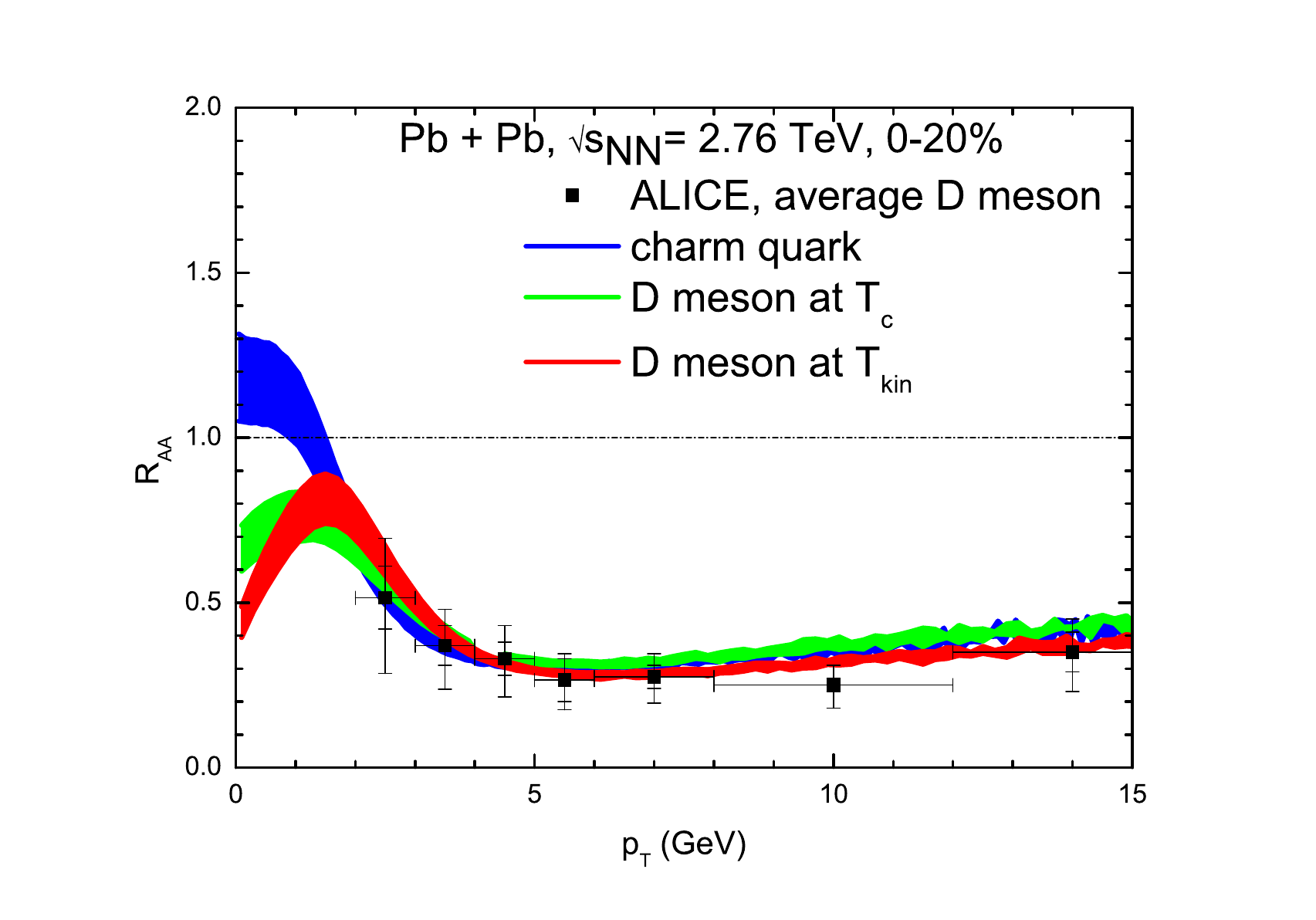}
\includegraphics[scale=0.28]{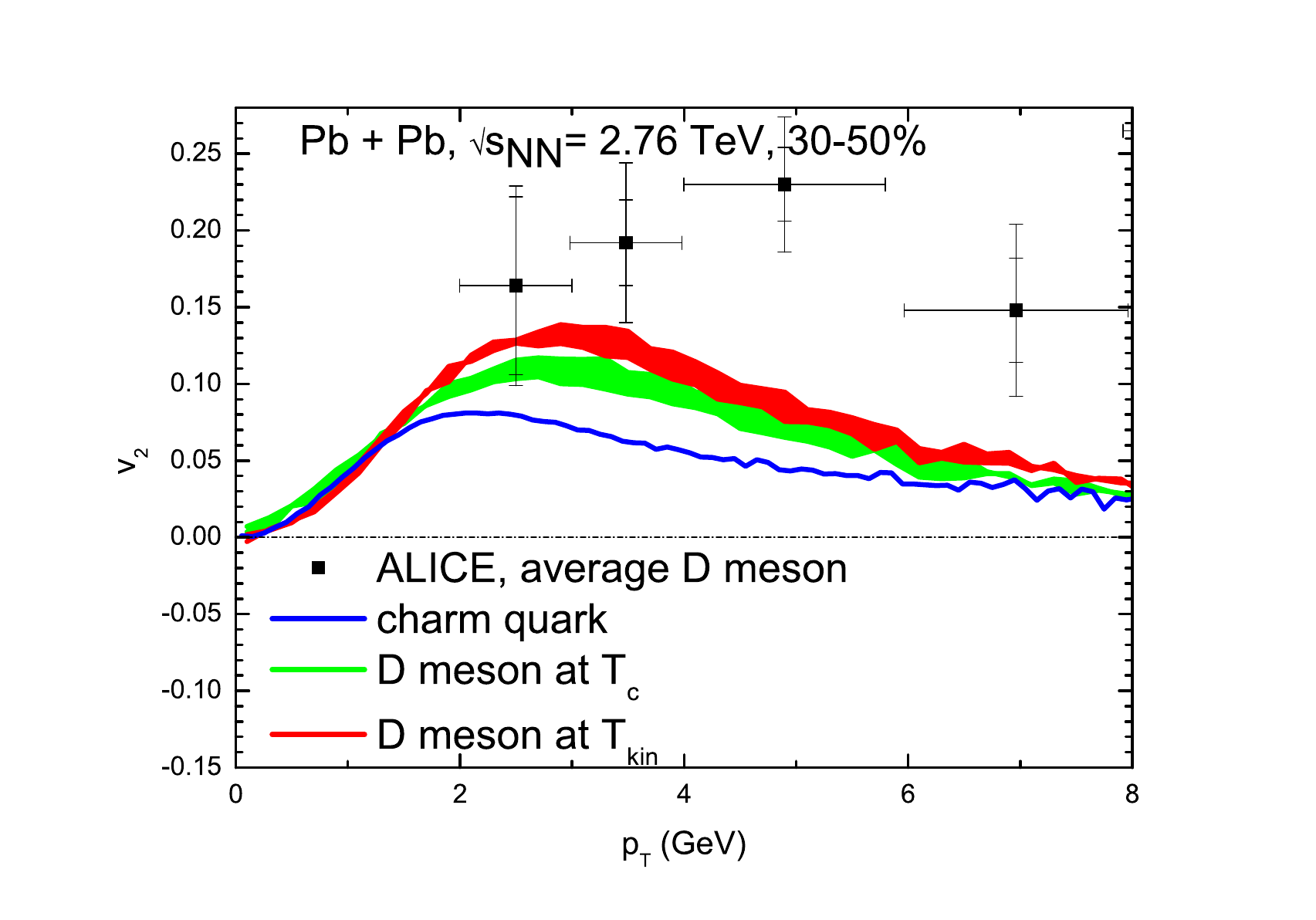}
\caption{Nuclear modification factor (left panel) and elliptic flow (right panel) of charm quarks (blue line) and $D$-mesons (red and green lines) at LHC, as computed within the TAMU transport approach; the red and green lines correspond to the $D$-meson results with and without the contribution of hadronic diffusion, respectively. In the nuclear modification factor, the bands indicate uncertainties due to the shadowing of charm production. In the elliptic flow, the bands indicate uncertainties due to the total charm-quark coalescence probability. Figures adapted from Ref.~\cite{He:2014cla}.}
\label{fig:THRaav2_LHC}
\end{figure}

A similar study has been conducted at LHC energies within the same model. In Fig.~\ref{fig:THRaav2_LHC} (left panel), the $\raa$ as a function of $\pT$ is shown for the charm quarks, $D$ mesons just after hadronization and $D$ mesons after the kinetic freeze-out in central Pb+Pb collisions at $\sqrts$=2.76\,TeV. The HQ diffusion in the QGP is calculated by via Langevin dynamics with transport coefficients from a thermodynamic $T$-matrix approach~\cite{Riek:2010fk}, and hadronization is realized in the RRM, and the hadronic diffusion utilizes the hadronic coefficients from Fig.~\ref{fig:A_D-tamu}~\cite{He:2011yi}. Again, a relatively minor impact on the $\raa$, the larger radial flow at the LHC leads to a slight enhancement and shift to higher $\pT$ of the maximum structure (``flow bump'') at low $\pT$. From the right panel of Fig.~\ref{fig:THRaav2_LHC}, which shows the $\pT$ dependence of the elliptic flow coefficient, one finds that the $D$-meson $v_2$ just after hadronization is enhanced due to the hadronic diffusion until kinetic freeze-out by a maximum of up to $\sim$25\%. 
The observed increase in elliptic flow due to diffusion in the hadronic phase appears to play an important factor in the simultaneous description of both observables, the $\raa$ and $v_2$.

%%%%%%%%%%%%%%%%%%%%%%%%%%%%%%%%%%
\subsubsection{PHSD model}
\label{sssec_phsd}
%%%%%%%%%%%%%%%%%%%%%%%%%%%%%%%%%%%%%%%
\begin{figure} 
\centering
\includegraphics[scale=0.35]{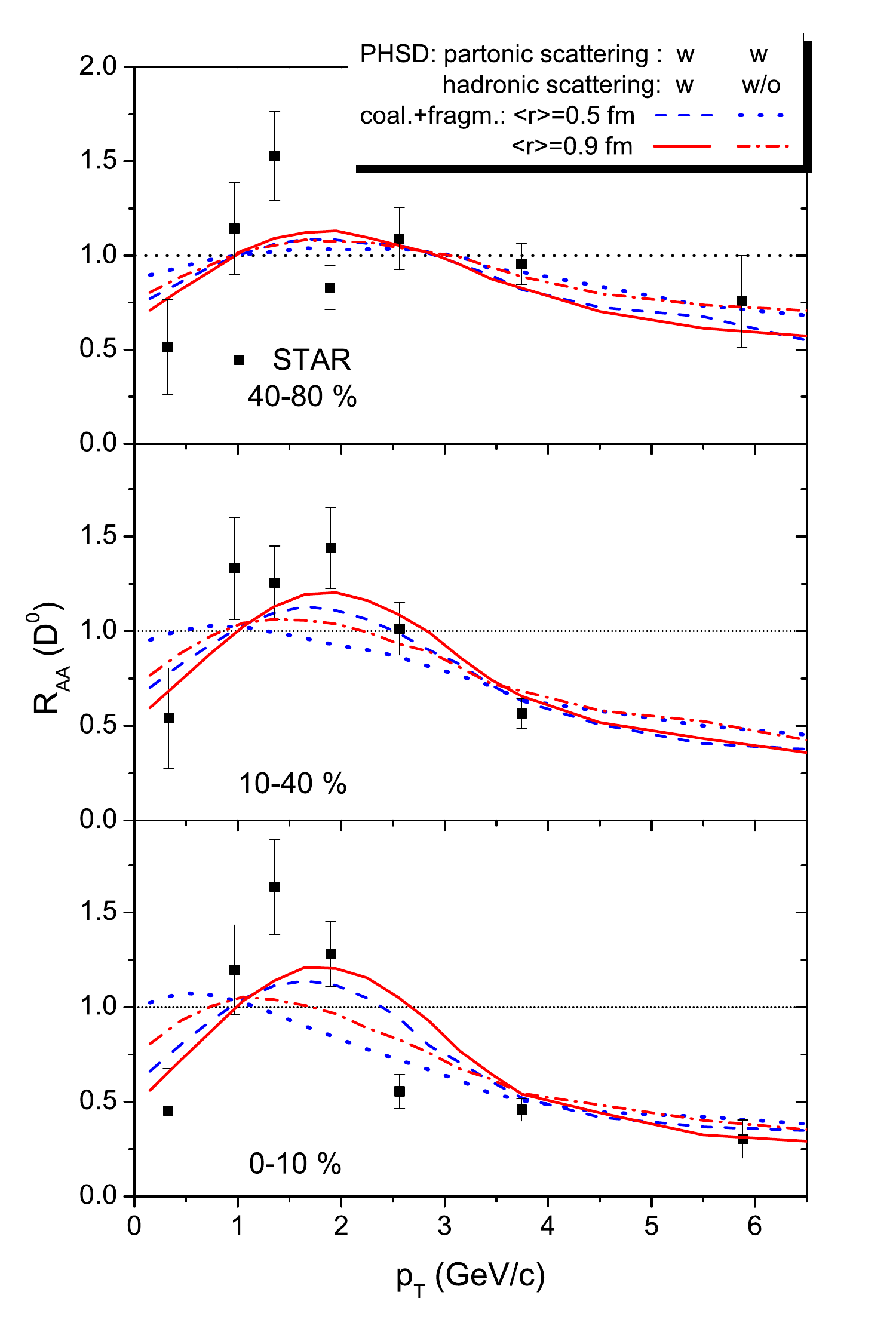}
\caption{ Nuclear modification factor of $D$-mesons at RHIC, as computed within the PHSD transport approach; the red solid line and blue dashed line corresponds to results with contribution of hadronic diffusion. The red dot-dashed line and blue dotted line represents the results without contribution of hadronic diffusion.  Figures adapted from Ref.~\cite{Song:2015sfa}.}
\label{fig:FHRaa_RHIC}
\end{figure}
The production of charm quarks and hadrons in ultra-relativistic heavy-ion collisions has been investigated through the Parton-Hadron-String Dynamics (PHSD) transport approach~\cite{Song:2015sfa}. The PHSD model~\cite{Cassing:2008sv,Cassing:2009vt} is based on a microscopic off-shell transport methodology designed for elucidating the behavior of strongly interacting hadronic and partonic matter, both in and out-of equilibrium. It relies on solving the Kadanoff–Baym equations within a first-order gradient expansion in phase space. This framework enables the depiction of the time evolution of nonperturbative interacting systems.

The transport of HF particles in HICs considers the evolution in both the partonic and hadronic phases. Charm quarks generated in primordial NN collisions undergo scattering in the QGP with off-shell partons whose  masses and widths are determined by the Dynamical Quasi-Particle Model (DQPM)~\cite{Cassing:2008nn,Cassing:2007nb} via fits to reproduce the LQCD equation-of-state in thermal equilibrium. The interactions of charm quarks with the partons in the QGP have been obtained using DQPM propagators and couplings~\cite{Berrehrah:2013mua}.
In PHSD, charm quarks undergo hadronization into $D$ mesons through coalescence and fragmentation processes near the critical energy density of the phase transition. Subsequently, the hadronized $D$ mesons interact with various hadrons in the hadronic phase, with cross sections calculated using a unitarized approach effective theory approach that incorporates heavy-quark spin symmetry~\cite{Abreu:2011ic,Tolos:2013kva}, cf.~Sec.~\ref{ssssec:unitarity}.

In Fig.~\ref{fig:FHRaa_RHIC}, the $D$-meson $\raa$ is displayed as a function of $\pT$ with and without hadronic scattering in Au-Au collisions at different centrality classes for the highest RHIC energy obtained within PHSD transport approach. The results for charm-quark hadronization through coalescence are presented for two different radii. It is found that hadronic rescattering plays a significant role in the $\raa$ at low $\pT$, by enhancing the flow bump by up to 20\% and shifting it to higher $\pT$ by up to $\sim$ 1 GeV. The suppression observed around $p_T \sim $ 6 GeV due to hadronic interactions could be a result of the enhancement at $p_T \sim $ 2 GeV, driven by charm number conservation.

\begin{figure} 
\centering
\includegraphics[scale=0.35]{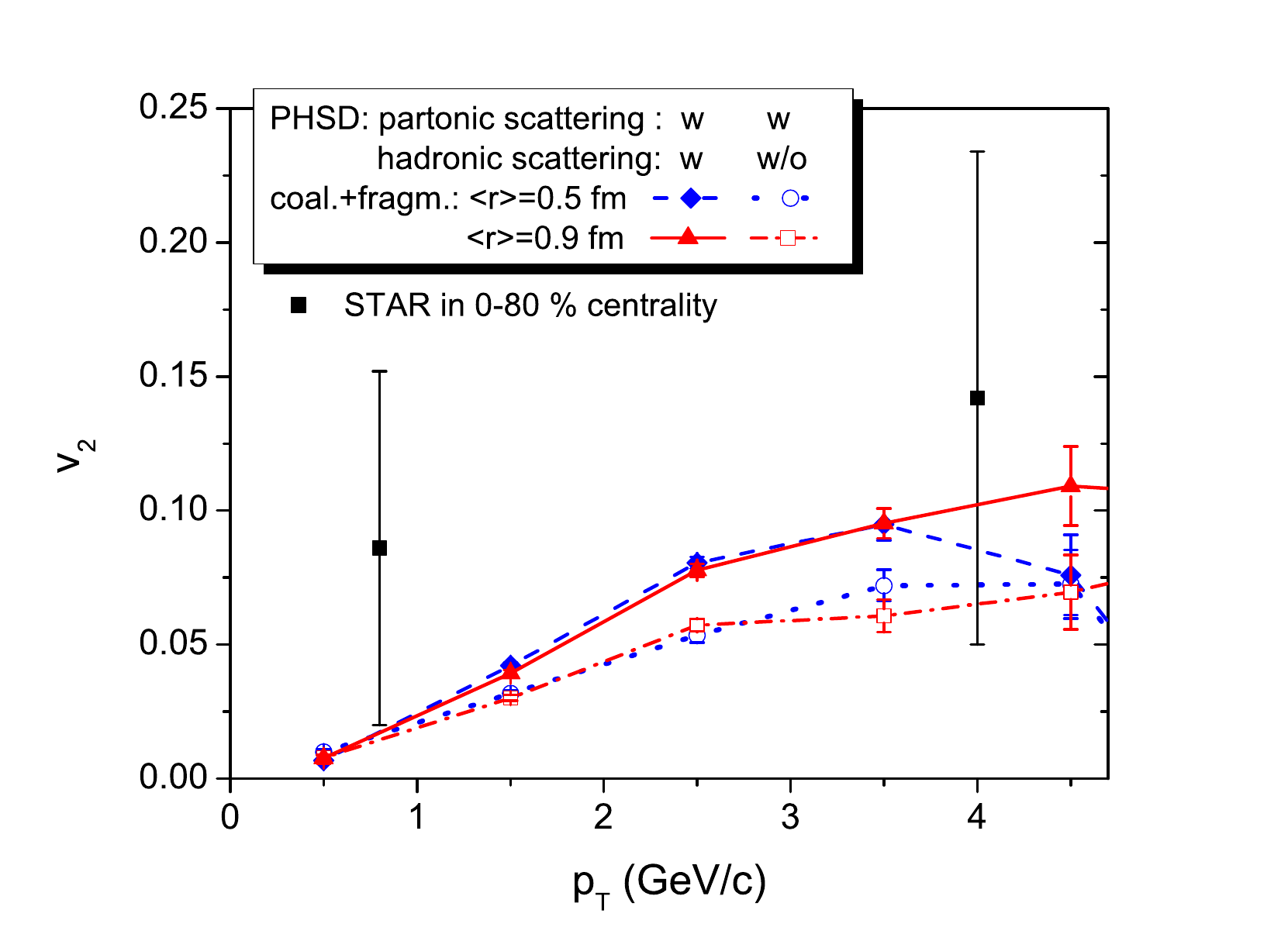}
\caption{The elliptic flow of of $D$-mesons at RHIC, as computed within the PHSD transport approach; the red solid line and blue dashed line corresponds to results with contribution of hadronic diffusion. The red dot-dashed line and blue dotted line represents the results without contribution of hadronic diffusion.  Figures adapted from Ref.~\cite{Song:2015sfa}.}
\label{fig:FHv2_RHIC}
\end{figure}
In Fig.~\ref{fig:FHv2_RHIC},  the $D$-meson $v_2$ is displayed as a function of $\pT$ with and without hadronic scattering for 0-80\% Au-Au collisions at the highest RHIC energy, as  obtained from PHSD. The hadronic rescattering has a very sizable effect, producing an enhancement of up to a maximum of near 40$\%$.

%%%%%%%%%%%%%%%%%%%%%%%%%%%%%%%%%%
\subsubsection{Nantes model}
\label{sssec_nantes}
%%%%%%%%%%%%%%%%%%%%%%%%%%%%%%%%%%%%%%%
\begin{figure} 
\centering
\includegraphics[scale=0.45]{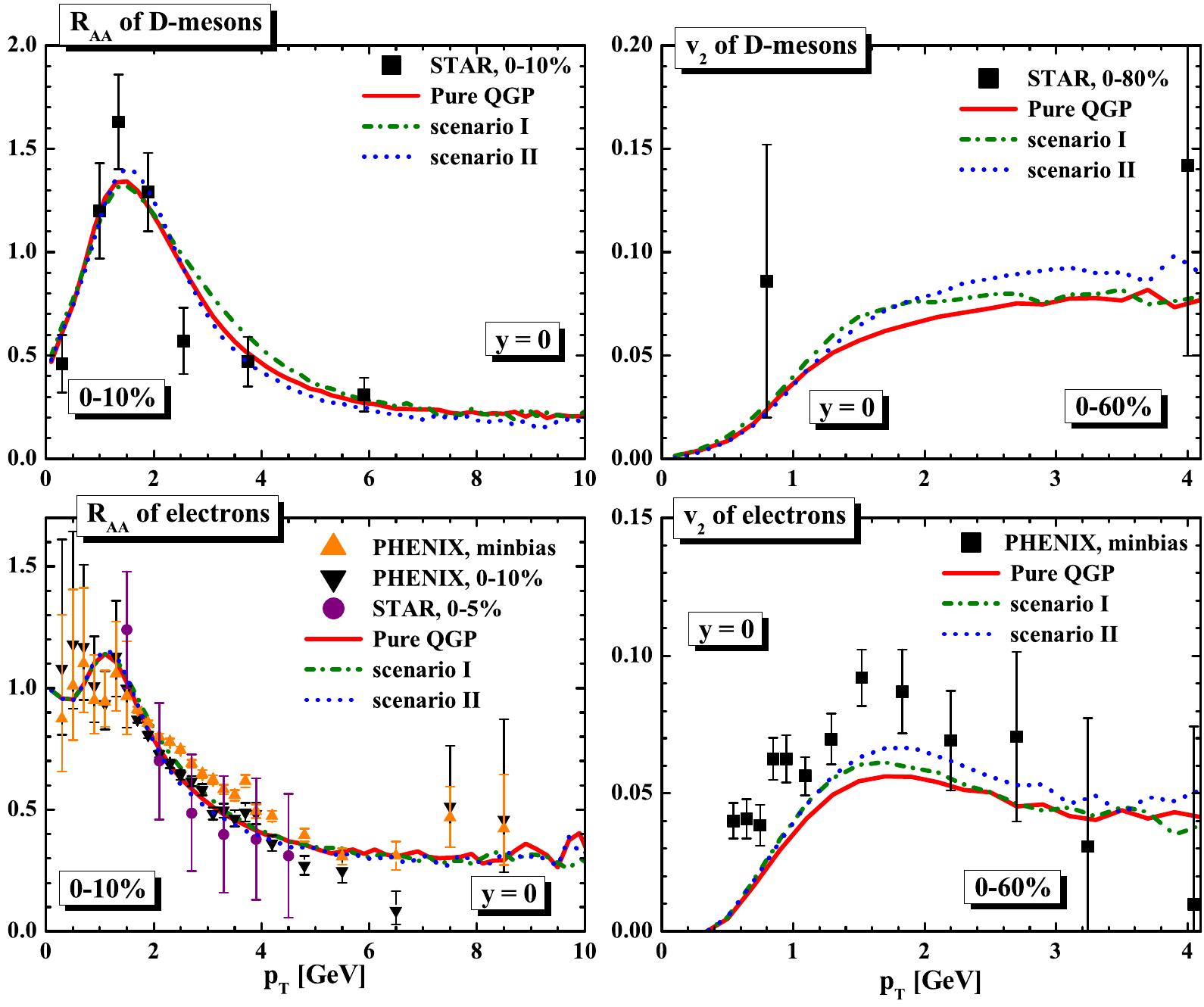}
\caption{Nuclear modification factor (left panel) and elliptic flow (right panel) of $D$-mesons (top) and single non-photonic electrons (bottom) originating from the decays of HF mesons at RHIC, as computed within Nantes transport approach;  green dot-dashed and blue dotted lines corresponds to results with contribution of hadronic diffusion. The red solid line represents the results without contribution of hadronic diffusion. Figures adapted from Ref.~\cite{Ozvenchuk:2014rpa}.}
\label{fig:SHRaav2_RHIC}
\end{figure}
The effects of hadronic phase rescattering on the propagation of $D$ mesons within the hadronic medium, along with its implications for heavy-flavor observables, have further been studied by the Nantes group~\cite{Ozvenchuk:2014rpa}. In their investigation, the authors employ Monte Carlo propagation of heavy quarks, denoted as MC@sHQ, within a 2+1D expanding ideal hydro medium~\cite{Gossiaux:2008jv,Gossiaux:2010yx}. The plasma evolution starts from smooth initial conditions and an equation of state featuring a first-order phase transition. The interaction between heavy quarks and plasma partons occurs through elastic and radiative collisions. The transport of heavy quarks is carried out with  the Boltzmann equation, solved using the test particle method, employing Monte Carlo techniques. The hadronization of heavy quarks is performed when the energy density of a fluid cell falls below a critical value, taken as $\epsilon_c$ = 0.45 GeV/fm$^3$. This process involves a combination of coalescence and fragmentation mechanisms~\cite{Gossiaux:2009mk}. The propagation of $D$ mesons in the hadronic medium is treated using the Fokker-Planck (FP) equation~\cite{Svetitsky:1987gq} using elastic scattering of $D$ mesons. The cross sections, needed to compute the transport coefficients, for $D\pi\rightarrow D\pi$, $D\eta\rightarrow D\eta$, and $DK(\bar K)\rightarrow DK(\bar K)$, as well as for $DN(\bar N)\rightarrow DN(\bar N)$ and $D\Delta(\bar\Delta)\rightarrow D\Delta(\bar\Delta)$, are taken from Ref.~\cite{Tolos:2013kva}, recall Sec.~\ref{ssssec:unitarity}. In addition, elastic scattering processes involving $D$ mesons with higher excited mesons $m$ (not listed above), $Dm\rightarrow Dm$, are simulated with a constant cross section of $\sigma=10~{\rm mb}$, motivated by the work of He {\it et al.}~\cite{He:2011yi}.
The authors then conducted calculations in two different scenarios for the hadronic diffusion~\cite{Ozvenchuk:2014rpa}. In scenario-I, they implemented the drag and diffusion coefficients from their calculations without imposing the Einstein relation. Consequently, the asymptotic solution of the FP equation does in general not agree with the equilibrium distribution function. In scenario-II, they utilized the drag coefficient obtained from their calculations and imposed the Einstein relation for the momentum- diffusion coefficients. The Einstein relation ensures that the asymptotic distribution function conforms to the Boltzmann distribution function. In Fig.~\ref{fig:SHRaav2_RHIC}, the variation of the final $D$-meson  $\raa$ and $v_2$ for $D$ mesons is illustrated as a function of $\pT$, along with the corresponding quantities for single non-photonic electrons (which also include bottom decay contributions), and compared with experimental data from the PHENIX and STAR collaboration in Au-Au collisions at  maximum RHIC energy.  
Based on these results it has been concluded that the rescattering of $D$ mesons in the hadronic gas has a negligible impact on the $\raa$ but leads to an increase of up to $\sim$20\% of the elliptic flow for both $D$ mesons and their decay electrons.

%%%%%%%%%%%%%%%%%%%%%%%%%%%%%%%%%%
\subsubsection{Duke model}
\label{sssec_duke}
%%%%%%%%%%%%%%%%%%%%%%%%%%%%%%%%%%%%%%%
\begin{figure} 
\centering
\includegraphics[scale=0.32]{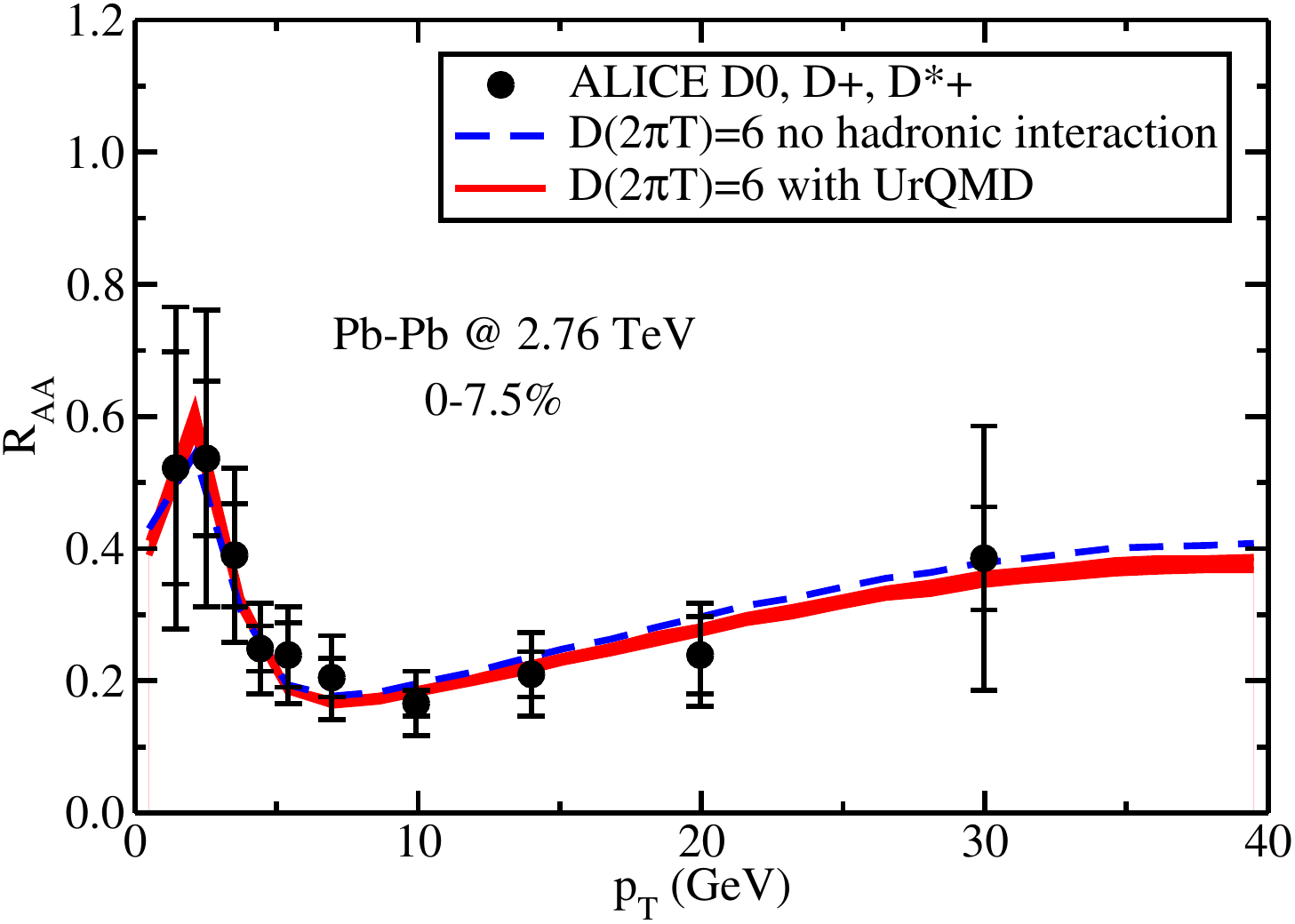}
\includegraphics[scale=0.32]{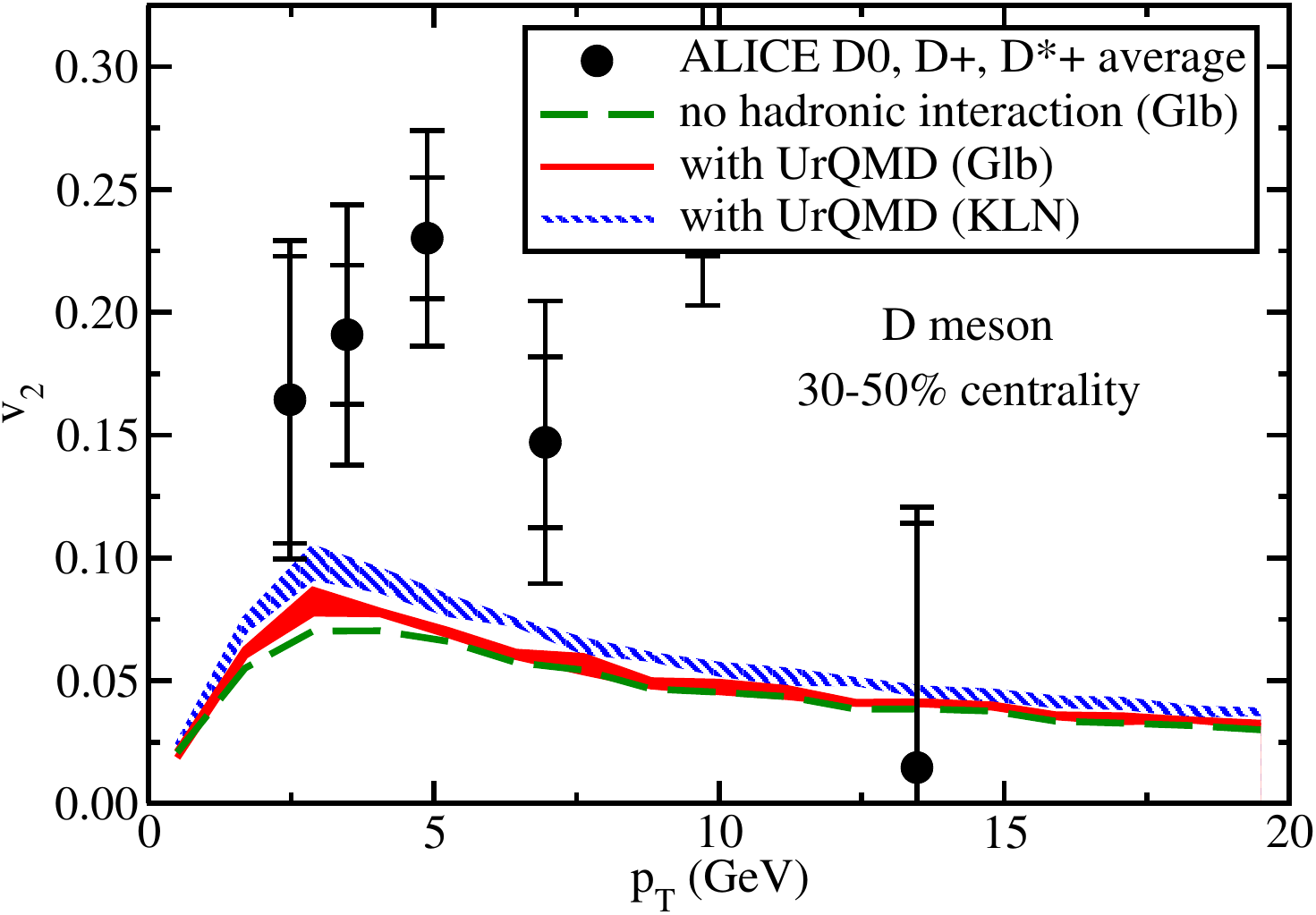}
\caption{Nuclear modification factor (left flow) and elliptic flow (right panel) of $D$ mesons at LHC, as computed within the Duke transport approach; blue dashed and red lines correspond to the nuclear modification factor with and without the contribution of hadronic diffusion, respectively. The green dashed line represents the elliptic flow without the contribution of hadronic diffusion, while the red and blue lines represent the elliptic flow with the contribution of hadronic diffusion under two different initial conditions. The error bands characterize the uncertainties introduced in the choice of the cutoff parameter in the hadron form factors when calculating the heavy meson scattering cross sections. Figures adapted from Ref.~\cite{Cao:2015hia}.}
 \label{fig:DHRaav2_LHC}
\end{figure}
In Ref.~\cite{Cao:2015hia} the influence of hadronic interactions on the $\raa$ and $v_2$ of $D$ mesons is addressed at both RHIC and LHC energies. Here a modified Langevin equation~\cite{Cao:2013ita} is employed to study the momentum evolution of the heavy quarks as they traverse the QGP medium:  in addition to the drag and diffusion terms, an extra term is incorporated to account for the recoil force acting on heavy quarks when they undergo medium-induced gluon radiation, with a lower gluon energy cutoff of $\pi T$ to mitigate the violation of detailed balance~\cite{Cao:2013ita,Cao:2015hia}. 
In order to attain the best description of HF data from RHIC and the LHC, a spatial diffusion coefficient of $\Ds (2\pi T)=6$ is employed. For the radiated gluons, a distribution function from higher-twist calculations is adopted~\cite{Zhang:2003wk}. The modified Langevin setup is implemented into an expanding QGP medium, which is simulated using a (2+1)D viscous hydrodynamic model~\cite{Qiu:2011hf}. Close to the critical temperature, both the bulk matter of the QGP and heavy quarks undergo hadronization, forming color-neutral bound states. To characterize the hadronization process of the bulk matter, they utilized the numerical tool ISS~\cite{Shen:2014vra}, which is based on the Cooper-Frye formula~\cite{Cooper:1974mv}. This tool is employed to obtain soft hadrons from the hydrodynamic medium. For heavy quarks, they employed a hybrid model combining fragmentation and coalescence to describe their hadronization.

After hadronization, the $D$ mesons  undergo rescattering with the hadrons of the bulk medium, which is modeled through the UrQMD transport model~\cite{Bass:1998ca,Bleicher:1999xi}. 
The latter utilizes microscopic cross-sections of hadronic scatterings are the basic inputs, taken from data as available and assuming constituent quark-scaling otherwise. The rescattering of $D$ mesons within a hadron gas is incorporated to the UrQMD framework using scattering cross sections with $\pi$ and $\rho$ mesons taken from Refs.~\cite{Lin:2000jp,Lin:1999ve} which are based on a hadronic Lagrangian generated from local flavor SU(4) gauge symmetry. In this calculation, there is appreciable uncertainty related to the choice of the cutoff parameter in the hadron form factors, which is varied to obtain an estimate of the uncertainty in the subsequent calculations of heavy-meson observables. In Fig.~\ref{fig:DHRaav2_LHC}, the $D$-meson $\raa$ (left panel) is depicted as a function of $\pT$ in Pb-Pb collisions at LHC energy, $\sqrts=2.76$ TeV. It is observed that the additional interaction experienced by $D$ mesons within the hadron gas leads to a mild additional suppression of their $\raa$  at large transverse momentum, while at the low momentum, the impact of hadronic rescattering is essentially unnoticeable. The impact of hadronic interactions on the elliptic flow of $D$ mesons at the LHC is illustrated in the right panel of Fig.~\ref{fig:DHRaav2_LHC}. Owing to additional scatterings of $D$ mesons in the hadron phase, their $v_2$ is further enhanced by up to approximately 20$\%$, most notably in the peak region around $\pT\simeq 3$ GeV. The authors also showcase the effect of different initial conditions for staring the hydrodynamic simulations. The KLN model yields larger eccentricity in the initial entropy density profiles compared to the Glauber model, entailing an additional 20$\%$ difference in the $v_2$ of heavy mesons following their evolution through the QGP, hadronization and the hadron gas.

\begin{figure}
\centering
\includegraphics[scale=0.32]{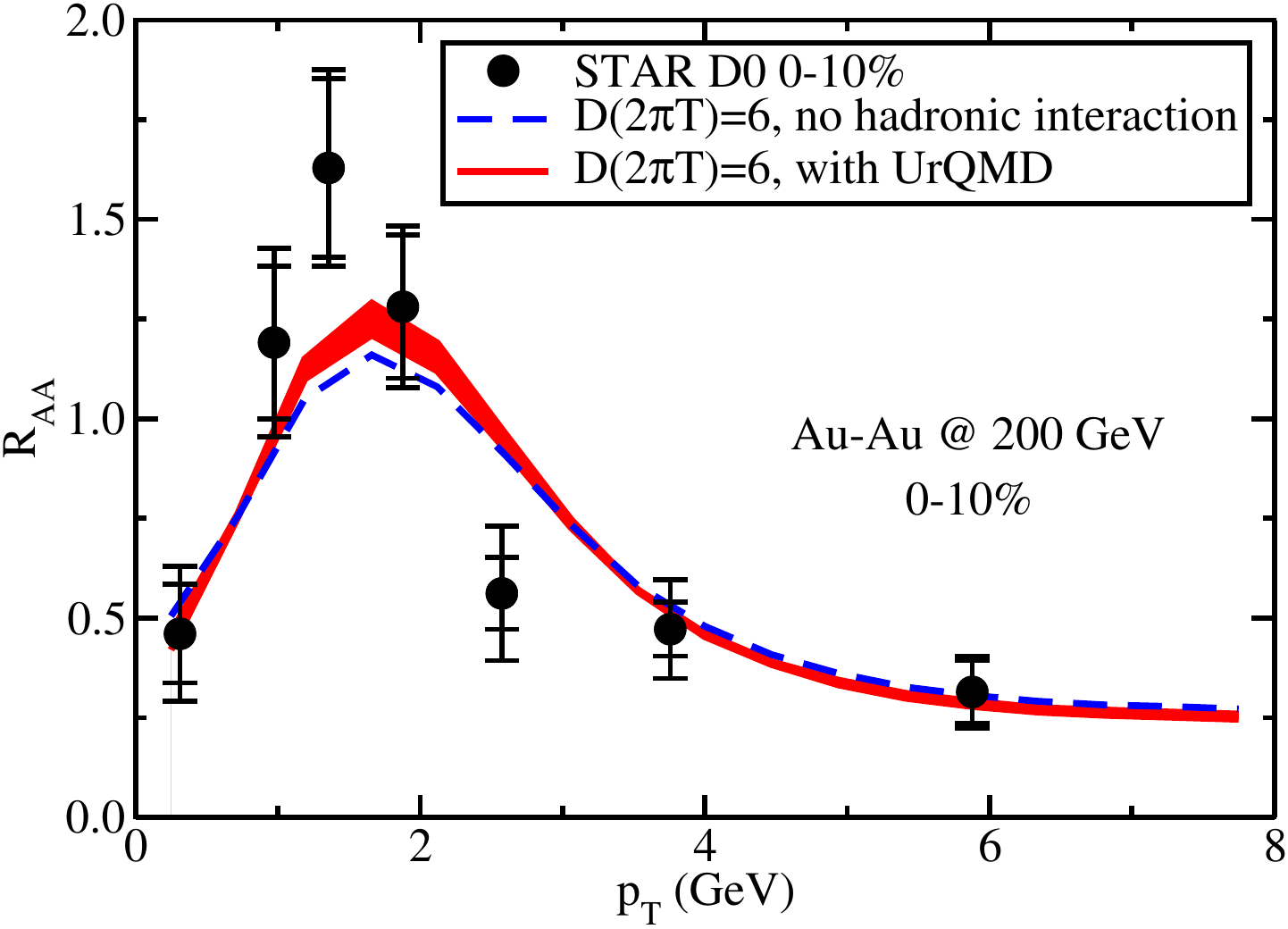}
\includegraphics[scale=0.32]{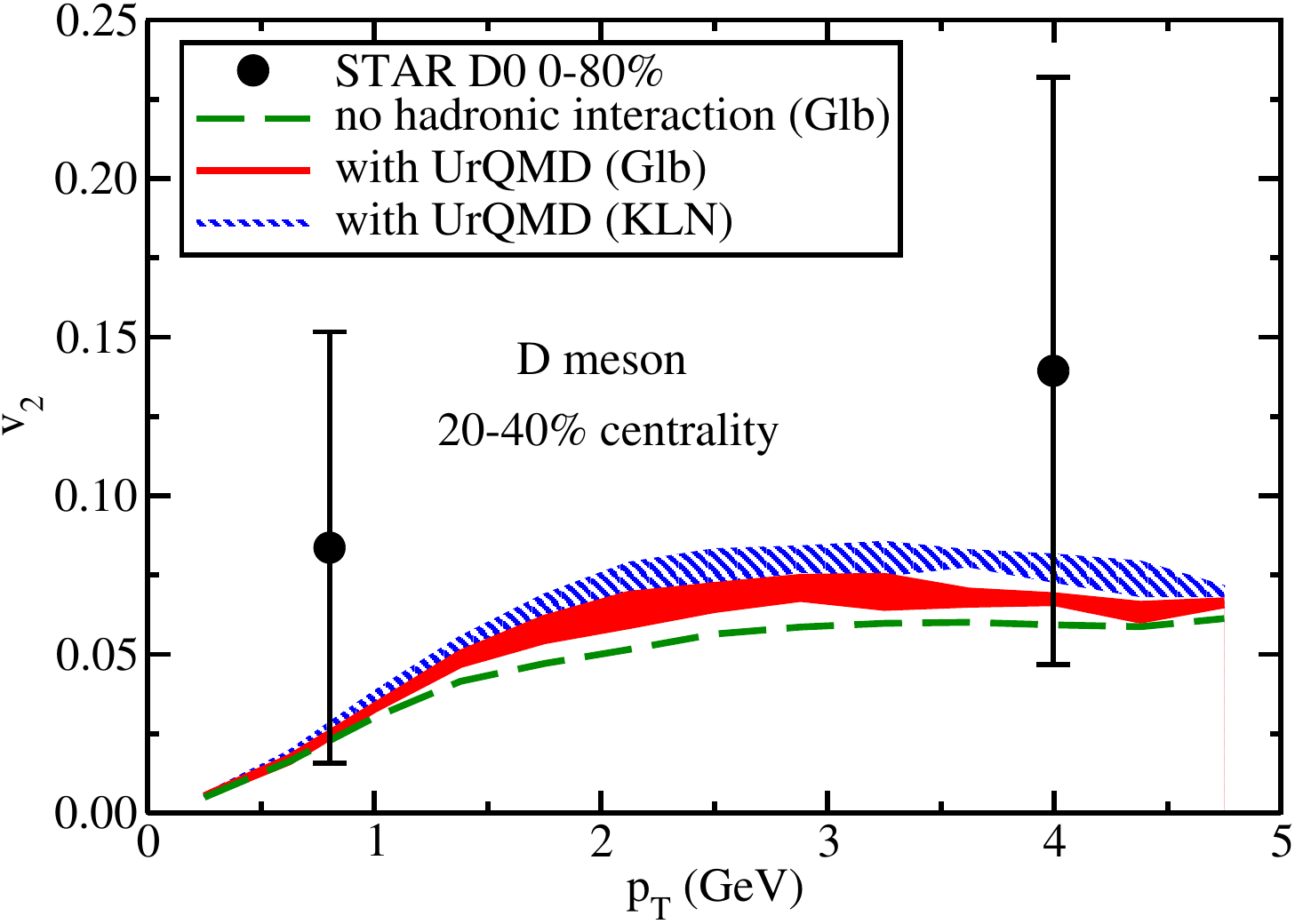}
\caption{Same as Fig.~\ref{fig:DHRaav2_LHC}, but for collisions at the RHIC. Figures adapted from Ref.~\cite{Cao:2015hia}.}\label{fig:DHRaav2_RHIC}
\end{figure}

Fig.~\ref{fig:DHRaav2_RHIC}  illustrates the variation of the $D$-meson $\raa$ (upper panel) and $v_2$ at the top RHIC energy of 0.2 GeV. Similar to case at the LHC, the inclusion of hadronic interactions using the UrQMD model leads to a slight suppression of $D$ meson $\raa$ at large $\pT$ and significantly enhanced elliptic flow of up to almost $\sim$40\%, not unlike the results in the PHSD model.\\

In summary of the sections on hadronic $D$-meson diffusion in HICs, Secs.~\ref{sssec_tamu}-\ref{sssec_phsd}, it seems fair to say that all of the phenomenological studies of the $D$-meson $\raa$ and $v_2$ that have considered rescattering in the hadronic medium (in addition to the QGP and hadronization effects) indicate that its impact on the nuclear modification factor is small. This is mainly due to the fact that the softening (thermalization) of the additional scattering in the hadronic phase is largely compensated by the additional radial flow as they are dragged along with the collective expansion. However, at the same time all calculations observe a very significant impact on the elliptic flow, that typically amounts to an increase of up to 30-40\% at RHIC energies and around 20$\%$ at LHC energies, most notably in the region of $\pT \simeq 3$ GeV, where the effect of thermalization can be expected to be largest. This is in large part due to the fact that the elliptic flow of the expanding medium is already near its maximum once the hadronic phase commences.    

%%%%%%%%%%%%%%%%%%%%%%%%%%%%%%%%%%%%%%%%%%%
\subsubsection{Femtoscopy studies}
\label{ssec:femtoscopy}
%%%%%%%%%%%%%%%%%%%%%%%%%%%%%%%%%%%%%%%%%%%%%%
\begin{figure}[th]
\centering
\includegraphics[scale=0.3]{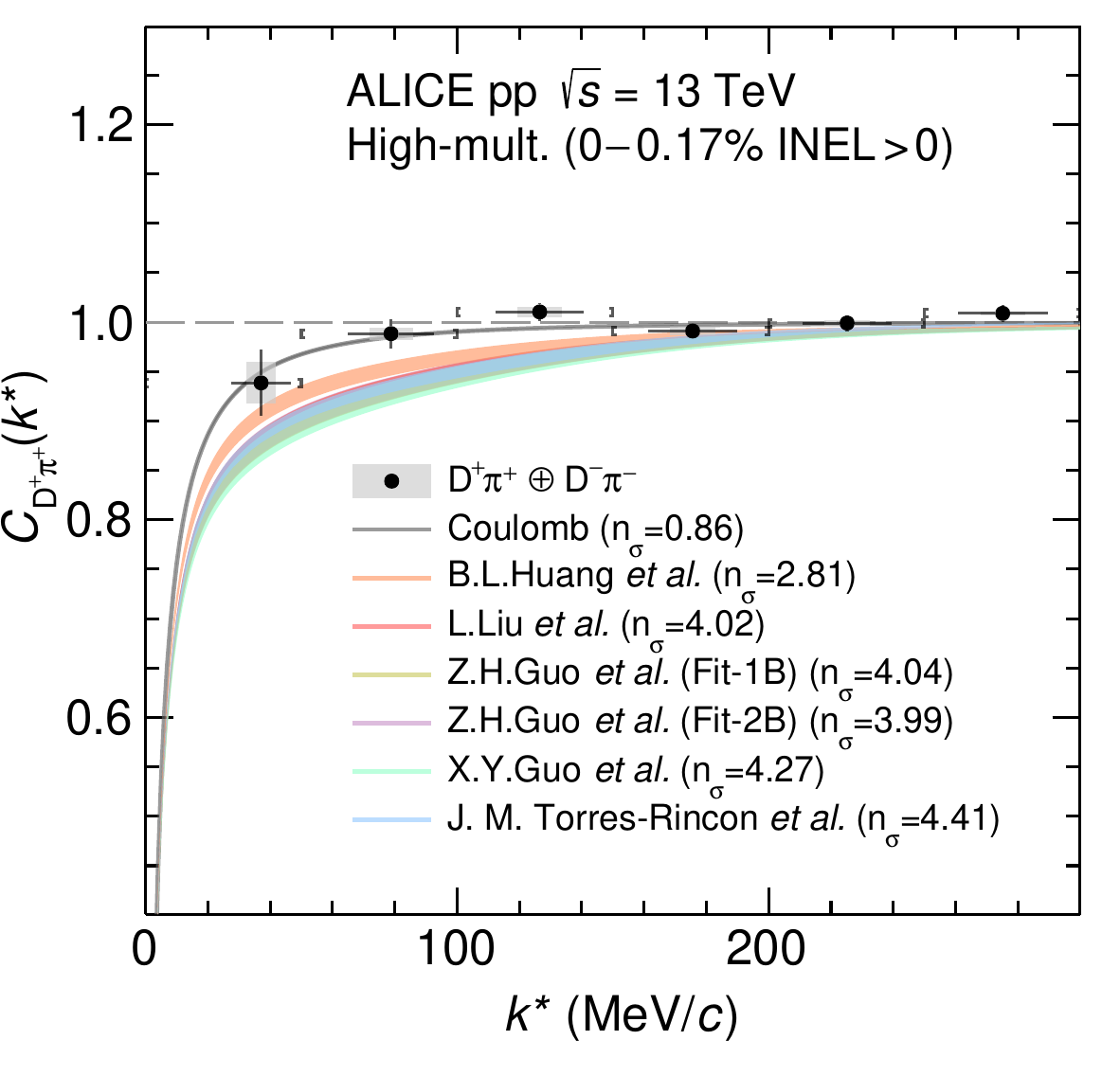}
\includegraphics[scale=0.3]{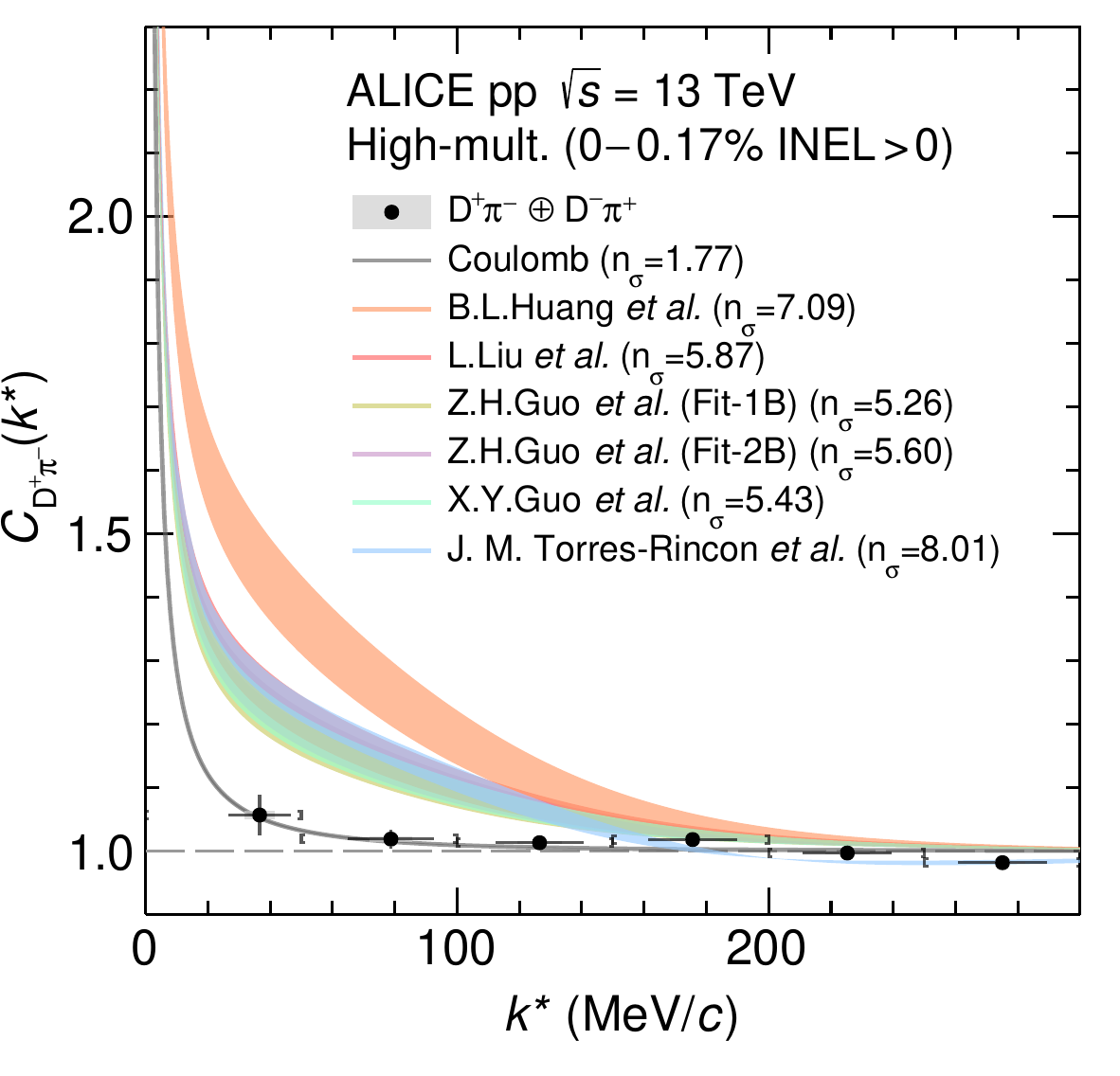}
\includegraphics[scale=0.3]{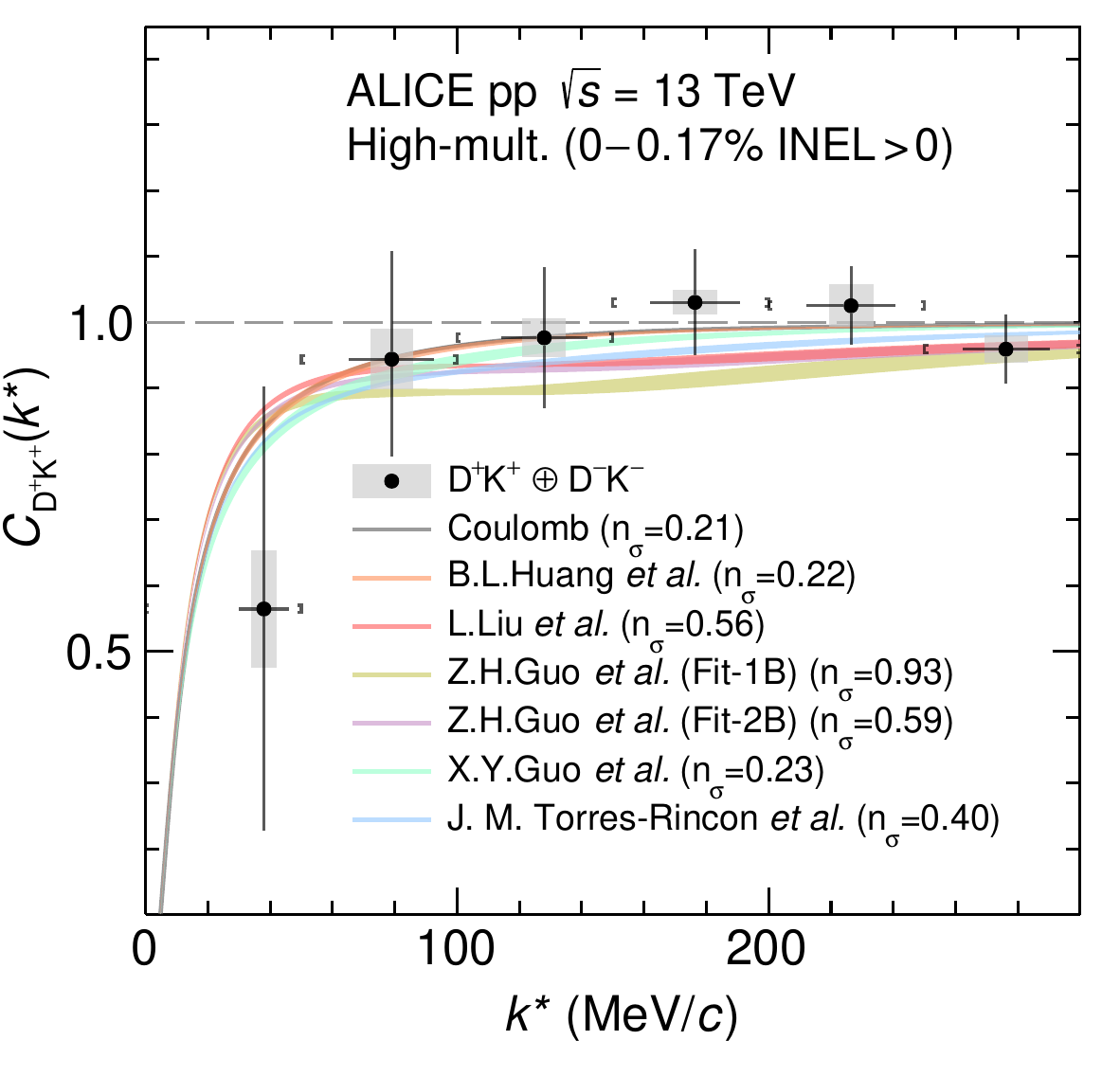}
\includegraphics[scale=0.3]{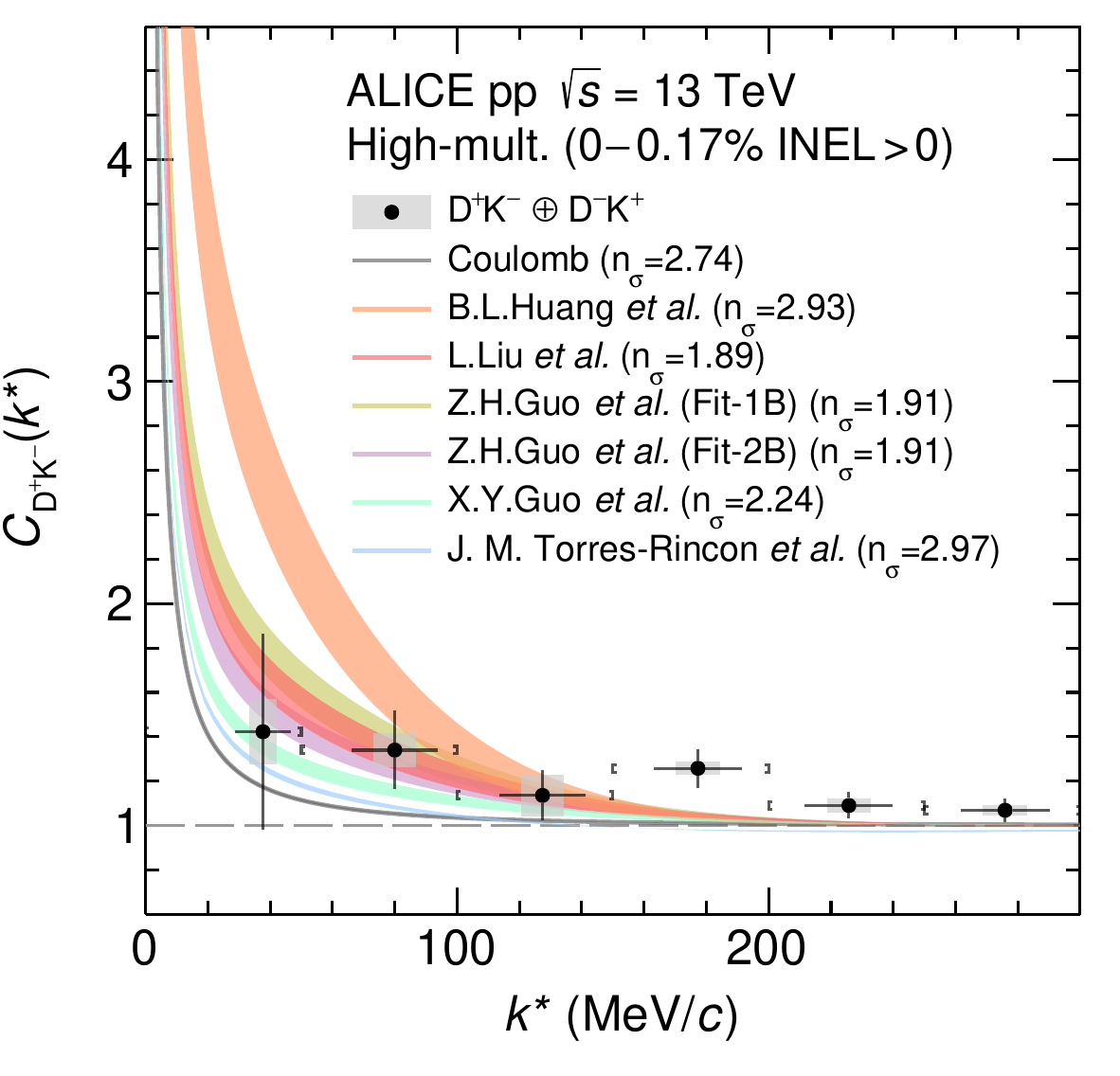}
\caption{Femtoscopic correlation functions of $D^+\pi^+$ + c.c. (top, left),
$D^+ \pi^-$ + c.c. (top, right), $D^+ K^+$ + c.c. (bottom, left) and $D^+ K^-$ + c.c. (bottom, right), measured by the ALICE collaboration in high-multiplicity pp($\sqrt{s} =13$\,TeV) collisions. Predictions of several theoretical models using low-energy scattering data are also displayed
(B.L.Huang {\it et al.}~\cite{Huang:2021fdt}, L.Liu {\it et al.}~\cite{Liu:2012zya},
Z.H.Guo {\it et al.}~\cite{Guo:2018tjx}, X.Y.Guo {\it et al.}~\cite{Guo:2018kno} and J.M.Torres-Rincon {\it et al.}~\cite{Torres-Rincon:2023qll}). Figures taken from Ref.~\cite{ALICE:2023bsp}.} 
\label{fig:Dmesonfemto}
\end{figure}
Two-particle correlation measurements in nuclear collisions have a long history. Originally, the idea was to utilize them as a way to characterize spatial properties of the particle emission source~\cite{Pratt:1995,Wiedemann:1999qn,Lisa:2005dd}, by using the most abundant, weakly-interacting particles, \ie, pions. In general, at small relative momentum, however, they can carry information about the nature of the final-state interactions, such as scattering lengths~\cite{Gyulassy:1979yi, Lednicky:1981su,Heinz:1999rw,Fabbietti:2020bfg}, which has been utilized to study a diverse set of two-hadron systems. 

Given two hadrons, 1 and 2, the femtoscopy correlation function $C(\bm{p}_1,\bm{p}_2)$ is given in terms of their respective momenta $\bm{p}_i$ and energy $E_i$ as,
\begin{equation}
    C(\bm{p}_1,\bm{p}_2) = \frac{E_1 E_2 \frac{d^6N}{d \bm{p}_1 d \bm{p}_2}}{E_1 \frac{d^3N}{d \bm{p}_1} E_2 \frac{d^3N}{d \bm{p}_2}} = \frac{N(\bm{p}_1,\bm{p}_2)}{N(\bm{p}_1) N(\bm{p}_2)} \ ,
\end{equation}
where $N(\bm{p}_1,\bm{p}_2)$ is the number of hadron pairs with momenta $\bm{p}_1$ and $\bm{p}_2$, and $N(\bm{p}_i)$ is the number of hadrons of type $i$ with 3-momentum ${\bm p}_i$. The correlation function measures deviations from the fully uncorrelated assumption, $N(\bm{p}_1,\bm{p}_2) = N(\bm{p}_1) N(\bm{p}_2)$, given by quantum or final state interactions between the two hadrons. In the center-of-mass frame of the pair (also known as pair-rest frame, and denoted by an asterisk $^*$) the correlation function becomes a function of the relative momentum of the pair $\bm{k}^*$,
\begin{equation}
    C(k^*)= \xi(k^*) \frac{ N_{\textrm{same}} (k^*)}{ N_{\textrm{mixed}} (k^*)} \ ,
\end{equation}
where $N_{\textrm{same}} (k^*)$ represents the number of pairs at a given relative momentum $k^*$ measured in the same collision event, while $N_{\textrm{mixed}} (k^*)$ is the same quantity for pairs detected in different (or mixed) events so that the pairs are fully uncorrelated. Experimentally, a normalization factor---possibly depending on $k^*$---is added just to correct for the different conditions and normalizations of the ``same'' and ``mixed'' collision events.

In Ref.~\cite{ALICE:2024bhk}, the ALICE collaboration has presented results for correlation functions of $D^{(*)+}\pi^+, D^{(*)+}\pi^-, D^{(*)+} K^+$ and $D^{(*)+} K^-$ in high-multiplicity pp collisions at $\sqrt{s}=13$ TeV, together with various theoretical predictions including several of the EFT calculations discussed in Sec.~\ref{sec:charm}. In most of the models, only the information at low energy (scattering length and scattering range) is accounted for, along with the long-range Coulomb interaction via the Koonin-Pratt formula~\cite{Pratt:1995,Heinz:1999rw,Lisa:2005dd}. As shown in Fig.~\ref{fig:Dmesonfemto}, which reproduces the results from Ref.\cite{ALICE:2024bhk}, the theoretical approaches show a common trend and reasonable agreement with the ALICE data  (within the rather large errors bars) in most channels ~\cite{ALICE:2024bhk,Torres-Rincon:2023qll}. The most striking discrepancy between the data and the theory occurs in the $D^{+}\pi^-$ channel (top, right panel of Fig.~\ref{fig:Dmesonfemto}), which only shows an agreement with the pure Coulomb interaction (gray line). All the theoretical models accounting for the strong interaction present a degree of attraction, indicated by the extra positive correlation as compared to Coulomb. In terms of low-energy scattering parameters, the theory calculations predict an $s$-wave scattering length in the $I=1/2$ and $I=3/2$ channels of around $a_0^{D\pi} (I=1/2) \simeq 0.5$ fm and $a_0^{D\pi} (I=3/2) \simeq -0.1$ fm, providing a net positive scattering length for the $D^+ \pi^-$ case (we use the convention in which a positive scattering length indicates attraction, while a negative one means repulsion, or the presence of a bound state). On the other hand, since the experimental data is described by a pure Coulomb interaction, the strong scattering length appears to be compatible with zero (or, according to the analysis of Ref.~\cite{ALICE:2024bhk}, even a small positive value for $a_0^{D\pi} (I=3/2)$)---in clear contradiction with all EFT predictions. This difference is yet to be understood. A caveat in this puzzle is the possible existence of a broad state decaying 
into the $D^+ \pi^-$ final state, as manifest in the calculation of~\cite{Torres-Rincon:2023qll}, which generates the $D_0^*(2300)$, decaying into $D^+\pi^-$. In the top, right panel of Fig.~\ref{fig:Dmesonfemto} the effect of this resonance can be traced to a shallow minima below $C(k^*)=1$ around $k^*=250$ MeV in the theoretical prediction of~\cite{Torres-Rincon:2023qll} (light blue line in the figure). Should this resonance have any visible effect on the $D\pi$ final state, the source function would need to be revisited to account for this extra emission contribution. In the meson-baryon sector we should mention that preliminary femtoscopy results for $DN$ interactions are reported in Ref.~\cite{ALICE:2022enj} by the ALICE collaboration also in high-multiplicity pp collisions at $\sqrt{s}=13$ TeV.

%%%%%%%%%%%%%%%%%%%%%%%%%%%%%%%%%%%%%%%%%%%%%%%%%%%%%%%%%%%%%%
\subsection{$B$ mesons and heavy baryons in the hadronic phase}
\label{ssec:B+baryon-obs}
%%%%%%%%%%%%%%%%%%%%%%%%%%%%%%%%%%%%%%%%%%%%%%%%%%%%%%%%%%%%%
The majority of the analysis regarding the impact of hadronic medium scattering on heavy-hadron observables is centered around the $D$ meson, as elaborated in the preceding section.
We now turn to reviewing and analyzing the information on available from the literature on hadronic HF diffusion in HICs of hadrons other than the $D$ meson, specifically $B$ mesons in Sec.~\ref{sssec:B-obs} and HF baryons ($\Lambda_c$ and $\Lambda_b$) in Sec.~\ref{sssec:bar-obs}.

%%%%%%%%%%%%%%%%%%%%%%%%%%%%%
\subsubsection{$B$-meson $\raa$ and $v_2$}
\label{sssec:B-obs}
%%%%%%%%%%%%%%%%%%%%%%%%%%%%%%%%%%%
The rescattering of $B$ mesons in the hadronic medium of HICs has been evaluated in the Catania transport model~\cite{Das:2016llg} in Au-Au ($\sqrt{s_{NN}} = 200$ GeV) collisions at RHIC.  The dynamics governing the time evolution of heavy quarks in the QGP and $B$ mesons in the hadronic phase were simulated using Langevin dynamics. The drag and diffusion coefficients for heavy quarks in the QGP are determined through calculations based on the quasi-particle model (QPM)~\cite{Das:2015ana,Scardina:2017ipo}. The quasi-particle model incorporates non-perturbative dynamics by introducing temperature-dependent quasi-particle masses for both light quarks and gluons. The quasi-particle model can reproduces LQCD thermodynamics~\cite{Plumari:2011mk} by appropriately fitting the strong-coupling parameter, $g(T)$. 
The initial distribution of heavy quarks in momentum space is determined based on the bottom-quark distribution in pp collisions, based on the Fixed-Order-Next-to-Leading-Log (FONLL) framework~\cite{Cacciari:2005rk,Cacciari:2012ny}. In coordinate space, the distribution is assumed to follow the profile of binary nucleon-nucleon collisions ($\Ncoll$) derived from the Glauber model.  The bulk medium expansion to determine cooling collective flow of the bulk matter is simulated with  a (3+1)D relativistic transport code. This model allows for the description of fluid evolution with a constant $\eta/s$ in a manner analogous to viscous-hydrodynamics simulation. In this way, one can simulate the dynamical evolution of a fluid with a specified $\eta/s$ ratio using the Boltzmann equation, see also Refs.~\cite{Ruggieri:2013bda,Ruggieri:2013ova,Ferini:2008he}. 
For the hadronization process, where bottom quarks transform into $B$-mesons, the Peterson fragmentation function is employed. Subsequently,  
the Langevin dynamics is employed to simulate the propagation of $B$-mesons within the hadronic bath which includes $\pi$, $K$, $\bar K$, and $\eta$ particles. The interaction between the $B$-mesons and the bath is evaluated using unitarized interactions derived from effective field theories that uphold both chiral and heavy-quark symmetries. Specifically, the interactions of $B$ mesons with $\pi$, $K$, $\bar K$, and $\eta$ mesons are obtained through a unitarized coupled-channel model that incorporates heavy-quark spin symmetry~\cite{Garcia-Recio:2008rjt,Gamermann:2010zz,Romanets:2012hm,Garcia-Recio:2012lts,Garcia-Recio:2013gaa, Tolos:2013gta}. We remind that in the absence of a net baryon density, the drag and diffusion coefficients of $\bar{B}$ mesons are the same as those for $B$ mesons.

\begin{figure} 
\centering
\includegraphics[scale=0.32]{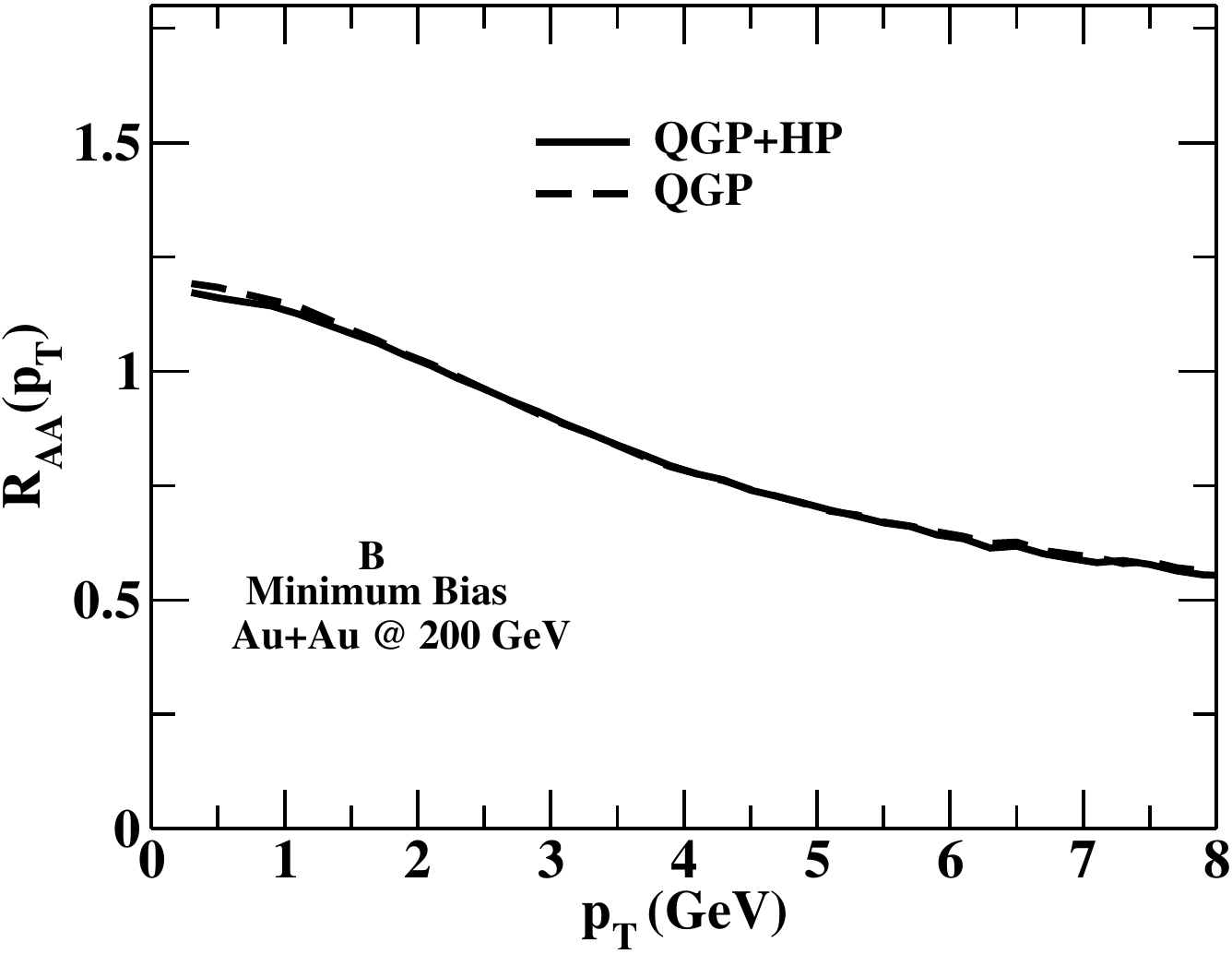}
\includegraphics[scale=0.32]{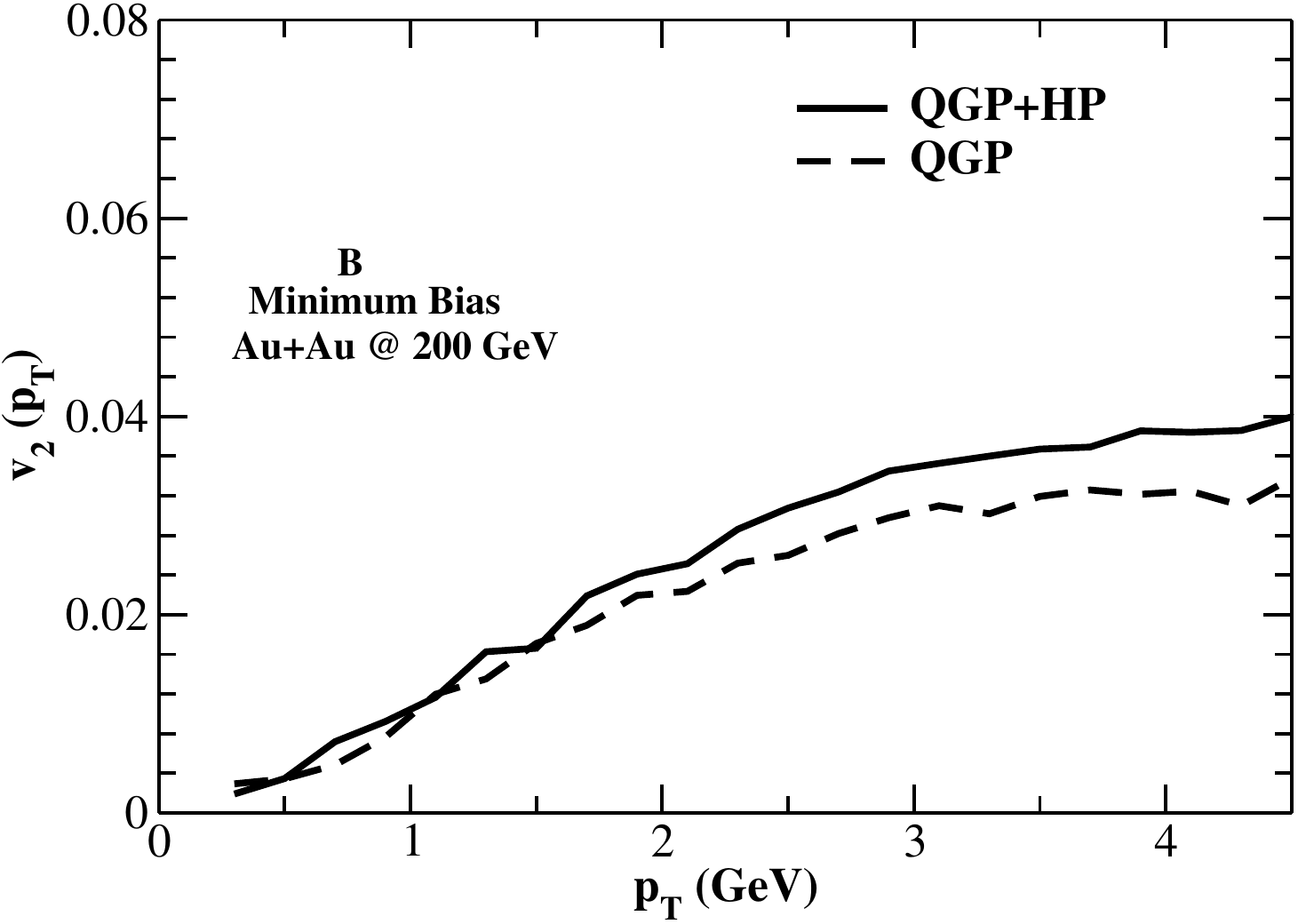}
\caption{Results of the Catania transport model for $B$-meson observables in in Au+Au collisions at $\sqrt{s_{NN}} = 200$ GeV. The nuclear modification factor $\raa$ (left panel) and elliptic flow (right panel) are shown considering either both QGP and hadronic phase interactions (black solid lines) or only QGP phase interactions (black dashed lines). Figures adapted from Ref.~\cite{Das:2016llg}.} 
\label{fig:CBRaav2_RHIC}
\end{figure}
Figure~\ref{fig:CBRaav2_RHIC}  depicts the $\pT$ dependence of the $\raa$ (left panel) and $v_2$ (right panel) of the $B$ meson in minimum-bias Au-Au collisions at RHIC with and without hadronic medium rescattering, as obtained in Ref.~\cite{Das:2016llg}. 
In qualitative analogy to the $D$-meson case, the impact of hadronic phase rescattering on the nuclear modification factor is negligible, while
the influence on the elliptic flow ($v_2$) is quite significant, amounting to a maximum of approximately 20\% at $\pT \simeq m_B$.

%%%%%%%%%%%%%%%%%%%%%%%%%%%%%%%%%%%%%%%%%%%%%%
\subsubsection{Heavy-flavor baryon $\raa$ and $v_2$}
\label{sssec:bar-obs}
%%%%%%%%%%%%%%%%%%%%%%%%%%%%%%%%%%%%%%%%%%%%%%
Heavy-flavor baryons ($\Lambda_c$ and $\Lambda_b$) play a key role in probing HF dynamics in QCD matter beyond the realm of heavy mesons ($D$ and $B$). Most notably, this concerns the hadro-chemistry in the hadronization process (which will be discussed in Sec.~\ref{ssec:chem}
below) but is also of interest for re-interactions in the hadronic phase. On the one hand, the baryons provide an additional mass scale which affects the kinetic diffusion properties of HF particles, but on the other hand one may also ask in how far the presence of two light quarks (rather than one in mesons) affects the interactions and thus the dynamics of the hadronic re-interactions of HF baryons.

The nuclear modification factor of the $\Lambda_c$ has been recently measured in Pb-Pb collisions at the LHC~\cite{ALICE:2018hbc}. The increased presence of heavy baryons~\cite{Lee:2007wr,Oh:2009zj,Ghosh:2014oia}, possibly triggered by quark coalescence processes, can impact the nuclear modification factor of the heavy mesons in terms of a reduction in their yields. Additionally, the baryon-to-meson ratio~\cite{Greco:2003vf}, $\Lambda_c$/$D$ and $\Lambda_b$/$B$, when measured as a function of $\pT$, can provide insights into in-medium hadronization, in particular on the additional collective flow (both radial and elliptic) that is imparted on the HF baryons through the presence of an extra light quark in the coalescence process. Recent experimental measurements of the charm-baryon-to-meson ratio~\cite{ALICE:2018hbc,STAR:2019ank, ALICE:2021bib} have provided intriguing data that may become a valuable tool for disentangling various hadronization mechanisms~\cite{Das:2016llg,Plumari:2017ntm,Cao:2019iqs,He:2019vgs,He:2022tod}. Again, this renders compelling motivation, if not a mandatory task, to quantify the impact of hadronic rescattering of $\Lambda_c$ and $\Lambda_b$ on the pertinent experimental observables. Transport coefficients of the heavy baryons, $\Lambda_c$ and $\Lambda_b$, in the hadronic medium have been reported in Refs.~\cite{Ghosh:2014oia,Tolos:2016slr} (see Sec.~\ref{ssec:baryon-trans}).

Let us now turn to the available phenomenological studies.
In Ref.~\cite{Das:2016llg}  the influence of rescattering within the hadronic medium on the $\raa$ and $v_2$ of the  $\Lambda_c$ has been investigated. The dynamics governing the time evolution of heavy quarks in the QGP and heavy baryons in the hadronic phase were modeled using Langevin dynamics. Hadronization was performed using Peterson fragmentation functions~\cite{Peterson:1982ak} whose functional form given by
\be
f(z) \propto 
\frac{1}{ z \left[ 1- \frac{1}{z}- \frac{\epsilon_c}{1-z} \right]^2 } \ ,
\ee
where $\epsilon_c$ serves as a free parameter used to calibrate the shape of the fragmentation function for individual charm hadrons to be consistent with experimental data obtained in $pp$ collisions. 
Specifically, in the absence of $pp$  data, the shape of the $\Lambda_c$ fragmentation was determined by utilizing electron-positron annihilation data. 
In electron-positron annihilation, the value $\epsilon_c$=0.2 for the $\Lambda_c$ baryon turns out to be approximately twice as large as that for the $D$ meson. This is in line with the common expectation that the fragmentation function for heavier hadrons is softer compared to that of lighter ones since extra energy has to taken out of the parent quark. This is even more evident for the production of baryons where the extra (pair) production (\eg, from the splitting of a radiated gluon) can be expected to require even more energy, thereby further reducing the energy available for the momentum of the produced baryon. 

For the $\Lambda_b$, an even smaller value of $\epsilon_c$=0.006 has been used. 
This approach has been applied to study $\Lambda_c$ and $\Lambda_b$ production at RHIC and LHC energies using Langevin simulations for both QGP and hadronic matter. The drag and diffusion coefficients for heavy quarks in the QGP are determined through calculations based on the quasi-particle model (QPM)~\cite{Das:2015ana,Scardina:2017ipo} and for $\Lambda_{b,c}$ baryons in hadronic matter within  a unitarized meson-baryon coupled-channel model that incorporates heavy-quark spin symmetry~\cite{Garcia-Recio:2008rjt,Gamermann:2010zz,Romanets:2012hm,Garcia-Recio:2012lts,Garcia-Recio:2013gaa, Tolos:2013gta} for interactions with $\pi$, $K$, $\bar K$, and $\eta$ particles, as discussed in the preceding section on $B$-meson observables (which also contains the information on the initial conditions and the Boltzmann background medium). 

\begin{figure} 
\centering
\includegraphics[scale=0.32]{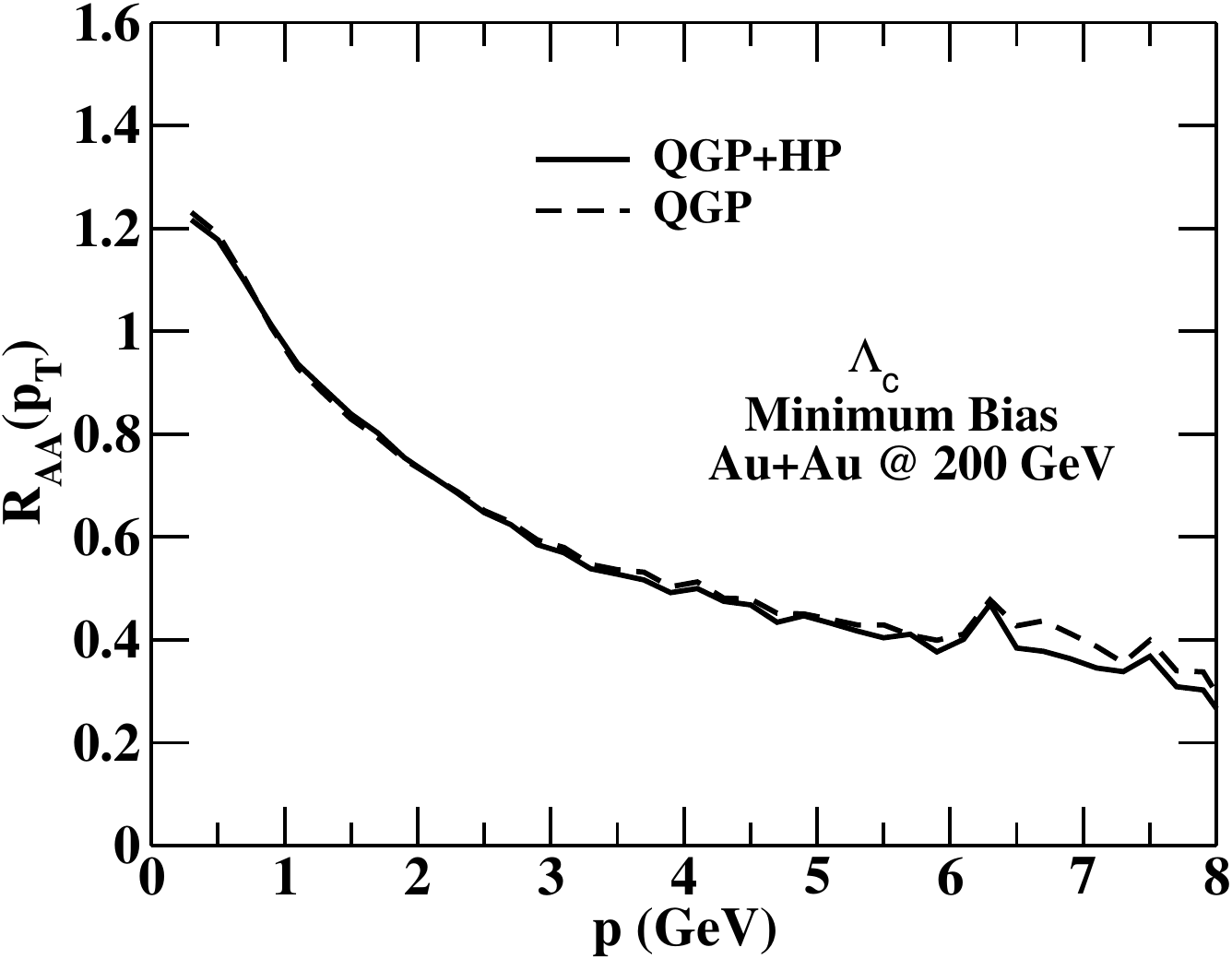}
\includegraphics[scale=0.32]{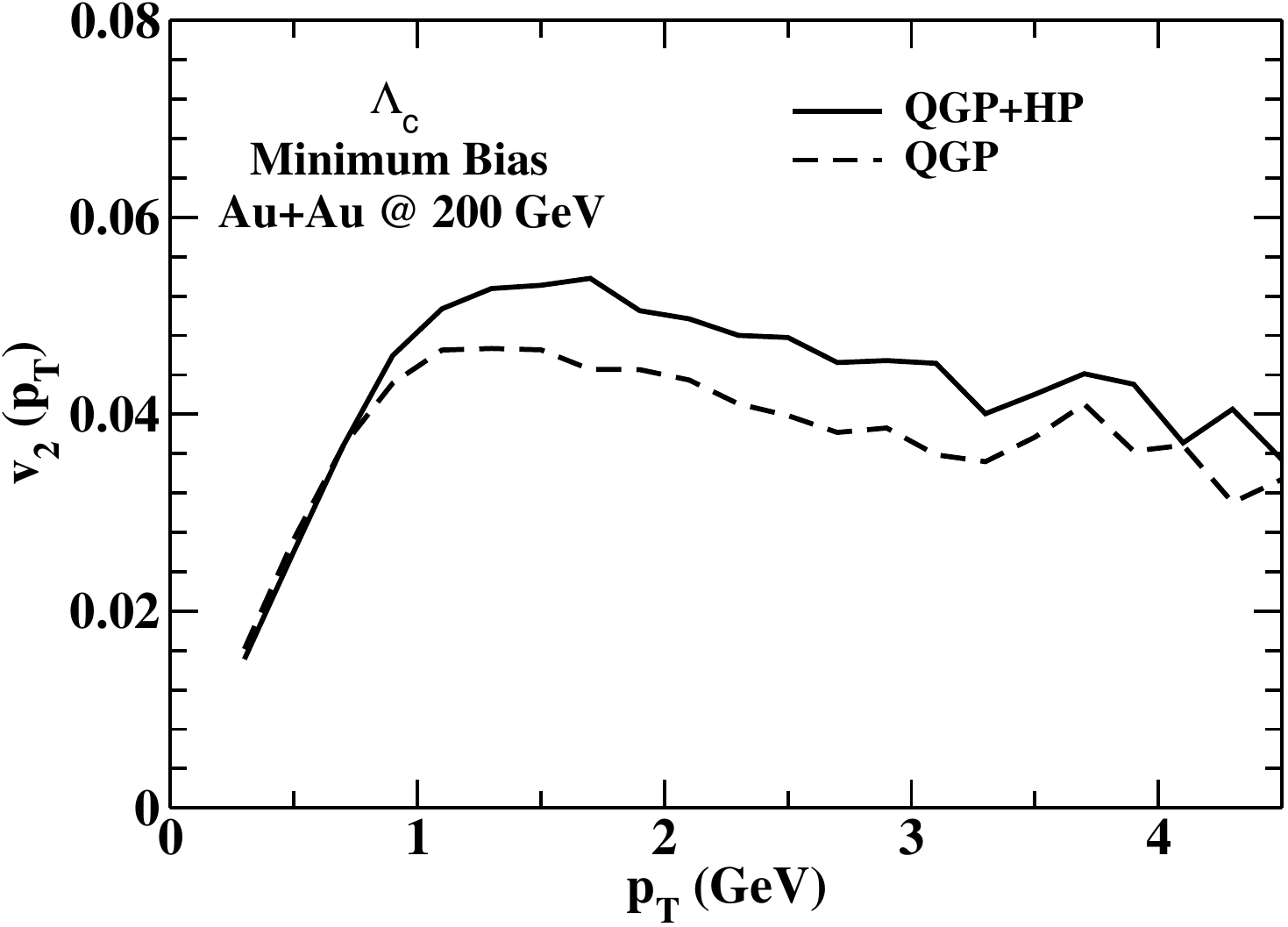}
\caption{Nuclear modification factor (left panel)  and elliptic flow (right panel) of $\Lambda_c$  baryons in minimum-bias Au+Au collisions at RHIC, as computed within the Catania transport approach; solid and dashed lines correspond to the results with and without the contribution of hadronic diffusion, respectively. Figures adapted from Ref.~\cite{Das:2016llg}.}
\label{fig:CRaav2_RHIC}
\end{figure}
Fig.~\ref{fig:CRaav2_RHIC} depicts the $\raa$ (left panel) and $v_2$ (right panel) for direct $\Lambda_c$ production (i.e., without feed-down contributions) as a functions of $\pT$ following from $c$-quark diffusion in the QGP plus fragmentation, and for additional hadronic phase diffusion  at top RHIC energy. The influence of the hadronic phase on the $\raa$ is almost imperceptible, while the $v_2$ is enhanced due to the presence of the hadronic phase, with an increase of up to approximately 20\%.

\begin{figure} 
\centering
\includegraphics[scale=0.35]{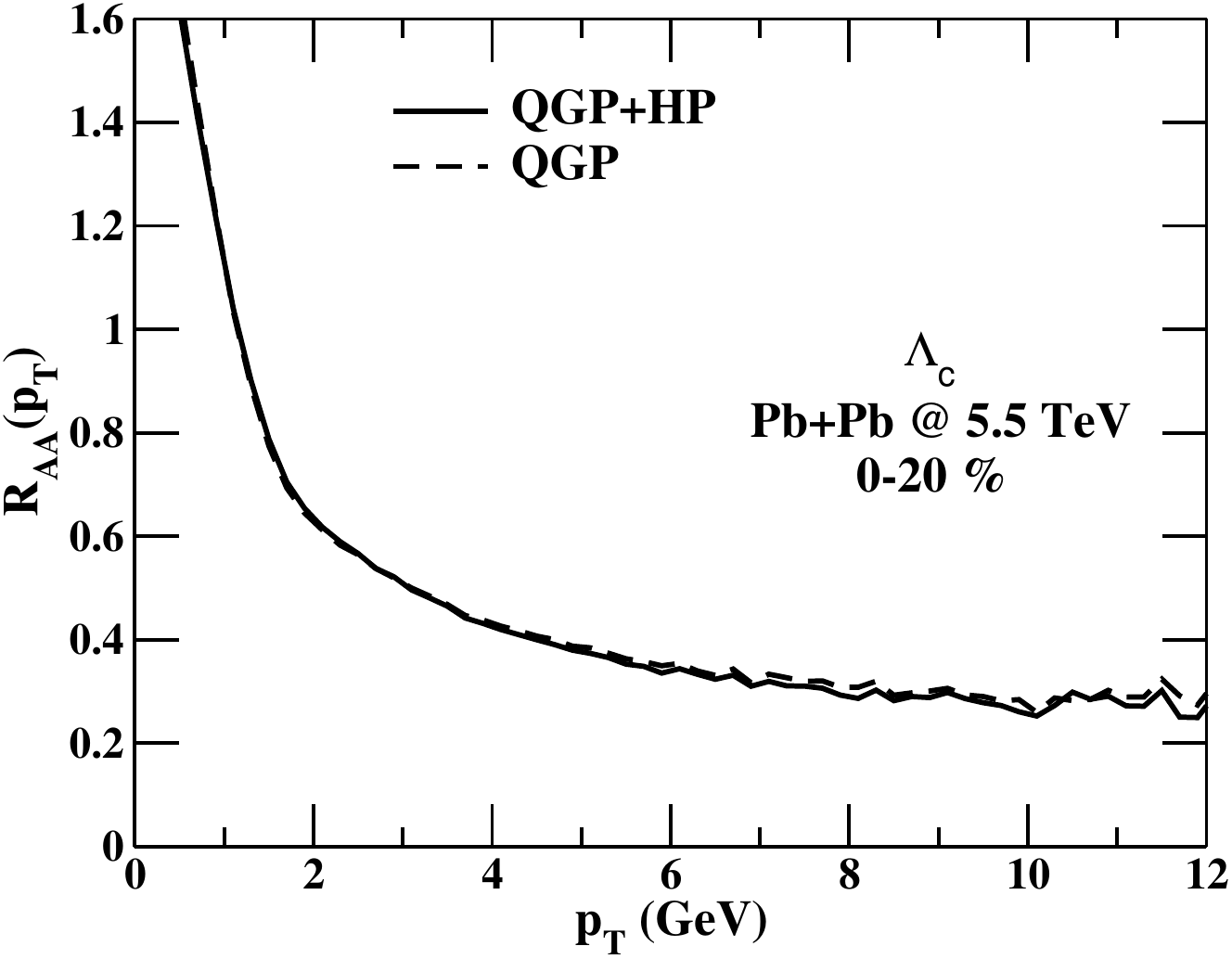}
\includegraphics[scale=0.35]{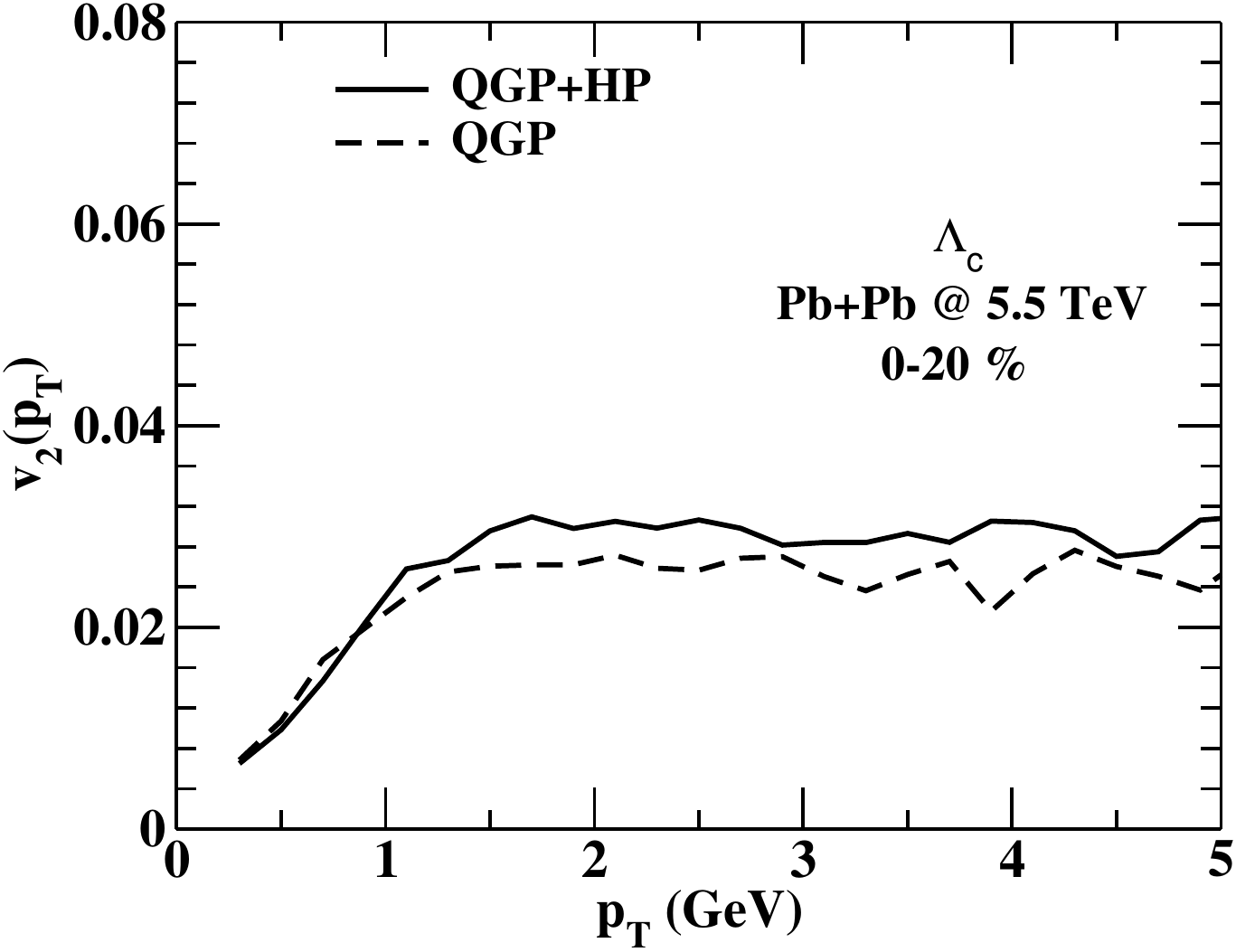}
\caption{Same as Fig.~\ref{fig:CRaav2_RHIC}, but for 0-20\% central Pb-Pb (5.5~TeV) collisions at the LHC. Figures adapted from Ref.~\cite{Das:2016llg}.}
\label{fig:CRaav2_LHC}
\end{figure}
Within the same setup, the results for $\Lambda_c$ in central Pb-Pb collisions at the LHC are shown in Fig.~\ref{fig:CRaav2_LHC}.
Again, the  impact of the hadronic phase on the $\raa$ is minimal, while $v_2 $ exhibits a noticeable. The absolute magnitude of the effects is slightly diminished compared to the results at RHIC energy, although the absolute values of the $v_2$ (with and without hadronic diffusion) are smaller than at RHIC.
However, the trend that the hadronic phase is  more important at RHIC than at the LHC persists.

\begin{figure}[thb] 
\centering
\includegraphics[scale=0.32]{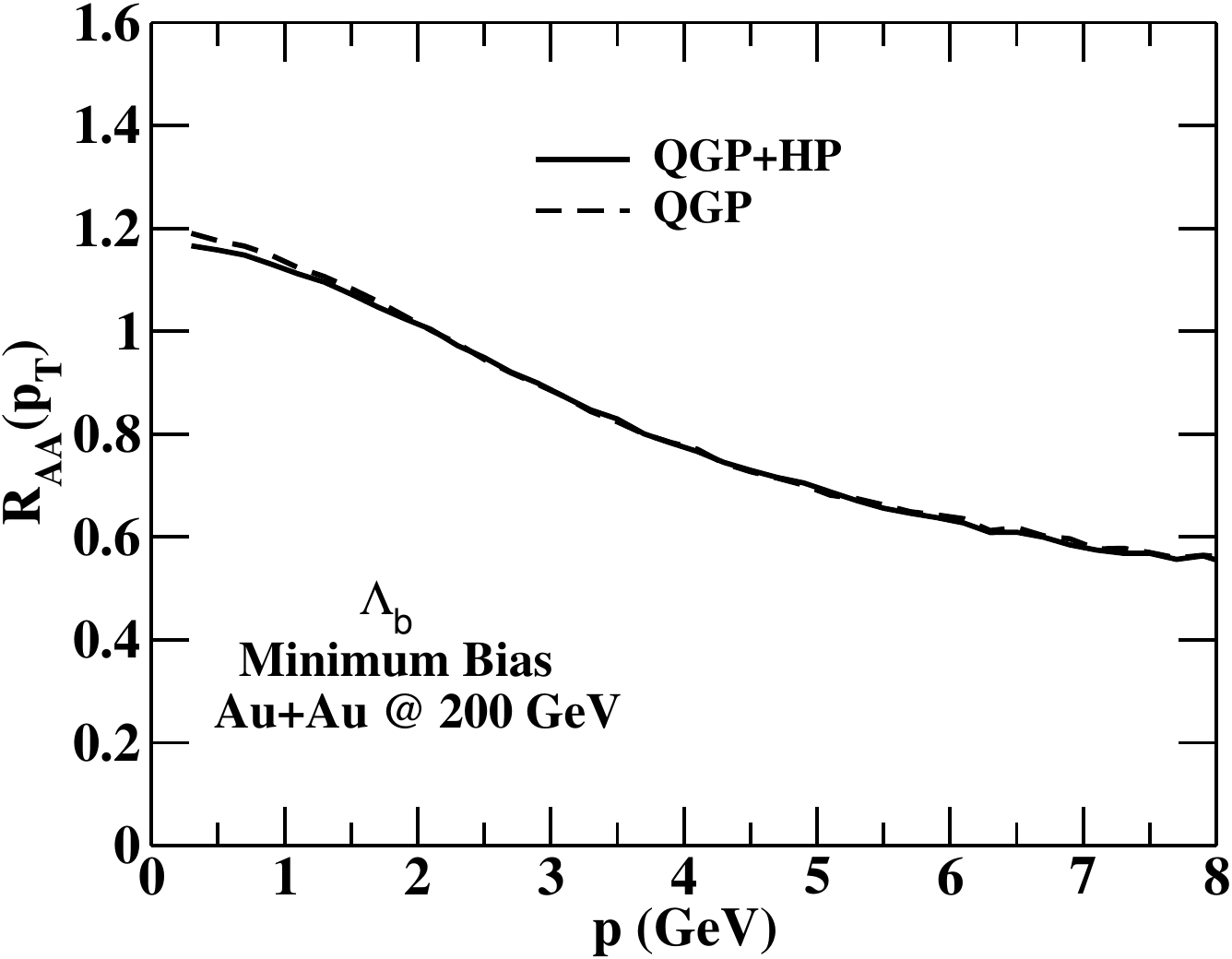}
\includegraphics[scale=0.32]{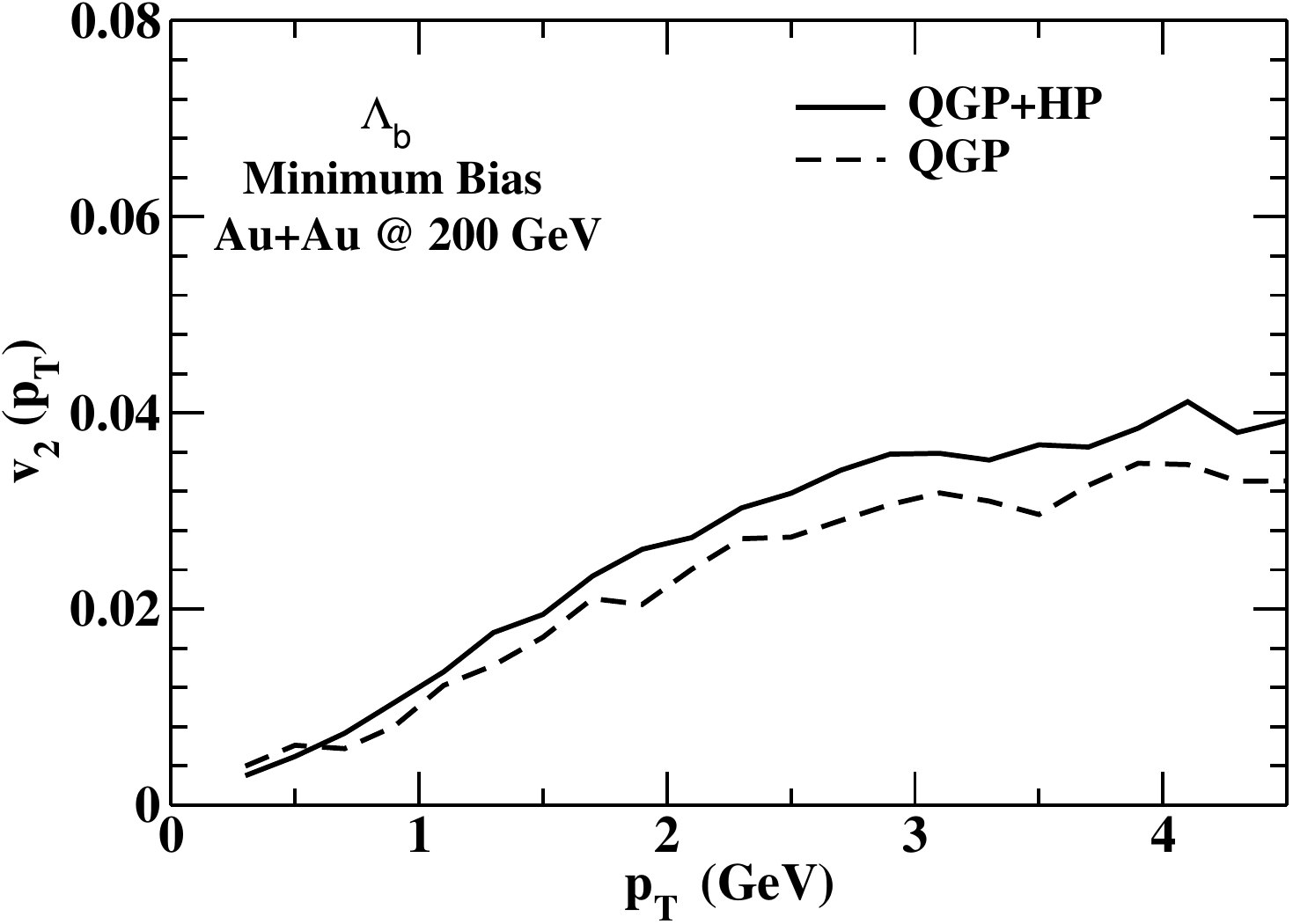}
\caption{Nuclear modification factor (left panel)  and elliptic flow (right panel) of $\Lambda_b$  baryons in central Au+Au collisions at RHIC, as computed within the Catania transport approach; solid and dashed lines correspond to the results with and without the contribution of hadronic diffusion. Figures adapted from Ref.~\cite{Das:2016llg} )}\label{fig:C1Raav2_RHIC}
\end{figure}
\begin{figure} 
\centering
\includegraphics[scale=0.35]{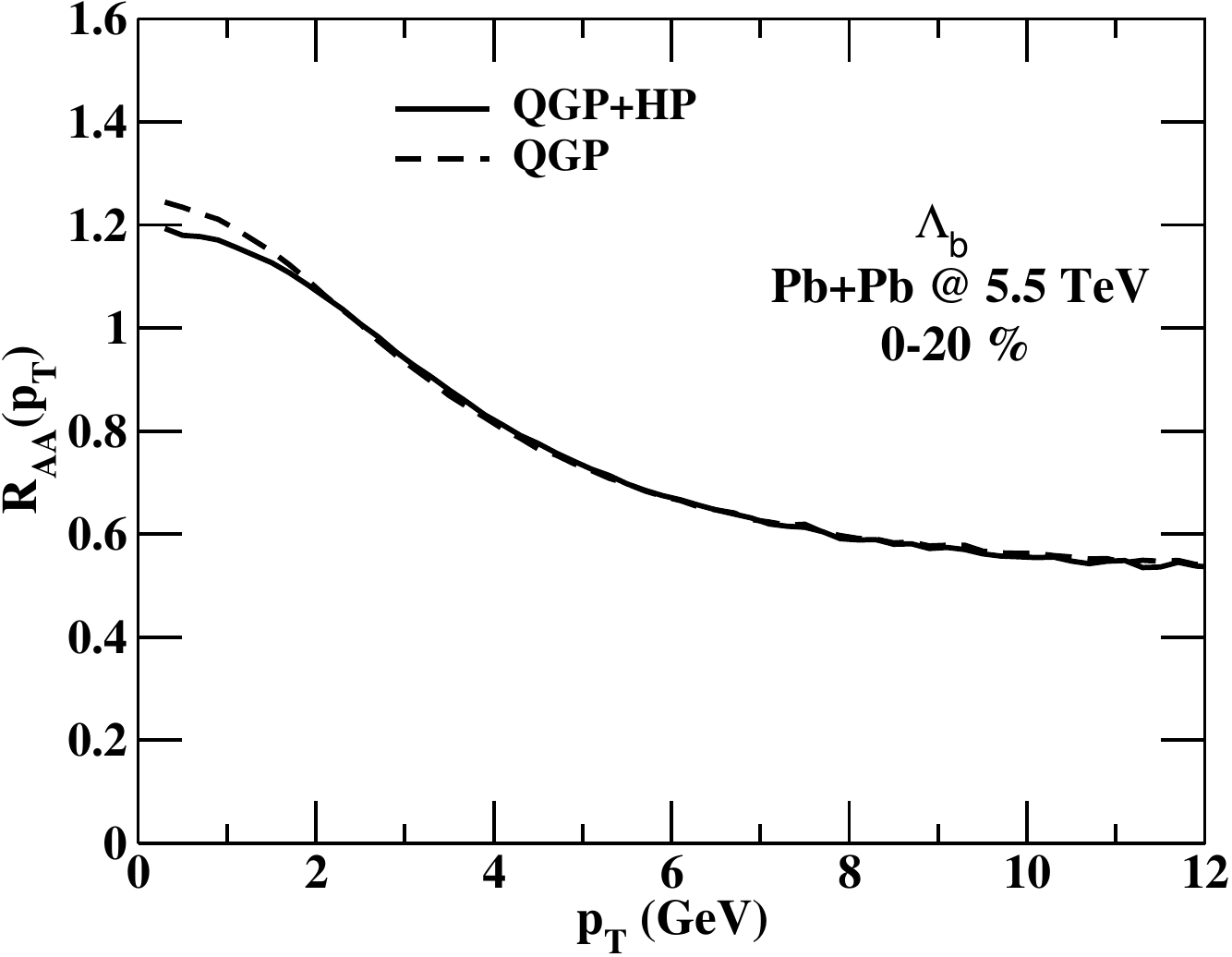}
\includegraphics[scale=0.35]{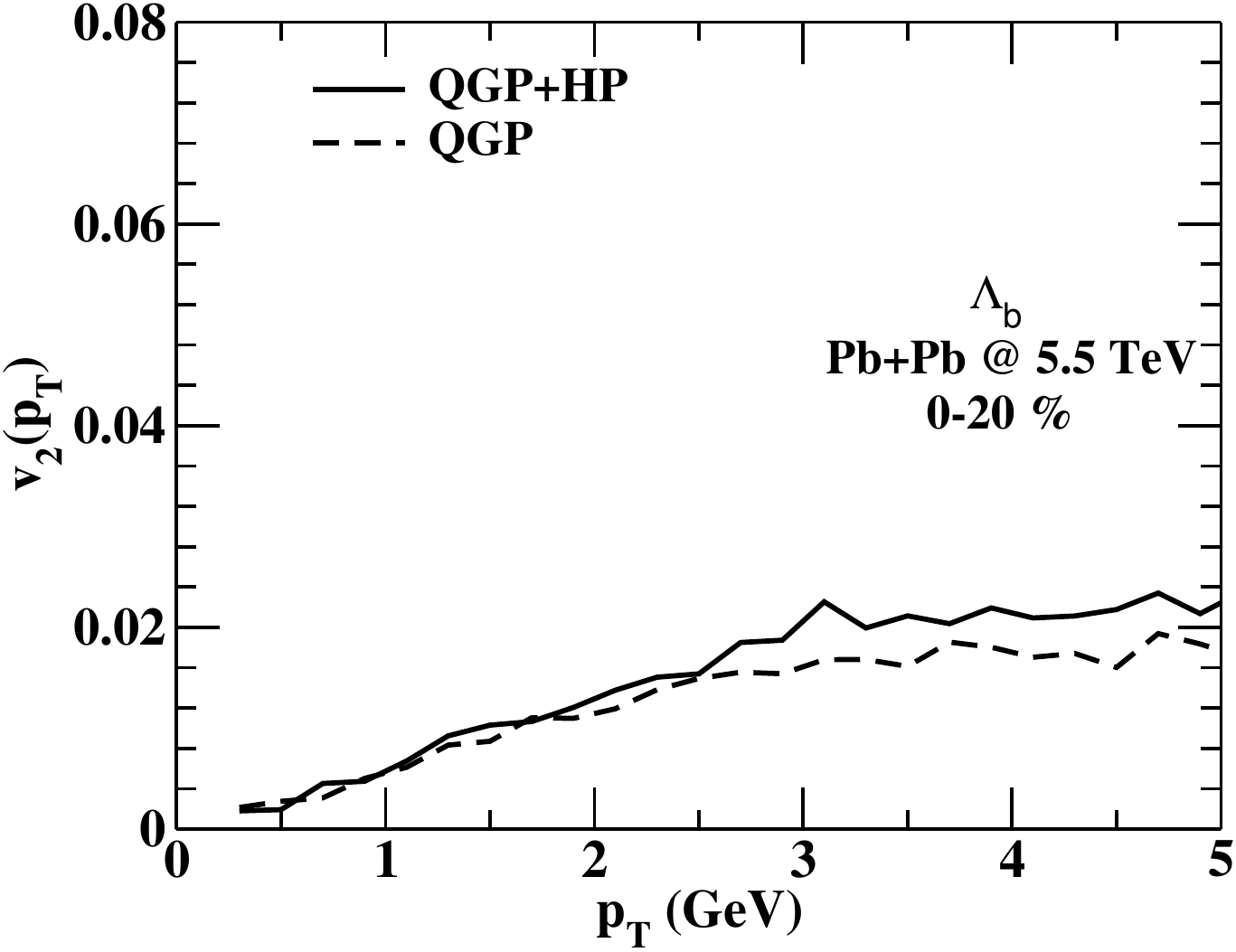}
\caption{Same as Fig.~\ref{fig:C1Raav2_RHIC}, but for 0-20\% central Pb-Pb (5.5~TeV) collisions at the LHC. Figures adapted from Ref.~\cite{Das:2016llg}.}
\label{fig:C1Raav2_LHC}
\end{figure}

The Catania group has conducted the same type of study, within the same setup as for the $\Lambda_c$, for the $\Lambda_b$, with appropriately updated transport coefficients for QGP and hadronic phase using the same framework for the interactions, as well as accordingly adapted fragmentation functions.
In Figs.~\ref{fig:C1Raav2_RHIC} and \ref{fig:C1Raav2_LHC} the $\raa$ and $v_2$ results are compiled for Au-Au at RHIC and Pb-Pb at the LHC. 
By and large, the same findings have been reported: negligible effects from the hadronic phase on the $\raa$ (left panels) but noticeable ones on the $v_2$ (right panels), more pronounced at RHIC, although overall quantitatively slightly smaller. There is, however, a small hint of an enhancement of the $\raa$ at the lowest $\pT\lsim 1$ GeV.

%%%%%%%%%%%%%%%%%%%%%%%%%%%%%%%%%%%%%%%%%%%%%%%%
\subsection{Heavy-flavor hadro-chemistry}
\label{ssec:chem}
%%%%%%%%%%%%%%%%%%%%%%%%%%%%%%%%%%%%%%%%%%%%%%%%%
\begin{figure} 
\centering
\includegraphics[scale=0.35]{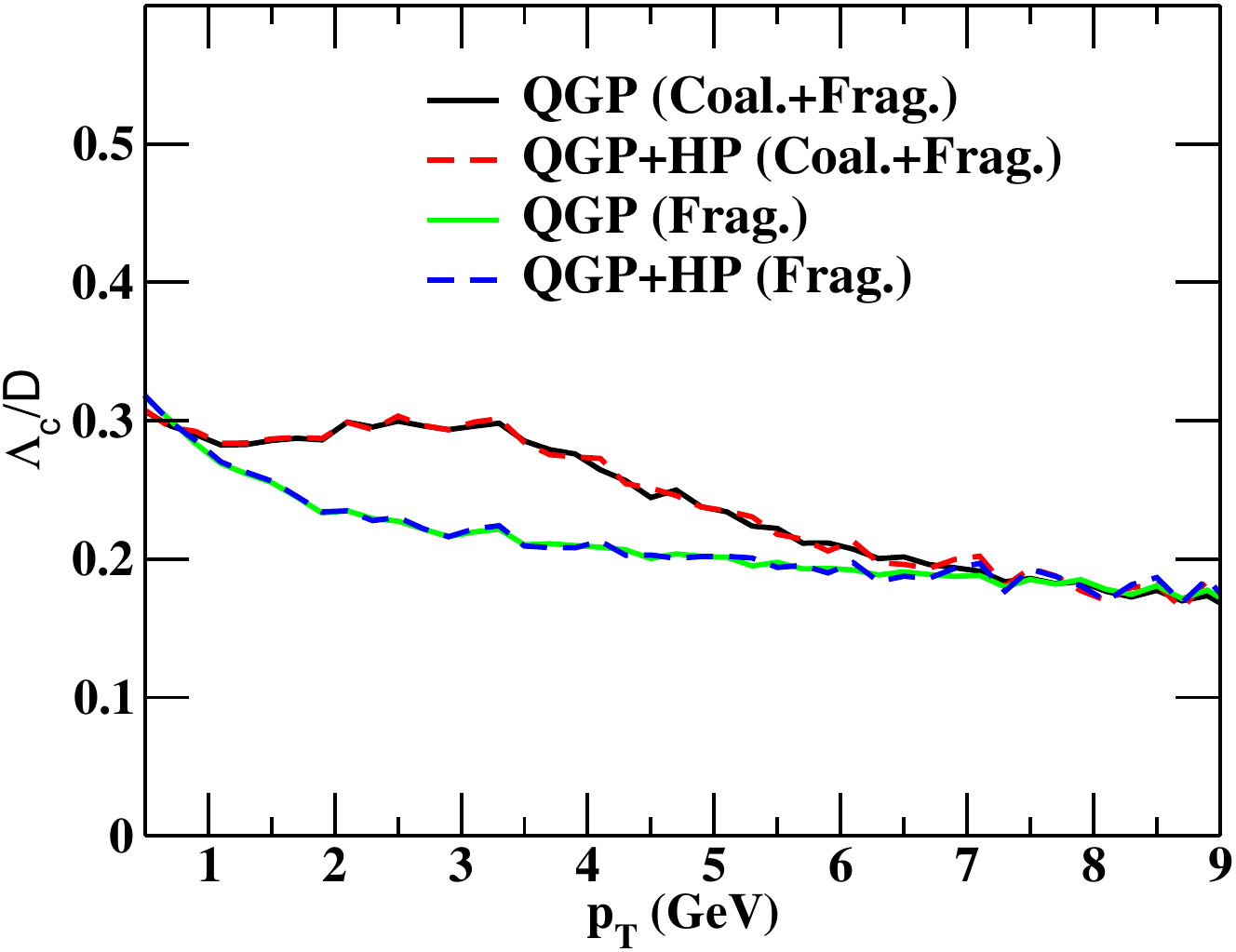}
\includegraphics[scale=0.35]{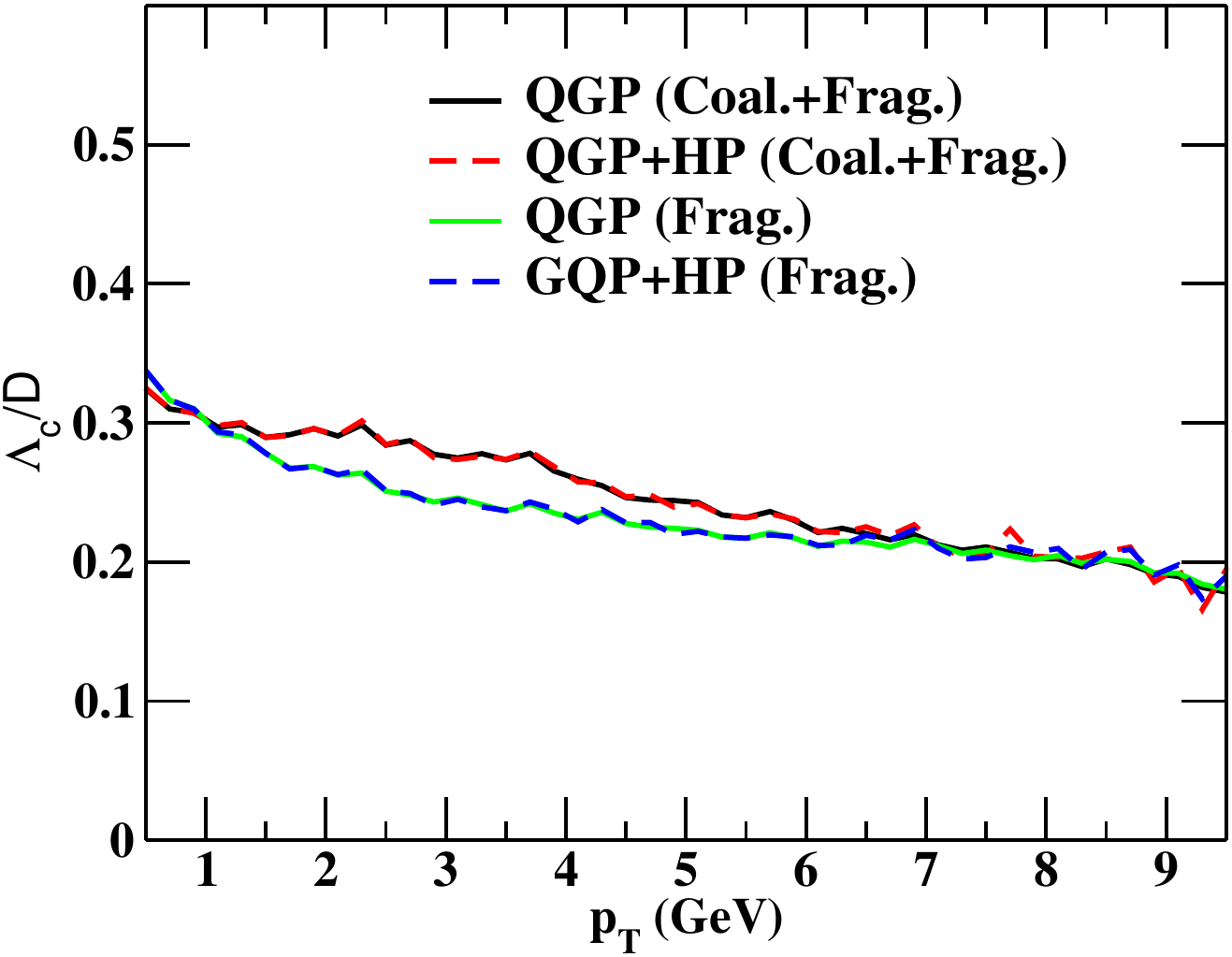}
\caption{The ratio $\Lambda_c/D$ as a function of $\pT$ is shown for minimum bias Au+Au collisions at RHIC (left panel) and for 0-20\% central Pb-Pb (5.5 TeV) collisions at the LHC (right panel), as computed within the Catania transport approach using coalescence plus fragmentation and fragmentation-only. The dashed and solid lines correspond to the results with and without the contribution of hadronic diffusion, respectively. Figures adapted from Ref.~\cite{Das:2016llg}.}
\label{fig:CBMR}
\end{figure}
The presence of a deconfined medium in HICs is expected to affect the hadronization process of heavy quarks into open heavy-flavor hadrons, relative to elementary reactions with small hadron multiplicities in the final state. In the light-flavor sector, the enhancement observed in HICs for baryon-over-meson ratios (such as $p/\pi$ or $\Lambda/K$) at intermediate $\pT$ has been interpreted in terms of coalescence processes~\cite{Fries:2008hs}, although their manifestation is not always straightforwardly disentangled  from the effects of radial flow (which leads to a stronger ``blue-shift'' for more massive particles); in addition, the $\pT$-integrated ratios, which are the more relevant quantity for the hadro-chemistry, did not exhibit significant changes from pp to  AA collisions at the same collision energy.
The classic example for a chemistry effect is the strangeness enhancement, or more precisely, the hadro-chemical equilibration of strange particles in AA collisions relative to an under-saturation in pp collisions. This occurs already at relatively low collision energies, such as at the fixed-target energies of the  SPS ($\sqrts$=17.3 GeV).
The heavy-baryon-to-meson ratios, $\Lambda_c$/$D$ and $\Lambda_b$/$B$, are providing a fresh perspective on these questions, and arguably a better controlled one, as it is not the total $c \bar c$ number that is expected to equilibrate (since secondary pair production or annihilation is strongly suppressed at the typical medium temperatures), but rather the {\em relative} abundance of charm and bottom hadrons. 

Recent experimental measurements, reported in Refs.~\cite{ALICE:2018hbc,STAR:2019ank,ALICE:2017thy}, have highlighted the heavy baryon-to-meson ratio as a valuable probe for elucidating different hadronization mechanisms. The observed enhancement in the $\Lambda_c^+/D^0$ yield ratio at midrapidity in pp and pPb collisions at the LHC, as reported by ALICE~\cite{ALICE:2020wla,ALICE:2017thy,ALICE:2020wfu} and CMS experiments~\cite{CMS:2019uws}, is substantial compared to baseline value of $\sim$0.1 in $e^-e^+$ and $e^-p$ collisions~\cite{ZEUS:2005pvv,ZEUS:2013fws,ZEUS:2010cic}. This enhancement can reach up to a factor of 2–5 for $\pT$ below 8 GeV. While the $\Lambda_c^+/D^0$ ratio in Pb-Pb collisions, as measured by ALICE~\cite{ALICE:2018hbc,ALICE:2021bib}, was observed to reach a maximum value of nearly one at $\pT\simeq 5$ GeV, it is important to note that the integrated ratio maybe be very similar to the one in pp and pPb collisions. 
These findings can be described by theoretical calculations incorporating hadronization through both coalescence and fragmentation mechanisms~\cite{Plumari:2017ntm,He:2019vgs}, although based on different micro-physics. While Ref.~\cite{Plumari:2017ntm} largely relies on controlling the production abundances through hadron wave function properties (in particular, different hadron radii), the approach in Ref.~\cite{He:2019vgs} essentially invokes large decay feed-down contributions from excited states (in the HF sector, many of these excited states have not been discovered yet, but are expected to exist based on the analogous states in the strange and light sectors and have been predicted by, \eg, relativistic quark models~\cite{Ebert:2011kk}). The population of these states has been originally put forward to explain the large  $\Lambda_c^+/D^0$ in pp collisions at the LHC at mid-rapidity (and is also consistent with pPb measurements)~\cite{He:2019tik}; the idea (and assumption) is that these states can be effectively populated once the particle multiplicity is large enough to justify a statistical treatment; once the multiplicity becomes too small, the ratio should decrease, which is indeed observed in both charm~\cite{ALICE:2021npz,LHCb:2018weo} and bottom sectors~\cite{LHCb:2023wbo}, and in agreement with model predictions~\cite{Chen:2020drg} through a canonical-ensemble induced suppression (due to enforcing exact quantum-number conservation).

In practice, the influence of the hadronic phase on the $\Lambda_c/D$ ratio has been studied in Ref.~\cite{Das:2016llg}, within the same setup (Langevin simulations in QGP and hadronic phase) as described in the context of Figs.~\ref{fig:CRaav2_RHIC}-\ref{fig:C1Raav2_LHC} of the preceding section. However, for hadronization, a scenario with additional recombination with light partons at $\Tpc$ has been considered, based on the instantaneous coalescence model of Ref.~\cite{Minissale:2015zwa}.
It is worth mentioning that the contribution from resonance decays impacts the heavy-baryon-to-meson ratios, as it involves the ratio of two distinct hadron species with varying contributions from resonance decays~\cite{Oh:2009zj}. The influence of resonance decay on the $\raa$ is usually less pronounced (especially if the chemistry does not change significantly), as it pertains to ratios involving the same hadrons.
Fig.~\ref{fig:CBMR} illustrates the variation of $\Lambda_c/D$ as a function of $\pT$ at top RHIC (left panel) and LHC energy (right panel) for both the QGP and QGP plus hadronic phase. The effects of coalescence are mostly relevant in the low and intermediate-$\pT$  regions, where they lead to a significant increase in the ratio (basically since the charm baryon pick up two light quarks from the thermal medium vs.~one for the charm meson).  However, the  ratio remains essentially unaffected by the rescattering in the hadronic phase. This observation reiterates the previously discussed findings, \ie, that the $\raa$ (and hence, the momentum spectra figuring in tit numerator) are affected very little, if any, by rescattering in the hadronic phase. This implies that the heavy-baryon-to-meson ratios are a rather clean probe of the hadronization dynamics (recall that hadronization into different species is based on the same underlying HQ distribution as resulting from diffusion through the QGP).

%%%%%%%%%%%%%%%%%%%%%%%%%%%%%%%%%%%%%%%%%%%
\subsection{Heavy-flavor diffusion: QGP vs. hadronic phase}
\label{ssec:assess}
%%%%%%%%%%%%%%%%%%%%%%%%%%%%%%%%%%%%%%%%%%%%%%
\begin{figure} 
\centering
\includegraphics[scale=0.35]{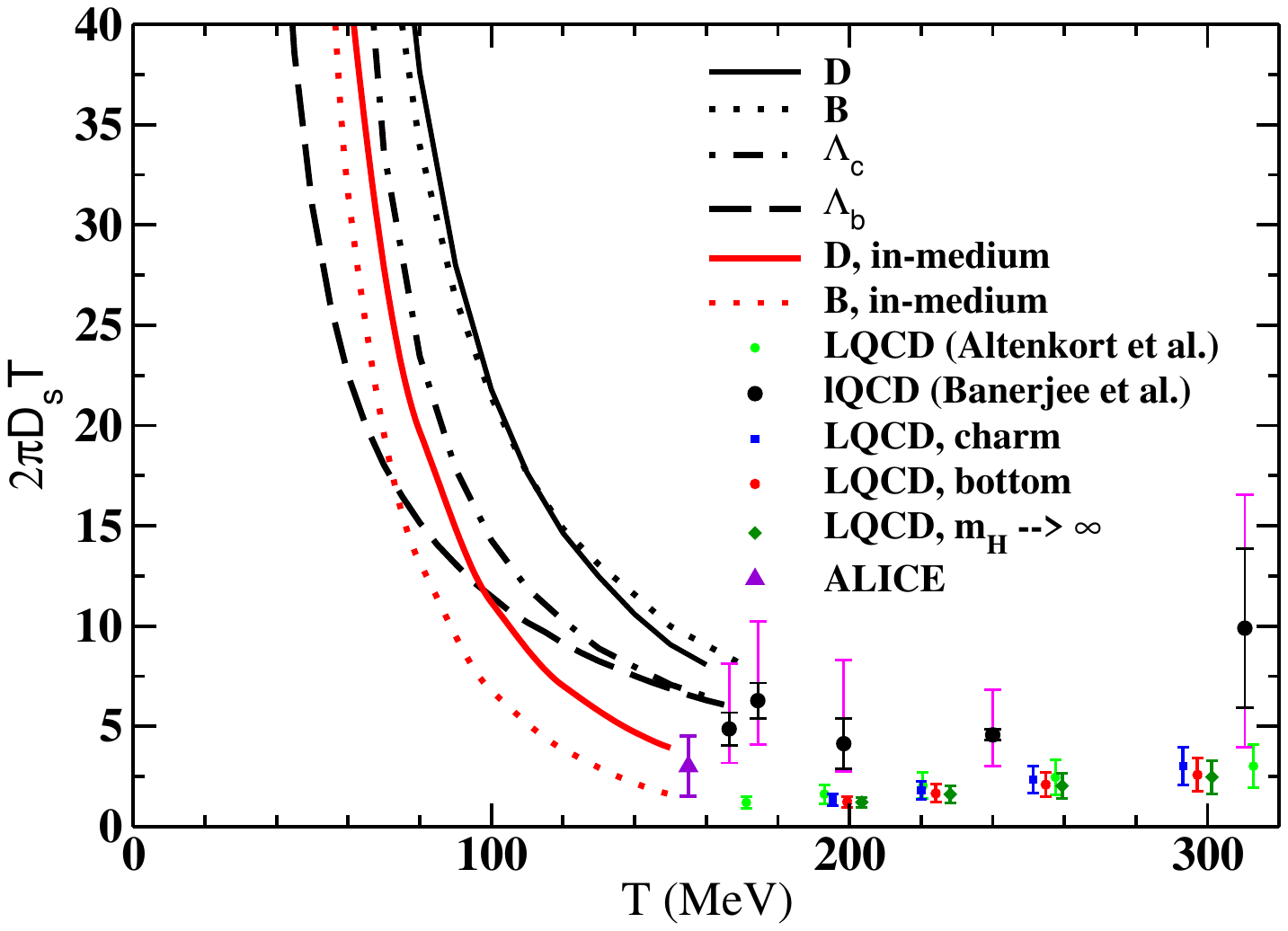}
\caption{Spatial diffusion coefficient as a function of temperature for both the QGP (only from LQCD computations) and hadronic phase obtained within different models. The $D$-meson and $B$-meson results are take from Ref.~\cite{Tolos:2013kva} and Ref.~\cite{Abreu:2012et}, respectively. The $\Lambda_c$ and $\Lambda_b$ results are taken from Ref.~\cite{Tolos:2016slr}. The in-medium results for $D$-mesons and $B$-mesons are taken from Ref.~\cite{Torres-Rincon:2021yga} and Ref.~\cite{Montana:2023sft}, respectively. The LQCD results on charm, bottom, and $m_H \rightarrow \infty$ are taken from Ref.~\cite{Altenkort:2023eav}. The LQCD results from Banerjee {\it et al.} are taken from Ref.~\cite{Banerjee:2011ra}. The LQCD results from Altenkort {\it et al.} are taken from Ref.~\cite{Altenkort:2023oms}. The ALICE Collaboration’s model fits to their data are taken from Ref.~\cite{ALICE:2021rxa}.} \label{fig:Dsall}
\end{figure}
A primary objective of all phenomenological investigations concerning open heavy-flavor observables is to extract the heavy-flavor spatial diffusion coefficient, $\Ds$~\cite{He:2012df}. 
As discussed in Appendix~\ref{app:DTR}, this quantity can be defined through the drag coefficient in the zero-momentum limit,
\be \label {eq:Dsall}
\Ds=\frac{T}{A (p \rightarrow 0) m_H}, 
\ee
where $m_H$ is the mass of the heavy quark in the QGP phase or the mass of the heavy hadron in the hadronic phase.
The diffusion coefficient serves to quantify the interaction between heavy quarks and the surrounding bulk medium. For example, in pQCD it is of the parametric form $\Ds \propto 1/(\alpha_s^2 T)$, \ie, does not, to lowest order, depend on the mass of the HF particle.
This illustrates why the HF diffusion coefficient contains fundamental information about the QCD medium. As a result, it is thought to be closely connected to other transport coefficients, such as shear viscosity or electric conductivity. To facilitate comparison and interpretation as indicators of the interaction strength within the medium, these coefficients are commonly scaled to dimensionless quantities. For instance, the  HF diffusion coefficient is scaled by the inverse ``thermal wavelength'' of the medium, $1/(2\pi T)$, resulting in $2\pi T \Ds$. Similarly, the shear viscosity is scaled by the entropy density ($s$) to obtain the dimensionless ratio $\eta/s$, while the electric conductivity is scaled by temperature ($T$), yielding 
$\sigma_{\rm el}/T$. In pQCD, these quantities are all proportional to the square of the inverse coupling constant, $1/\alpha_s^2$.

Transport models~\cite{He:2012df,Scardina:2017ipo} have proven to be highly effective in reproducing experimentally measured observables such as the nuclear modification factor and elliptic flow of $D$-mesons. These models rely on inputs such as drag and momentum diffusion coefficients, which encapsulate microscopic details regarding the interactions of heavy quarks with the QGP medium. In contrast, the Boltzmann equation relies on inputs such as cross-sections. After comparing the results obtained within the transport model with experimental data, different groups estimate the spatial diffusion coefficient, $\Ds$, from the transport coefficients that allow them to describe the experimental data effectively using Eq.~\eqref{eq:Dsall}.
Experimental data on $D$-meson production, obtained from both RHIC and the LHC, have reached a level of precision that permits quantitative constraints on the spatial diffusion coefficient. These constraints can be established by comparing the data with theoretical models. This has been carried out, \eg, by the ALICE collaboration using their recent measurements of the $\raa$ and $v_2$ of $D$ mesons~\cite{ALICE:2021rxa}. 
By conducting $\chi^2$ analysis of the available model calculations, the ones with a reasonably good confidence were selected and used to extract the underlying values of the charm quark diffusion constant in QGP near the $\Tpc$ as 1.5~$\le (2\pi T)\Ds \le \,4.5$. Several of these models did not account for hadronic diffusion.

The spatial diffusion coefficient of heavy quarks has also been computed using LQCD~\cite{Banerjee:2011ra,Francis:2015daa, Banerjee:2022uge}. 
For $N_f$=2+1-flavor QCD with light dynamical quarks corresponding to a pion mass of around 320 MeV recent computations have been carried out in Ref.~\cite{Altenkort:2023oms}. It has been found that the $c$-quark diffusion coefficient is quite close to the conjectured lower bound at temperatures just above $\Tpc$, $(2\pi T)\Ds \simeq 1.2\pm 0.25$ and rises only moderately with temperature, to about 3$\pm$1 at 2$\Tpc$; these results are not inconsistent with the phenomenological extraction from heavy-ion data.
As a further advancement, its mass dependence has been reported, which indeed turned out to be weak, with a 20-30\% reduction when going from $c$-quarks to the static limit ($m_H\to \infty$)~\cite{Altenkort:2023eav}.
 
This once again underscores the significance of open HF hadrons as probes of the transport properties of hot QCD matter, providing essential connections between LQCD and experiments via the phenomenology of transport models.

The spatial diffusion coefficient $\Ds$ is directly related to the HF thermalization time. One can estimate the thermalization time of heavy quarks as $\tau_{\rm{th}}$, defined as the inverse of the drag coefficient, $ \tau_{\rm{th}} = A^{-1}$. Hence, 
\be
\label{eq:th}
\tau_{\rm{th}} = \frac{m_H \Ds}{T}  = \frac{m_H}{2\pi T^2} \ (2\pi T \Ds) \ .
\ee

In Fig.~\ref{fig:Dsall}, we compile the temperature dependence 
$\Ds$ as a function of temperature in both the QGP (where we limit our selection to LQCD results) and the hadronic phase, as thoroughly discussed in this review. The hadronic model calculations have a clear tendency to drop to rather small values when extrapolated to temperatures in the vicinity of $\Tpc$, suggestive for a continuous transition from the confined to the deconfined medium through a minimum near the quark-hadron transition.  This indicates the strength of the interaction is maximum near the quark-hadron transition temperature. As discussed above, in the QGP phase there is again a tendency for $D_s$ to increase but the values remain overall very small, far from a weakly coupled gas of quarks and gluons. 

If the thermalization time remains constant, it implies that the drag coefficient will be independent of temperature, meaning $2\pi T \Ds \propto T^2$. Therefore, a constant value of 
$2\pi T \Ds$ corresponds to a thermalization time that is proportional to $1/T^2$. The LQCD results for the QGP indicate a (mild) linear increase in $2\pi T \Ds$ with temperature, implying that the interaction strength decreases. One might wonder whether the marked drop of $\Ds (2\pi T)$ with $T$ in the hadronic phase indicates a strong increase in the interaction strength; this is probably not the case, but rather due to the fact that the total hadronic density increases with a rather large power in $T$, due to numerous and rather massive excited states that are increasingly populated toward higher temperatures. As a result, the friction coefficient has a stronger temperature scaling; for example, if $A(0)\propto T^4$ (\ie, two additional powers compared to pQCD), one would obtain a parametric dependence of the diffusion coefficient as $\Ds (2\pi T) \propto 1/T^2$).  

In phenomenological applications (and extractions of $\Ds$), one has to keep in mind that the 3-momentum dependence of the friction coefficient plays an important role. This is due to the fact that thermal motion and the collective flow of the expanding medium will always result in a smearing of the 3-momentum dependence of the transport coefficient relative to the observed $\pT$ in the lab-frame of experiment. In particular, it has been shown through model comparisons that there is a direct correlation between a larger value of $\Ds$ along with a harder 3-momentum dependence of the corresponding friction coefficient, $A(p)$, to yield a comparable description of $D$-meson data, see also Ref.~\cite{He:2022ywp} for additional details.

%%%%%%%%%%%%%%%%%%%%%%%%%%%%%%%%%%%
\section{Summary and Outlook}
\label{sec:sum}
%%%%%%%%%%%%%%%%%%%%%%%%%%%%%%%%%%%%
The theoretical investigation of the transport properties of open heavy-flavor particles in hot hadronic matter essentially commenced a little more than a decade ago, originally driven by the phenomenological need to study the impact on their transverse-momentum spectra in heavy-ion collisions. While this is essential for a quantitative extraction of the transport properties of heavy quarks in the strongly coupled quark-gluon plasma and its hadronization, it is also of interest in its own right in a comprehensive understanding of hot QCD matter and its underlying quantum many-body physics. In addition, the construction and applications of effective hadronic interactions have been mutually beneficial to other sub-fields, such as the properties of HF hadrons in the cold nuclear medium or spectroscopic applications in particular to exotic states.         

In this article, we have attempted to review the progress made in the evaluation of open HF diffusion in hot hadronic matter, specifically for the pertinent drag and diffusion coefficients, and to illuminate its quantitative role in transport model simulations of the fireball in heavy-ion collisions. 
Most of this research to date, for both microscopic calculations and applications to heavy-ion collisions, has focused on the transport coefficients of $D$ mesons.
While initial approaches have utilized  LO approximations in heavy-meson effective theory, it quickly became clear that for the temperatures encountered in experiment, resummed interactions -- in particular those involving resonance excitations -- are required for realistic descriptions of $D$-meson interactions with light mesons in a heat bath. This led to the development of systematic approaches where symmetry constraints from both the heavy-quark and the light-quark sectors have been combined, and pertinent diagrams utilized via unitarization techniques which enable to extend the calculations to higher temperatures. We have found the different approaches to be in fair agreement when compared under the same assumptions and/or inputs, lending robustness to the results for the transport coefficients. A bigger uncertainty originates from the proliferation of states in the hadron resonance medium when approaching temperatures near the pseudo-critical one.
Here, constituent-quark scaling arguments for effective cross sections have been put forward, but it is not known how accurate they are especially if higher excited resonances are involved that are not necessarily accessible through interactions based on symmetry arguments (but are rather rooted in the quark-level spectroscopy). These arguments become  even more pertinent if one considers the diffusion of HF baryons.

Quantitatively, theoretical estimates of the collisional widths of $D$-mesons turn out to be at least $\sim$50 MeV from known interactions at temperatures $T\simeq$\,150\,MeV, with corresponding drag coefficient of about 0.05 fm$^{-1}$. In reality they are probably larger, \eg, due to further excited states (or $t$-channel interactions) in the hadron resonance gas that are currently not taken into account. Nevertheless, these widths are significant, especially when compared to the threshold and mass gaps in $D$-meson spectra that are often rather small, \eg, a $D$-$D^*$ mass splitting which essentially amounts to the pion mass. Thus, quantum many-body treatments are in order and have been carried out and may lead to further enhanced effects, \eg, by rendering bound states accessible in $2\to2$ scatterings. Additional investigations in these directions are required.

In principle, the above outlined approaches directly translate into bottom sector, although the empirical constraints on the interactions (and excited states) are more scarce. On general grounds, one expects the thermalization rate (or drag coefficient) for $B$-mesons to be smaller than that for $D$ mesons by their mass ratio, $m_D/m_B$. In the definition of the spatial diffusion coefficient the leading mass dependence is taken out, and therefore they should be quite comparable.
%(this has been confirmed in  recent LQCD computations for bottom and charm quarks in the quark-gluon plasma), 

To assess the impact of the hadronic phase on various heavy-hadron observables, we have conducted a review of the results obtained within different transport models. Again, almost all pertinent works have focused on observables related to $D$ mesons. Basic arguments derived from the typical time evolution of the nuclear modification factor and elliptic flow of heavy quarks in the QGP already suggest that latter will be significantly stronger affected in the hadronic phase than the former (since the $\raa$ saturates much earlier than the $v_2$). This has been corroborated by studies that terminate the HQ transport at different ``critical temperatures'' which found large (small) variations on the $v_2$ ($\raa$), which, of course, implies an incomplete simulation. 
Quantitative studies in various transport models have confirmed these expectations. More concretely, the relative increase of the maximal $v_2$ value due to the inclusion of hadronic diffusion can reach up to $\sim$30\% in Au-Au collisions at RHIC, and possibly up to $\sim$20\% in Pb-Pb collisions at the LHC. In either case, this is not a small effect, but it also suggests that a better testing ground could be toward lower collision energies or more peripheral collisions since the QGP phase is expected to live shorter in these configurations. Caveats include that the bulk-medium $v_2$ tends to be lower at smaller collision energies (perhaps because the mechanism of converting the initial spatial geometry into momentum anisotropy is less effective, \eg, due to longer thermalization times), but, on the other hand, the higher concentration of baryons could enhance the hadronic diffusion effects. Existing results from lower-energy runs at RHIC tend to indicate a lower overall heavy-flavor $v_2$, but its precision and interpretation remains inconclusive to date. Precision charm-hadron measurements at SPS energies ($\sqrts\simeq$~6-17\,GeV) as planned by the NA60+ experiment will be highly illuminating. In fact, a stronger proliferation of excited states in the baryon sector compared to a meson-dominated medium could have a sizable effect on $D$-meson interactions in high-$\mu_B$ matter.
Another observable that could be sensitive to hadronic-diffusion effects is the splitting of the $v_2$ of $D$ and $D_s$ mesons, based on the idea that the latter is largely inert to hadronic interactions as it does not contain a light valence quark. Finally, we reiterate the idea that the limited kinematics for charm production at lower collision energies could lead to an enhancement (even divergence) in the nuclear modification factor of $D$-mesons if they were to diffuse to momenta that cannot be produced in $pp$ collisions.

With the current best estimates of $D$-meson transport coefficients, values of down to $\sim$5 for the (scaled) spatial diffusion coefficient, $\Ds (2\pi T)$, at temperatures of $T\simeq$160\,MeV seem attainable. This is still significantly larger than the results of recent lattice-QCD computations on the QGP side, where values near 1-2 were found. Two possible reasons for this are that (i) the minimum of  $\Ds (2\pi T)$ as a function of temperature is significantly above $\Tpc$, and (ii) current hadronic calculations are still incomplete. Nevertheless, it is encouraging that the discrepancy is not that large, and we are confident that future theoretical efforts will be able to make significant progress on this, as well as on the quantitative implementations of hadronic diffusion into heavy-ion phenomenology.

\section{Acknowledgements}
\label{sec:acknowledgments}
%%%%%%%%%%%%%%%%%%%%%%%%%%%%%%%%%%%
SKD acknowledges Jan-e Alam, Sourav Sarkar, Sabyasachi Ghosh, and Mohammad Yousuf Jamal for their valuable discussions. SKD acknowledges the support from DAE-BRNS, India, Project No. 57/14/02/2021-BRNS. The work of JMT-R has been supported by the project number
CEX2019-000918-M (Unidad de
Excelencia ``Mar\'ia de Maeztu''), PID2020-118758GB-I00, PID2023-147112NB-C21, financed by the Spanish MCIN/ AEI/10.13039/501100011033/, the
EU STRONG-2020 project, under the program H2020-
INFRAIA-2018-1 grant agreement no. 824093, and DFG project no. 315477589 - TRR 211 (Strong-interaction matter under extreme conditions). The work of RR has been supported by the US National Science Foundation under grant no. PHY-2209335 and the U.S.~Department of Energy, Office of Science, Office of Nuclear Physics through the Topical Collaboration in Nuclear Theory on {\it Heavy-Flavor Theory for QCD Matter (HEFTY)} under award no. DE-SC0023547. 

\appendix

%%%%%%%%%%%%%%%%%%%%%%%%%%%%%%%%%%%%%%%%%%%%%%%%%%%%%%%%%%%%%
\section{Appendix: Heavy-flavor transport coefficients}
\label{app:DTR}
%%%%%%%%%%%%%%%%%%%%%%%%%%%%%%%%%%%%%%%%%%%%%%%%%%%%%%%%%%%
In this Appendix we recall the formal derivation of the Fokker-Planck equation from the master (Boltzmann) equation, and the definition of the relevant transport coefficients, as well as their connection to the Langevin description of a relativistic Brownian particle. For further details we refer we readers to the Refs.~\cite{pitaevskii2012physical,Svetitsky:1987gq,van1992stochastic,risken1996fokker,Rapp:2009my,Abreu:2011ic}.

We consider the elastic collisions of a generic $D$-meson (denoted by $H$, for heavy hadrons in general) with a light particle (denoted by $l$) from the thermal bath. The thermal bath might consist of hadrons ($\pi$, $K$, $\bar{K}$, $\eta$, nucleons, $\Delta$, etc.), characterized by their thermal phase space distributions. 

The momentum evolution of the heavy hadron's with momentum, ${\bm p}$, $f_H (t,{\bm p})$, in the thermal bath can be studied using the Boltzmann equation,
\bea 
\label{eq:Bolt} 
\frac{\pa f_H (t , \bm{p})}{\pa t} = \left[ \frac{\pa f_H (t , \bm{p})}{\pa t} \right]_{ \textrm{coll}} \ . 
\eea
The right-hand side of Eq.~\eqref{eq:Bolt} represents the collision integral, wherein the phase-space distribution function of the bulk medium appears as an integrated quantity. This form of the Boltzmann equation  assumes no external forces, indicating that the evolution of $f_H(t, {\bm p})$ is solely governed by collisions. Given the low density of heavy mesons, collisions between them can be safely disregarded. Therefore our focus can be directed solely towards the interaction of the heavy hadrons with the light particles in the thermal bath. Heavy hadrons can enter and exit the momentum phase space element $d{\bm p}$ around ${\bm p}$ through collisions with the thermal particles. Consequently, the collision term consists of two parts associated with a gain and a loss term. 
If we define $w(p,k)$ as the rate of collisions that change the momentum of $H$ from $p$ to $p-k$, then we can write
\bea 
\label {eq:meq}
\left[ \frac{\pa f_H (t , {\bm p})}{\pa t} \right]_{ \textrm{coll}}=\int d^3 {\bm k} \left[ f_H(t, {\bm p}+{\bm k}) w ( {\bm p}+ {\bm k}, {\bm k}) -  f_H(t,{\bm p}) w ({\bm p}, {\bm k} ) \right]  \ .
\eea
The first term on the right-hand side of Eq. ~(\ref{eq:meq}) represents the gain of probability through collisions, which cause $H$ to enter the volume element of momentum space at ${\bm p}$. Conversely, the second term represents the loss of probability out of that element; $w$ is the aggregate of contributions arising from heavy-hadron scattering off all thermal particles. In essence, the Boltzmann equation should be treated as a quantum Boltzmann-Uehling-Uhlenbeck equation. This involves considering the Bose enhancement effect in the final state, where factors like $1+f_H$ encode the increased probability of a $D$ meson scattering into an already occupied state. As long as the number of $D$-mesons is very small, we can approximate $1+f_H \sim 1$.  However, this approximation may not hold for the light thermal particles.
Expanding $f_H(t,{\bm p}+{\bm k}) w ({\bm p}+ {\bm k}, {\bm k})$ around ${\bm p}$,
\bea 
\label {eq:meq2}
f_H(t,{\bm p}+{\bm k}) w ({\bm p}+ {\bm k}, {\bm k}) \simeq  f_H(t,{\bm p}) w ({\bm p}, {\bm k}) + k_i \frac{\pa }{\pa p_i}[f_H(t,{\bm p}) w ({\bm p}+ {\bm k})]+
\frac{1}{2}k_ik_j \frac{\pa ^2}{\pa p_i \pa p_j}[f_H(t,{\bm p}) w ({\bm p}+ {\bm k})]+ \cdots  \ , 
\eea 
and substituting into this into Eq.~\eqref{eq:meq}, we obtain:
\bea 
\label {eq:FP0}
\left[ \frac{\pa f_H (t , {\bm p})}{\pa t} \right]_{ \textrm{coll}}= \frac{\pa}{\pa p_i} 
\left\{ A_i({\bm p})f_H+\frac{\pa}{\pa p_j}\left[ B_{ij}({\bm p})f_H \right] \right\} \ ,
\eea
where the kernels are defined as

\bea
A_i ({\bm p})= \int d^3{\bm k} \ w({\bm p},{\bm k}) k_i     \nn
B_{ij} ({\bm p})= \int d^3{\bm k} \ w({\bm p},{\bm k}) k_i k_j \ . 
\eea
The classical interpretation of these kernels is that they represent drag and momentum-space diffusion coefficients affecting the kinematics of $H$. For a collection of different notations used in the literature, cf.~Table~\ref{tab:notation}.
\begin{table}[ht!]
\centering
\vspace*{2mm}
\begin{tabular}{|c|c|}
\hline
\multicolumn{2}{|c|}{Drag Force} \\ 
\hline
$A(p)$ & \cite{He:2011yi},\cite{Ghosh:2011bw},\cite{Rapp:2018qla},\cite{Torres-Rincon:2021yga},\cite{He:2022ywp}  \\
$F(p)$ & \cite{Abreu:2011ic},\cite{Tolos:2013kva},\cite{Tolos:2016slr}   \\
$\gamma(p)$ & \cite{Ozvenchuk:2014rpa}  \\
$\eta_D$  & \cite{Laine:2011is} \\
\hline
\multicolumn{2}{|c|}{Momentum-space diffusion coefficients} \\ 
\hline
$B_0(p),B_1(p)$ &  \cite{Ghosh:2011bw},\cite{Ozvenchuk:2014rpa},\cite{Rapp:2018qla},\cite{Torres-Rincon:2021yga},\cite{He:2022ywp} \\
$\Gamma_0,\Gamma_1$ &  \cite{Abreu:2011ic},\cite{Tolos:2013kva},\cite{Tolos:2016slr} \\
$\kappa/2$  & \cite{Laine:2011is} \\
\hline
\multicolumn{2}{|c|}{Coordinate-space diffusion coefficient} \\ 
\hline
${\cal D}_s, D_s$ & \cite{Rapp:2018qla},\cite{Torres-Rincon:2021yga},\cite{He:2022ywp} \\
$D_x$ & \cite{Ghosh:2011bw}, \cite{Abreu:2011ic} ,\cite{Tolos:2013kva}, \cite{Ozvenchuk:2014rpa},\cite{Tolos:2016slr}\\
$D$ & \cite{Laine:2011is}  \\
\hline
\end{tabular}
\caption{List of notations for the different transport coefficients used in the works referenced in this review.} 
\label{tab:notation}
\end{table}
The function $w({\bm p},{\bm k})$ is given by
\bea 
\label {eq:ome}
w({\bm p},{\bm k})=g_l \int \frac{d^3q}{(2\pi)^3} f_l(q)v_{q,p}\sigma_{p,q \rightarrow p-k,q+k} \ ,
\eea
where $f_l(q)$ represents the phase space distribution of the light particles in the thermal bath, $g_l$ is their statistical degeneracy, $v_{q,p}$ indicates the relative velocity between the collision partners and $\sigma_{p,q \rightarrow p-k,q+k}$ stands for the interaction cross-section of $H$ with the light particles in the heat bath. 

With these approximations, the Boltzmann Eq.~\eqref{eq:Bolt} reduces to the Fokker-Planck equation, 
\bea \label {eq:FP}
\frac{\pa f_H (t , {\bm p})}{\pa t}= \frac{\pa}{\pa p_i} 
\left\{ A_i({\bm p})f_H+\frac{\pa}{\pa p_j}\left[ B_{ij}({\bm p})f_H\right] \right\} \ .
\eea

In this context, $A_i ({\bm p})$ acts like a friction term, indicating the average momentum change of $H$, while $B_{ij}({\bm p})$ acts as a diffusion coefficient in momentum space,  responsible for the broadening of the average momentum distribution of $H$. 
Since $A_i$ and $B_{ij}$ solely depend on $p_i$, in a isotropic background medium they can be represented by three scalar functions of $p$,
\begin{align} 
\label {eq:Dec}
A_i(p) & =A(p^2)p_i      \nn \\
B_{ij} (p)  & = B_0 (p^2) \Delta_{ij} + B_1 (p^2) \frac{p_i p_j}{p^2} \ ,
\end{align}
where
\bea 
\Delta_{ij} \equiv \delta_{ij} - \frac{p_i p_j}{p^2} . 
\label{eq:projector}
\eea
The explicit expressions of these three scalar coefficients,  as given in \eqref{eq:Dec} in terms of $w({\bm p},{\bm k})$, are as follows:
\bea 
\label{eq:TP}
A(p) & = & \frac{p^i A_i}{p^2} = \int d^3{\bm k}\  w({\bm p},{\bm k})  \ \frac{k_ip^i}{p^2} \ , \\ 
B_0 (p) & = & \frac{1}{2} \Delta_{ij} B^{ij} = \frac{1}{4} \int d^3{\bm k}\ w({\bm p},{\bm k}) \left[ {\bm k}^2 - \frac{(k_i p^i)^2}{p^2} \right] \ , \\ 
 B_1(p) & = & \frac{p_i p_j}{p^2} B^{ij} = \frac{1}{2} \int d^3{\bm k}\  w({\bm p},{\bm k}) \ \frac{(k_i p^i)^2}{p^2} \ , 
\eea
where pertinent scattering matrix elements, $|\mathcal{M}_{Hl}|$, dictate the dynamics through the cross-section given in Eq.~\eqref{eq:ome}. 

The three coefficients, $A(p)$, $B_{0}(p)$ and $B_{1}(p)$, in the Fokker-Planck equation are not independent; instead, they are related through the fluctuation-dissipation relation, which can be derived from the Fokker-Plank equation (see, \eg, Ref.\cite{vanHees:2004gq}. One should expect that in the long-time limit, the heavy hadrons reach the same thermal distribution as the medium constituents, $f_H={\rm e}^{-E_H/T}$. However, if the drag and diffusion coefficients are directly taken from their calculations, this is generally not the case. Since the drag and diffusion coefficients represent the first two moments of the full collisional integral, not all momentum transfers  can be accurately captured solely by these two moments. Establishing a connection between the drag and diffusion coefficients is necessary to ensure that the correct equilibrium distribution $f_{eq}=e^{-E/T}$  is attained as the asymptotic distribution. In equilibrium, the right-hand side of Eq.~(\ref{eq:FP}) vanishes, and a connection between the drag and diffusion coefficients can be established. At $ p \rightarrow 0$, when $B_0=B_1=B$, Eq.~\eqref{eq:FP} simplifies to Einstein’s famous fluctuation-dissipation relation, $B=m_H A T$. Further details can be found in Refs.~\cite{Walton:1999dy,Rapp:2009my}. Employing the fluctuation-dissipation relation reduces the problem to a single transport coefficient. This highlights that the Fokker-Planck equation serves as a consistent approximation to the Boltzmann equation, as it accurately captures the proper equilibrium limit.  It is essential to include both friction and diffusion terms to maintain the principle of detailed balance.
However, one should keep in mind that if the momentum transfer is significant too large, \ie, does not satisfy $k \ll m_H$, the Fokker-Planck equation may not serve as an appropriate approximation of the Boltzmann equation, and it may deviate from it~\cite{Das:2013kea}.

The fluctuation-dissipation relation suggests that a pure diffusion equation, $A=0$, is only valid in the limit of $T \rightarrow \infty$  for a particle with finite mass. On the other hand, in phenomenological applications to heavy-ion collisions, the evolution of high-$\pT$ particles is often approximated using an energy-loss treatment to compute the nuclear modification factor, which involves neglecting the diffusion term, $B = 0$. This implies that only momentum or energy-degrading processes are considered, a behavior akin to the  $T \rightarrow 0$  limit in the fluctuation-dissipation relation. The absence of momentum diffusion implies that both momentum randomization and energy gain processes are neglected, thus preventing particles from achieving equilibration.

The diffusion diffusion in momentum space or drag coefficient can be translated into a spatial diffusion coefficient, which quantifies the variance in the particle's position as it evolves over time. The spatial diffusion coefficient, represented as $\Ds$,  can be expressed as follows:
\be
\langle ( {\bm r}- {\bm r}_0)^2\rangle = 6 \Ds t \ .
\ee
The spatial diffusion coefficient can be linked to the drag and momentum-diffusion coefficient through the following relation,
\be \label {eq:Ds}
{\cal D}_s=\frac{T}{m_H A}=\frac{T^2}{B} \ .
\ee
The conventional method for studying the HQ momentum evolution within the QGP phase and the momentum evolution of open HF hadrons within the hadronic medium through the Fokker-Planck equation is a stochastic one using the Langevin equations. The relativistic Langevin equations governing the temporal changes in the position and momentum of heavy quarks or heavy hadrons can be expressed as follows:
\be 
\left\{
\begin{array}{rcl}
dx_i & = & \frac{p_i}{E}dt \ ,  \nn \\
dp_i & = & -A(p) p_i dt+C_{ij}(p)\rho_j\sqrt{dt} \ ,  \label{eq:LV}
\end{array}
\right. 
\ee
where for each discrete time step $dt$, the change in the coordinates and momenta, denoted as $dx_i$ and $dp_i$ respectively, represent the shifts in position and momentum. $A(p)$ represents the drag force, while $C_{ij}(p)$ denotes the so-called covariance matrix; $\rho$ represents a stochastic noise, which adheres to the probability distribution of independent Gaussian-normal distributed random variables, described by $P(\rho) = (2\pi)^{-3/2} e^{-\rho^2/2}$. Additionally, it satisfies the relations $ \langle \rho_i \rho_j \rangle = \delta_{ij}$ and $ \langle \rho_i \rangle = 0$. The covariance matrix is associated with the transverse and longitudinal diffusion coefficients through the equation,
\bea
C_{ij} (p)=\sqrt{2B_0(p)} \ \Delta_{ij}+\sqrt{2B_1(p)} \  \frac{p_i p_j}{p^2} \ ,
\label{cm}
\eea
where $\Delta_{ij}$ is given by Eq.~(\ref{eq:projector}). Under the assumption that $B_0(p) = B_1(p) = B(p)$, Eq.~(\ref{cm}) simplifies to $C_{ij} = \sqrt{2 B(p)} \ \delta_{ij}$. This assumption, which is strictly valid for $p\rightarrow 0$, is commonly employed for the heavy-quark/hadron momentum evolution at finite momentum $p$. The Langevin equation can be effectively solved by incorporating the drag and diffusion coefficients as inputs, as they encapsulate the microscopic intricacies of the interactions between heavy hadrons and the thermal bath. In addition, one needs to specify the initial distribution of coordinates and momenta of the heavy hadrons (usually taken from the Glauber model) and a realistic space-time of the bulk medium. Subsequently, the results obtained from solving the Langevin equation can be utilized to compute various observables that are measured in experiments, such as the nuclear modification factor and flow coefficients. The Fokker-Planck and Langevin equations are equivalent when considering the first two moments, which correspond to the drag and diffusion coefficients in the Fokker-Planck equation, see Eq.~\eqref{eq:FP}.

%Appendix sections are coded under \verb+\appendix+.
%\verb+\printcredits+ command is used after appendix sections to list 
%author credit taxonomy contribution roles tagged using \verb+\credit+ 
%in frontmatter.

\printcredits

%\bibliographystyle{cas-model2-names}
%\bibliographystyle{plain}
\bibliographystyle{unsrt}

\bibliography{cas-refs}


%Author biography without author photo.
%Author biography. Author biography. Author biography.

%\bio{figs/pic1}
%Author biography with author photo.
%Author biography. Author biography. Author biography.

%\bio{figs/pic1}
%Author biography with author photo.
%Author biography. Author biography. Author biography.

\end{document}